\documentclass{article}

\usepackage[utf8]{inputenc} 
\usepackage[T1]{fontenc}    

\usepackage{bbm, authblk}
\usepackage{hyperref}       
\usepackage{url}            
\usepackage{booktabs}       
\usepackage{amsfonts}       
\usepackage{microtype}      
\usepackage{xcolor}         
\usepackage{natbib}
\usepackage{float}
\usepackage[ruled,vlined]{algorithm2e}
 \usepackage{tikz}
\usetikzlibrary{arrows}

\usepackage[accepted]{icml2023}

\usepackage{./Latex/macros_01_proof_envs}
\usepackage{./Latex/macros_02_operators_commands}
\usepackage{./Latex/macros_03_symbols}
\usepackage{./Latex/macros_04_propernames}
\usepackage{./Latex/macros_05_reviewer}
\allowdisplaybreaks

\usepackage{csquotes}

\newcommand{\p}{\mathbb P}

 \newcommand{\op}{\text{op}}

 \newcommand{\s}{ { \mathfrak{s} } }

\newcommand{\I}{{\mathcal I }}

  \newcommand{\J}{{  \mathcal J } }

 \newcommand{\Q}{ { \mathcal Q} }
  \newcommand{\Bn}{ { \mathcal B_n} }
 \newcommand{\G}{ { \mathfrak{G} } }
  \newcommand{\E}{ { \mathbb{E} } }
  
\newcommand{\Bt}{ \mathcal{B}_n}
\newcommand{\tr}{ { \mathcal B_n^{-1}   \Delta_{\min} } }
\newcommand{\htr}{ { \mathcal B_n^{-1/2}   \Delta_{\min} } }

\usepackage{array}
\newcommand{\PreserveBackslash}[1]{\let\temp=\\#1\let\\=\temp}
\newcolumntype{C}[1]{>{\PreserveBackslash\centering}p{#1}}
\newcolumntype{R}[1]{>{\PreserveBackslash\raggedleft}p{#1}}
\newcolumntype{L}[1]{>{\PreserveBackslash\raggedright}p{#1}}

\newcommand{\bftheta}{ { \bm \theta } }
\newcommand{\bfbeta}{ { \bm \beta } }
\newcommand{\bfmu}{ { \bm \mu } }
\newcommand{\bfeta}{ { \bm \eta } }
\newcommand{\bfepsilon}{ { \bm \epsilon } }
\newcommand{\bfSigma}{ { \bm \Sigma } }
\newcommand{\bfOmega}{ { \bm \Omega } }

\icmltitlerunning{DCDP: an almost linear time change point detection methodology in high-dimensions}

\begin{document}

\twocolumn[
\icmltitle{Divide and Conquer Dynamic Programming: 
An Almost Linear Time  Change Point Detection Methodology in High Dimensions}




\begin{icmlauthorlist}
\icmlauthor{Wanshan Li}{x}
\icmlauthor{Daren Wang}{y}
\icmlauthor{Alessandro Rinaldo}{x}
\end{icmlauthorlist}

\icmlaffiliation{x}{Department of Statistics and Data Science, Carnegie Mellon University}
\icmlaffiliation{y}{Department of ACMS, University of Notre Dame}

\icmlcorrespondingauthor{Wanshan Li}{wanshanl@andrew.cmu.edu}

\icmlkeywords{Change point detection, high-dimensional statistics}

\vskip 0.3in
]
\printAffiliationsAndNotice{}


\begin{abstract}
    We develop a novel, general and computationally efficient framework, called Divide and Conquer Dynamic Programming (DCDP), for localizing change points in time series data with high-dimensional features.  DCDP deploys a class of greedy algorithms that are applicable to a broad variety of high-dimensional statistical models and can enjoy almost linear computational complexity. 
    We investigate the performance of DCDP in three commonly studied change point settings in high dimensions: the mean model, the Gaussian graphical model, and the linear regression model. 
       In all three cases, we derive non-asymptotic bounds for the accuracy of the DCDP change point estimators. We demonstrate that the DCDP procedures consistently estimate the change points with sharp, and in some cases, optimal rates while incurring significantly smaller computational costs than the best available algorithms. Our findings are supported by extensive numerical experiments on both synthetic and real data.  
\end{abstract}

\section{Introduction}
\label{sec:introduction}
Change point analysis is a well-established topic in statistics that is concerned with identifying abrupt changes in the data, typically observed as a time series, that are due to structural changes in the underlying distribution. Initially introduced in the 1940s \citep{wald1945, page1954}, change point analysis has been the subject of a rich statistical literature and has produced a host of well-established methods for statistical inference. Despite their popularity, most existing change point methods available to practitioners are ill-suited or computationally   costly to handle high-dimensional    complex data.  In this paper, we develop a general and flexible framework for high-dimensional change point analysis that enjoys very favorable statistical and computational properties.

We adopt a standard offline change point analysis set-up, whereby we observe a sequence $\{\bfZ_i\}_{i\in [n]}$ of independent data points, where $[n]:=\{1,\ldots, n\}$. We assume that each $\bfZ_i$ follows a high-dimensional parametric distribution $\mathbb{P}_{\bftheta^*_i}$ specified by an unknown parameter $\bftheta^*_i$, and that sequence of parameters $\{\bftheta_i\}_{i \in [n]}$ is piece-wise constant over time. For example, in the mean change point model (see ~\Cref{sec: result mean} below), $\E(\bfZ_i)   = \bftheta^*_i   \in \mathbb R^p$, where $\bftheta^*_i$ is a vector in $\mathbb R^p$. In the regression change point model (see ~\Cref{sec: result regression}),   $\bfZ_i = (\bfX_i, y_i) \in \mathbb R^p\times \mathbb R$ satisfying  $\E(y_i|\bfX_i) =    \bfX_i^\top \bftheta ^*_i $  where $\bftheta^*_i$  is a vector of regression parameters.

We postulate that there exists an unknown sub-sequence 
of {\it change points} $ 1=\eta_0 <  \eta_1  < \eta_2 < \ldots <\eta_K < \eta _{K+1}= n+1$ such that $\bftheta^*_{i}\neq \bftheta^*_{i - 1}$ if and only if $i \in \{\eta_{k}\}_{k\in [K]}$. For each $k\in [K]=\{ 1,\ldots,K\}$, define the local spacing parameter  and local jump size parameter as
\begin{equation}
    \Delta_k = \eta_{k} -\eta_{k-1} \quad  \text{and} \quad   \kappa_k 
	  : =   \|\bftheta^*_{\eta_{k}} - \bftheta^*_{\eta_k - 1}\| 
\end{equation}
respectively, where $\|\cdot \|$ is some appropriate  norm that is problem specific.  Throughout the paper, we will allow the parameters of the data generating distributions, the spacing and jump sizes to change with $n$, though we will require $K$ to be bounded.
Our goal is to estimate the number and locations of the change points sequence $\{ \eta_k\}_{k\in[K]}$. We will deem any estimator $\{ \widehat \eta_k\}_{k\in[ \widehat{K}]}$ of the change point sequence {\it consistent} if, with probability tending to $1$ as $n\rightarrow \infty$, 
\begin{equation} 
\label{eq: consistency general}
     \widehat{K} = K \quad \text{and} \quad   \max_{k\in [K]}|\widehat{\eta}_k - \eta_k|   = o(\Delta_{\min}),
\end{equation}
where $\Delta_{\min} = \min_{ k\in[K] } \Delta_k$.

Recent years have witnessed significant advances in the fields of high-dimensional change point analysis,  both in terms of methodological developments and theoretical advances. Most change point estimators for high-dimensional problems can be divided into two main categories: those based on variants of the binary segmentation algorithm and those relying on the penalized likelihood. See below for a brief summary of the relevant literature.

In this paper, we aim to develop a comprehensive framework for estimating change points in high-dimensional models using an $\ell_0$-penalized likelihood approach. 
While $\ell_0$-based change point algorithms have demonstrated excellent -- in fact, often optimal -- localization rates, their computational costs remain a significant challenge. Indeed, optimizing the $\ell_0$-penalized objective function using a dynamic programming (DP) approach requires quadratic time complexity \citep{friedrich2008complexitypenalizedmestimation} and, therefore, is often impractical. 

To overcome this computational bottleneck, we propose a novel class of algorithms for high-dimensional multiple change point estimation problems called \textit{divide and conquer dynamic programming} (DCDP) -  see \Cref{algorithm:DCDP}.
The DCDP framework is very versatile and can be applied to a wide range of high-dimensional change point problems. At the same time, it yields a substantial reduction in computational complexity compared to the vanilla DP. In particular, when the minimal spacing $\Delta_{\min}$ between consecutive change points is of order $n$, DCDP exhibits almost linear time complexity.

Moreover, the DCDP algorithm retains a high degree of statistical accuracy. Indeed, we show that DCDP delivers minimax optimal localization error rates for   change point localization in the sparse high-dimensional mean model, the Gaussian graphical model  and the sparse linear regression model. 
To the best of our knowledge, DCDP is the first near-linear time procedure that can provide optimal statistical guarantees in these three different models. See \Cref{remark:optimal mean} and \Cref{remark:optimal regression} for more detailed discussions on optimality.
 


{\bf Structure of the paper.} 
Below we provide a selective review of the recent relevant literature on high-dimensional change point analysis. In \Cref{sec:method}, we describe the DCDP framework. In \Cref{section:main}, we provide detailed theoretical studies to demonstrate that DCDP    achieves minimax optimal localization errors in the three models.  In \Cref{sec:experiment}, we conduct extensive numerical experiments on synthetic and real data to illustrate the superior numerical performance of DCDP compared to existing procedures.

{\bf Relevant litearture.} 
Binary Segmentation(BS) is a greedy iterative approach that breaks the multiple-change-point problem down into a sequence of single change-point sub-problems. Originally introduced by \cite{Scott1974} to handle the case of one change point, the BS algorithm was later shown by \cite{Venkatraman1992} to be effective also in the multiple-change-point senerios.   
Modern computationally efficient variants of the original BS algorithms include wild-binary segmentation of \cite{Fryzlewicz2014} and   \textit{Seeded Binary Segmentation} (SBS) algorithm of \citep{kovacs2020seeded}. 
Binary Segmentation  procedures have been designed for various change point problems, including high-dimensional mean models \citep{eichinger2018mosum,wang_samworth2018}, graphical models \cite{londschien2021change}, covariance models {\citep{wang_yi_rinaldo_2021_cov}, network models \cite{wang2021network_cpd}, functional models \cite{oscar2022nonpara} and many more. 

Penalized likelihood-based approaches are also popular in the change point literature. Broadly, these approaches 
 segment the time series by maximizing a likelihood function with an appropriate penalty to avoid over-segmentation.
 \cite{yao_au1989}  
 showed that  $\ell_0$-penalized  likelihood-based methods yield consistent estimators of change points.  
 Relaxing the $\ell_0$-penalty to the  $\ell_1$-penalty results in the  Fused Lasso algorithm, whose theoretical and computational properties have been analyzed by many, including  \citep{lin2017fused} for the mean setting and by \citep{qian2016fused} for the linear regression setting. More recently, \cite{unify_sinica2022} proposed a unified framework to analyze Fused-Lasso-based change point estimators in linear models.


Few recent notable contributions in the literature have focused on designing \textit{unified methodological frameworks} for offline change point analysis. 
\cite{verzelen2020hd_mean} developed  a general approach based on local two-sample tests to detect changes in means, but their approach can only consistently estimate the number of change points and the localization accuracy of the estimators is unspecified. 
\citep{changeforest2022} proposed
 a novel multivariate nonparametric multiple change point detection method based on the likelihood ratio tests.
\cite{unify_sinica2022} studied a general framework based on the Fused Lasso to deal with change points in mean and linear regression models,  but their detection boundary is sub-optimal and it is computationally demanding to numerically optimize the  Fused Lasso objective function for high-dimensional time series. Until now, a unified framework for offline change point localization with optimal statistical guarantees and low computational complexity is still missing in the literature.

{\bf Notation.} For   $n \in \mathbb Z^+$, denote $[n]:=\{1,\cdots, n\}$. For a vector $\bfv\in \mathbb{R}^p$,   denote the $i$-th entry as $v_i$, and similarly, for a matrxi $\bfA\in \mathbb{R}^{m\times n}$, we use $A_{ij}$ to denote its element at the $i$-th row and $j$-th column. We use $\mathbb{S}^p_+$ to denote the cone of positive semidefinite matrices in $\mathbb{R}^{p\times p}$. For two real numbers $a,b$, we denote $a\vee b:=\max\{a,b\}$.

$\|\cdot\|_1,\|\cdot\|_2$ refer to the $\ell_1$ and $\ell_2$ norm of vectors, i.e., $\|\bfv\|_1 = \sum_{i\in [p]}|v_i|$ and $\|\bfv\|_2 = (\sum_{i\in [p]}v_i^2)^{1/2}$. For a square matrix $A\in \mathbb{R}^{n\times n}$, we use $\|\bfA\|_F$ to denote its Frobenius norm, ${\rm Tr}(\bfA) = \sum_{i\in [n]}A_{ii}$ to denote its trace, and $|\bfA|$ to denote its determinant. For a random variable $X\in \mathbb{R}$, we denote $\|X\|_{\psi_2}$ as the subgaussian norm \citep{vershynin2018high}: $\|X\|_{\psi_2}:=\inf \{t>0: \mathbb{E} \psi_2(|X| / t) \leq 1\}$ where $\psi_2(t) = e^{t^2} - 1$.

For asymptotics, we denote $x_n\lesssim y_n$
or $x_n = O(y_n)$ if $\forall n$,
$x_n \leq c_1 y_n$ for some universal constant $c_1>0$. $a_n = o(b_n)$ means $a_n/b_n\rightarrow 0$ as $n\rightarrow \infty$, and $X_n = o_p(Y_n)$ if $X_n/Y_n\rightarrow 0$ in probability. We call a positive sequence $\{a_n\}_{n\in \mathbb{Z}^+}$ a diverging sequence if $a_n\rightarrow \infty$ as $n\rightarrow \infty$.

\section{Methodology}
\label{sec:method}
In this section, we introduce the DCDP framework and analyze its computational complexity.
We assume that  we observe a time series of independent data  $\{\bfZ_i\}_{i\in [n]}$ sampled from the unknown sequence of distributions $ \{\mathbb P_{\bftheta_i^*}\}_{i\in [n]}$. For a time interval $ \mclI \subset [1, n]$  comprised of integers, let $\mclF(\bftheta,\mclI) $ denote the value of an appropriately chosen goodness-of-fit function of the subset $\{\bfZ_i\}_{i\in  \mclI }$, and for a fixed and common value of the parameter $\bftheta$. The choice of the goodness-of-fit function is problem dependent.

Next, we use  $\widehat{\bftheta}_{\mclI}$ to denote the penalized or unpenalized maximum likelihood estimator of $\bftheta^*$ within the interval $\mclI$. Intuitively, 
  $  \mathcal F( \widehat \bftheta_\mathcal I ,\mathcal I) $
   can be considered a  local statistic to test for the existence of one or more change points in $\mathcal I$.

DCDP is a two-stage algorithm that entails a divide step and an   conquer step; see \Cref{algorithm:DCDP} for details. In the divide step, described in \Cref{algorithm:DP},  DCDP first computes preliminary estimates of the change point locations by running DDP, a dynamic programming algorithm over a uniformly-spaced grid of time points $\{s_i=\lfloor i\cdot n/(\mclQ + 1)\rfloor\}_{i\in [\mclQ]}$. (DDP can also take as input a random collection of time points, but there are no computational or statistical advantages in randomizing this choice). In the subsequent conquer step, detailed in \Cref{algo:local_refine_general},  the localization accuracy of the initial estimates is improved using a penalized local refinement (PLR) methodology.

{\bf Computational complexity of DCDP.} 
The DCDP procedure achieves substantial computational gains by using a coarse, regular grid of time points $\{s_i\}_{i \in \mclQ} \subset [n]$ during the divide step. Additionally, the PLR procedure in the conquer step is a local algorithm and is easily parallelizable. The number of grid points $\mclQ$ to be given as input to DDP in the divide step should be chosen to be of smaller order than the length of the time course $n$, but large enough to identify the number and the approximate positions of the true change points.


\begin{algorithm}[h]
 \caption{Divide and Conquer Dynamic Programming. DCDP $(\{\bfZ_i\}_{i \in [n]}, \gamma, \zeta ,  \mclQ)$}
\label{algorithm:DCDP}
	\textbf{Input:} Data $\{ \bfZ_i \}_{i \in [n ]  }$, tuning parameters $\gamma,\zeta,  \mclQ  > 0$.

  Set grid points  $s_i=\lfloor\frac{i\cdot n}{\mclQ+1}\rfloor$ for $i\in [\mclQ]$.
\vskip 0.2cm 	  
(Divide Step) Compute the proxy estimators  $\{\widehat{\eta}_k\}_{k\in [\widehat {K}]}$  using DDP $(\{\bfZ_i\}_{i \in [n]},\{s_i\}_{i\in [\mclQ]},\gamma )$ in \Cref{algorithm:DP}.
\vskip 0.2cm 
(Conquer Step) Compute the final estimators $\{\widetilde {\eta}_k\}_{k\in [\widehat {K}]}$ using $ {\rm PLR}(\{\widehat {\eta}_k\}_{k\in [\widehat{K}]},\zeta )$ in \Cref{algo:local_refine_general}. 
\vskip 0.2cm 	
\textbf{Output:}  The   change point estimators $\{\widetilde {\eta}_k\}_{k\in [\widehat{K}]}$.
\end{algorithm}

\begin{algorithm}[h]

\caption{  Divided Dynamic Programming 
 DDP $(\{\bfZ_i\}_{i \in [n]},\{s_i\}_{i\in [\mclQ]},\gamma )$: 
  the divide step. }
\label{algorithm:DP}
	\textbf{Input:} Data $\{ \bfZ_i \}_{i  \in [n ]}$, ordered integers $\{s_i\}_{i\in [\Q]}\subset (0,n)$, tuning parameter $\gamma>0$.

 	 Set $ \widehat {\mathcal P}= \emptyset$,  $\mathfrak{ p} =\underbrace{(-1,\ldots, -1)}_{ n }  $,  ${\bm B} =\underbrace{( \gamma, \infty , \ldots,  \infty)}_{n } $.  
   \\
 	\For {$r  $ in $\{s_i\}_{i\in [\Q]}$} {
 	 \For {$l  $ in $\{s_i\}_{i\in [\Q]}$, $ l<r  $} {
   \vskip 0.2cm 	
         $\mclI \leftarrow[l , r] \cap \{1, \ldots, n\}$;
 	  \\
 	  compute $\widehat{\bftheta}_{\mclI} $ and $\mathcal F (\widehat{\bftheta}_\mclI, \mclI) $ based on $\{\bfZ_i\}_{i \in \I}$  ;
    
 	 $ b \leftarrow B_{l}+ \gamma + \mathcal F (\widehat{\bftheta}_\mclI, \mclI) ;$
   \vskip 0.2cm 	
 	 \If {$b <  B_r $} {
 	 $B_r \leftarrow  b $;
   \\
 	 $\mathfrak{p}_r  \leftarrow l  $.
 	 } 
 	}
 	}
 To compute $  \mathfrak{p}  \in \mathbb N^n $, 	set 
 $ k \leftarrow  n  $.

 	  \While {$k >1$} {
  $ h  \leftarrow \mathfrak{p}_ k  $;
  \\
   $\widehat {\mathcal P} \leftarrow   \widehat {\mathcal P} \  \cup \   \{h\}$;
\\ 
  $ k  \leftarrow h $.
 	  }
	\textbf{Output:}  The set of estimated change points 
	$\widehat {\mathcal P}$. 
\end{algorithm} 

\begin{algorithm}[h] 
\caption{Penalized Local Refinement  ${\rm PLR}(\{\widehat{\eta}_k\}_{k\in [\widehat{K}]},\zeta)$: the conquer step.  
}
\label{algo:local_refine_general}
	\textbf{Input:} Data $\{ \bfZ_i \}_{i\in[n]}$,  estimated change points $\{\widehat{\eta}_k\}_{k \in[\widehat{K}]}$ from \Cref{algorithm:DP}, tuning parameter  $\zeta > 0$.

	Let $(\widehat{\eta}_0, \widehat{\eta}_{\widehat{K} + 1}) \leftarrow (0, n)$.
	
	\For{$k = 1, \ldots, \widehat{K}$} { 
	$(s_k, e_k) \leftarrow (\frac{2}{3} \widehat{\eta}_{k-1} + \frac{1}{3}\widehat{\eta}_{k} , \ \  \frac{1}{3}\widehat{\eta}_{k}  
 + \frac{2}{3}\widehat{\eta}_{k+1} )$
\begin{align*}\bigg(\check{\eta}_k,\widehat{\bftheta}^{(1)},&\widehat{\bftheta}^{(2)}\bigg) \leftarrow \argmin_{\eta, \bftheta^{(1)}, \bftheta^{(2)}} \{\mclF(\bftheta^{(1)},[s_k,\eta)) + 
\\
  &\mclF(\bftheta^{(2)}, [\eta,e_k)) + \zeta R(\bftheta^{(1)}, \bftheta^{(2)},\eta; s_k, e_k )\}
\end{align*}
	$$\widetilde{\eta}_k \leftarrow \argmin_{\substack{\eta}} \left\{\mclF(\widehat \bftheta^{(1)},[s_k,\eta)) + \mclF(\widehat \bftheta^{(2)}, [\eta,e_k))  \right\}$$
	}
	\textbf{Output:} The refined estimators $\{\widetilde{\eta}_k\}_{k \in [\widehat{K}]}$.
\end{algorithm} 

In detail, let   $\mclC_1(|\mclI|, p)$ denote the time complexity of  computing the goodness-of-fit function  $ \mathcal F (\widehat{\theta}_\mclI, \mclI)  $. Naively, the time  complexity of \Cref{algorithm:DP} is $O(\mclQ^2 \cdot \mclC_1(n, p))$, where $\Q$ is the size of the grid $\{ s_i\}_{i \in [\mclQ]}$ in \Cref{algorithm:DP}. 
With the memorization technique proposed in \citep{yu2022temporal}, we show in \Cref{lem: complexity divide step} that the complexity of the divide step can be reduced to $O(n\mclQ\cdot \mclC_2(p))$, and in \Cref{lem: complexity local refine} that the conquer step can be computed with time complexity $O(n\cdot \mclC_2(p))$, where $\mclC_2(p)$ is independent of $n$. Furthermore, as shown later in \Cref{section:main} and \Cref{sec: fundamental lemma}, setting $\mclQ = \frac{4n}{\Delta_{\min}}\log^2(n)$ ensures consistency of \Cref{algorithm:DP}. Therefore, the complexity of DCDP is
$$O\left( \frac{n^2}{\Delta_{\min}}\cdot\log^2(n)\cdot \mclC_2(p) \right).$$
When $\Delta_{\min}$ is of the same order as $n$, the complexity of DCDP becomes $O(n\log^2(n)\cdot \mclC_2(p))$. To the best of our knowledge, DCDP is the first multiple-change-point detection algorithm that can provably achieve near-linear time complexity in the three models presented in \Cref{section:main}.

{\bf Statistical accuracy.} As we will show below, though the DDP procedure in the divide step may already be sufficiently accurate to deliver consistent estimates as defined in \eqref{eq: consistency general}, its error rate is suboptimal. Sharper, even optimal, localization errors can be achieved through the PLR algorithm in the conquer step (see \Cref{algo:local_refine_general}). The PLR  procedure takes as input the preliminary change points estimates from the divide step\footnote{More generally, it can be shown that the PLR procedure remains effective as long as it is given as input any change point estimates whose Hausdorff distance from the true change points is bounded by $\Delta_{\min}$. Thus, the preliminary estimates need not even be consistent.}, and provably reduces their localization errors --   for some of the models considered in the next section, down to the minimax optimal rates. The effectiveness of local refinement methods to enhance the precision of initial change point estimates has been well-documented in the recent literature on change point analysis \citep{rinaldo2021cpd_reg_aistats, li2022cpd_btl}.
In \Cref{algo:local_refine_general}, 
the additional penalty function $R(\bftheta^{(1)}, \bftheta^{(2)},\eta; s, e)$ in \Cref{algo:local_refine_general} is introduced to ensure numerical stability of the parameter estimates in high dimensions and, possibly, to reproduce desired structural properties, such as sparsity. Its choice is, therefore, problem specific. For example, in the sparse mean and linear change point model in \Cref{sec: result mean},  $\bftheta^{(1)}, \bftheta^{(2)} \in \mathbb{R}^{p}$ and we consider the group lasso penalty function
\begin{equation}
\label{eq:group lasso penalty}
    R(\cdot) =   \sum_{i \in [p]} \sqrt{(\eta - s)(\bftheta^{(1)} )_{i}^2 + (e - \eta)(\bftheta ^{(2)})_{i} ^2}.
\end{equation}
\bnrmk[Penalization] In     \Cref{algorithm:DP}, $\gamma$  is a tuning parameter to control the number of selected change points and to avoid false discoveries. In \Cref{algo:local_refine_general}, the tuning parameter $\zeta$ is used to modulate the impact of the penalty function  $R$. We derive theoretically valid choices of tuning parameters in \Cref{section:main}, and provide practical guidance on how to select them in a data-driven way in  \Cref{sec:experiment}.
 \enrmk



\section{Main Results }
\label{section:main}

We investigate the theoretical performance of DCDP in three different high-dimensional change point models. For each of the models examined, we first derive localization rates for the DDP algorithm in the divide step and find that, though they imply consistency, they are worse than the corresponding rates afforded by the computationally costly vanilla DP algorithm \citep{wang2018univariate, rinaldo2021cpd_reg_aistats}. This suboptimal performance reflects the trade-off between computation efficiency and statistical accuracy and should not come as a surprise. 
  Next, we demonstrate that, by using the PLR algorithm in the conquer step, the estimation accuracy increases and the final localization rates become comparable to the (often minimax) optimal rates.

Throughout the section, we will consider the following high-dimensional offline change point analysis framework of reference.

\bnassum[]
\label{assp: DCDP_general}
We observe independent data points $\{\bfZ_{i}\}_{i\in[n]}$ such that, for each $i$, $\bfZ_{i}$ is a draw from a parametric distribution $\mathbb{P}_{\bftheta^*_i}$ specified by an unknown parameter vector $\bftheta^*_i$.  There exists an unknown  collection of change points $ 1=\eta_0 <  \eta_1  < \eta_2 < \ldots <\eta_K < \eta _{K+1}= n+1$ such that $\bftheta^*_{i} \neq \bftheta^*_{i-1}$ if and only if $i \in \{\eta_k\}_{k\in[K]}$. For each change point $\eta_k$, we will let $\kappa_k = \| \bftheta^*_{\eta_{k}} - \bftheta_{\eta_k - 1}^* \|$ be the size of the corresponding change, where $\|\cdot\|$ is an appropriate norm (to be specified, depending on the model). For simplicity, we further assume that the magnitudes of the changes are of the same order: there exists a $\kappa>0$ such that 
	 $  \kappa _k \asymp  \kappa $ for all $k\in [K]$. We denote the spacing between $\eta_{k}$ and $\eta_{k-1}$ with 
  $\Delta_k =  \eta_{k} -\eta_{k-1} $ and let  $ \Delta_{\min} = \min_{k\in[K]}\Delta_k$ denote the minimal spacing. All the model parameters are allowed to change with $n$, with the exception of $K$.
\enassum

   

{

\subsection{ Changes in  means}
\label{sec: result mean}

Change point detection and localization of a piece-wise constant mean signal is arguably the most traditional and well-studied change point model.  Initially developed in the 1940s for univariate data, the model  has recently been generalized under various high-dimensional settings and thoroughly investigated: see, e.g., 
\cite{wang_samworth2018, Gao2019, verzelen2020hd_mean, unify_sinica2022}. Below, we show that, for this model,  DCDP achieves the sharp detection boundary and delivers the minimax optimal localization error rate.
\bnassum[Mean model]
\label{assp: DCDP_mean main}
   Suppose that for each $i\in [n]$, $\bfZ_i= \bfX_{i }$ satisfies the mean model $\bfX_{i} =    \bfmu _i^*    + \bfepsilon_i\in \mathbb{R}^p$ and \Cref{assp: DCDP_general} holds with $\bftheta^*_i = \bfmu^*_i$ and $\|\cdot\| = \|\cdot\|_2$.   

{\bf (a)} The measurement errors $\{\bfepsilon_i\}_{i\in[n]}$ are independent mean-zero random vectors with independent subgaussian entries such that $0<\sigma_{\epsilon} =\sup_{i\in[n]}\sup_{j\in[p]} \|(\bfepsilon_i)_j\|_{\psi_2}<\infty$.

{\bf (b)} For each $i \in [n]$,    there exists a collection of subsets  $  S_i \subset [p]$,  such that 
	  $  (\bfmu^*_{i})_j =0 \text{ if } j \not \in S_i.$
	  In addition, the cardinality of the support satisfies $|S_i|\leq \mathfrak{s} $. 
\enassum

 Conditions {\bf (a)} and {\bf (b)} above are standard assumptions for the high-dimensional linear regression time series models \citep{Basu2015,unify_sinica2022}.  In our first result, we establish consistency of the divide step. The proof of the following theorem is in   \Cref{sec: main proof mean}.
 
\bnthm \label{thm:DCDP mean}
Suppose  that \Cref{assp: DCDP_mean main} holds and that 
  \begin{align}\label{eq:dp mean snr 1} \Delta_{\min} \kappa^2  \ge \Bt \sigma_{\epsilon}^2 \s \log(p\vee n),
  \end{align}
  for some slowly diverging sequence $\{ \mclB_n\}_{n\in \mathbb{Z}^+}$.
  For sufficiently large constants $C_\gamma$ and $C_\mclF$, let 
 $\{\widehat { \eta }_k\}_{k\in [\widehat K]}  $ denote the output of  \Cref{algorithm:DP} with $\mclQ = \frac{4n}{\Delta_{\min}}\log^2(n)$,
 \begin{equation*}
    \mathcal F  (\widehat{\bfmu}_\I,\I) := \begin{cases}
    \sum_{i \in \I } \|\bfX_i - \widehat \bfmu_\I \|_2 ^2  &\text{if } |\I|\geq C_\mclF \s \log(p\vee n),\\
    0  &\text{otherwise},
    \end{cases}
\end{equation*}
 and   $\gamma = C_\gamma \mathcal B_n^{-1/2} \Delta_{\min}\kappa^2$. Here 
\begin{equation}
    \widehat{\bfmu}_{\I} = \argmin_{\bfmu\in \mathbb{R}^p}\|\bfX_i - \bfmu\|_2^2 + \lambda \sqrt{|\I|}\|\bfmu\|_1,
\end{equation}
with $\lambda = C_{\lambda}\sqrt{\log(p\vee n)}$ and $C_\lambda $ a
  sufficiently large constant. 
 Then, 
 with probability  $1 - n^{-3}$, $\widehat K = K$ and 
 \begin{equation*}
     \max_{ k \in [K] } |\eta_k-\widehat \eta_k|   \lesssim   \frac{  \sigma_{\epsilon}^2\log(p\vee n)}{\kappa^2 } + \mathcal{B}_n^{-1/2} \Delta_{\min}.
 \end{equation*}
 \enthm

The \textit{signal-to-noise-ratio} (SNR) condition \eqref{eq:dp mean snr 1} assumed in \Cref{thm:DCDP mean} is frequently used in the change point detection literature \citep{unify_sinica2022,wang_samworth2018}. Recently,  \cite{verzelen2020hd_mean} showed that, if $\s\le \sqrt p$, condition \eqref{eq:dp mean snr 1} is indeed necessary, in the sense that if 
$$ \frac{\Delta_{\min} \kappa^2 }{\sigma_{\epsilon}^2 \s  \log(p\vee n)} =o(1) ,$$
then there exists a setting for which no change point estimator is consistent.    
The localization error of DCDP estimator $\{ \widehat \eta_k\}_{k\in[\widehat K] } $ returned by   \Cref{algorithm:DP} satisfies 
$$
 \frac{\max_{k\in [K]} |   \eta_k-\widehat{\eta}_k|}{\Delta_{\min}} \lesssim    \frac{ \sigma_{\epsilon}^2 \log(p\vee n)}{\Delta_{\min}\kappa^2} +   \mclB_n^{-1/2},
$$
with high probability.
   Thus, using  \eqref{eq:dp mean snr 1}, 
it follows  that the resulting estimator is consistent:
$$ \frac{\max_{k\in [K]}|   \eta_k-\widehat{\eta}_k|}{\Delta_{\min}} \lesssim   
\mclB_n^{-1} + \mclB_n^{-1/2}
    =o_p(1). $$ 

\bnrmk[Grid size]
In \Cref{thm:DCDP mean} and in all the results of this section, we choose a  value for the grid size $\mclQ$ that, while coarse, ensures consistency. Any finer grid can yield the same error rate, at an additional computational cost. 
\enrmk

Compared to the localization error of the vanilla DP, the localization error of Divided DP \Cref{algorithm:DP} picks up an additional term $\htr$. As remarked above, this is to be expected, as \Cref{algorithm:DP} only deploys a subset of the data indices.    Starting with the coarse (but still consistent)   preliminary estimators from the divide step  \Cref{algorithm:DP}, the local refinement algorithm   \Cref{algo:local_refine_general}  further improves its accuracy and, in fact, yields an optimal error rate.
 
\bnthm
\label{cor:mean local refinement}  Let $\{ \mclB_n\}_{n\in \mathbb{Z}^+}$ be any slowly diverging sequence and suppose  that $\Delta_{\min} \kappa ^2 \geq \Bt \sigma_{\epsilon}^2 \mathfrak{s}^2\log^3(p\vee n) $.
	 Let 
 $\{ \widetilde \eta_k\}_{k\in[\widehat K]}$ be the output of 
 \Cref{algo:local_refine_general} with $\zeta = C_{\zeta} \sqrt{\log(p\vee n)}$ for sufficiently large constant $C_{\zeta}$ and $R(\theta^{(1)}, \theta^{(2)},\eta; s, e)$ be specified in \eqref{eq:group lasso penalty}. Then under \Cref{assp: DCDP_mean main}, for any $\alpha \in (0,1) $, with probability at least $1-  (  \alpha \vee n^{-1} ) $  it holds that $\widehat{K} = K$ and
    \begin{equation}
    \label{eq: mean rate op1}
        \max_{k  \in [K]} |\eta_k-\widetilde \eta_k| \lesssim  { \frac{\sigma_{\epsilon}^2}{\kappa^2} (1+\log ( {1}/{\alpha} )  }).
    \end{equation}
\enthm
  The proof of \Cref{cor:mean local refinement} can be found in \Cref{sec:mean op1}.
\bnrmk    \label{remark:optimal mean}
 The localization error bound \eqref{eq: mean rate op1} is the tightest in the literature. It improves the existing bounds by \cite{wang_samworth2018}  and   \cite{unify_sinica2022}
 by a   factor of $\s \log(p) $.  It also matches the lower bound established in \cite{wang_samworth2018}, showing that    $O_p(1/\kappa^2)$ is the optimal error order and can not be further improved. 
\enrmk

\subsection{Changes in  regression coefficients}
\label{sec: result regression}
We now consider the more complex high-dimensional regression change point model in which the regression coefficients are sparse and change in a piecewise constant manner. Recently, various approaches and methods have been proposed to address this challenging scenario; see, in particular, \cite{rinaldo2021cpd_reg_aistats,wang2021_jmlr,unify_sinica2022,yu2022temporal}. Below, we will show that DCDP yields optimal localization errors also for this class of change point models. 

\bnassum[High-dimensional linear model]\label{assp:dcdp_linear_reg main}
   Let  the  observed data   $\{\bfX_{i }, y_i  \}_{i\in [n] }  \subset \mathbb R^p \times \mathbb R $ be such that   
	 $y_{i  } =   \bfX_i^\top \bfbeta ^* _i  +\epsilon_i   $ and let \Cref{assp: DCDP_general} hold  with $\bftheta^*_i = \bfbeta^*_i\in\mathbb{R}^p$ and $\|\cdot\| = \|\cdot\|_2$. In addition,

{\bf (a)} Suppose that $\{ \bfX_i\}_{i\in [n]} \overset{i.i.d.}{\sim} N_p(0, \bfSigma )$  and that the minimal and the maximal eigenvalues of  $\bfSigma$ satisfy 
	  $\Lambda_{\min}   (\bfSigma)\geq c_X$ and $\Lambda_{\max} (\bfSigma) \le C_X$, with universal constants $c_X, C_X\in(0,\infty)$.
   In addition, suppose that $ \{ \epsilon_i \}_{i\in [n]}  \overset{i.i.d.}{\sim}  N(0, \sigma^2_ \epsilon)$  and is  independent of $\{ \bfX_i\}_{i\in [n]} $.

{\bf (b)}  For each $i \in [n]$,    there exists a collection of indices $  S_i \subset [p]$,  such that 
	  $  (\bfbeta^* _ { i})_j =0 \text{ if } j \not \in S_i.$
	  In addition, the cardinality of the support satisfies $|S_i|\leq \mathfrak{s} $. 
\enassum
  
We note that   \Cref{assp:dcdp_linear_reg main}  {\bf (a)} and {\bf (b)}   are standard assumptions for Lasso estimators. Similarly to the case of the mean change point model, we first analyze the performance of the divide step of DCDP and find it to be consistent, albeit at a sub-optimal rate.

 \bnthm \label{thm:DCDP regression}   Suppose  \Cref{assp:dcdp_linear_reg main} holds  and  that 
 \begin{align}\label{eq:snr divide linear regression}
 \Delta_{\min }  \kappa ^2 \geq \Bt  \sigma_{\epsilon}^2 \mathfrak{s}\log(p\vee n) 
 \end{align}for some diverging sequence $\{ \mclB_n\}_{n\in \mathbb{Z}^+}$.
Let  
 $\{\widehat { \eta }_k\} _{k\in [\widehat K]}  $ be the  output of  \Cref{algorithm:DP}  with $\mclQ = \frac{4n}{\Delta_{\min}}\log^2(n)$,    $\gamma = C_\gamma  \mathcal B_n^{-1/2} \Delta_{\min}\kappa^2$ and
\begin{equation*}
     \mathcal F  (\widehat{\beta}_{\I},\mclI) :  = \begin{cases} 
 0 \quad \quad\quad \  \text{if } |\mclI|< C_\mclF \s \log(p\vee n);
 \\ 
 \sum_{i \in \mclI } (y_i - \bfX_i^\top \widehat \bfbeta _\mclI ) ^2 \quad \text{otherwise,}  
 \end{cases} 
\end{equation*}
 for sufficiently large constants $ C_\gamma $ and $C_\mclF$ and $\widehat{\bfbeta}_{\I}$ given by
\begin{equation}
    \widehat{\bfbeta}_{\I} = \argmin_{\bfbeta\in \mathbb{R}^p}(y_i - \bfX_i^\top \bfbeta)^2 + \lambda \sqrt{|\I|}\|\bfbeta\|_1,
\end{equation}
with $\lambda = C_{\lambda}\sqrt{\log(p\vee n)}$, for  $C_{\lambda}$  a sufficiently large constant. 
 Then,
 with probability $1 - n ^{-3}$, $\widehat K = K$ and that
  $$ \max_{ k\in [K]} |\eta_k-\widehat \eta_k|   \lesssim    \frac{  \sigma_{\epsilon}^2\mathfrak{s} \log(p\vee n)}{\kappa^2 }+  \htr.$$
 \enthm

The proof of \Cref{thm:DCDP regression} is deferred to \Cref{sec: main proof linear}. It is immediate to verify that, under the SNR condition \eqref{eq:snr divide linear regression} and given the choice of $\gamma$, estimators satisfy that $ \max_{k\in [K]}  |   \eta_k-\widehat{\eta}_k|  
    =o_p(\Delta_{\min}) $ and are therefore consistent.  


With a slightly stronger SNR condition than \eqref{eq:snr divide linear regression},  
statistically optimal change point estimators can be obtained  in the conquer step.   
 
\bnthm
 \label{cor:regression local refinement}
 Let $\{ \mclB_n\}_{n\in \mathbb{Z}^+}$ be any slowly diverging sequence and suppose  that $\Delta_{\min} \kappa ^2 \geq \Bt  \sigma_{\epsilon}^2 \mathfrak{s}^2\log^3(p\vee n) $.
  Let 
 $\{ \widetilde \eta_k\}_{k\in[\widehat K]}$ be the output of 
 \Cref{algo:local_refine_general} with $\zeta = C_{\zeta} \sqrt{\log(p\vee n)}$ for sufficiently large constant 
 $C_\zeta $ and $R(\bftheta^{(1)}, \bftheta^{(2)},\eta)$ specified in \eqref{eq:group lasso penalty}
 Then under \Cref{assp:dcdp_linear_reg main}, for any $\alpha\in (0,1)$, with probability at least  $1 - (\alpha\vee n^{-1})$, it holds that $\widehat {K} = K$ and
	\begin{equation}
      \max_{k  \in [K]} | \eta_k-\widetilde \eta_k |   \lesssim     ( 1+\frac{\sigma_\epsilon^2}{\kappa ^2}\log^2({1}/{\alpha}) ).
	 \label{eq: linear reg rate op1}
	\end{equation}
\enthm
The proof of \Cref{cor:regression local refinement} can be found in \Cref{sec:regression op1}.
\bnrmk \label{remark:optimal regression}
The localization  error \eqref{eq: linear reg rate op1} matches the existing lower bound 
 established in \cite{rinaldo2021cpd_reg_aistats}  and, therefore, it is rate minimax optimal. To the best of our knowledge, the only other existing change point algorithm that can achieve optimal localization errors in the high-dimensional linear regression setting is the one developed in  \cite{yu2022temporal}, which allows for dependent observations. However, the approach by \cite{yu2022temporal} requires quadratic time complexity. 
 It is worth mentioning that both \cite{rinaldo2021cpd_reg_aistats} and \cite{yu2022temporal} also assume the SNR condition we use in \Cref{thm:DCDP regression} and \Cref{cor:regression local refinement}. 
\enrmk

\subsection{Changes in precision matrices}
\label{sec: result covariance}


For our third and final example, we specialize the general change point framework of  \Cref{assp: DCDP_general} to the case of Gaussian graphical models, in which the distributional changes are induced by a sequence of temporally piece-wise constant precision matrices, with the magnitude of the changes measured in Frobenius norm. 

\bnassum[Gaussian graphical model]
\label{assp:DCDP_covariance main} 
  Suppose  for each $i\in [n]$, $\bfX _{i }$ is a mean-zero Gaussian vector in $\mathbb{R}^p$ with covariance matrix $\bfSigma^*_i = \mathbb{E}[\bfX_i\bfX_i^\top]$, and \Cref{assp: DCDP_general} holds with $\bftheta^*_i = (\bfSigma_i^*)^{-1}$  with $\|\cdot\| = \|\cdot\|_F$. Assume that for each $i\in [n]$, the minimal and maximal eigenvalues of  $\bfSigma_i^*$ satisfy 
	  $\Lambda_{\min}   (\bfSigma_i^*)\geq c_X$ and $\Lambda_{\max} (\bfSigma _i^*) \le C_X$, with universal constants $c_X, C_X\in(0,\infty)$.
\enassum 

 Several contributions in  he recent literature  address  the problem of detecting change points in precision matrices; see, e.g., \cite{Gibberd2017ggm,Gibberd2017fusedGlasso,  Bybee2018jmlr,Keshavarz2020jmlr,londschien2021change,liu2021jmlr, unify_sinica2022}. Most of these studies focus on estimating a {\it single} change point. To the best of our knowledge, only \cite{unify_sinica2022} has provided theoretical guarantees for the multiple-change-point setting assuming sparse changes in the precision matrices. Below, we show that the divide step of the DCDP procedure is able to detect multiple change points in the precision matrices in the dense regime. 
 
\bnthm
\label{thm:DCDP covariance main} Suppose \Cref{assp:DCDP_covariance main} holds and that 
	  \begin{align}\label{eq:snr precision 1} \Delta_{\min} \kappa^2  \geq \Bt   p^2\log(n\vee p)
   \end{align}
   for some slowly diverging sequence $\{ \mclB_n\}_{n\in \mathbb{Z}^+}$. Let  $ \{ \widehat \eta_k\}_{k\in [\widehat K]}$ be the output of \Cref{algorithm:DP} with $\mclQ = \frac{4n}{\Delta_{\min}}\log^2(n)$,  $\gamma = C_\gamma  \mclB_n^{-1/2}\Delta_{\min}\kappa^2$ and
\begin{equation*}
    \mathcal F  (\widehat{\bfOmega}_\I,\I) 
     = \begin{cases}  0  
\quad \quad\quad \quad \quad\quad \quad  \text{ if } |\I|< C_\mclF p\log(p\vee n);\\
       \sum_{i \in \I } {\rm Tr}[\widehat{\bfOmega}_{\mclI}^\top \bfX_i\bfX_i^\top] - |\I|\log|\widehat{\bfOmega}_{\mclI}|   
       \ \text{otherwise}.
    \end{cases}
\end{equation*}
for sufficiently large constants $ C_\gamma $ and $C_\mathcal F$. Here $\widehat{\bfOmega}_{\I}$ is
\begin{equation}
\label{eq:goodness-of-fit.precision}
    \widehat{\bfOmega}_{\I} = \argmin_{\bfOmega\in \mathbb{S}^p_+} \sum_{i \in \I } {\rm Tr}[{\bfOmega}^\top \bfX_i\bfX_i^\top] - |\I|\log|\bfOmega|. 
\end{equation}
 Then
 with probability  at least $1 - n ^{-3}$, $\widehat  K = K$ and that
 \begin{equation} 
 \label{eq:loc.rate.precison}
  \max_{ k\in [K] } |\eta_k-\widehat \eta_k|   \lesssim    \frac{ p^2 \log(p\vee n)}{\kappa^2 } + \mclB_n^{-\frac{1}{2}}{\Delta_{\min}}.
 \end{equation}
\enthm
The proof of \Cref{thm:DCDP covariance main} is deferred to  \Cref{sec: main proof covariance}.


Under the assumption of the theorem, the localization rate \eqref{eq:loc.rate.precison} implies consistency, as defined in \eqref{eq: consistency general}; indeed, it is easy to see that $ \max_{k\in [K]}  |   \eta_k-\widehat{\eta}_k|  
    =o_p( \Delta_{\min}   ). $ 
 
An analogous condition to Condition \eqref{eq:snr precision 1} is used in \cite{unify_sinica2022} under the slightly different settings of sparse changes. More precisely, the authors  there requires that    $\Delta_{\min}\kappa^2\geq \mclB_n d\log(n\vee p)$, where $d $   is the maximal number of nonzero entries in the precision matrices. When applied to our dense settings, their SNR condition matches  \eqref{eq:snr precision 1}.

Under a slightly stronger SNR condition, we further obtain that the local refinement algorithm in the conquer step improves the localization rate to match the sharpest rate known for this problem.

\bnthm[]
\label{cor:covariance local refinement main}Let $\Bt$ be an arbitrary slowly diverging sequence and suppose $ \Delta_{\min} \kappa^2  \geq \Bt   p^4\log^2(n\vee p)$.
 Let 
 $\{ \widetilde \eta_k\}_{k\in[\widehat K]}$ be the output of 
 \Cref{algo:local_refine_general} with $R(\theta^{(1)}, \theta^{(2)},\eta) =0$. 
Then under \Cref{assp:DCDP_covariance main},  it holds that  with probability at least $1 - n^{-1}$ 
\begin{equation}\label{eq:covariance final rate}
        \max_{k \in {[K]}} |\eta_k-\widetilde \eta_k|\lesssim \frac{1}{\kappa^2}\log(n).
    \end{equation}
\enthm
The proof of \Cref{cor:covariance local refinement main} is in \Cref{sec:cov op1}.
 The localization error bound obtained for DCDP in
\Cref{cor:covariance local refinement main} matches the sharpest error bounds obtained for the precision matrices change point model \cite{liu2021jmlr, unify_sinica2022} and does not require the precision matrices to be sparse. To the best of our knowledge, DCDP is the first linear time algorithm that can optimally estimate multiple change points in the precision matrices in high dimensions.

\section{Numerical Experiments}
\label{sec:experiment}
We evaluate the numerical performance of DCDP through examples of synthetic and real data. The tuning parameters  $ \gamma$ and  $\zeta$ of DCDP are chosen using cross-validation. The implementations of our numerical experiments   are available online
\footnote{\url{https://github.com/MountLee/DCDP}}.
More details, including the implementation for cross-validation and additional numerical  results, can be found in   \Cref{sec:detail experiment}  due to space constraints.

\subsection{Time complexity  and accuracy of DCDP}
\label{sec:experiment_time_error_divide}
We generate  i.i.d.   Gaussian random variables $\{y_i\}_{i\in[n]} \subset  \mathbb R $ with $y_i = \mu^*_i + \epsilon_i$ and  $\sigma_\epsilon   =1$. We set $n=4\Delta$ where $\Delta $ will be specified in each setting.   The three population change points of $\{\mu^*_i\}_{i\in[n]}$ are set to be $\mu_{\eta_0}^* = 0$, $\mu_{\eta_1}^* = \delta$, $\mu_{\eta_2}^* = 0$, $\mu_{\eta_3}^* = \delta$, where $\eta_{k} = k\Delta + \delta_k$ with $\delta_k\sim{\rm Unif[-\frac{3}{10}\Delta,\frac{3}{10}\Delta]}$ for $ k = 1,2,3$. We use the Hausdorff distance $H(\{\widehat{\eta}_k\}_{k\in [\widehat{K}]},\{\eta_k\}_{k\in [K]})$ to quantify the difference between the estimators and the true change points.

\begin{figure}[ht]
    \centering
    \includegraphics[width = 0.48
    \textwidth]{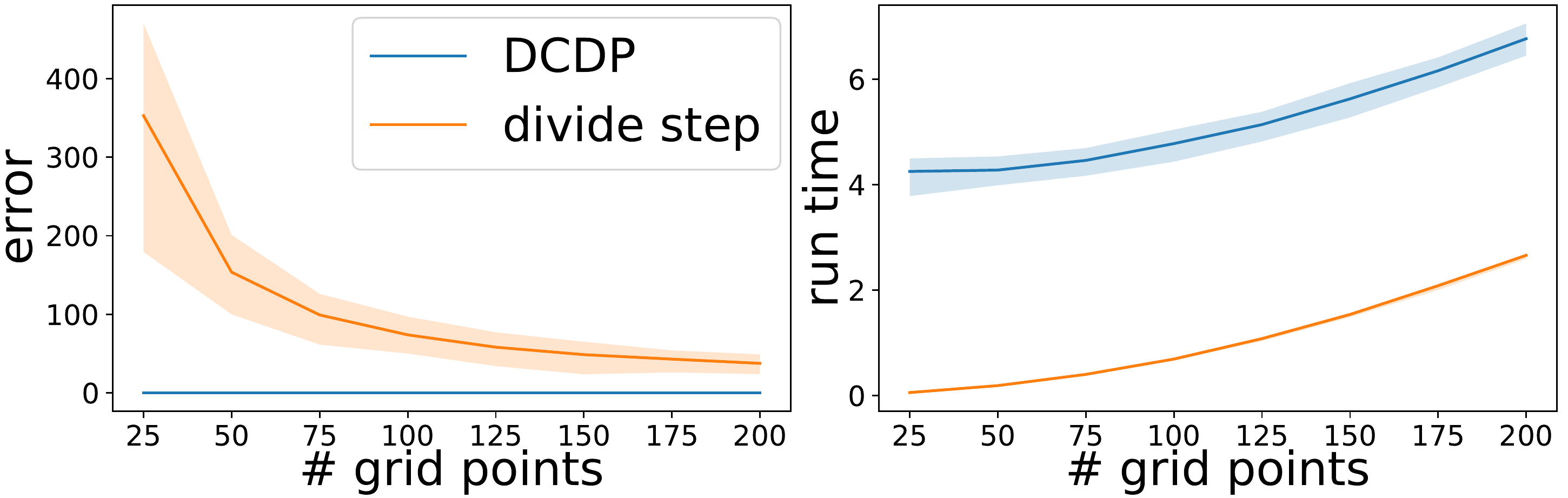}
    \caption{Average localization error   and average run time  versus  the number of grid points $\mclQ$ over 100 trials. The shaded area indicates  the upper and lower 0.1 quantiles of the corresponding quantities. }
\label{fig:loc_runtime_vs_n_grid}
\end{figure}
In the first set of experiments, we set $ \Delta =5000,\delta = 5$ and vary $\mathcal Q$  from $25$ to $200$, and summarize results in \Cref{fig:loc_runtime_vs_n_grid}. The left plot of the figure shows that while the localization errors of the divide step are sensitive to the choice of $\mathcal Q$, the additional conquer step (Algorithm 3) greatly improves the numerical accuracy of the final estimators of DCDP. The right plot of the figure demonstrates that the time complexity of DCDP is quadratic in  $\mathcal Q$, which is in line with the complexity analysis presented in \Cref{sec:method}.

In the second set of experiments, we fix $\mathcal Q = 100. \delta = 5 $ and let $\Delta$ range from 1000 to 6000. The results are summarized in \Cref{fig:error_runtime_on_n}. The left plot of the figure shows that while the localization errors of the dive step change with $\Delta$, the accuracy of DCDP is consistently small for all the different values of $\Delta$. The right plot of the figure shows that the time complexity is linear in $n$, and this observation matches the findings presented in \Cref{sec:method}.

\begin{figure}[ht]
    \centering
    \includegraphics[width = 0.48 \textwidth]{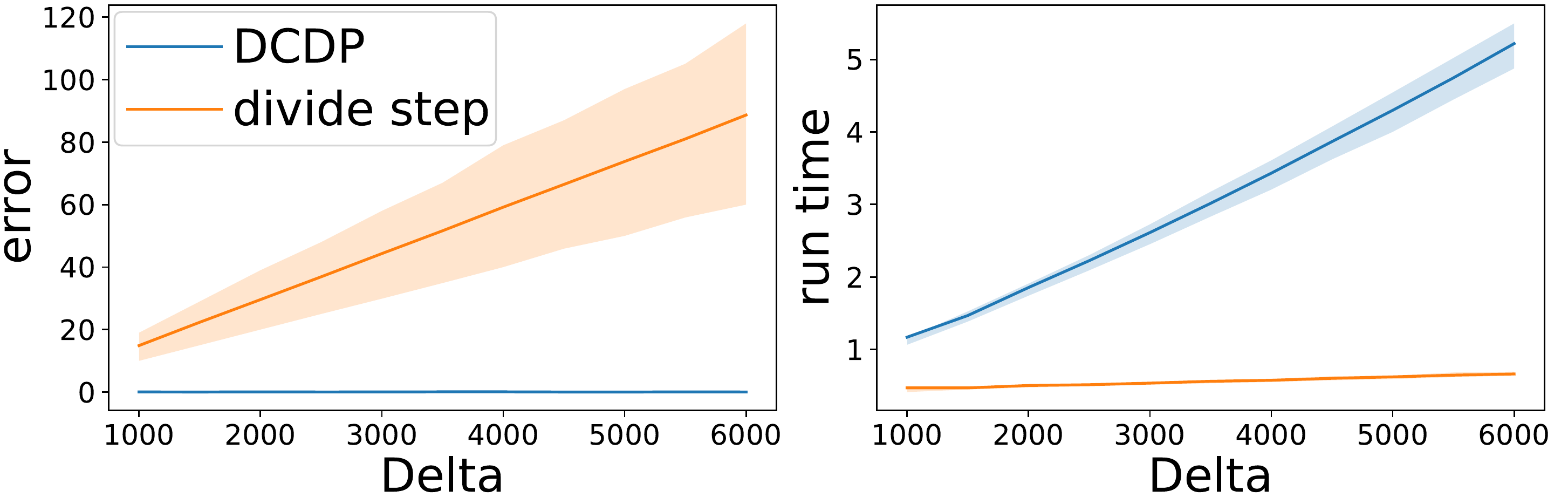}
    \caption{
    Average localization error and average run time v.s. $\Delta$ over 100 trials. }
    \label{fig:error_runtime_on_n}
\end{figure}

\begin{figure}[H]
    \centering
    \includegraphics[width = 0.48 \textwidth]{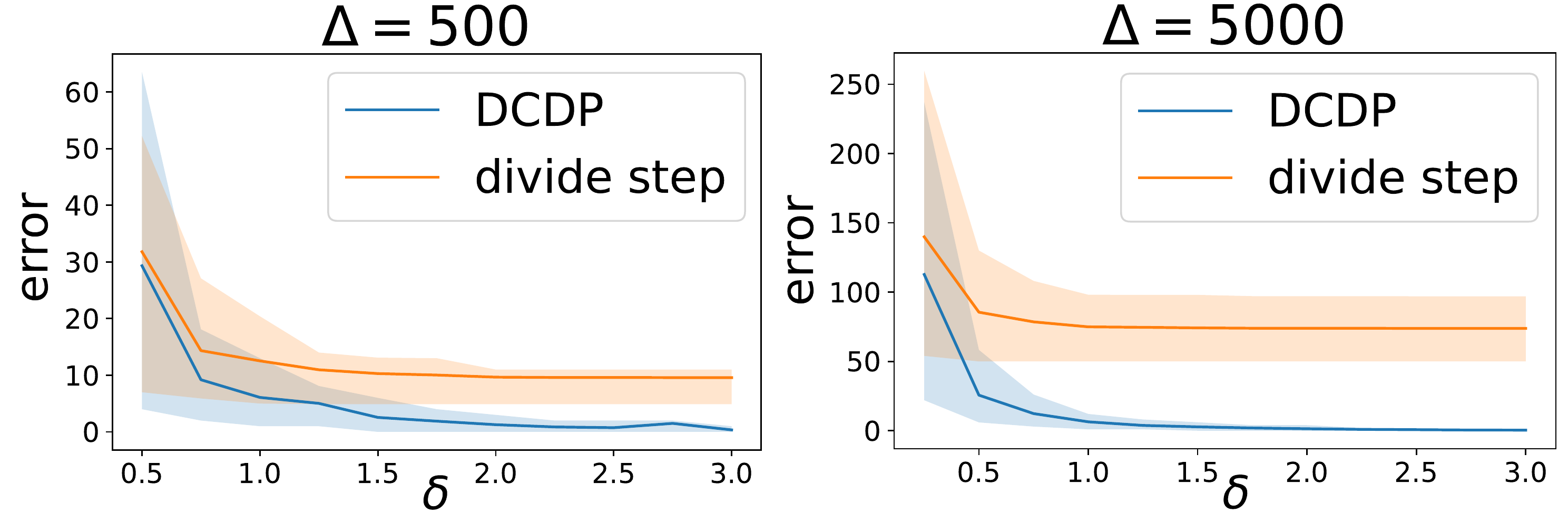}
    \caption{Localization error when varying $\delta$, the magnitude of nonzero signals.}
    \label{fig:snr}
\end{figure}

Next, we fix $\mclQ = 100$ and $\Delta\in \{500, 5000\}$ and vary $\delta$, the strength of signals, to illustrate the performance of DCDP under different SNR levels. The results are summarized in \Cref{fig:snr}. More discussions on the accuracy under small $\delta$ are included in \Cref{sec:snr}.

\subsection{Numerical performance of DCDP}
\label{sec: experiment comparison}

Below we report  the outcome of various simulation studies in which we compare the numerical performance of DCDP with that of several other state-of-the-art methods,  for each  of the three models presented in \Cref{section:main}. 

In the following experiments,
for each specific  $\Delta$  we set the total number of observations $n = (K + 1)\Delta$ and the locations of true change points $\eta_k = k\Delta + \delta _k$, where $\delta_k$ is a random variable sampled from  the uniform distribution ${\rm Unif}[-\frac{3}{10}\Delta,\frac{3}{10}\Delta]$.
In each setting, we conduct  
100 trials and report the average execution time,   the average  Hausdorff distance between true and estimated change points,  and the frequency of  cases in which $\widehat{K}=K$, for each method.

{\bf The mean model}
\\
We set $K = 3$  and, for $k=0,\cdots, K$ and $\delta \in \{1, 5 \}$, we assume  a population mean vector of the form 
\begin{align*}
\bfmu^*_{\eta_k} = (  \underbrace{ 0, \ldots, 0}_{5k},  \underbrace{ \delta ,  \ldots, \delta }_{5} ,   \underbrace{ 0,\ldots,0}_{p-5k-5})^\top   \in \mathbb R^{p}.
\end{align*}

We compare DCDP with Change-Forest
(CF) \cite{changeforest2022}, Block-wise Fused Lasso (BFL) \cite{unify_sinica2022}, and Inspect \citep{wang_samworth2018}. The results are summarized in \Cref{tab:mean compare_methods}. On average, DCDP outputs the most accurate change point estimators while remaining computationally efficient.

\begin{table}[h]
 \centering
 \begin{tabular}{l l l c}
 \hline
 Method & $H(\hat{\bfeta},\bfeta)$ & Time & $\widehat{\mathbb{P}}[\widehat{K}=K]$  \\
 \hline
   \multicolumn{4}{c}{$n = 200, p=100, K =3, \delta = 5$}
   \\
  DCDP & 0.00 (0.00) & 0.6s (0.0) & 1.00 \\
 Inspect & 0.40 (3.50) &  0.0s (0.0) & 0.91 \\
  CF & 1.84 (6.27) &  0.8s (0.2) & 0.90  \\
  BFL &  47.84 (6.69) &  1.4s (0.2) & 0.00 \\
 \hline
 \multicolumn{4}{c}{$n = 200, p=100, K =3, \delta = 1$}  
  \\
  DCDP & 0.83 (0.87) & 0.8s (0.2) & 1.00 \\
 Inspect  &  2.65 (5.16) &   0.0s (0.0) & 0.86  \\
  CF &  6.29 (9.57) &  1.1s (0.3) & 0.78 \\
  BFL &  47.19 (6.48) &  1.1s (0.2) & 0.00 \\
 \hline
 \end{tabular}
  
 \caption{Numerical comparison of different methods in the high-dimensional mean shift models. The numbers in the cells indicate the averages over 100 trials and the numbers in the brackets  indicate the corresponding standard errors.}
 \label{tab:mean compare_methods}
\end{table}

{\bf The linear regression model}

We set $K = 3$  and, for $k=0,\cdots, K$,  assume population regression coefficients of the form
\begin{align*}
\bfbeta^*_{\eta_k} = (  \underbrace{ 0, \ldots, 0}_{5k},  \underbrace{ \delta ,  \ldots, \delta }_{5} ,  \underbrace{ 0,\ldots,0}_{p-5k-5})^\top \in \mathbb R^{p},
\end{align*}
where $\delta \in \{1, 5 \}$.

We compare the numerical performance of DCDP with   Variance-Projected Wild Binary Segmentation (VPBS) \citep{wang2021_jmlr} and vanilla Dynamic Programming (DP) \citep{rinaldo2021cpd_reg_aistats}. The results are summarized in \Cref{tab:linear compare_methods}. On average, DCDP is the most efficient algorithm with compelling numerical accuracy.

\begin{table}[h]
 \centering
\begin{tabular}{l l l c}
  \hline
Method & $H(\hat{\bfeta},\bfeta)$ & Time & $\widehat{\mathbb{P}}[\widehat{K}=K]$ 
 \\
 \hline
 \multicolumn{4}{c}{$n = 200, p=100, K =3, \delta = 5$} \\
  DCDP & 0.13 (0.39) & 18.4s (1.1) & 1.00 \\
  DP & 0.01 (0.10) & 220.3s (16.8) & 0.98 \\
  VPWBS & 15.44 (17.99) &  120.1s (13.1) & 0.70 \\
 \hline
 \multicolumn{4}{c}{$n = 200, p=100, K =3, \delta = 1$} \\
  DCDP & 1.45 (8.59) & 8.8s (0.7) & 0.98 \\
  DP & 0.22 (2.00) & 84.4s (5.7) & 0.99 \\
  VPWBS & 11.54 (11.23) &  120.4s (14.5) & 0.65 \\
 \hline
 \end{tabular}
\caption{Numerical comparison of different methods in the high-dimensional regression coefficient shift models.}
 \label{tab:linear compare_methods}
\end{table}

{\bf The Gaussian graphical model}
\\
We set $K=3$ and the population  covariance matrix matrices as    $\bfSigma^*_{\eta_0}= \bfSigma^*_{\eta_2} = {\bm I}_p$ and 
$\bfSigma^*_{\eta_1}= \bfSigma^*_{\eta_3}$ where
\begin{equation*}
(\bfSigma^*_{\eta_1})_{ij} =(\bfSigma^*_{\eta_3})_{ij} =\begin{cases}
    \delta_1, & i=j;\\
    \delta_2, & |i-j|=1;\\
    0, & \text{otherwise},
\end{cases}
\end{equation*}
with $\delta_1 = 5, \delta_2 = 0.3$.

We compare the numerical performance of DCDP with Change-Forest
(CF) \cite{changeforest2022} and  Block-wise Fused Lasso (BFL) \cite{unify_sinica2022}.  
Note that the BFL algorithm produces empty set in all trials, so we only report DCDP and CF in \Cref{tab:covariance compare_methods}.
It can be seen that on average DCDP outputs  the most accurate change point estimates and  is highly computationally efficient.

\begin{table}[h]
 \centering
\begin{tabular}{l l l c}
  \hline
 Method & $H(\hat{\bfeta},\bfeta)$ & Time & $\widehat{\mathbb{P}}[\widehat{K}=K]$ 
 \\
 \hline
 \multicolumn{4}{c}{$n = 400, p=10, K =3, \delta_1 = 5,\delta_2 = 0.3$} \\
  DCDP & 0.42 (0.64) & 0.5s (0.0) & 1.00  \\
  CF &  5.54 (14.71) &  0.6s (0.1) & 0.88  \\
 \hline
 \multicolumn{4}{c}{$n = 400, p=20, K =3, \delta_1 = 5,\delta_2 = 0.3$} \\
  DCDP & 0.66 (4.37) & 0.9s (0.3) & 1.00 \\
  CF &  7.37 (18.76) &  1.0s (0.0) & 0.85  \\
 \hline
 \end{tabular}
 \caption{Numerical comparison of different methods in the precision matrix shift models. }
 \label{tab:covariance compare_methods}
\end{table}

\subsection{Real data analysis}
\label{sec: application}
In this section, we apply DCDP to three popular real data examples and compare it with state-of-the-art methods.



\textbf{Bladder tumor micro-array data.}
This dataset contains the micro-array records of 43 patients with bladder tumor, collected and studied by \cite{acgh2006}. The result is visualized in \Cref{fig:acgh}, where we only show the data of 10 patients for the ease of presentation and reading. While there is no accurate ground truth of locations of change points, the 37 change points spotted by DCDP align well with previous research \citep{ecp2015, wang_samworth2018}. \Cref{fig:acgh} provides virtual support for the findings by DCDP.

\begin{figure}[ht]
    \centering
    \includegraphics[width = 0.48 \textwidth]{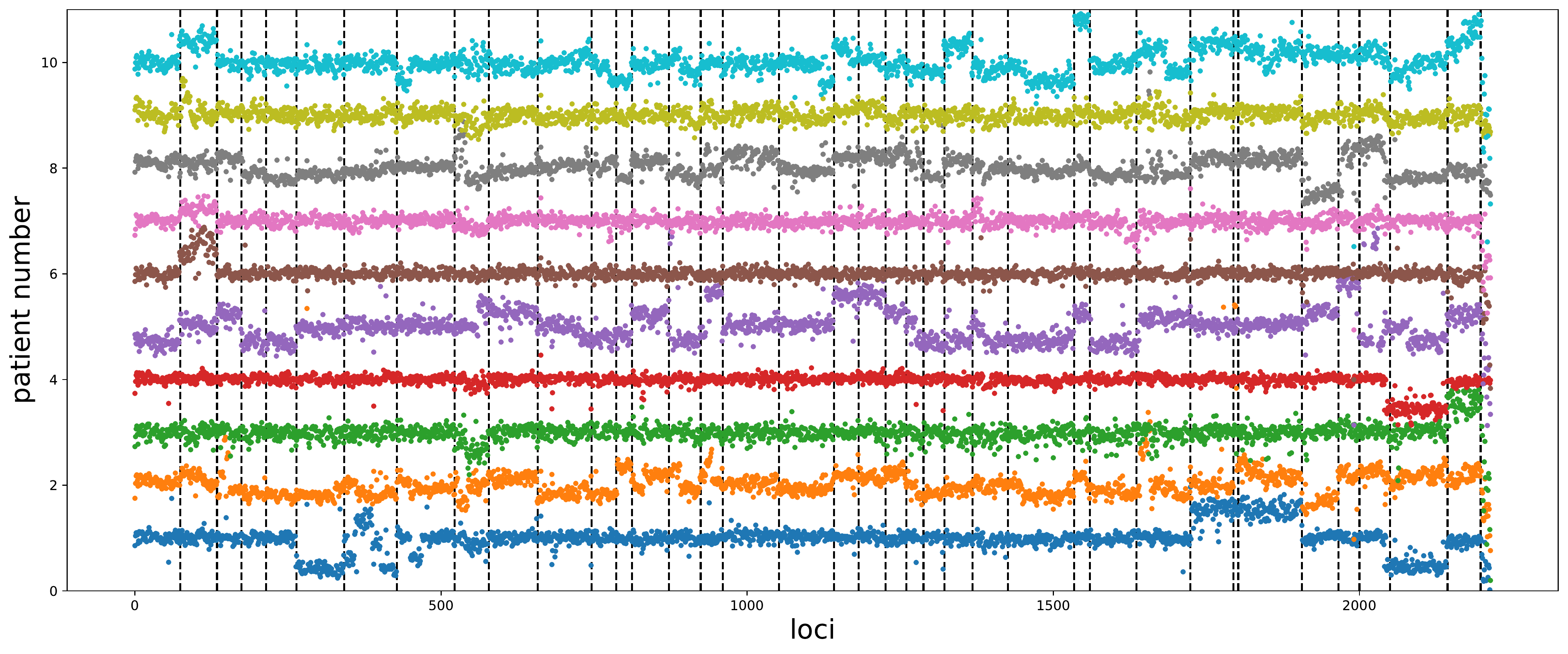}
    \caption{Estimated change points in the micro-array data. The result is based on the data of all 43 patients, while only the data of 10 patients is presented. The estimated change points are indicated by dashed vertical lines.}
    \label{fig:acgh}
\end{figure}

\textbf{Dow Jones industrial average index.}
We apply DCDP to the weekly log return of the 29 companies composing the \textit{Dow Jones Industrial Average(DJIA)} from April, 1990 to January, 2012, to detect changes in the covariance structure. We use the version of the data provided in \cite{ecp2015}. Two change points at September 22, 2008 and May 4, 2009 are detected, which correspond to the months during which the market was impacted by the financial crisis in 2008. The estimates by DCDP match well with previous research \cite{ecp2015} on this data.

To give a virtual evaluation on estimated change points, in \Cref{fig:djia} we show the estimated precision matrices on the three segments of the data split by the estimated change points.

\begin{figure}[ht]
    \centering
    \includegraphics[width = 0.48 \textwidth]{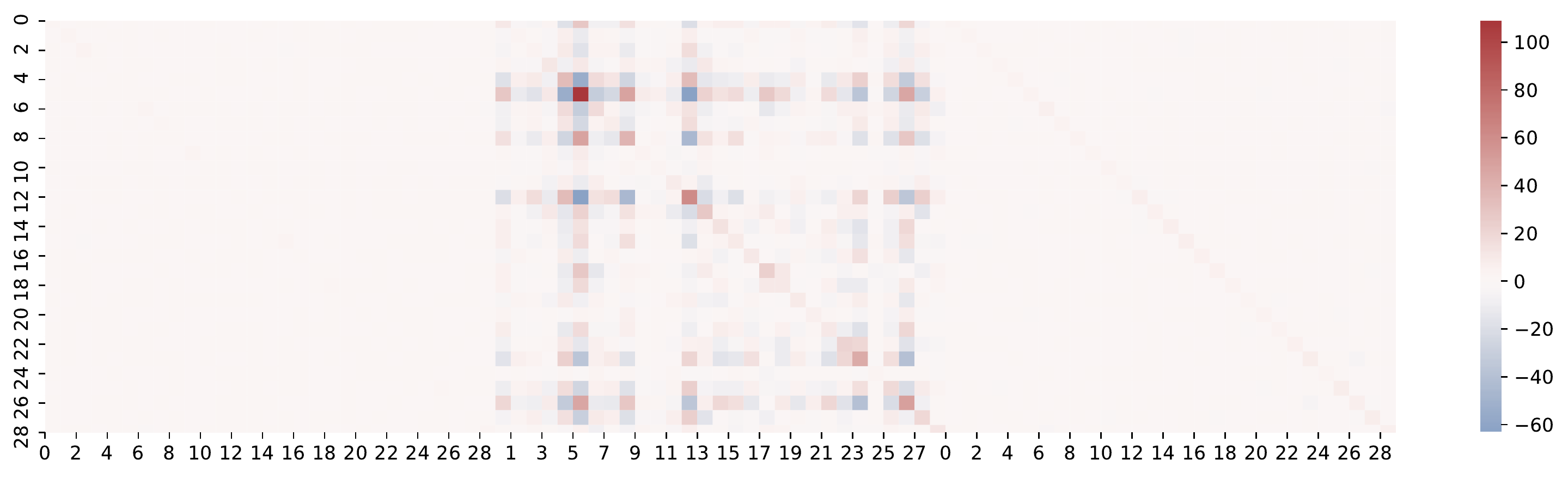}
    \caption{Estimated change points in the DJIA data.}
    \label{fig:djia}
\end{figure}

\textbf{FRED data.}
We also apply DCDP to \textit{Federal Reserve Economic Database (FRED)} data.\footnote{The dataset is publicly available at \url{https://research.stlouisfed.org/econ/mccracken/fred-databases}.} We use the subset of monthly data spanning from January 2000 to December 2019, which consists of $n = 240$ samples. The original data has 128 features. We use the R package \texttt{fbi}\citep{fbi} to transform the raw data to be stationary and remove outliers, as is suggested by the data collector \citep{fred2016}. After preprocessing, there are 118 features, including the date.

We use logarithm of the monthly growth rate of the US industrial production index (named as INDPRO in the FRED data set) as the response variable, and other 116 macroeconomic variables as predictors. Previous research \citep{wang2022testing_linear, yu2022temporal} suggests that there exist change points in the association between INDPRO and predictors.

DCDP spots a change point at January 2008, which is consistent with previous research on this data \citep{wang2022testing_linear,yu2022temporal}.

\section{Discussion}
\label{sec:discussion}
In this paper, we propose  a novel framework called DCDP   for offline change point detection that can efficiently localize multiple change points for a broad range of high-dimensional models. DCDP improves the computational efficiency of vanilla dynamic programming while preserving the accuracy of change point estimation. DCDP serves as a unified methodology for a large family of change point models and theoretical guarantees for the localization errors of DCDP under three specific models are established. Extensive numerical experiments are conducted to compare the performance of DCDP with other popular methods to support our theoretical findings.

There are two main limitations in this paper. First, although the methodology itself is model-agnostic, we only consider linear-type models in the theoretical analysis. Thus, an important future direction is to generalize the theoretical analysis to other models like non-parametric families or artificial neural networks. Moreover, in our theoretical results, the sharpest localization error rates require stronger SNR conditions, as is discussed in \Cref{sec: proof local refinement}. Since there is no existing work in the literature achieving the same error rate with weaker assumptions, weakening the SNR conditions for the sharp error rate will be another important direction for future work.

\noindent{\bf Acknowledgments}\label{subsec:acknowledgments}

We would like to thank the anonymous
reviewers for their feedback which greatly helped improve our exposition. Wanshan Li and Alessandro Rinaldo acknowledge partial support from NSF grant DMS-EPSRC 2015489.

\bibliographystyle{apalike}
\bibliography{refs}

\begin{thebibliography}{}

\bibitem[Adamczak, 2015]{hanson-wright-general}
Adamczak, R. (2015).
\newblock {A note on the Hanson-Wright inequality for random vectors with
  dependencies}.
\newblock {\em Electronic Communications in Probability}, 20(none):1 -- 13.

\bibitem[Bai and Safikhani, 2022]{unify_sinica2022}
Bai, Y. and Safikhani, A. (2022).
\newblock A unified framework for change point detection in high-dimensional
  linear models.
\newblock {\em Arxiv:2207.09007}.

\bibitem[Basu and Michailidis, 2015]{Basu2015}
Basu, S. and Michailidis, G. (2015).
\newblock {Regularized estimation in sparse high-dimensional time series
  models}.
\newblock {\em The Annals of Statistics}, 43(4):1535 -- 1567.

\bibitem[Bybee and Atchad{{\'e}}, 2018]{Bybee2018jmlr}
Bybee, L. and Atchad{{\'e}}, Y. (2018).
\newblock Change-point computation for large graphical models: A scalable
  algorithm for gaussian graphical models with change-points.
\newblock {\em Journal of Machine Learning Research}, 19(11):1--38.

\bibitem[Chao, 2019]{Gao2019}
Chao (2019).
\newblock Phase transitions in approximate ranking.
\newblock {\em arXiv:1711.11189}.

\bibitem[Eichinger and Kirch, 2018]{eichinger2018mosum}
Eichinger, B. and Kirch, C. (2018).
\newblock A mosum procedure for the estimation of multiple random change
  points.
\newblock {\em Bernoulli}, 24(1):526--564.

\bibitem[Friedrich et~al., 2008]{friedrich2008complexitypenalizedmestimation}
Friedrich, F., Kempe, A., Liebscher, V., and Winkler, G. (2008).
\newblock Complexity penalized {$M$}-estimation: fast computation.
\newblock {\em J. Comput. Graph. Statist.}, 17(1):201--224.

\bibitem[Fryzlewicz, 2014]{Fryzlewicz2014}
Fryzlewicz, P. (2014).
\newblock {Wild binary segmentation for multiple change-point detection}.
\newblock {\em The Annals of Statistics}, 42(6):2243 -- 2281.

\bibitem[Gibberd and Nelson, 2017]{Gibberd2017fusedGlasso}
Gibberd, A.~J. and Nelson, J. D.~B. (2017).
\newblock Regularized estimation of piecewise constant gaussian graphical
  models: The group-fused graphical lasso.
\newblock {\em Journal of Computational and Graphical Statistics},
  26(3):623--634.

\bibitem[Gibberd and Roy, 2017]{Gibberd2017ggm}
Gibberd, A.~J. and Roy, S. (2017).
\newblock Multiple changepoint estimation in high-dimensional gaussian
  graphical models.

\bibitem[James and Matteson, 2015]{ecp2015}
James, N.~A. and Matteson, D.~S. (2015).
\newblock ecp: An r package for nonparametric multiple change point analysis of
  multivariate data.
\newblock {\em Journal of Statistical Software}, 62(7):1–25.

\bibitem[Keshavarz et~al., 2020]{Keshavarz2020jmlr}
Keshavarz, H., Michaildiis, G., and Atchade, Y. (2020).
\newblock Sequential change-point detection in high-dimensional gaussian
  graphical models.
\newblock {\em Journal of Machine Learning Research}, 21(82):1--57.

\bibitem[Kovács et~al., 2020]{kovacs2020seeded}
Kovács, S., Li, H., Bühlmann, P., and Munk, A. (2020).
\newblock Seeded binary segmentation: A general methodology for fast and
  optimal change point detection.

\bibitem[Li et~al., 2022]{li2022cpd_btl}
Li, W., Rinaldo, A., and Wang, D. (2022).
\newblock Detecting abrupt changes in sequential pairwise comparison data.
\newblock In Oh, A.~H., Agarwal, A., Belgrave, D., and Cho, K., editors, {\em
  Advances in Neural Information Processing Systems}.

\bibitem[Lin et~al., 2017]{lin2017fused}
Lin, K., Sharpnack, J., Rinaldo, A., and Tibshirani, R.~J. (2017).
\newblock A sharp error analysis for the fused lasso, with application to
  approximate changepoint screening.
\newblock In {\em Proceedings of the 31st International Conference on Neural
  Information Processing Systems}, NIPS'17, page 6887–6896, Red Hook, NY,
  USA. Curran Associates Inc.

\bibitem[Liu et~al., 2021]{liu2021jmlr}
Liu, B., Zhang, X., and Liu, Y. (2021).
\newblock Simultaneous change point inference and structure recovery for high
  dimensional gaussian graphical models.
\newblock {\em Journal of Machine Learning Research}, 22(274):1--62.

\bibitem[Loh and Wainwright, 2012]{Loh2012}
Loh, P.-L. and Wainwright, M.~J. (2012).
\newblock {High-dimensional regression with noisy and missing data: Provable
  guarantees with nonconvexity}.
\newblock {\em The Annals of Statistics}, 40(3):1637 -- 1664.

\bibitem[Londschien et~al., 2022]{changeforest2022}
Londschien, M., Bühlmann, P., and Kovács, S. (2022).
\newblock Random forests for change point detection.
\newblock {\em Arxiv:2205.04997}.

\bibitem[Londschien et~al., 2021]{londschien2021change}
Londschien, M., Kov{\'a}cs, S., and B{\"u}hlmann, P. (2021).
\newblock Change-point detection for graphical models in the presence of
  missing values.
\newblock {\em Journal of Computational and Graphical Statistics},
  30(3):768--779.

\bibitem[Madrid~Padilla et~al., 2022]{oscar2022nonpara}
Madrid~Padilla, O.~H., Yu, Y., Wang, D., and Rinaldo, A. (2022).
\newblock Optimal nonparametric multivariate change point detection and
  localization.
\newblock {\em IEEE Transactions on Information Theory}, 68(3):1922--1944.

\bibitem[McCracken and Ng, 2016]{fred2016}
McCracken, M.~W. and Ng, S. (2016).
\newblock Fred-md: A monthly database for macroeconomic research.
\newblock {\em Journal of Business \& Economic Statistics}, 34(4):574--589.

\bibitem[Page, 1954]{page1954}
Page, E.~S. (1954).
\newblock {Continuous Inspection Schemes}.
\newblock {\em Biometrika}, 41(1-2):100--115.

\bibitem[Pilliat et~al., 2020]{verzelen2020hd_mean}
Pilliat, E., Carpentier, A., and Verzelen, N. (2020).
\newblock Optimal multiple change-point detection for high-dimensional data.
\newblock {\em ArXiv:2011.07818}.

\bibitem[Qian and Su, 2016]{qian2016fused}
Qian, J. and Su, L. (2016).
\newblock Shrinkage estimation of regression models with multiple structural
  changes.
\newblock {\em Econometric Theory}, 32(6):1376--1433.

\bibitem[Rinaldo et~al., 2021]{rinaldo2021cpd_reg_aistats}
Rinaldo, A., Wang, D., Wen, Q., Willett, R., and Yu, Y. (2021).
\newblock Localizing changes in high-dimensional regression models.
\newblock In Banerjee, A. and Fukumizu, K., editors, {\em Proceedings of The
  24th International Conference on Artificial Intelligence and Statistics},
  volume 130 of {\em Proceedings of Machine Learning Research}, pages
  2089--2097. PMLR.

\bibitem[Scott and Knott, 1974]{Scott1974}
Scott, A. and Knott, M. (1974).
\newblock A cluster analysis method for grouping means in the analysis of
  variance.
\newblock {\em Biometrics}, 30:507.

\bibitem[Stransky et~al., 2006]{acgh2006}
Stransky, N., Vallot, C., Reyal, F., Bernard-Pierrot, I., de~Medina, S. G.~D.,
  Segraves, R., de~Rycke, Y., Elvin, P., Cassidy, A., Spraggon, C., Graham, A.,
  Southgate, J., Asselain, B., Allory, Y., Abbou, C.~C., Albertson, D.~G.,
  Thiery, J.~P., Chopin, D.~K., Pinkel, D., and Radvanyi, F. (2006).
\newblock {\em Nature genetics}, 38(12):1386—1396.

\bibitem[Venkatraman, 1992]{Venkatraman1992}
Venkatraman, E.~S. (1992).
\newblock {Consistency results in multiple change-point problems}.
\newblock {\em PhD thesis, Stanford University}.

\bibitem[Vershynin, 2018]{vershynin2018high}
Vershynin, R. (2018).
\newblock {\em High-dimensional probability: An introduction with applications
  in data science}.
\newblock Cambridge University Press, Cambridge.

\bibitem[Wald, 1945]{wald1945}
Wald, A. (1945).
\newblock {Sequential Tests of Statistical Hypotheses}.
\newblock {\em The Annals of Mathematical Statistics}, 16(2):117 -- 186.

\bibitem[Wang et~al., 2020]{wang2018univariate}
Wang, D., Yu, Y., and Rinaldo, A. (2020).
\newblock {Univariate mean change point detection: Penalization, CUSUM and
  optimality}.
\newblock {\em Electronic Journal of Statistics}, 14(1):1917 -- 1961.

\bibitem[Wang et~al., 2021a]{wang2021network_cpd}
Wang, D., Yu, Y., and Rinaldo, A. (2021a).
\newblock {Optimal change point detection and localization in sparse dynamic
  networks}.
\newblock {\em The Annals of Statistics}, 49(1):203 -- 232.

\bibitem[Wang et~al., 2021b]{wang_yi_rinaldo_2021_cov}
Wang, D., Yu, Y., and Rinaldo, A. (2021b).
\newblock {Optimal covariance change point localization in high dimensions}.
\newblock {\em Bernoulli}, 27(1):554 -- 575.

\bibitem[Wang and Zhao, 2022]{wang2022testing_linear}
Wang, D. and Zhao, Z. (2022).
\newblock Optimal change-point testing for high-dimensional linear models with
  temporal dependence.
\newblock {\em Arxiv.2205.03880}.

\bibitem[Wang et~al., 2021c]{wang2021_jmlr}
Wang, D., Zhao, Z., Lin, K.~Z., and Willett, R. (2021c).
\newblock Statistically and computationally efficient change point localization
  in regression settings.
\newblock {\em Journal of Machine Learning Research}, 22(248):1--46.

\bibitem[Wang and Samworth, 2018]{wang_samworth2018}
Wang, T. and Samworth, R.~J. (2018).
\newblock High dimensional change point estimation via sparse projection.
\newblock {\em Journal of the Royal Statistical Society: Series B (Statistical
  Methodology)}, 80(1):57--83.

\bibitem[Xu et~al., 2022]{yu2022temporal}
Xu, H., Wang, D., Zhao, Z., and Yu, Y. (2022).
\newblock Change point inference in high-dimensional regression models under
  temporal dependence.
\newblock {\em ArXiv:2207.12453}.

\bibitem[Yankang (Bennie)~Chen, 2022]{fbi}
Yankang (Bennie)~Chen, Serena~Ng, J.~B. (2022).
\newblock fbi: Factor-based imputation and fred-md/qd data set.
\newblock {\em R package version 0.6.0.}

\bibitem[Yao and Au, 1989]{yao_au1989}
Yao, Y.-C. and Au, S.~T. (1989).
\newblock Least-squares estimation of a step function.
\newblock {\em Sankhyā: The Indian Journal of Statistics, Series A
  (1961-2002)}, 51(3):370--381.

\bibitem[Željko Kereta and Klock, 2021]{kereta2021ejs}
Željko Kereta and Klock, T. (2021).
\newblock {Estimating covariance and precision matrices along subspaces}.
\newblock {\em Electronic Journal of Statistics}, 15(1):554 -- 588.

\end{thebibliography}


\clearpage


\onecolumn

\pagestyle{plain}

\appendix

\begin{center}
\textbf{\LARGE Appendix}
\end{center}

The appendix contains seven parts. The first five parts present the proof of main results in \Cref{section:main} and the last two parts show some additional results of experiments on synthetic and real data. In detail,

\begin{enumerate}
    \item \Cref{sec:detail experiment} contains supplementary materials to numerical experiments in \Cref{sec:experiment}.
    \item \Cref{sec: fundamental lemma} contains key properties that make the proof of DCDP different from that of the vanilla DP. The computation complexity of the divide step is discussed in \Cref{lem: complexity divide step}.
    \item \Cref{sec: main proof mean} contains proof of \Cref{thm:DCDP mean} for the divide step under the mean model in \Cref{sec: result mean}.
    \item \Cref{sec: main proof linear} contains proof of \Cref{thm:DCDP regression} for the divide step under the linear model in \Cref{sec: result regression}.
    \item \Cref{sec: main proof covariance} contains proof of \Cref{thm:DCDP covariance main} for the divide step under the Gaussian graphical model in \Cref{sec: result covariance}.
    \item \Cref{sec: proof local refinement} contains proof of Theorem \ref{cor:mean local refinement}, \ref{cor:regression local refinement}, \ref{cor:covariance local refinement main} for the conquer step.
    
\end{enumerate}

\clearpage
\section{Additional Experiments}
\label{sec:detail experiment}
This section serves as a supplement to \Cref{sec:experiment}. In \Cref{sec:add_experiment}, we discuss the selection of $\gamma$. In \Cref{sec:experiment supp results}, we present full results of numerical experiments in \Cref{sec: experiment comparison}.

\subsection{Selection of $\gamma$}
\label{sec:add_experiment}
In the theory of DCDP, we need $\gamma = C_{\gamma}\mclB_n^{-1/2}\Delta_{\min}\kappa^2$, which involves unknown population parameter $\Delta_{\min}$ and $\kappa^2$. It is common in the change point literature and even broader literature that theoretically best tuning parameters involve unknown quantities, and a typical practical solution is to perform cross validation to select the best tuning parameter from a list of candidates.

Suppose we have data $\{\bfZ_i\}_{i\in [n]}$ with $\bfZ_i\sim \mathbb{P}_{\bftheta_i}$. Without loss of generality, suppose $n = 2m$ for some $m\in\mathbb{Z}^+$. We split the data by indices, such that data with odd indices $\{\bfZ_{2i - 1}\}_{i\in [m]}$ is the training set and data with even indices $\{\bfZ_{2i}\}_{i\in [m]}$ is the test set. This is a common way to conduct cross validation in the change point literature. Given a set of candidate parameters $G = \{(\gamma_i,\zeta_i)\}_{i\in [l]}$, for each $i\in [l]$, the CV has three steps: 
\begin{enumerate}
    \item Run DCDP on $\{\bfZ_{2i - 1}\}_{i\in \mclI_k}$ with $(\gamma_i,\zeta_i)$ to get a segmentation $\widetilde{P} = \{\mclI_k\}_{k\in[\widehat{K}+1]}$ of $[1,m]$ where $\mclI_k = [\widetilde{\eta}_{k-1}, \widetilde{\eta}_{k})$.
    \item Calculate $\{\widehat{\bftheta}_k\}_{k\in [\widehat K+1]}$ from $\{\{\bfZ_{2i - 1}\}_{i\in \mclI_k}\}_{k\in [\widehat K+1]}$ and $$R_i = \sum_{k\in [\widehat{K} + 1]}\mclF(\widehat{\bftheta}_k,\mclI_k)$$ from $\{\{\bfZ_{2i}\}_{i\in \mclI_k}\}_{k\in [\widehat K + 1]}$.
    \item Select $(\gamma_{i_{cv}},\zeta_{i_{cv}})$ with the index $i_{cv}=\argmin_{i\in[l]}R_i$.
\end{enumerate}

\subsection{Impact of SNR}
\label{sec:snr}
In \Cref{sec:experiment_time_error_divide}, we illustrate the performance of DCDP with varying SNR levels. As is shown in \Cref{fig:snr}, the localization error gets larger when $\delta$, the signal strength, becomes smaller. In this section, we show that the localization errors of DCDP for small $\delta$ are in fact reasonably good. The data generating mechanism is the same as in \Cref{sec:experiment_time_error_divide}.

We set $\Delta = 500$. In the left panel of \Cref{fig:snr 2}, we set $n = 2\Delta$ and allow the estimator to know that there is a single change point, which is the simplest setting of change point detection. In this setting, the optimal estimator is to simply pick the extreme point of the CUSUM statistic. It can be seen that with similar SNR, the localization error of DCDP under the (much more difficult) multiple change point setting is only twice of the error of the most powerful method in the simplest case. This demonstrate that DCDP performs well in low SNR scenarios.

\begin{figure}[H]
    \centering
    \includegraphics[width = 0.4 \textwidth, height = 0.28 \textwidth ]{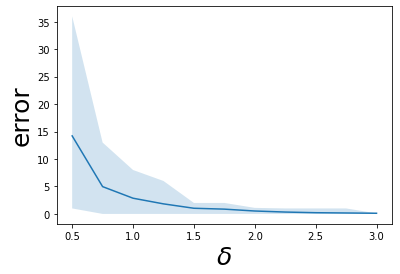}
    \includegraphics[width = 0.43 \textwidth, height = 0.28 \textwidth ]{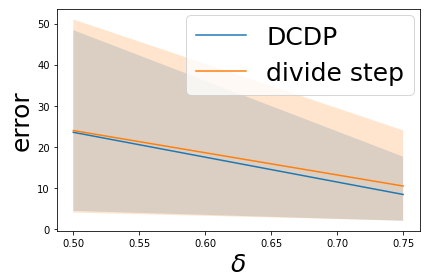}
    \caption{Left: localization error of the extreme point of the CUSUM statistic when $n = 2\Delta$ and it is known that there is only one change point; right: localization error of DCDP when $\mclQ =n$ under $n = 4\Delta$ and $\delta\in \{0.50,0.75\}$.}
    \label{fig:snr 2}
\end{figure}

In the right panel of \Cref{fig:snr 2}, we set $n = 4\Delta$ (i.e., there are 3 change points) and let $\mclQ = n$, $\delta\in \{0.50,0.75\}$. In this setting, the "divide step" corresponds to the vanilla DP and "DCDP" corresponds to vanilla DP + local refinement. Theoretically, this would lead to more accurate estimates, but with a much higher computational price. However, comparing the resulted errors with those in \Cref{fig:snr}, it can be seen that the improvement on the localization error against that of $\mclQ = 100$ is fairly small, while the actual run time is more than 200 times longer. This demonstrates that DCDP is efficient and accurate, even when the SNR is low.

\clearpage
\subsection{More results on comparisons}
\label{sec:experiment supp results}
In this section we present full results of comparisons between DCDP and other methods in \Cref{tab:mean compare_methods full}, \Cref{tab:linear compare_methods full}, and \Cref{tab:covariance compare_methods full}, as a supplement to \Cref{sec: experiment comparison}. Among all involved methods, DCDP is implemented in Python, ChangeForest is implemented in Rust and provides Python API, Inspect, Variance-Projected WBS, vanilla DP, and Block-Fused-Lasso are implemented in R based on \texttt{Rcpp}. For fair comparison, we first generate data in Python and then load the data in R for R-based methods. All experiments for DCDP and ChangeForest are run on a virtual machine of Google Colab with Intel(R) Xeon(R) CPU of 2 cores 2.30
GHz and 12GB RAM (one setting at a time). All other experiments are run on a personal computer with Intel Core i7 8850H CPU of 6 cores 2.60GHz and 64GB RAM (one setting at a time). Notice that programs implemented by \texttt{Rcpp} is usually faster than Python, and the machine to run \texttt{Rcpp}-based methods has better parameters than the virtual machine to run DCDP and ChangeForest, the comparison of execution time would not be unfair against \texttt{Rcpp}-based methods.

\Cref{tab:mean compare_methods full} shows full results of the comparison under the mean shift model.

\begin{table}[H]
 \centering
 \begin{tabular}{l l l l l l l}
  \hline
 Setting & Method & $H(\hat{\eta},\eta)$ & Time & $\hat{K}<K$ &$\hat{K}=K$ &$\hat{K}>K$
 \\
 \hline
 \multirow{4}{3.6cm}{$n = 200, p=20, K =3, \delta = 5$} 
 & DCDP & 0.00 (0.00) & 0.7s (0.2) & 0 & 100 & 0 \\
 & Inspect & 0.54 (4.46) &  0.0s (0.0) & 0 & 96 & 4 \\
 & CF &  3.59 (10.10) &  0.3s (0.0) & 0 & 84 & 16 \\
 & BFL &  42.56 (6.95) &  3.5s (0.6) & 100 & 0 & 0 \\
 \hline
 \multirow{4}{3.6cm}{$n = 200, p=20, K =3, \delta = 1$} 
 & DCDP & 0.51 (0.77) & 0.7s (0.2) & 0 & 100 & 0 \\
 & Inspect &  3.13 (5.50) &  0.0s (0.0) & 0 & 67 & 33 \\
 & CF &  4.38 (10.13) &  0.4s (0.1) & 0 & 81 & 19 \\
 & BFL &  43.30 (8.25) &  2.9s (0.6) & 100 & 0 & 0 \\
 \hline
 \multirow{4}{3.6cm}{$n = 200, p=20, K =3, \delta = 0.5$} 
 & DCDP & 8.30 (12.90) & 0.4s (0.0) & 8 & 90 & 2 \\
 & Inspect & 6.85 (7.53) &  0.0s (0.0) & 0 & 78 & 22 \\
 & CF &  7.15 (9.57) &  0.4s (0.1) & 1 & 78 & 21 \\
 & BFL &  54.48 (20.98) &  2.8s (1.1) & 100 & 0 & 0 \\
 \hline
 \multirow{4}{3.6cm}{$n = 200, p=100, K =3, \delta = 5$} 
 & DCDP & 0.0 (0.0) & 0.6s (0.0) & 0 & 100 & 0 \\
 & Inspect & 0.40 (3.50) &  0.0s (0.0) & 0 & 91 & 9 \\
 & CF & 2.85 (7.50) &  0.8s (0.2) & 0 & 85 & 15 \\
 & BFL &  47.80 (6.66) &  1.5s (0.3) & 100 & 0 & 0 \\
 \hline
 \multirow{4}{3.6cm}{$n = 200, p=100, K =3, \delta = 1$} 
 & DCDP & 0.83 (0.87) & 0.8s (0.2) & 0 & 100 & 0 \\
 & Inspect &  2.65 (5.16) &  0.0s (0.0) & 0 & 86 & 14 \\
 & CF &  3.28 (7.01) &  1.3s (0.1) & 0 & 85 & 15 \\
 & BFL &  47.59 (6.08) &  1.1s (0.2) & 100 & 0 & 0 \\
 \hline
 \multirow{4}{3.6cm}{$n = 800, p=100, K =3, \delta = 0.5$} 
 & DCDP &  9.36 (29.96) &  2.1s (0.3) & 3  &  97 & 0 \\
 & Inspect & 12.55 (22.14) &  0.1s (0.0) & 0 & 77 & 23 \\
 & CF &  14.73 (30.50) &  5.5s (0.3) & 0 & 82 & 18 \\
 & BFL &  80.10 (137.33) &  15.7s (3.8) & 28 & 71 & 1 \\
 \hline
 \end{tabular}
 \caption{Comparison of DCDP and other methods under the mean model with different simulation settings. 100 trials are conducted in each setting. For the localization error and running time (in seconds), the average over 100 trials is shown with standard error in the bracket. The three columns on the right record the number of trials in which $\hat{K}<K$, $\hat{K}=K$, and $\hat{K}>K$ respectively.}
 \label{tab:mean compare_methods full}
\end{table}

\Cref{tab:linear compare_methods full} shows full results of the comparison under the linear regression coefficient shift model.

\begin{table}[H]
 \centering
\begin{tabular}{l l l l l l l}
  \hline
 Setting & Method & $H(\hat{\eta},\eta)$ & Time & $\hat{K}<K$ &$\hat{K}=K$ &$\hat{K}>K$
 \\
 \hline
 \multirow{4}{3.6cm}{$n = 200, p=20, K =3, \delta = 5$} 
 & DCDP & 0.03 (0.17) & 5.1s (0.3) & 0 & 100 & 0 \\
 & DP & 0.04 (0.20) & 17.0s (0.5) & 0 & 100 & 0 \\
 & VPWBS & 7.69 (15.53) &  28.4s (3.5) & 1 & 71 & 28 \\
 & BFL &  84.45 (15.33) &  4.2s (0.7) & 100 & 0 & 0 \\
 \hline
 \multirow{4}{3.6cm}{$n = 200, p=20, K =3, \delta = 1$} 
 & DCDP & 0.94 (5.17) & 2.3s (0.2) & 2 & 98 & 0 \\
 & DP & 0.05 (0.22) & 12.8s (0.5) & 0 & 100 & 0 \\
 & VPWBS & 11.71 (19.82) &  30.4s (2.2) & 21 & 73 & 6 \\
 & BFL &  43.31 (8.82) &  3.1s (0.8) & 100 & 0 & 0 \\
 \hline
 \multirow{4}{3.6cm}{$n = 200, p=100, K =3, \delta = 5$} 
 & DCDP & 0.13 (0.39) & 18.4s (1.1) & 0 & 100 & 0 \\
 & DP & 0.01 (0.10) & 220.3s (16.8) & 0 & 98 & 2 \\
 & VPWBS & 15.44 (17.99) &  120.1s (13.1) & 18 & 70 & 12 \\
 & BFL &  47.84 (6.69) &  1.4s (0.2) & 100 & 0 & 0 \\
 \hline
 \multirow{4}{3.6cm}{$n = 200, p=100, K =3, \delta = 1$} 
 & DCDP & 1.45 (8.59) & 8.8s (0.7) & 2 & 98 & 0 \\
 & DP & 0.22 (2.00) & 84.4s (5.7) & 0 & 99 & 1 \\
 & VPWBS & 11.54 (11.23) &  120.4s (14.5) & 3 & 65 & 32 \\
 & BFL &  47.19 (6.48) &  1.1s (0.2) & 100 & 0 & 0 \\
 \hline
 \end{tabular}
 \caption{Comparison of DCDP and other methods under the linear model with different simulation settings. 100 trials are conducted in each setting. For the localization error and running time (in seconds), the average over 100 trials is shown with standard error in the bracket. The three columns on the right record the number of trials in which $\hat{K}<K$, $\hat{K}=K$, and $\hat{K}>K$ respectively.}
 \label{tab:linear compare_methods full}
\end{table}

\Cref{tab:covariance compare_methods full} shows full results of the comparison under the precision shift model. In \Cref{tab:covariance compare_methods full}, we didn't present the results of BFL because it produces empty set in all trials, for some unknown reason. We tried to fine tune the parameters in BFL, but didn't manage to produce nonempty sets, probably because the precision matrices under our setting are not sparse enough for BFL to perform well.

\begin{table}[H]
 \centering
\begin{tabular}{l l l l l l l}
  \hline
 Setting & Method & $H(\hat{\eta},\eta)$ & Time & $\hat{K}<K$ &$\hat{K}=K$ &$\hat{K}>K$
 \\
 \hline
 \multirow{2}{3.6cm}{$n = 2000, p=5, K =3, \delta_1 = 2,\delta_2 = 0.3$} 
 & DCDP & 5.16 (6.52) & 0.7s (0.3) & 0 & 100 & 0 \\
 & CF &  58.25 (151.74) &  1.8s (0.3) & 2 & 69 & 29 \\
 \hline
 \multirow{2}{3.6cm}{$n = 2000, p=10, K =3, \delta_1 = 5,\delta_2 = 0.3$} 
 & DCDP & 0.27 (0.49) & 0.7s (0.1) & 0 & 100 & 0 \\
 & CF &  42.5 (137.92) &  2.9s (0.2) & 0 & 84 & 16 \\
 \hline
 \multirow{2}{3.6cm}{$n = 2000, p=20, K =3, \delta_1 = 5,\delta_2 = 0.3$} 
 & DCDP & 0.03 (0.17) & 1.2s (0.2) & 0 & 100 & 0 \\
 & CF &  27.68 (97.20) &  4.8s (0.4) & 0 & 86 & 14 \\
 \hline
 \multirow{2}{3.6cm}{$n = 400, p=10, K =3, \delta_1 = 5,\delta_2 = 0.3$} 
 & DCDP & 0.42 (0.64) & 0.5s (0.0) & 0& 100& 0 \\
 & CF &  5.54 (14.71) &  0.6s (0.1) & 0 & 88 & 12 \\
 \hline
 \multirow{2}{3.6cm}{$n = 400, p=20, K =3, \delta_1 = 5,\delta_2 = 0.3$} 
 & DCDP & 0.66 (4.37) & 0.9s (0.3) & 0 & 100 & 0 \\
 & CF &  7.37 (18.76) &  1.0s (0.0) & 0 & 85 & 15 \\
 \hline
 \end{tabular}
 \caption{Comparison of DCDP and other methods under the covariance model with different simulation settings. 100 trials are conducted in each setting. For the localization error and running time (in seconds), the average over 100 trials is shown with standard error in the bracket. The three columns on the right record the number of trials in which $\hat{K}<K$, $\hat{K}=K$, and $\hat{K}>K$ respectively.}
 \label{tab:covariance compare_methods full}
\end{table}

\clearpage

\section{Fundamental lemma}
\label{sec: fundamental lemma}

In the proof of localization error of the vanilla dynamic programming, we frequently compare the goodness-of-fit function $\mclF( \widehat \theta_\mclI , \mclI)$ over an interval $\mclI = (s,e]$ with
\[
\mclF(\widehat \theta_{(s,\eta_{i + 1}] },   (s,\eta_{i + 1}]) + \cdots + \mclF(\widehat \theta_{  (\eta_{i + m},e]},(\eta_{i + m},e]) + m\gamma
\]
where $\{\eta_{i + j}\}_{j\in [m]} = \{\eta_{\ell}\}_{\ell\in [K]}\cap \mclI$ is the collection of true change points within interval $\mclI$ and $\gamma$ is the penalty tuning parameter of the DP. 

However, for DCDP, we only search over the rough grid $\{s_{i} = \lfloor \frac{i\cdot n}{\mclQ + 1}\rfloor \}_{i\in [\mclQ]}$ that may or may not  contain any true change points. Therefore, we need to   {\bf a)}  guarantee the existence of some reference points (contained in   $\{s_{i}\}_{i\in [\mclQ]}$)  that are close enough to true change points, and {\bf b)} quantify  the deviation of the goodness-of-fit function  evaluated at   the  reference points    compared to that evaluated at the true change points.

\paragraph{Reference points.} The grid is given by points $s_q = \lfloor\frac{q\cdot n}{\mclQ + 1}\rfloor$ for $q\in[\Q]$.
 Let 
$ \{ \eta_k\}_{k\in[K]}$ be the collection of change points and denote 
\begin{align*} 
  \mathcal L _k (\delta)  : = \bigg  \{    \{ s_q\}_{q\in[\Q]} \bigcap   [\eta_k -\delta ,\eta_k  ]  \not  =\emptyset \bigg  \} , \quad \text{and} \quad    \mathcal R  _k (\delta)  : = \bigg  \{    \{ s_q\}_{q\in[\Q]} \bigcap   [\eta_k ,\eta_k + \delta   ]  \not  =\emptyset \bigg  \} .
\end{align*} 
Intuitively, if $ s_q \in [\eta_k -\delta ,\eta_k  ]$ and $ s_{q'} \in [\eta_k ,\eta_k + \delta   ]$ , then $s_q, s_{q'}$ can serve as reference points of  the true change point $\eta_k$.  
Denote 
\begin{align} 
\label{eq:left and right approximation of change points}  \mathcal L(\delta)  : = \bigcap_{k=1}^{K}   \mathcal L _k  \big  (   \delta   \big  )  
\quad \text{and} \quad 
\mathcal R (\delta)  : = \bigcap_{k=1}^{K}   \mathcal R _k  \big  (  \delta  \big  ) .   
\end{align}  

Then it is straightforward to see that both events $\mathcal L(\delta)$ and $\mathcal R(\delta)$ will hold as long as $\min_{q\in [\mclQ + 1]}|s_q - s_{q-1}|<\frac{\delta}{2}$, which is guaranteed if $\mclQ > 3\frac{n}{\delta}$. For the proofs in \Cref{sec: main proof mean}, \Cref{sec: main proof linear}, and \Cref{sec: main proof covariance}, we require that $\mathcal L(\mclB_n^{-1}\Delta_{\min})$ and $\mathcal R(\mclB_n^{-1}\Delta_{\min})$ hold. Therefore, for the theoretical results in \Cref{section:main} to hold, $\mclQ$ should satisfy that
\begin{equation*}
    \mclQ > \frac{3n}{\Delta_{\min}}\mclB_n.
\end{equation*}
Since in our paper, $\{\mclB_n\}_{n\in \mclZ^+}$ is a slowly diverging sequence, we can take it as $\mclB_n = \log(n)$ and then it suffices to take $\mclQ = \frac{4n}{\Delta_{\min}}\log^2(n)$.

Under the fixed-$K$ setting of paper and when $\{\Delta_k\}_{k\in [K]}$ are of the same order, the existence of reference points will be guaranteed as long as $\mclQ > 4\log^2(n)$.

\paragraph{Goodness-of-fit.} The deviation of goodness-of-fit functions at reference points are different from the one that occurs in the proof of the vanilla DP, because the fitted parameters would have some bias since reference points may not locate at true change points. For different models, the deviation of the goodness-of-fit has different orders. We need to analyze each model separately. The deviations are described in \Cref{lem:mean one change deviation bound}, \Cref{lem:regression one change deviation bound}, and \Cref{lem:cov one change deviation bound}.

\paragraph{Complexity analysis.} In \Cref{lem: complexity divide step} we analyze the complexity of the divide step.

\bnlem[Complexity of the divide step]
\label{lem: complexity divide step}
Under all three models in \Cref{section:main}, with a memorization technique, the computation complexity of \Cref{algorithm:DP} would be $O(n\mclQ\cdot \mclC_2(p))$.
\enlem
\bprf
For generality, suppose $\{s_i\}_{i\in [\mclQ]}$ is an arbitrary grid of integers over $(0,n)$, i.e., $0<s_1<s_2<\cdots<s_{\mclQ}<n$, and denote $s_0 = 0$, $s_{\mclQ + 1} = n$, $\delta_i = s_{i} - s_{i - 1}$ for $i\in [\mclQ + 1]$. 

Under the three models in \Cref{section:main}, calculating $\widehat{\theta}_{\mclI}$ only involves summations like $\sum_{i\in \mclI}X_i$, $\sum_{i\in \mclI}X_iX_i^\top$, $\sum_{i\in \mclI}X_iy_i$. In $l$-th step ($l\geq 1$) of the inner loop of \Cref{algorithm:DP} at the right end point $s_r$, it suffices to remove $\delta_l$ terms from the summation. Thus, the complexity for the inner loop at $s_r$ would be $O(s_r\cdot \mclC_2(p))$, and the total complexity would be
$$
\sum_{r\in [\mclQ]} O(s_r\cdot \mclC_2(p)) = O(n\mclQ\cdot \mclC_2(p)).
$$
\eprf

\clearpage

\section{Mean model} 
\label{sec: main proof mean}

In this section we show the proof of \Cref{thm:DCDP mean}. Throughout this section, for any generic interval $\I\subset [1,n]$, denote $\mu^*_{\I} = \frac{1}{|\I|}\sum_{i\in \I}\mu^*_i$ and
$$ \widehat \mu_\I = \argmin_{\mu\in \mathbb{R}^p}\frac{1}{|\mclI|}\sum_{i\in \mclI}\|X_i - \mu\|_2^2 + \frac{\lambda}{\sqrt{|\mclI|}}\|\mu\|_1. $$
Also, unless specially mentioned, in this section, we set the goodness-of-fit function $\mathcal F  (\I)$ in \Cref{algorithm:DCDP} to be
\begin{equation}
    \mathcal F  (\I) := \begin{cases}
    \sum_{i \in \I } \|X_i - \widehat \mu_\I \|_2 ^2, &\text{ when } |\I|\geq C_\mclF\sigma_{\epsilon}^2\s\log(n\vee p),\\
    0, &\text{otherwise},
    \end{cases}.
\end{equation}
where $C_\mclF$ is a universal constant.

\paragraph{Assumptions.} For the ease of presentation, we combine the SNR condition we will use throughout this section and \Cref{assp: DCDP_mean main} into a single assumption.

\bnassum[Mean model] 
\label{assp: DCDP_mean}
Suppose that \Cref{assp: DCDP_mean main} holds. In addition, suppose that $ \Delta_{\min} \kappa^2  \geq \Bt   \frak{s}\log(n\vee p)$ as is assumed in \Cref{thm:DCDP mean}.
\enassum

\begin{proof}[Proof of \Cref{thm:DCDP mean}] 
By  \Cref{prop:mean local consistency}, $K \leq |\widehat{\mathcal{P}}| \leq 3K$.  This combined with \Cref{prop:mean change points partition size consistency} completes the proof.
\end{proof}

 \bnprop\label{prop:mean local consistency}
 Suppose  \Cref{assp: DCDP_mean} holds. Let 
 $\widehat { \mathcal P }  $ denote the output of  \Cref{algorithm:DCDP}.
 Then with probability  at least $1 - C n ^{-3}$, the following conditions hold.
	\begin{itemize}
		\item [(i)] For each interval  $ \I = (s, e] \in \widehat{\mathcal{P}}$    containing one and only one true 
		   change point $ \eta_k $, it must be the case that
 $$\min\{ \eta_k -s ,e-\eta_k \}  \lesssim  \sigma_{\epsilon}^2\bigg( \frac{  \s\log(n\vee p) +\gamma }{\kappa_k^2 }\bigg) +  \tr      .$$
		\item [(ii)] For each interval  $  \I = (s, e] \in \widehat{\mathcal{P}}$ containing exactly two true change points, say  $\eta_ k  < \eta_ {k+1} $, it must be the case that
			\begin{align*} 
 \eta_k -s    \lesssim      \htr    \quad \text{and} \quad 
 e-\eta_{k+1}  \lesssim   \htr     . 
 \end{align*} 
		 
\item [(iii)] No interval $\I \in \widehat{\mathcal{P}}$ contains strictly more than two true change points.

	\item [(iv)] For all consecutive intervals $ \I_1 $ and $ \I_2 $ in $\widehat{ \mathcal P}$, the interval 
		$ \I_1 \cup  \I_2 $ contains at least one true change point.
				
	 \end{itemize}
\enprop 

\bprf
The four cases are proved in \Cref{lem:mean single cp}, \Cref{lem:mean two change points}, \Cref{lem:mean three or more cp}, and \Cref{lem:mean two intervals}, respectively.
\eprf

\bnprop\label{prop:mean change points partition size consistency}
Suppose  \Cref{assp: DCDP_mean} holds. Let 
 $\widehat { \mathcal P }  $ be the output of  \Cref{algorithm:DCDP}.
 Suppose 
$ \gamma \ge C_\gamma K\tr \kappa^2 $  for sufficiently large constant $C_\gamma$. Then
 with probability  at least $1 - C n ^{-3}$,  $| \widehat { \mathcal P} | =K $.
\enprop

\begin{proof}[Proof of   \Cref{prop:mean change points partition size consistency}] 
Denote $\mathfrak{G} ^*_n = \sum_{ i =1}^n \|X_i  -  \mu^*_i\|_2^2$.  Given any collection $\{t_1, \ldots, t_m\}$, where $t_1 < \cdots < t_m$, and $t_0 = 0$, $t_{m+1} = n$, let 
\begin{equation}\label{eq: mean sn-def}
	\G _n(t_1, \ldots, t_{m}) = \sum_{k=1}^{m} \mclF(\widehat \mu_  {(t_{k}, t_{k+1}]}, (t_{k}, t_{k+1}]).
\end{equation}
For any collection of time points, when defining \eqref{eq: mean sn-def}, the time points are sorted in an increasing order.

Let $\{ \widehat \eta_{k}\}_{k=1}^{\widehat K}$ denote the change points induced by $\widehat {\mathcal P}$.  Suppose we can justify that 
	\begin{align}
		\G^*_n + K\gamma  \ge  &\G _n(s_1,\ldots,s_K)   + K\gamma - C_1 ( K +1)\sigma_{\epsilon}^2 \s  \log(n\vee p) - C_1\sum_{k\in [K]}\kappa_k^2\tr  \label{eq:K consistency step 1} \\ 
		\ge & \G_n (\widehat \eta_{1},\ldots, \widehat \eta_{\widehat K } ) +\widehat K \gamma - C_1 ( K +1)  \sigma_{\epsilon}^2 \s  \log(n\vee p) - C_1\sum_{k\in [K]}\kappa_k^2\tr  \label{eq:K consistency step 2} \\ 
		\ge &  \G _n ( \widehat \eta_{1},\ldots, \widehat \eta_{\widehat K } , \eta_1,\ldots,\eta_K ) + \widehat K \gamma    - 2C_1 ( K +1)  \sigma_{\epsilon}^2 \s  \log(n\vee p)- C_1\sum_{k\in [K]}\kappa_k^2\tr  \label{eq:K consistency step 3}
	\end{align}
	and that 
	\begin{align}\label{eq:K consistency step 4}
		\G ^*_n   -\G _n ( \widehat \eta_{1},\ldots, \widehat \eta_{\widehat K } , \eta_1,\ldots,\eta_K ) \le C_2 (K + \widehat{K} + 2) \sigma_{\epsilon}^2 \s  \log(n\vee p).
	\end{align}
	Then it must hold that $| \mathcal{P} | = K$, as otherwise if $\widehat K \ge K+1 $, then  
	\begin{align*}
		C _2 (K + \widehat{K} + 2)  \sigma_{\epsilon}^2 \s  \log(n\vee p)  & \ge   \G ^*_n  -\G _n ( \widehat \eta_{1},\ldots, \widehat \eta_{\widehat K } , \eta_1,\ldots,\eta_K ) \\
		& \ge      (\widehat K - K)\gamma  -2C_1 ( K +1)  \sigma_{\epsilon}^2 \s  \log(n\vee p) - 2C_1\sum_{k\in [K]}\kappa_k^2\tr.
	\end{align*} 
	Therefore due to the assumption that $| \widehat p|  =\widehat K\le 3K $, it holds that 
	\begin{align} \label{eq:Khat=K} 
		[C_2 (4K + 2) + 2C_1(K+1)] \sigma_{\epsilon}^2 \s  \log(n\vee p) +2C_1\sum_{k\in [K]}\kappa_k^2\tr \ge  (\widehat K - K)\gamma \geq \gamma,
	\end{align}
	Note that \eqref{eq:Khat=K} contradicts the choice of $\gamma$.

\
\\
{\bf Step 1.} Note that \eqref{eq:K consistency step 1}  is implied by 
	\begin{align}\label{eq:step 1 K consistency}  
		\left| 	\G ^*_n - \G_n(s_1,\ldots,s_K)    \right| \le  C_3(K+1) \sigma_{\epsilon}^2 \s  \log(n\vee p) + C_3\sum_{k\in [K]}\kappa_k^2\tr,
	\end{align}
	which is  an  immediate consequence  of \Cref{lem:mean one change deviation bound}. 
	\
	\\
	\\
	{\bf Step 2.} Since $\{ \widehat \eta_{k}\}_{k=1}^{\widehat K}$ are the change points induced by $\widehat {\mathcal P}$, \eqref{eq:K consistency step 2} holds because $\widehat {\mathcal P}$ is a minimizer.
\\
{\bf Step 3.} For every $ \I  =(s,e]\in \widehat p$, by \Cref{prop:mean local consistency}, we know that $\I$ contains at most two change points. We only show the proof for the two-change-points case as the other case is easier. Denote
	\[
		 \I =  (s ,\eta_{q}]\cup (\eta_{q},\eta_{q+1}] \cup  (\eta_{q+1} ,e]  = \J_1 \cup \J_2 \cup \J_{3},
	\]
where $\{ \eta_{q},\eta_{q+1}\} =\I \, \cap \, \{\eta_k\}_{k=1}^K$. 

For each $ m=1,2,3$, if $|\J_m|\geq C_\mclF\sigma_{\epsilon}^2\s\log(n\vee p)$, then by \Cref{lem:mean one change deviation bound}, it holds that
$$
\sum_{ i \in \J_m }\|y_ i - \widehat\mu_{\J_m}\|_2^2   \leq \sum_{ i \in \J_ m } \|y_ i - \mu^*_i\|_2^2 + C\sigma_{\epsilon}^2\s\log(n\vee p). 
$$
Thus, we have
\begin{equation}
    \mclF(\widehat\mu_{\J_m},\J_m)\leq \mclF(\mu^*_{\J_m},\J_m) + C\sigma_{\epsilon}^2\s\log(n\vee p).
\end{equation}
On the other hand, by \Cref{lem:mean loss deviation no change point}, we have
$$
\mclF(\widehat\mu_{\I},\J_m)   \geq \mclF(\mu^*_{\J_m},\J_m) - C\sigma_{\epsilon}^2\s\log(n\vee p). 
$$
Therefore the last two inequalities above imply that 
\begin{align}   
\mclF(\widehat \mu_\I,\I) \geq & \sum_{m=1}^{3} \mclF(\widehat \mu_\I,\J_m) \nonumber \\
\geq & \sum_{m=1}^{3}   \mclF(\widehat{\mu}_{\J_m},\J_m) - 6C\sigma_{\epsilon}^2\s\log(n\vee p).
\label{eq:K consistency step 3 inequality 3}
\end{align}
Note that \eqref{eq:K consistency step 3} is an immediate consequence of   \eqref{eq:K consistency step 3 inequality 3}.

{\bf Step 4.}
Finally, to show \eqref{eq:K consistency step 4},   let  $\widetilde { \mathcal P}$ denote the partition induced by $\{\widehat \eta_{1},\ldots, \widehat \eta_{\widehat K } , \eta_1,\ldots,\eta_K\}$. Then 
$| \widetilde { \mathcal P} | \le K + \widehat K+2   $ and that $\mu^*_i$ is unchanged in every interval $\I \in \widetilde { \mathcal P}$. 
So  \Cref{eq:K consistency step 4} is an immediate consequence of   \Cref{lem:mean one change deviation bound}.
\end{proof}

\subsection{Fundamental lemmas}

\bnlem[Deviation, mean model]
\label{lem:mean one change deviation bound}
Let $\mathcal I =(s,e] \subset (0, n] $ be any generic interval and $$\widehat \mu_\I = \argmin_{\mu}  \frac{1}{|\mclI|}\sum_{i\in \mclI}\|X_i - \mu\|_2^2 + \frac{\lambda}{\sqrt{|\mclI|}}\|\mu\|_1   .$$  

{\bf a.} If $\I$ contains no change points, then it holds that 
$$\p \bigg( \bigg|   \sum_{ i  \in \I } \|X_i   -\widehat \mu_\I   \|_2^2  - \sum_{ i   \in \I } \|X_i      - \mu^*_i   \|_2^2  \bigg|  \ge C \sigma_{\epsilon}^2\mathfrak{s}\log(n\vee p)  \bigg)  \le  (n\vee p)^{-3}. $$

{\bf b.}  Suppose that the interval $ \I $ contains one and only one change point $ \eta_k $. Denote
$$  \mathcal J = (s,\eta_k] \quad \text{and} \quad  \mathcal J' =  (\eta_k, e]  .$$   Then it holds that 
$$\p \bigg( \bigg|   \sum_{ i  \in \I } \|X_i   -\widehat \mu_\I   \|^2  - \sum_{ i   \in \I } \|X_i      - \mu^*_i    \|^2  \bigg|  \geq 2\frac{ | \J||\J'| }{ |\I| }  \kappa_k ^2  +C \sigma_{\epsilon}^2\mathfrak{s}\log(n\vee p)  \bigg)   \le  (n\vee p)^{-3}. $$
\enlem
\begin{proof} 
We show {\bf b} as {\bf a} immediately follows from {\bf b} with $ |\J'| =0$. Denote 
$$  \mathcal J = (s,\eta_k] \quad \text{and} \quad  \mathcal J' =  (\eta_k, e]  .$$ 
Denote  $\mu_\I = \frac{1}{|\I | } \sum_{i\in \I } \mu^*_i $.
The it holds that 
\begin{equation*}
    \begin{split}
        &\sum_{i\in \I} \|X_i -\widehat \mu_\I\|_2^2  - \sum_{i\in \I} \|X_i - \mu^*_i\|_2^2  \\
        =& \sum_{i \in \I } \|\widehat \mu_\I-\mu^*_i \|_2^2  -2 \sum_{i \in \I } \epsilon_i^\top  (\widehat \mu_\I  - \mu^*_i)\\
        \leq& 2\sum_{i \in \I } \|\widehat \mu_\I-\mu^*_\I \|_2^2 + 2\sum_{i \in \I } \|\mu^*_\I-\mu^*_i \|_2^2  -2 \sum_{i \in \I } \epsilon_i^\top  (\widehat \mu_\I  - \mu^*_\I) - 2 \sum_{i \in \I } \epsilon_i^\top  (\mu^*_\I  - \mu^*_i).
    \end{split}
\end{equation*} 
Observe that 
\begin{equation*}
    \p \bigg(\|\sum_{i\in \I } \epsilon_i \|_{\infty}  \geq C\sigma_{\epsilon}\sqrt {\log(n\vee p)|\I |   }\bigg) \leq (n\vee p)^{-5 } 
\end{equation*}
Suppose this good  event holds.

{\bf Step 1.} By the event and \Cref{lem:estimation_high_dim_mean}, we have
\begin{equation*}
    \begin{split}
        \sum_{i \in \I } \|\widehat \mu_\I-\mu^*_\I \|_2^2 &\leq C\sigma_{\epsilon}^2\s\log(n\vee p),\\
        |2 \sum_{i \in \I } \epsilon_i^\top  (\widehat \mu_\I  - \mu^*_\I)| &\leq |\I|\|\sum_{i\in \I}\epsilon_i\|_{\infty}\|\widehat \mu_\I-\mu^*_\I\|_1\leq C\sigma_{\epsilon}^2\s\log(n\vee p).
    \end{split}
\end{equation*}
{\bf Step 2.} Notice that 
\begin{align*} 
    \sum_{   i \in \I } \|\mu^*_\I-\mu^*_i\| ^2  =  
        & \sum_{   i \in \I }  \| \frac{ | \J | \mu^*_\J + | \J'| \mu^*_{ \J ' }}{|\I| } -\mu^*_i  \|_2 ^2\\
 =  &     \sum_{i\in \J }   \| \frac{ |\J '|  (\mu^*_\J  - \mu^*_{\J'} ) }{|\I| }  \|_2^2 + 
     \sum_{i\in \J' }   \| \frac{ |\J  |  (\mu^*_\J  - \mu^*_{\J'} ) }{|\I| }  \|_2^2
  \\ = &  \frac{|\J | | \J'| }{|\I|  }\|\mu^*_\J  - \mu^*_{\J'}\|_2 ^2 = \frac{|\J | | \J'| }{|\I|  }    \kappa_k  ^2.
    \end{align*}
Meanwhile, it holds that 
\begin{align*} 
\sum_{  i     \in \I } \epsilon_i ^\top (\mu^*_\I  - \mu^*_i  ) =
& \sum_{   i \in \I }  \epsilon_i^\top  \bigg ( \frac{ | \J | \mu^*_\J + | \J'| \mu^*_{ \J ' }}{|\I| }-\mu^*_i  \bigg )  
\\
=&  \frac{|\J'|}{|\I|}\sum_{i\in \J }   \epsilon_i^\top(    \mu^*_{\J'}-\mu^*_\J  ) + 
   \frac{|\J|}{|\I|}\sum_{i\in \J' }   \epsilon_i^\top(    \mu^*_{\J}-\mu^*_{\J'}  )
   \\
   \le &    C_2\sigma_{\epsilon}\sqrt {   \frac{|\J | | \J'| }{|\I|  }     \kappa_k  ^2  \log(n\vee p) }
   \le \frac{|\J | | \J'| }{|\I|}   \kappa_k  ^2    +  C\sigma_{\epsilon}^2\log(n\vee p)  ,
\end{align*}
where the first inequality follows from the fact that the variance is upper bounded by
$$\sum_{i\in \J }  \sigma_{\epsilon}^2 \frac{ |\J '|^2 }{|\I|^2 } \|\mu^*_\J  - \mu^*_{\J'} \|_2^2 + \sum_{i\in \J' }  \sigma_{\epsilon}^2 \frac{ |\J|^2 }{|\I|^2 } \|\mu^*_\J  - \mu^*_{\J'} \|_2^2 =    \frac{|\J | | \J'| }{|\I|  }\sigma_{\epsilon}^2 \kappa_k^2 . $$
\end{proof}

\bnlem
\label{lem:estimation_high_dim_mean}
For any interval $\mclI\subset [1, n]$ with $|\mclI|\geq C_0\mathfrak{s}\log(n\vee p)$ that   contains finitely many  change points. Let
\begin{equation*}
    \widehat {\mu}_{\mclI}\defined \argmin_{\mu\in \mathbb{R}^p}\frac{1}{|\mclI|} \sum_{i\in \I }\|X_i - \mu\|_2^2 + \frac{\lambda}{\sqrt{|\mclI|}}\|\mu\|_1,
\end{equation*}
for $\lambda = C_{\lambda}\sigma_{\epsilon}\sqrt{\log(n\vee p)}$ for sufficiently large constant $C_{\lambda}$. Then it holds with probability at least $1 - (n\vee p)^{-5}$ that
\begin{equation}
    \begin{split}
    \|\widehat{\mu}_{\I}-\mu^*_{\I}\|_{2}^{2} &\leq \frac{C\sigma_{\epsilon}^2 \s \log (n\vee p)}{\I} \\
\left\|\widehat{\mu}_{\I}-\mu^*_{\I}\right\|_{1} &\leq C \sigma_{\epsilon}\s \sqrt{\frac{\log (n\vee p)}{|\I|}} \\
\|\left(\widehat{\mu}_{\I}-\mu^*_{\I}\right)_{S^{c}}\|_{1} &\leq 3\|\left(\widehat{\mu}_{\I}-\mu^{*}_{\I}\right)_{S}\|_{1},
    \end{split}
\end{equation}
where $\mu^*_\mclI = \frac{1}{|\I|}\sum_{i\in \I}\mu^*_i$.
\enlem

\begin{proof}
By definition, we have $L(\widehat{\mu}_{\I},\I)\leq L({\mu}_{\I},\I)$, that is
\begin{align}
\nonumber
    &\sum_{i\in \I}\|Y_i - \widehat{\mu}_{\I}\|_2^2 + \lambda\sqrt{|\I|}\|\widehat{\mu}_{\I}\|_1\leq \sum_{i\in \I}\|Y_i - {\mu}^*_{\I}\|_2^2 + \lambda\sqrt{|\I|}\|{\mu}^*_{\I}\|_1\\
\nonumber
    \Rightarrow & \sum_{i\in \I}(\widehat{\mu}_{\I} - \mu^*_{\I})^\top (2Y_i - \mu^*_{\I} - \widehat{\mu}_{\I}) + \lambda\sqrt{|\I|}[\|{\mu}^*_{\I}\|_1 - \|\widehat{\mu}_{\I}\|_1]\geq 0 \\
    \nonumber
    \Rightarrow & (\widehat{\mu}_{\I} - \mu^*_{\I})^\top (\sum_{i\in \I}\epsilon_i) + 2(\widehat{\mu}_{\I} - \mu^*_{\I})^\top\sum_{i\in \I}(\mu^*_i - \mu^*_{\I}) - |\I|\sum_{i\in \I}\|\widehat{\mu}_{\I} - \mu^*_{\I}\|_2^2 + \lambda\sqrt{|\I|}[\|{\mu}^*_{\I}\|_1 - \|\widehat{\mu}_{\I}\|_1]\geq 0\\
    \Rightarrow & \|\widehat{\mu}_{\I} - \mu^*_{\I}\|_1 \|2\sum_{i\in \I}\epsilon_i\|_{\infty} + \lambda\sqrt{|\I|}[\|{\mu}^*_{\I}\|_1 - \|\widehat{\mu}_{\I}\|_1]\geq |\I|\sum_{i\in \I}\|\widehat{\mu}_{\I} - \mu^*_{\I}\|_2^2.
    \label{tmp_eq:mean_concentration}
\end{align}
By a union bound, we know that for some universal constant $C>0$, with probability at least $1 - (n\vee p)^{-5}$,
$$
\|\sum_{i\in \I}\epsilon_i\|_{\infty}\leq C\sigma_{\epsilon}\sqrt{|\I|\log(n\vee p)}\leq \frac{\lambda}{4}\sqrt{|\I|},
$$
as long as $C_{\lambda}$ is sufficiently large. Therefore, based on the sparsity assumption in \Cref{assp: DCDP_mean}, it holds that
\begin{align*}
    &\frac{\lambda}{2}\|\widehat{\mu}_{\I} - \mu^*_{\I}\|_1 + \lambda[\|{\mu}^*_{\I}\|_1 - \|\widehat{\mu}_{\I}\|_1]\geq 0 \\
    \Rightarrow & \frac{\lambda}{2}\|\widehat{\mu}_{\I} - \mu^*_{\I}\|_1 + \lambda[\|({\mu}^*_{\I})_S\|_1 - \|(\widehat{\mu}_{\I})_S\|_1]\geq \lambda \|(\widehat{\mu}_{\I})_{S^c}\|_1\\
    \Rightarrow & \frac{\lambda}{2}\|\widehat{\mu}_{\I} - \mu^*_{\I}\|_1 + \lambda\|({\mu}^*_{\I}-\widehat{\mu}_{\I})_S\|_1 \geq \lambda \|(\mu^*_{\I} - \widehat{\mu}_{\I})_{S^c}\|_1\\
    \Rightarrow & 3\|({\mu}^*_{\I}-\widehat{\mu}_{\I})_S\|_1\geq \|({\mu}^*_{\I}-\widehat{\mu}_{\I})_{S^c}\|_1.
\end{align*}
Now from \Cref{tmp_eq:mean_concentration} we can get
\begin{align*}
    |\I|\|\widehat{\mu}_{\I} - \mu^*_{\I}\|_2^2\leq &\frac{3\lambda}{2}\sqrt{|\I|}\|\widehat{\mu}_{\I} - \mu^*_{\I}\|_1\\
    \leq & \frac{12\lambda}{2}\sqrt{|\I|}\|(\widehat{\mu}_{\I} - \mu^*_{\I})_S\|_1\\
    \leq & 6\lambda \sqrt{\s} \sqrt{|\I|}\|(\widehat{\mu}_{\I} - \mu^*_{\I})_S\|_2\\
    \leq & 6\lambda \sqrt{\s} \sqrt{|\I|}\|\widehat{\mu}_{\I} - \mu^*_{\I}\|_2,
\end{align*}
which implies that
$$
\|\widehat{\mu}_{\I} - \mu^*_{\I}\|_2\leq 6C_{\lambda}\sigma_{\epsilon}\sqrt{\frac{\s\log(n\vee p)}{|\I|}}.
$$
The other inequality follows accordingly.
\end{proof}

\subsection{Technical lemmas}
Throughout this section, let $\widehat { \mathcal P }  $ denote the output of  \Cref{algorithm:DCDP}.

\bnlem[No change point]
\label{lem:mean loss deviation no change point}
Let $\I\subset [1,\ldots, n]$ be any interval that contains no change point. Then for any interval $\J \supset \I$, it holds with probability at least $1 - (n\vee p)^{-5}$ that
\begin{equation*}
    \mclF(\mu^*_{\I},\I)\leq \mclF(\widehat{\mu}_\J,\I) + C\sigma_{\epsilon}^2\s\log(n\vee p).
\end{equation*}
\enlem
\begin{proof}
{\bf Case 1}. If $|\I|< C_\mclF\sigma_{\epsilon}\s\log(n\vee p)$, then by definition, we have $\mclF(\mu^*_{\I},\I) = \mclF(\widehat \mu^*_{\J},\I) = 0$ and the inequality holds.

{\bf Case 2}. If $|\I|\geq  C_\mclF\sigma_{\epsilon}\s\log(n\vee p)$, then take difference and we can get
\begin{align*}
    &\sum_{i\in \I}\|X_i - \mu^*_i\|_2^2 - \sum_{i\in \I}\|X_i - \widehat{\mu}_\J\|_2^2 \\
    =& 2(\widehat \mu_{\J} - \mu^*_{\I})^\top\sum_{i\in \I}\epsilon_i -|\I|\|\mu^*_{\I} - \widehat \mu_{\J}\|_2^2\\
    \leq & 2(\|(\widehat \mu_{\J} - \mu^*_{\I})_S\|_1 + \|(\widehat \mu_{\J} - \mu^*_{\I})_{S^c}\|_1)\|\sum_{i\in \I}\epsilon_i\|_{\infty} -|\I|\|\mu^*_{\I} - \widehat \mu_{\J}\|_2^2\\
    \leq& c_1\|\widehat \mu_{\J} - \mu^*_{\I}\|_2\sigma_{\epsilon}\sqrt{\s|\I|\log(n\vee p)} + c_2\sigma_{\epsilon}\s\sqrt{\frac{\log(n\vee p)}{|\I|}}\cdot c_1\sigma_{\epsilon}\sqrt{|\I|\log(n\vee p)} - |\I|\|\mu^*_{\I} - \widehat \mu_{\J}\|_2^2 \\
    \leq & \frac{1}{2}|\I|\|\mu^*_{\I} - \widehat \mu_{\J}\|_2^2 + 2c_1^2\sigma_{\epsilon}^2\s\log(n\vee p) + c_2\sigma_{\epsilon}^2\s\log(n\vee p) - |\I|\|\mu^*_{\I} - \widehat \mu_{\J}\|_2^2\\
    \leq & C\sigma_{\epsilon}^2\s\log(n\vee p),
\end{align*}
where in the second inequality we use the definition of the index set $S$ and \Cref{lem:estimation_high_dim_mean}.
\end{proof}

\bnlem[Single change point]
\label{lem:mean single cp}
Suppose the good events 
$\mathcal L (\tr ) $ and  $\mathcal R (\tr) $ defined  in \Cref{eq:left and right approximation of change points} hold. 
Let   $ \I=(s,e] \in \mathcal {\widehat P}  $  be such that $\I$ contains exactly  one    change point $ \eta_k $. 
 Then  with probability  at least $1-(n\vee p)^{-3}$,   it holds that 
 $$\min\{ \eta_k -s ,e-\eta_k \}  \lesssim    \sigma_{\epsilon}^2\bigg( \frac{ \s\log(n\vee p) +\gamma }{\kappa_k^2 }\bigg)  +  \tr    .$$
  \enlem
\begin{proof}   
If either $ \eta_k -s \le \tr  $ or $e-\eta_k\le \tr$, then there is nothing to show. So assume that 
$$  \eta_k -s  > \tr \quad \text{and} \quad   e-\eta_k  > \tr . $$
By event $\mathcal R (\tr )$, there exists  $ s_ u  \in \{ s_q\}_{q=1}^\Q$ such that 
$$0\le s_u - \eta_k \le \tr. $$
 So 
$$ \eta_k \le s_ u \le e .$$
Denote 
$$ \I_ 1 = (s,s_u] \quad \text{and} \quad \I_2 = (s_u, e] .$$
Since 
$s, e,  s_u \in  \{ s_q\}_{q=1}^\Q  $,   it follows that 
\begin{align}\nonumber
\sum_{i \in \I }\|X_i - \widehat \mu_\I \|_2 ^2 \le &  \sum_{i \in \I_1 }\|X_i - \widehat \mu_{\I_1}  \|_2 ^2  + \sum_{i \in \I_2 }\|X_i - \widehat \mu_{\I_2}  \|_2 ^2   + \gamma 
\\\nonumber 
\le &  \sum_{i \in \I_1 }\|X_i -  \mu^*_i \|_2 ^2 + C_1 \big (  \sigma_{\epsilon}^2\s\log(n\vee p) + (s_u -\eta_k) \kappa_k^2 \big )  
\\ \nonumber
& + \sum_{i \in \I_1 }\|X_i -  \mu^*_i \|_2 ^2 + C_1\sigma_{\epsilon}^2\s\log(n\vee p)  + \gamma
\\ \nonumber 
=&  \sum_{i \in \I }\|X_i -  \mu^*_i  \|_2 ^2  +C_2 \big (  \sigma_{\epsilon}^2\s\log(n\vee p) + (s_u -\eta_k)   \kappa_k^2 \big )+ \gamma 
\\
\le &  \sum_{i \in \I }\|X_i -  \mu^*_i \|_2 ^2  +C_2 \big (  \sigma_{\epsilon}^2\s\log(n\vee p) +  \tr    \kappa_k^2   \big ) + \gamma  , \label{eq:one change point step 1}
\end{align}
where the first inequality follows from the fact that $ \I =(s ,e ] \in \widehat { \mathcal P} $ and so it is the local minimizer, the second inequality follows from \Cref{lem:mean one change deviation bound} {\bf a} and {\bf b} and the observation that 
$$  \eta_k -s > \tr  \ge s_u -\eta_k $$ 
Denote 
$$ \J_1  = (s,\eta_k ] \quad \text{and} \quad \J_2 = (\eta_k , e] .$$
\Cref{eq:one change point step 1}  gives 
\begin{align*}
\sum_{i \in \J_1 }\|X_i - \widehat \mu_\I \|_2 ^2 +\sum_{i \in \J_2 }\|X_i - \widehat \mu_\I \|_2 ^2  
 \le \sum_{i \in \J_1  }\|X_i -  \mu^*_{ \J_1 } \|_2 ^2   +\sum_{i \in \J_2  }\|X_i -  \mu^*_{ \J_2  } \|_2 ^2  +C_2  \big (\sigma_{\epsilon}^2\s\log(n\vee p) +   \tr \kappa_k^2  \big )  + \gamma ,
\end{align*}
which leads to 
\begin{align*}
&\sum_{i \in \J_1 }\|\widehat \mu_\I  - \mu^*_{\J_1 }\|_2^2 +\sum_{i \in \J_2 }\|\widehat \mu_\I  - \mu^*_{\J_2  } \|_2^2  
\\
 \leq  & 2 \sum_{i \in \J_1  } \epsilon_i^\top (\widehat \mu_\I  -  \mu^*_{ \J_1 } )     +2 \sum_{i \in \J_2  } \epsilon_i^\top (\widehat \mu_\I  -  \mu^*_{ \J_2 } )      +C_2  \big (\sigma_{\epsilon}^2\s\log(n\vee p) +  \kappa_k^2  \tr    \big )  + \gamma 
 \\
 \leq & 2\sigma_{\epsilon}\sum_{j = 1,2}\|\widehat \mu_\I  -  \mu^*_{ \J_j }\|_2 \sqrt{|\J_j|\log(n\vee p)}+   C_2  \big( \sigma_{\epsilon}^2\s\log(n\vee p) +  \kappa_k^2\tr\big ) + \gamma  \\
 \leq & \frac{1}{2} \sum_{j = 1,2}|\J_j|\|\widehat \mu_\I  -  \mu^*_{ \J_j }\|_2^2 + C_3  \big (\sigma_{\epsilon}^2\s\log(n\vee p) +  \kappa_k^2  \tr    \big )  + \gamma,
\end{align*} 
where the second inequality holds because the Orlicz norm $\|\cdot\|_{\psi_2}$ of $\sum_{i \in \J_1  } \epsilon_i^\top (\mu_\I  -  \mu^*_{ \J_1 } )$ is upper bounded by $|\J_1|\sigma^2_{\epsilon}\|\mu_\I  -  \mu^*_{ \J_1 }\|_2^2$.

It follows that 
$$ |\J_1 |\|  \widehat \mu_\I  - \mu^*_{\J_1 }\|_2 ^2 + |\J_2 |\|  \widehat \mu_\I  - \mu^*_{\J_2 }\|_2 ^2  =  \sum_{i \in \J_1 }\|  \widehat \mu_\I  - \mu^*_{\J_1 }\|_2 ^2 +\sum_{i \in \J_2 }\| \widehat \mu_\I  - \mu^*_{\J_2  }  \|_2^2    \le C_4\big ( \sigma_{\epsilon}^2\s\log(n\vee p) +\tr \kappa_k^2    \big )+ 2\gamma  .$$
Note that 
$$ \inf_{ a \in \mathbb R  }  |\J_1 |\|  a - \mu^*_{\J_1 }\|_2 ^2 + |\J_2 |\| a  - \mu^*_{\J_2 }\| ^2  = \kappa_k ^2  \frac{|\J_1| |\J_2|}{| \I| }   \ge \frac{ \kappa_k^2 }{2} \min\{ |\J_1| ,|\J_2| \}    . $$ 
This leads to 
$$  \frac{ \kappa_k^2 }{2}\min\{ |\J_1| ,|\J_2| \}    \le  C_4\big ( \sigma_{\epsilon}^2\s\log(n\vee p) +\tr  \kappa_k^2   + \gamma  \big )  ,$$
which is 
$$ \min\{ |\J_1| ,|\J_2| \}   \le C_5 \bigg( \frac{\sigma_{\epsilon}^2\s\log(n\vee p) + \gamma  }{\kappa_k^2 } +  \tr  \bigg) .$$
 \end{proof}

 \bnlem [Two change points]
 \label{lem:mean two change points}
Suppose the good events 
$\mathcal L (\tr ) $ and  $\mathcal R ( \tr ) $ defined  in  \Cref{eq:left and right approximation of change points} hold. 
Let   $ \I=(s,e] \in \mathcal {\widehat P}  $  be an interval that contains  exactly two  change points $ \eta_k,\eta_{k+1}  $.  Suppose in addition that 
\begin{align} \label{eq:1D two change points snr}
\Delta_{\min} \kappa^2 \ge C   \Bn^{1/2}  \big(\sigma_{\epsilon}^2\s\log(n\vee p) + \gamma)    
\end{align} 
for sufficiently large constant $C $. 
 Then  with probability  at least $1-(n\vee p)^{-3}$,   it holds that 
\begin{align*} 
 \eta_k -s    \lesssim      \htr    \quad \text{and} \quad 
 e-\eta_{k+1}  \lesssim   \htr     . 
 \end{align*} 
 \enlem 
\begin{proof} 
Since the events $\mathcal L ( \tr  ) $ and  $\mathcal R ( \tr ) $ hold, let $ s_u, s_v$ be such that 
 $\eta_k \le s_u \le s_v \le \eta_{k+1} $ and that 
 $$ 0 \le s_u-\eta_k \le   \tr ,  \quad  0\le  \eta_{k+1} - s_v \le \tr   .    $$

 \begin{center}
 \begin{tikzpicture} 
\draw[ - ] (-10,0)--(1,0);
 \node[color=black] at (-8,-0.3) {\small s};
 \draw[  (-, ultra thick, black] (-8,0) -- (-7.99,0);
  \draw[ { -]}, ultra thick, black] (-1.0002,0) -- (-1,0);

   \node[color=black] at (-7,-0.3) {\small $\eta_k$};
\draw(-7 ,0)circle [radius=2pt] ;

  \node[color=black] at (-6.5,-0.3) {\small $s_u$};
\draw plot[mark=x, mark options={color=black, scale=1.5}] coordinates {(-6.5,0)  };

  \node[color=black] at (-2.3,-0.3) {\small $\eta_{k+1}$};
\draw(-2.3 ,0)circle [radius=2pt] ;
  \node[color=black] at (-3 ,-0.3) {\small $s_v$};
\draw plot[mark=x, mark options={color=black, scale=1.5}] coordinates {(-3,0)  }; 
  
   \node[color=black] at (-1,-0.3) {\small $e$};

\end{tikzpicture}
\end{center} 

Denote 
$$ \mathcal I_ 1= (  s, s _u], \quad \I_2  =(s_u, s_v] \quad \text{and} \quad \I_3  = (s_v,e]. $$
In addition, denote 
$$ \J_1  = (s,\eta_k ], \quad \J_2=(\eta_k, \eta_{k} + \frac{ \eta_{k+1}   -\eta_k  }{2}], \quad \J_3 =  ( \eta_k+ \frac{ \eta_{k+1}   -\eta_k  }{2},\eta_{k+1 } ]  \quad \text{and} \quad \J_4 = (\eta_{k+1}  , e] .$$
Since 
$s, e,  s_u ,s_v  \in  \{ s_q\}_{q=1}^\Q  $, then it follows from the definition of $\widehat {\mclP}$ that 
\begin{align}\nonumber
& \sum_{i \in \I }\|X_i - \widehat \mu_\I \|_2^2
\\  \nonumber 
\le &  \sum_{i \in \I_1 }\|X_i - \widehat \mu_{\I_1} \|_2 ^2  + \sum_{i \in \I_2 }\|X_i - \widehat \mu_{\I_2}\|_2^2  + \sum_{i \in \I_3 }\|X_i - \widehat \mu_{\I_3}\|_2^2  +2 \gamma 
\\\nonumber 
\le &  \sum_{i \in \I_1 }\|X_i -  \mu^*_i\|_2^2 + C_1 \bigg ( \sigma_{\epsilon}^2 \s\log(n\vee p) +  \frac{|\J_1| (s_u -\eta_k ) }{ |\J_1| + (s_u -\eta_k )  }\kappa_k^2 \bigg )  +
 \sum_{i \in \I_2 }\|X_i -  \mu^*_i\|_2^2    +C_1   \sigma_{\epsilon}^2 \s\log(n\vee p)
 \\ \nonumber 
  + &  \sum_{i \in \I_3 }\|X_i -  \mu^*_i\|_2^2 + C_1 \bigg( \sigma_{\epsilon}^2 \s\log(n\vee p)+  \frac{|\J_4 | (\eta_{k+1} -s_v   )  }{ |\J_4 | + (\eta_{k+1} -s_v   )   } \kappa_{k+1} ^2   \bigg)   +2  \gamma
\\ 
\le  &  \sum_{i \in \I }\|X_i -  \mu^*_i\|_2^2  +C_1' \bigg (  \sigma_{\epsilon}^2 \s\log(n\vee p) +  \frac{|\J_1| (s_u -\eta_k ) }{ |\J_1| + (s_u -\eta_k )  }\kappa_k^2     +  \frac{|\J_4 | (\eta_{k+1} -s_v   )  }{ |\J_4 | + (\eta_{k+1} -s_v   )   }  \kappa_{k+1} ^2        \bigg )+2  \gamma 
  \label{eq:two change points step 1}
\end{align}
where the first inequality follows from the fact that $ \I =(s ,e ] \in \widehat { \mathcal P} $, the second inequality follows from \Cref{lem:mean one change deviation bound} {\bf a} and {\bf b}. 
\Cref{eq:two change points step 1}  gives 
\begin{align} \nonumber 
& \sum_{i \in \J_1 }\|X_i - \widehat \mu_\I\|_2^2 +\sum_{i \in \J_2 }\|X_i - \widehat \mu_\I\|_2^2  +\sum_{i \in \J_3 }\|X_i - \widehat \mu_\I\|_2^2 +\sum_{i \in \J_4 }\|X_i - \widehat \mu_\I\|_2^2   
\\ \nonumber  
 \le& \sum_{i \in \J_1  }\|X_i -  \mu^*_{ \J_1 }\|_2^2   +\sum_{i \in \J_2  }\|X_i -  \mu^*_{ \J_2  }\|_2^2 +\sum_{i \in \J_3  }\|X_i -  \mu^*_{ \J_3 }\|_2^2   +\sum_{i \in \J_4  }\|X_i -  \mu^*_{ \J_4  }\|_2^2  \\
 + & C_1' \bigg (\sigma_{\epsilon}^2 \s\log(n\vee p) +   \frac{|\J_1| (s_u -\eta_k ) }{ |\J_1| + (s_u -\eta_k )  }\kappa_k^2     + \frac{|\J_4 | (\eta_{k+1} -s_v   )  }{ |\J_4 | + (\eta_{k+1} -s_v   )   }  \kappa_{k+1}    ^2  \bigg )  + 2\gamma.  \label{eq:1D two change points first}
\end{align}
Note that for $\ell\in \{1,2,3,4 \}$,
\begin{align}\nonumber  
  &\sum_{i \in \J_\ell }\|X_i - \widehat \mu_\I\|_2^2  
 -   \sum_{i \in \J_\ell   }\|X_i -  \mu^*_{ \J_ \ell }\|_2^2  -\sum_{i \in \J_ \ell   }\|\widehat \mu_\I  - \mu^*_{\J_ \ell }\|_2^2 
\\
=& \nonumber 
 2 \sum_{i \in \J_ \ell   } \epsilon_i^\top ( \mu^*_{ \J_\ell  }-\widehat \mu_\I    )      
\\\nonumber   
\ge  & -  C\sigma_{\epsilon} \|\widehat \mu_\I  -  \mu^*_{ \J_\ell }\|_2 \sqrt{|\J_\ell|\log(n\vee p)}
\\
\ge  &  -\frac{1}{2}|\J_\ell| \|\widehat \mu_\I  -  \mu^*_{ \J_\ell }\|_2^2 - C'\sigma_{\epsilon}^2 \s\log(n\vee p). \nonumber  
\end{align} 
which gives 
\begin{align} \label{eq:1D two change points second}
  &\sum_{i \in \J_\ell }\|X_i - \widehat \mu_\I \|_2^2  
 -   \sum_{i \in \J_\ell   }\|X_i -  \mu^*_{ \J_ \ell }\|_2^2  \ge \frac{1}{2}\sum_{i \in \J_ \ell   }\| \widehat \mu_\I  - \mu^*_{\J_ \ell }\|_2^2
-   C_2\sigma_{\epsilon}^2 \s\log(n\vee p) .
\end{align} 
\Cref{eq:1D two change points first} and 
\Cref{eq:1D two change points second} together implies that
 
\begin{align} \label{eq:1D two change points third}   \sum_{l=1}^ 4 
   |\J_l | (\widehat \mu_\I  -  \mu^*_{ \J_\ell } ) ^2        \le C_3 \bigg ( \sigma_{\epsilon}^2 \s\log(n\vee p) +\frac{|\J_1| (s_u -\eta_k ) }{ |\J_1| + (s_u -\eta_k )  }  \kappa_k^2    +\frac{|\J_4 | (\eta_{k+1} -s_v   )  }{ |\J_4 | + (\eta_{k+1} -s_v   )   } \kappa_{k+1} ^2    \bigg ) +4 \gamma  .  
   \end{align}
Note that 
\begin{align}  \label{eq:1D two change points signal lower bound} \inf_{ a \in \mathbb R  }  |\J_1 |(  a - \mu^*_{\J_1 }) ^2 + |\J_2 |( a  - \mu^*_{\J_2 }) ^2  =&     \frac{|\J_1| |\J_2|}{ |\J_1| +  |\J_2| } \kappa_k ^2.
\end{align} 
Similarly
\begin{align} \label{eq:1D two change points signal lower bound 2} \inf_{ a \in \mathbb R  }  |\J_3 |(  a - \mu^*_{\J_3 }) ^2 + |\J_4 |( a  - \mu^*_{\J_4 }) ^2   =     \frac{|\J_3| |\J_4|}{|\J_3| + |\J_4| }       \kappa_{k+1}  ^2  ,\end{align} 
\Cref{eq:1D two change points third} together with 
\Cref{eq:1D two change points signal lower bound} and \Cref{eq:1D two change points signal lower bound 2} leads to 
\begin{align}
\label{eq:conclusion of two change points} 
  \frac{|\J_1| |\J_2|}{ |\J_1| +  |\J_2| }  \kappa_k ^2   +   \frac{|\J_3| |\J_4|}{|\J_3| + |\J_4| } \kappa_{k+1}  ^2  \le C_3 \bigg (\sigma_{\epsilon}^2 \s\log(n\vee p) +\frac{|\J_1| (s_u -\eta_k ) }{ |\J_1| + (s_u -\eta_k )  }  \kappa_k^2    +\frac{|\J_4 | (\eta_{k+1} -s_v   )  }{ |\J_4 | + (\eta_{k+1} -s_v   )   } \kappa_{k+1} ^2    \bigg ) +4 \gamma  .
  \end{align}
 Note that 
$$ 0\le s_u -\eta_k \le \tr \quad \text{and} \quad 
0 \le \eta_{k+1} -s_v \le \tr,
 $$ 
and so there are four possible cases.  

{\bf case a.} 
If 
$$|\J_1| \le \htr  \quad \text{and} \quad  
 |\J_4|   \le \htr , $$ 
 then the desired result  follows immediately. 

{\bf case b.}   $|\J_1|   >  \htr    $ and  $|\J_4|     \le \htr     $. Then since $|\J_2| \ge \Delta_{\min} /2  $, it holds that 
$$   \frac{|\J_1| |\J_2|}{ |\J_1| +  |\J_2| }  \ge  \frac{1}{2 }\min\{|\J_1| , |\J_2|  \} \ge  \frac{1}{2 }\htr.  $$
In addition,
$$\frac{|\J_1| (s_u -\eta_k ) }{ |\J_1| + (s_u -\eta_k )  }  \le s_u -\eta_k \le \tr  \quad   \text{and} \quad  \frac{|\J_4 | (\eta_{k+1} -s_v   )  }{ |\J_4 | + (\eta_{k+1} -s_v   )   }  \le \eta_{k+1} -s_v     \le \tr . 
$$
So  \Cref{eq:conclusion of two change points}  leads to 
\begin{align}\label{eq:conclusion of two change points case two} 
\frac{1}{2 }\htr   \kappa_k ^2   +   \frac{|\J_3| |\J_4|}{|\J_3| + |\J_4| } \kappa_{k+1}  ^2  \le C_3 \bigg (\sigma_{\epsilon}^2 \s\log(n\vee p) + \tr \kappa_k^2    + \tr \kappa_{k+1} ^2    \bigg ) +4 \gamma  .
  \end{align} 
Since $ \kappa_k \asymp \kappa$ and $ \kappa_{k+1}  \asymp \kappa$, \Cref{eq:conclusion of two change points case two} gives 
$$ \frac{1}{2 }\htr   \kappa  ^2  \le C_4 \bigg ( \sigma_{\epsilon}^2 \s\log(n\vee p) + \tr \kappa ^2    + \tr \kappa ^2    \bigg ) +4 \gamma   . $$
Since $\Bn $ is a diverging sequence, the above display gives 
$$ \Delta_{\min} \kappa ^2  \le C_5\Bn^{1/2 } (\log(n\vee p) + \gamma ).$$
This contradicts \Cref{eq:1D two change points snr}.

{\bf case c.}  $|\J_1|    \le   \htr    $ and 
$|\J_4|     >  \htr     $. Then the same argument  as that in  {\bf case b} leads to the same contradiction.

 {\bf case d.}     $|\J_1|   >  \htr    $ and  $|\J_4|     >  \htr     $. Then since $|\J_2|\ge \Delta_{\min} /2 , |\J_4| \ge \Delta_{\min} /2  $, it holds that 
$$   \frac{|\J_1| |\J_2|}{ |\J_1| +  |\J_2| }  \ge  \frac{1}{2 }\min\{|\J_1| , |\J_2|  \} \ge  \frac{1}{2 }\htr\quad \text{and} \quad 
 \frac{|\J_3| |\J_4|}{ |\J_3| +  |\J_4| }  \ge  \frac{1}{2 }\min\{|\J_3| , |\J_4|  \} \ge  \frac{1}{2 }\htr   $$
In addition, 
$$\frac{|\J_4 | (\eta_{k+1} -s_v   )  }{ |\J_4 | + (\eta_{k+1} -s_v   )   }  \le \eta_{k+1} -s_v     \le \tr  \quad  
\frac{|\J_1| (s_u -\eta_k ) }{ |\J_1| + (s_u -\eta_k )  }  \le s_u -\eta_k \le \tr . 
$$
So  \Cref{eq:conclusion of two change points}  leads to 
\begin{align}\label{eq:conclusion of two change points case four} 
\frac{1}{2 }\htr   \kappa_k ^2   +   \frac{1}{2 }\htr     \kappa_{k+1}  ^2  \le C_6 \bigg (\sigma_{\epsilon}^2 \s\log(n\vee p) + \tr \kappa_k^2    + \tr \kappa_{k+1} ^2    \bigg ) +4 \gamma  .
  \end{align} 
Note that $\Bn $ is a diverging sequence. So the above display gives 
$$ \Delta_{\min} \big  ( \kappa_{k}   ^2+ \kappa_{k+1}^2  \big)   \le C_ 7 \Bn^{1/2 } (\sigma_{\epsilon}^2 \s\log(n\vee p) + \gamma ) $$
Since   $ \kappa_k \asymp \kappa$ and  $ \kappa_{k+1}  \asymp \kappa$. This contradicts \Cref{eq:1D two change points snr}. 
 \end{proof}

 \bnlem[Three or more change points]
 \label{lem:mean three or more cp}
Suppose the good events 
$\mathcal L (\tr ) $ and  $\mathcal R ( \tr ) $ defined  in  \Cref{eq:left and right approximation of change points} hold. 
 Suppose in addition that 
\begin{align} \label{eq:1D three change points snr}
\Delta \kappa^2 \ge C   \big(\sigma_{\epsilon}^2 \s\log(n\vee p) + \gamma)    
\end{align} 
for sufficiently large constant $C $. 
 Then  with probability  at least $1-(n\vee p)^{-3}$,    there is no interval  $  \widehat { \mathcal P}  $  containing three or more true change points.   
 \enlem

\begin{proof} 
For  contradiction,  suppose    $ \I=(s,e] \in \mathcal {\widehat P}  $  be such that $ \{ \eta_1, \ldots, \eta_M\} \subset \I $ with $M\ge 3$. Throughout the proof, $M$ is assumed to be a parameter that can potentially change with $n$. 
Since the events $\mathcal L ( \tr  ) $ and  $\mathcal R ( \tr ) $ hold, by relabeling $\{ s_q\}_{q=1}^\Q  $ if necessary,   let $ \{ s_m\}_{m=1}^M $ be such that 
 $$ 0 \le s_m  -\eta_m  \le \tr    \quad \text{for} \quad 1 \le m \le M-1  $$ and that
 $$ 0\le \eta_M  - s_M \le \tr .$$ 
 Note that these choices ensure that  $ \{ s_m\}_{m=1}^M  \subset \I . $
 
 \begin{center}
 \begin{tikzpicture} 
\draw[ - ] (-10,0)--(1,0);
 \node[color=black] at (-8,-0.3) {\small s};
 \draw[  (-, ultra thick, black] (-8,0) -- (-7.99,0);
  \draw[ { -]}, ultra thick, black] (-1.0002,0) -- (-1,0);

   \node[color=black] at (-7,-0.3) {\small $\eta_1$};
\draw(-7 ,0)circle [radius=2pt] ;

  \node[color=black] at (-6.5,-0.3) {\small $s_1$};
\draw plot[mark=x, mark options={color=black, scale=1.5}] coordinates {(-6.5,0)  }; 

  \node[color=black] at (-5,-0.3) {\small $\eta_2$};
\draw(-5 ,0)circle [radius=2pt] ;
  \node[color=black] at (-4.5,-0.3) {\small $s_2$};
\draw plot[mark=x, mark options={color=black, scale=1.5}] coordinates {(-4.5,0)  };

  \node[color=black] at (-2.5,-0.3) {\small $\eta_3$};
\draw(-2.5 ,0)circle [radius=2pt] ;
  \node[color=black] at (-3 ,-0.3) {\small $s_3$};
\draw plot[mark=x, mark options={color=black, scale=1.5}] coordinates {(-3,0)  }; 
  
   \node[color=black] at (-1,-0.3) {\small $e$};

\end{tikzpicture}
\end{center}

{\bf Step 1.}
 Denote 
 $$ \mathcal I_ 1= (  s, s _1], \quad \I_m   =(s_{m-1} , s_m] \text{ for }  2 \le m \le M     \quad \text{and} \quad \I_{M+1}   = (s_M,e]. $$
Then since 
$ s , e,  \{ s_m \}_{m=1}^M  \subset   \{ s_q\}_{q=1}^\Q  $,   it follows that 
\begin{align}\nonumber
& \sum_{i \in \I }\|X_i - \widehat \mu_\I\|_2^2
\\  \nonumber 
\le &  \sum _{m=1}^{M+1 } \sum_{i \in \I_m}\|X_i - \widehat y_{\I_m }  \|_2^2   + M  \gamma 
\\  \label{eq: 1D three change points deviation term 1}
\le &  \sum_{i \in \I_1 }\|X_i -  \mu^*_i\|_2^2 + C_1 \bigg ( \sigma_{\epsilon}^2 \s\log(n\vee p) +  \frac{ (\eta_1 -s ) ( s_1-\eta_ 1)  }{  s_1-s  }\kappa_1^2 \bigg )
\\   \label{eq: 1D three change points deviation term 2} 
+ & 
 \sum_{m=2}^{M-1} \sum_{i \in \I_m  }\|X_i -  \mu^*_i\|_2^2    +C_1  \bigg(\sigma_{\epsilon}^2 \s\log(n\vee p)+     \frac{(\eta_m -s_{m-1} )(s_m-\eta_{m } ) }{ s _{m}-s _{m-1}  }  \kappa_m^2 \bigg ) 
 \\  \label{eq: 1D three change points deviation term 3}
  + &    C_1 \sigma_{\epsilon}^2 \s\log(n\vee p)
\\ \label{eq: 1D three change points deviation term 4}
+ & \sum_{i\in \I_{M+1}}\|X_i-\mu^*_i\|_2^2+ C_1  \bigg( \sigma_{\epsilon}^2 \s\log(n\vee p) + \frac{(\eta_M -s_{M} )(e-\eta_{M} ) }{ e-s _{M}  }  \kappa_{M} ^2 \bigg ) + M\gamma, 
 \end{align}  
 where Equations \eqref{eq: 1D three change points deviation term 1}, \eqref{eq: 1D three change points deviation term 2}
 \eqref{eq: 1D three change points deviation term 3} and \eqref{eq: 1D three change points deviation term 4} follow from \Cref{lem:mean one change deviation bound} and in particular,   \Cref{eq: 1D three change points deviation term 3} corresponds to the interval $\I _M = (s_{M-1},s_M] $ which by assumption  containing no change points. 
 Note that 
\begin{align*} & \frac{ (\eta_1 -s ) ( s_1-\eta_ 1)  }{  s_1-s  } \le s_1-\eta_1  \le \tr ,
\\ 
&   \frac{(\eta_m -s_{m-1} )(s_m-\eta_{m } ) }{ s _{m}-s _{m-1}  } \le  s_m-\eta_m  \le \tr ,   \ \text{ and } 
\\
&   \frac{(\eta_M -s_{M} )(e-\eta_{M} ) }{ e-s _{M}  } \le \eta_M-s_m  \le \tr  
 \end{align*} 
 and  that 
 $ \kappa_k \asymp \kappa $ for all $ 1\le k \le K$. Therefore 
\begin{equation}
    \label{eq:1D three change points step 1}
  \sum_{i \in \I }\|X_i - \widehat \mu_\I\|_2^2 \le  \sum _{ i \in \I } \|X_i - \mu^*_i\|_2^2 +  C_2  \bigg( M \sigma_{\epsilon}^2 \s\log(n\vee p)+   M \tr \kappa^2
+  M\gamma \bigg),
\end{equation}
where $ C_2$ is some large constant independent of $M$.

{\bf Step 2.}  Let 
$$  \J_1 =(s,  \eta_1], \ \J_m = (\eta_{m-1}, \eta_m]  \text{ for } 2 \le m \le M , \ \J_{M+1}  =(\eta_M, e]. $$
Note that $\mu^*_i$ is unchanged in any of $\{ \J_m\}_{m=0}^{M+1}$.  
So  for $ 1 \le m \le M+1  $,
\begin{align}\nonumber  
  &\sum_{i \in  \J_m  }\|X_i - \widehat \mu_\I\|_2^2  
 -   \sum_{i \in \J_  m    }\|X_i -  \mu^*_{ \J_  m }\|_2^2  -\sum_{i \in \J_  m   }\|\widehat \mu_\I  - \mu^*_{\J_  m }\|_2^2 
\\
=& \nonumber 
 2 \sum_{i \in \J_  m} \epsilon_i^\top (\mu^*_{\J_ m}-\widehat \mu_\I)   
\\\nonumber   
\ge  & - C\sigma_{\epsilon}\|\widehat \mu_\I  -  \mu^*_{ \J_ m  }\|_2\sqrt{|\J_m|\log(n\vee p)}  
\\
\ge &   -  C_3  \sigma_{\epsilon}^2 \s\log(n\vee p) - \frac{1 }{2 } |\J_  m |\|\widehat \mu_\I  -  \mu^*_{ \J_ m  }\|_2^2   \nonumber  
\end{align} 
which gives 
\begin{align} \label{eq:1D three change points second}
  &\sum_{i \in \J_m  }\|X_i - \widehat \mu_\I\|_2^2 
 -   \sum_{i \in \J_ m   }\|X_i -  \mu^*_{ \J_  m }\|_2^2  \geq \frac{1}{2}\sum_{i \in \J_  m   }\|\widehat \mu_\I  - \mu^*_{\J_  m  }\|_2 ^2 - C_3\sigma_{\epsilon}^2 \s\log(n\vee p).
\end{align} 
Therefore
\begin{align} \label{eq:1D three change points third} \sum_{ m=1}^{M+1}   |\J_m |\|\widehat \mu_\I  - \mu^*_{\J_  m  }\|_2^2 = \sum_{ m=1 }^{M+1}  \sum_{i \in \J_  m   }\|\widehat \mu_\I  - \mu^*_{\J_  m  }\|_2^2 \le  C_4  M\bigg(\sigma_{\epsilon}^2 \s\log(n\vee p)  +    \tr \kappa^2
+   \gamma \bigg),
\end{align}
where the equality follows from the fact that  $\mu^*_i$ is unchanged in any of $\{ \J_m\}_{m=0}^{M+1}$, and
 the inequality follows from \Cref{eq:1D three change points step 1} and 
\Cref{eq:1D three change points second}.
\\
\\
{\bf Step 3.}
For any $  m \in\{2, \ldots, M\}$, it holds that
\begin{align}  \label{eq:1D three change points signal lower bound} \inf_{ a \in \mathbb R  }  |\J_{m-1} |\|a - \mu^*_{\J_{m-1} }\|_2^2 + |\J_{m} |\|a  - \mu^*_{\J_{m} }\|_2^2  =&     \frac{|\J_{m-1}| |\J_m|}{ |\J_{m-1}| +  |\J_m| } \kappa_m ^2  \ge \frac{1}{2} \Delta_{\min} \kappa^2,
\end{align}  
where the last inequality follows from the assumptions that $\eta_k - \eta_{k-1}\ge \Delta_{\min}  $ and $ \kappa_k \asymp \kappa$  for all $1\le k \le K$. So
\begin{align}   \nonumber &2 \sum_{ m=1}^{M }   |\J_m |\|\widehat \mu_\I  - \mu^*_{\J_  m  }\|_2^2   
\\ 
\ge &  \nonumber    \sum_{m=2}^M  \bigg(  |\J_{m-1}  | \|\widehat \mu_\I  - \mu^*_{\J_ { m-1}  }\|_2^2   + |\J_m |\|\widehat \mu_\I  - \mu^*_{\J_  m  }\|_2^2   \bigg) 
\\ \label{eq:1D three change points signal lower bound two} 
\ge & (M-1)  \frac{ 1}{2} \Delta_{\min} \kappa^2  \ge \frac{M}{4} \Delta_{\min} \kappa^2 ,
\end{align} 
where the second inequality follows from  \Cref{eq:1D three change points signal lower bound} and the last inequality follows from $M\ge 3$. \Cref{eq:1D three change points third} and  \Cref{eq:1D three change points signal lower bound two} together imply that 
\begin{align}\label{eq:1D three change points signal lower bound three} 
 \frac{M}{4} \Delta_{\min} \kappa^2 \le 2  C_4  M\bigg(\sigma_{\epsilon}^2 \s\log(n\vee p) + \tr \kappa^2 + \gamma \bigg) .
\end{align}
Since $\Bn\to \infty $, it follows that for sufficiently large $n$,   \Cref{eq:1D three change points signal lower bound three} gives 
$$ \Delta_{\min}\kappa^2 \le C_5 \big(\sigma_{\epsilon}^2 \s\log(n\vee p) +\gamma),$$  
which contradicts \Cref{eq:1D three change points snr}.
 \end{proof}

 \bnlem[Two consecutive intervals]
 \label{lem:mean two intervals}
 Suppose $ \gamma \ge C_\gamma K\tr \kappa^2 $ for sufficiently large constant $C_\gamma $. 
With probability  at least $1-(n\vee p)^{-3}$,    there are no  two consecutive intervals $\I_1= (s,t ] \in \widehat {\mathcal  P}  $,   $ \I_2=(t, e]  \in   \widehat  {\mathcal  P}  $    such that $\I_1 \cup \I_2$ contains no change points.  
 \enlem 
 \begin{proof} 
 For contradiction, suppose that 
 $$ \I :  =\I_1\cup  \I_2 $$
 contains no change points. 
 Since 
 $ s,t,e \in \{ s_q\}_{q=1}^\Q $, it follows that
 $$ \sum_{i\in \I_1} \|X_i - \widehat \mu_{\I_1}\|^2  +\sum_{i\in \I_2} \|X_i - \widehat \mu_{\I_2}\|_2^2 +\gamma  \le \sum_{i\in \I } \|X_i - \widehat \mu_{\I }\|_2^2. $$
 By \Cref{lem:mean one change deviation bound}, it follows that 
\begin{align*} 
& \sum_{i\in \I_1} \|X_i - \mu^*_i\|_2^2 \le  C_1 \sigma_{\epsilon}^2\s\log(n\vee p) +  \sum_{i\in \I_1} \|X_i - \widehat \mu_{\I_1}\|_2^2 ,
 \\
 & \sum_{i\in \I_2} \|X_i - \mu^*_i\|_2^2 \le  C_1 \sigma_{\epsilon}^2\s\log(n\vee p) +  \sum_{i\in \I_2} \|X_i - \widehat \mu_{\I_2}\|_2^2
  \\
  & \sum_{i\in \I} \|X_i - \widehat{\mu}_{\I}\|_2^2 \le  C_1 \sigma_{\epsilon}^2\s\log(n\vee p) + \sum_{i\in \I} \|X_i - \mu_{i}^*\|_2^2.
 \end{align*} 
 So 
 $$\sum_{i\in \I_1} \|X_i - \mu^*_i\|_2^2  +\sum_{i\in \I_2}\|X_i - \mu^*_i\|_2^2 -2C_1 \sigma_{\epsilon}^2\s\log(n\vee p) +\gamma    \le \sum_{i\in \I }\|X_i - \mu^*_i\|_2^2 +C_1\sigma_{\epsilon}^2\s\log(n\vee p) .  $$
 Since $\mu^*_i$ is unchanged when $i\in \I$, it follows that 
 $$ \gamma \le 3C_1\sigma_{\epsilon}^2\s\log(n\vee p).$$
 This is a contradiction when $C_\gamma> 3C_1. $
 \end{proof}


\clearpage

\section{Linear model}
\label{sec: main proof linear}

In this section we show the proof of \Cref{thm:DCDP regression}. Throughout this section, 
for any generic interval $\I\subset [1,n]$, denote $\beta^*_{\I} = \frac{1}{|\I|}\sum_{i\in \I}\beta^*_i$ and
$$ \widehat \beta_\I = \argmin_{\beta\in \mathbb{R}^p}\frac{1}{|\mclI|}\sum_{i\in \mclI}(y_i - X_i^\top \beta)^2 + \frac{\lambda}{\sqrt{|\mclI|}}\|\beta\|_1. $$
Also, unless specified otherwise, for the output of \Cref{algorithm:DCDP}, we always set the goodness-of-fit function $\mclF(\cdot, \cdot)$ to be
\begin{equation}
\mathcal F  (\beta, \I) :  = \begin{cases} 
 \sum_{i \in \I } (y_i - X_i^\top \beta) ^2  &\text{if } |\I|\ge C_\mclF\s \log(n\vee p) ,
 \\ 
 0 &\text{otherwise,}
 \end{cases}
\end{equation} 
where $C_{\mclF}$ is a universal constant which is larger than $C_s$, the constant in sample size in \Cref{lemma:interval lasso} and \Cref{lemma:consistency}.

\paragraph{Assumptions.} For the ease of presentation, we combine the SNR condition we will use throughout this section and \Cref{assp:dcdp_linear_reg main} into a single assumption.

\bnassum[Linear model] 
\label{assp:dcdp_linear_reg}
Suppose that \Cref{assp:dcdp_linear_reg main} holds. In addition, suppose that $ \Delta_{\min} \kappa^2  \geq \Bt   \frak{s}\log(n\vee p)$ as is assumed in \Cref{thm:DCDP regression}.
\enassum

\begin{proof}[Proof of \Cref{thm:DCDP regression}] 
By  \Cref{prop:regression local consistency}, $K \leq |\widehat{\mathcal{P}}| \leq 3K$.  This combined with \Cref{prop:regression change points partition size consistency} completes the proof.
\end{proof}


 \bnprop\label{prop:regression local consistency}
 Suppose  \Cref{assp:dcdp_linear_reg} holds. Let 
 $\widehat { \mathcal P }  $ denote the output of  \Cref{algorithm:DCDP} with $\gamma =C_\gamma  K\tr\kappa^2$. Then with probability  at least $1 - n ^{-3}$, the following properties hold.
	\begin{itemize}
		\item [(i)] For each interval  $ \I = (s, e] \in \widehat{\mathcal{P}}$    containing one and only one true 
		   change point $ \eta_k $, it must be the case that
 $$\min\{ \eta_k -s ,e-\eta_k \}  \lesssim \frac{\sigma_{\epsilon}^2\vee 1  }{\kappa^2 }\bigg(  \s \log(n\vee p) +\gamma \bigg) +  \tr.$$
	\item [(ii)] For each interval  $  \I = (s, e] \in \widehat{\mathcal{P}}$ containing exactly two true change points, say  $\eta_ k  < \eta_ {k+1} $, it must be the case that
\begin{align*} 
 \eta_k -s    \lesssim   \frac{\sigma_{\epsilon}^2\vee 1  }{\kappa^2 }\bigg(  \s \log(n\vee p) +\gamma \bigg) +  \tr      \ \text{and} \ 
 e-\eta_{k+1}   \le  C   \frac{\sigma_{\epsilon}^2\vee 1  }{\kappa^2 }\bigg(  \s \log(n\vee p) +\gamma \bigg) +  \tr. 
 \end{align*} 
		 
\item [(iii)] No interval $\I \in \widehat{\mathcal{P}}$ contains strictly more than two true change points; and

	\item [(iv)] For all consecutive intervals $ \I_1 $ and $ \I_2 $ in $\widehat{ \mathcal P}$, the interval 
		$ \I_1 \cup  \I_2 $ contains at least one true change point.
				
	 \end{itemize}
\enprop

\bprf
The four cases are proved in \Cref{lemma: regression dcdp one change point}, \Cref{lemma:regression two change points}, \Cref{lem:regression three or more cp}, and \Cref{lem:regression two intervals} respectively.
\eprf

\bnprop\label{prop:regression change points partition size consistency}
Suppose  \Cref{assp:dcdp_linear_reg} holds. Let 
 $\widehat { \mathcal P }  $ denote the output of  \Cref{algorithm:DCDP}. Suppose 
$\gamma \ge  C_\gamma K \tr  \kappa^2$  for sufficiently large constant $C_\gamma$. Then
 with probability  at least $1 - C n ^{-3}$,  $| \widehat { \mathcal P} | =K $.
\enprop

\begin{proof}[Proof of   \Cref{prop:regression change points partition size consistency}] 
Denote $\mathfrak{G} ^*_n = \sum_{ i =1}^n (y_i  -  X_i^\top \beta^*_i)^2$.  Given any collection $\{t_1, \ldots, t_m\}$, where $t_1 < \cdots < t_m$, and $t_0 = 0$, $t_{m+1} = n$, let 
	\begin{equation}\label{regression eq-sn-def}
		\G _n(t_1, \ldots, t_{m}) = \sum_{k=1}^{m} \sum_{ i = t_k +1}^{t_{k+1}} \mclF(\widehat{\beta} _  {(t_{k}, t_{k+1}]}, (t_{k}, t_{k+1}]). 
	\end{equation}
For any collection of time points, when defining \eqref{regression eq-sn-def}, the time points are sorted in an increasing order.

Let $\{ \widehat \eta_{k}\}_{k=1}^{\widehat K}$ denote the change points induced by $\widehat {\mathcal P}$.  Suppose we can justify that 
	\begin{align}
		\G^*_n + K\gamma  \ge  &\G _n(s_1,\ldots,s_K)   + K\gamma - C_1 ( K +1)  \s \log(n\vee p) - C_1\sum_{k\in [K]}\kappa_k^2\tr  \label{eq:regression K consistency step 1} \\ 
		\ge & \G_n (\widehat \eta_{1},\ldots, \widehat \eta_{\widehat K } ) +\widehat K \gamma - C_1 ( K +1)  \s \log(n\vee p) - C_1\sum_{k\in [K]}\kappa_k^2\tr  \label{eq:regression K consistency step 2} \\ 
		\ge &  \G _n ( \widehat \eta_{1},\ldots, \widehat \eta_{\widehat K } , \eta_1,\ldots,\eta_K ) + \widehat K \gamma    - C_1 ( K +1)  \s \log(n\vee p)- C_1\sum_{k\in [K]}\kappa_k^2\tr  \label{eq:regression K consistency step 3}
	\end{align}
	and that 
	\begin{align}\label{eq:regression K consistency step 4}
		\G ^*_n   -\G _n ( \widehat \eta_{1},\ldots, \widehat \eta_{\widehat K } , \eta_1,\ldots,\eta_K ) \le C_2 (K + \widehat{K} + 2) \s \log(n\vee p) .
	\end{align}
	Then it must hold that $| \widehat p | = K$, as otherwise if $\widehat K \ge K+1 $, then  
	\begin{align*}
		C _2 (K + \widehat{K} + 2)  \s \log(n\vee p)  & \ge   \G ^*_n  -\G _n ( \widehat \eta_{1},\ldots, \widehat \eta_{\widehat K } , \eta_1,\ldots,\eta_K ) \\
		& \ge      (\widehat K - K)\gamma  -C_1 ( K +1)  \s \log(n\vee p) - C_1\sum_{k\in [K]}\kappa_k^2\tr.
	\end{align*} 
	Therefore due to the assumption that $| \widehat p|  =\widehat K\le 3K $, it holds that 
	\begin{align} \label{eq:regression Khat=K} 
		C_2 (4K + 2) \s \log(n\vee p)  + C_1(K+1) \s \log(n\vee p) +C_1\sum_{k\in [K]}\kappa_k^2\tr \ge  (\widehat K - K)\gamma \geq \gamma,
	\end{align}
	Note that \eqref{eq:regression Khat=K} contradicts the choice of $\gamma$.

\
\\
{\bf Step 1.} Note that \eqref{eq:regression K consistency step 1}  is implied by 
	\begin{align}\label{eq:regression step 1 K consistency}  
		\left| 	\G ^*_n   -   \G_n(s_1,\ldots,s_K)    \right| \le  C_3(K+1) \lambda^2 + C_3\sum_{k\in [K]}\kappa_k^2\tr ,
	\end{align}
	which is  an  immediate consequence  of \Cref{lem:regression one change deviation bound}. 
	\
	\\
	\\
	{\bf Step 2.} Since $\{ \widehat \eta_{k}\}_{k=1}^{\widehat K}$ are the change points induced by $\widehat {\mathcal P}$, \eqref{eq:regression K consistency step 2} holds because $\widehat p$ is a minimizer.
\\
\\
	{\bf Step 3.}
For every $ \I  =(s,e]\in \widehat p$, by \Cref{prop:regression local consistency}, we know that with probability at least $1 - (n\vee p)^{-5}$, $\I$ contains at most two change points. We only show the proof for the two-change-point case as the other case is easier. Denote
	\[
		 \I =  (s ,\eta_{q}]\cup (\eta_{q},\eta_{q+1}] \cup  (\eta_{q+1} ,e]  = \J_1 \cup \J_2 \cup \J_{3},
	\]
where $\{ \eta_{q},\eta_{q+1}\} =\I \, \cap \, \{\eta_k\}_{k=1}^K$. 

For each $m\in\{1,2,3\}$, by \Cref{lem:regression one change deviation bound}, it holds that
\[
\sum_{ i \in \J_m }(y_ i - X_i^\top \widehat{\beta} _{\J_m}    )^2   \leq \sum_{ i \in \J_ m } (y_ i  - X_i^\top \beta^*_i   )^2 + C\sigma_{\epsilon}^2\s \log(n\vee p).
\]
By \Cref{lem:regression loss deviation no change point}, we have
\[
\sum_{ i \in \J_m }(y_ i - X_i^\top\widehat {\beta} _\I    )^2   \ge \sum_{ i \in \J_ m } (y_ i  - X_i^\top \beta_i^*   )^2 - C\sigma_{\epsilon}^2\s \log(n\vee p).
\]
 Therefore the above inequality implies that 
	\begin{align} \label{eq:regression K consistency step 3 inequality 3}  \sum_{i \in \I }  (y_ i - X_i^*\widehat {\beta} _\I    )^2   \ge \sum_{m=1}^{3} \sum_{ i \in \J_m }(y_ i - X_i^\top\widehat {\beta} _{\J_m}    )^2  -C\sigma_{\epsilon}^2\s\log(n\vee p).  
	\end{align}
Note that   \eqref{eq:regression K consistency step 3} is an immediate consequence of   \eqref{eq:regression K consistency step 3 inequality 3}.

	{\bf Step 4.}
Finally, to show \eqref{eq:regression K consistency step 4},   let  $\widetilde { \mathcal P}$ denote    the partition induced by $\{\widehat \eta_{1},\ldots, \widehat \eta_{\widehat K } , \eta_1,\ldots,\eta_K\}$. Then 
$| \widetilde { \mathcal P} | \le K + \widehat K+2   $ and that $\beta^*_i$ is unchanged in every interval $\I \in \widetilde { \mathcal P}$. 
	So  \Cref{eq:regression K consistency step 4} is an immediate consequence of   \Cref{lem:regression one change deviation bound}.
\end{proof}

\subsection{Fundamental lemmas}

\bnlem \label{lem:regression one change deviation bound}
Let $\mathcal I =(s,e] $ be any generic interval.  
\\
{\bf a.} If $\I$ contains no change points and that 
$|\I| \ge C_s \s \log(n\vee p) $ where $C_s$ is the universal constant in \Cref{lemma:interval lasso}. Then it holds that 
$$\p \bigg( \bigg|   \sum_{ i  \in \I } (y_i   - X_i ^\top \widehat  \beta _\I   )^2  - \sum_{ i   \in \I } (y_i      -  X_i^\top \beta^*_\I    )^2  \bigg|  \ge C    \s  \log(n\vee p)  \bigg)  \le  n^{-4}. $$
\
\\
{\bf b.}  Suppose that   the   interval $ \I=(s,e]$ contains one  and only change point $ \eta_k $ and that 
$|\I| \ge C_s \s \log(n\vee p) $. Denote  $\widehat \mu_\I = \frac{1}{|\I | } \sum_{i\in \I } y_i $ and 
$$  \mathcal J = (s,\eta_k] \quad \text{and} \quad  \mathcal J' =  (\eta_k, e]  .$$   Then it holds that 
$$\p \bigg( \bigg|   \sum_{ i  \in \I } (y_i   - X_i^\top \widehat \beta_\I   )^2  - \sum_{ i   \in \I } (y_i      - X_i^\top \beta^* _i    )^2  \bigg|  \ge  C\bigg\{   \frac{ | \J||\J'| }{ |\I| }  \kappa_k ^2  +  \s \log(n\vee p)\bigg\}   \bigg)   \le  n^{-4}. $$
\enlem
\begin{proof} 
We show {\bf b} as {\bf a} immediatelly follows from {\bf b} with $ |\J'| =0$.
 Denote 
$$  \mathcal J = (s,\eta_k] \quad \text{and} \quad  \mathcal J' =  (\eta_k, e]  .$$ 
Denote  $  \beta _\I^*  = \frac{1}{|\I | } \sum_{i\in \I } \beta ^* _i $. Note that 
\begin{align}  \nonumber 
   \bigg|  \sum_{ i  \in \I } (y_i   - X_i^\top \widehat \beta_\I   )^2  - \sum_{ i   \in \I } (y_i      - X_i^\top \beta^* _i    )^2  \bigg|   
 = &\bigg| \sum_{i\in \I } \big\{  X_i^\top  ( \widehat \beta_\I  -   \beta^* _i )   \big\} ^2 -  2 \sum_{i\in \I }  \epsilon_i X_i^\top ( \widehat \beta_\I   -  \beta^*_i ) \bigg| 
 \\ \label{eq:regression one change point deviation bound term 1}
 \le & 2 \sum_{i\in \I } \big\{  X_i^\top  ( \widehat \beta_\I  -    \beta^*_\I  )   \big\}   ^2 
 \\ \label{eq:regression one change point deviation bound term 2}
 + & 2 \sum_{i\in \I } \big\{  X_i^\top  (    \beta^*_\I  -\beta^*_i  )    \big\} ^2  
 \\ \label{eq:regression one change point deviation bound term 3}
 +& 2 \bigg|    \sum_{i\in \I }  \epsilon_i X_i^\top ( \widehat \beta_\I   -  \beta^*_\I ) \bigg|  
 \\ \label{eq:regression one change point deviation bound term 4}
 +&  2 \bigg|    \sum_{i\in \I }  \epsilon_i X_i^\top (   \beta^*_\I -\beta^*_i  ) \bigg|  .
\end{align}
\
\\
Suppose all the good events in \Cref{lemma:interval lasso} holds. 
\\
\\
{\bf Step 1.} By \Cref{lemma:interval lasso},   $\widehat \beta_\I  -    \beta^*_\I  $ satisfies the cone condition that 
$$ \| (\widehat \beta_\I  -    \beta^*_\I  )_{S^c}\|_1 \le 3 \| (\widehat \beta_\I  -    \beta^*_\I )_S  \|_1   .$$
It follows from \Cref{corollary:restricted eigenvalues 2} that with probability at least $ 1-n^{-5}$,
\begin{align*} 
 \bigg| \frac{1}{|\I| } \sum_{i\in \I } \big\{  X_i^\top  ( \widehat \beta_\I  -    \beta^*_\I  )   \big\} ^2    - ( \widehat \beta_\I  -    \beta^*_\I    )^\top \Sigma  ( \widehat \beta_\I  -    \beta^*_\I  )  \bigg|  \le C_1  \sqrt { \frac{\s \log(n\vee p) }{|\I| }} \| \widehat \beta_\I  -    \beta^*_\I     \|_2 ^2  .
\end{align*}
The above display   gives
\begin{align*}    \bigg| \frac{1}{|\I| } \sum_{i\in \I } \big\{  X_i^\top  ( \widehat \beta_\I  -    \beta^*_\I  )    \big\}^2 \bigg| \le &  \| \Sigma\|_{\op}   \| \widehat \beta_\I  -    \beta^*_\I     \|_2 ^2   + C_1  \sqrt { \frac{\s \log(n\vee p) }{|\I| }} \| \widehat \beta_\I  -    \beta^*_\I     \|_2 ^2 
\\
\le & C_x \| \widehat \beta_\I  -    \beta^*_\I     \|_2 ^2 + C_1  \sqrt { \frac{\s \log(n\vee p) }{C_\zeta \s \log(n\vee p)  }} \| \widehat \beta_\I  -    \beta^*_\I     \|_2 ^2   
\\
\le &  \frac{ C_2\s\log(n\vee p)}{|\I| }   \end{align*} 
where the second inequality follows from the assumption that  $|\I| \ge C_\zeta \s \log(n\vee p) $ and the last inequality follows from  \Cref{eq:lemma:interval lasso term 1}  in \Cref{lemma:interval lasso}.
This gives 
$$    \bigg|   \sum_{i\in \I } \big\{  X_i^\top  ( \widehat \beta_\I  -    \beta^*_\I  )    \big\}^2 \bigg| \le  2 C_2\s\log(n\vee p) . $$
\
\\
{\bf Step 2.}  Observe that 
$ X_i^\top (\beta_\I ^* - \beta^*_i) $ is Gaussian with mean $0$ and variance 
\begin{align*} \omega_i ^2 = (\beta_\I ^* - \beta^*_i )^\top \Sigma  (\beta_\I ^* - \beta^*_i)  . 
\end{align*}
Since 
$$ \beta_\I ^* = \frac{|\J| \beta^*_\J +|\J'| \beta^*_{\J'}   }{|\I | },$$
it follows that 
$$ \omega_i ^2 = \begin{cases} 
\bigg(  \frac{ | \J'| (\beta^*_\J -\beta^*_{\J'}   ) }{ | \I| }  \bigg)^\top  \Sigma \bigg(  \frac{ | \J'| (\beta^*_\J -\beta^*_{\J'}   ) }{ | \I| }  \bigg)   \le   \frac{| \J'| ^2  \kappa_k^2 }{ |\I|^2 }   &\text{when }  i \in \J ,
\\
\bigg(  \frac{ | \J| (\beta^*_{\J'} -\beta^*_\J    ) }{ | \I| }  \bigg) ^\top  \Sigma \bigg(  \frac{ | \J'| (  \beta^*_{\J'}  -\beta^*_\J   ) }{ | \I| }  \bigg)   \le   \frac{| \J| ^2  \kappa_k^2 }{ |\I|^2 }   &\text{when }  i \in \J'.
\end{cases} $$
 Consequently,
$   \{ X_i^\top (\beta_\I ^* - \beta^*_i)\} ^2 $ is sub-Exponential with parameter $  \omega_i ^2$.  
By standard sub-Exponential tail bounds, it follows that 
\begin{align*}
 &\p \bigg(  \bigg|   \sum_{ i \in \I }\{ X_i^\top (\beta_\I ^* - \beta^*_i)\} ^2 -\E{\sum_{ i \in \I }\{ X_i^\top (\beta_\I ^* - \beta^*_i)\} ^2}    \bigg| \ge C_3 \tau    \bigg)  
 \\
 \le &  \exp\bigg (-c \min\bigg \{ \frac{  \tau^2}{\sum_{i\in \I } \omega_i ^4  }  ,  \frac{ \tau }{ \max_{i\in \I }   \omega_i ^2 } \bigg\}  \bigg)
 \\
 \le &  \exp\bigg (-c' \min\bigg \{ \frac{  \tau^2}{\sum_{i\in \I } \omega_i ^2  }  ,  \frac{ \tau }{ \max_{i\in \I }   |\omega_i | } \bigg\}  \bigg)
 \\
 \le & \exp\bigg (-c ''\min\bigg \{  \tau^2  \bigg( \frac{|\I |   }{| \J'| |\J|  } \kappa_k^{-2}  \bigg)    ,    \tau   \frac{|\I| }{ \max \{|\J|, |\J'|  \}}   \kappa_k ^{-1} \bigg\}  \bigg) ,
 \end{align*}
where the second inequality follows from the observation that 
$$  \omega_i^2     \le \kappa_k |\omega_i| \le  C_\kappa  |\omega_i| \text{ for all } i \in \I , $$
 and the last inequality follows from the observation that 
\begin{align*}
 \sum_{i\in \I} \omega_i ^2    
 \le  C_x | \J| \frac{| \J'| ^2  \kappa_k^2 }{ |\I|^2 } + C_x | \J ' | \frac{| \J| ^2  \kappa_k^2 }{ |\I|^2 }  
=   C_x \frac{| \J'| |\J| }{|\I | } \kappa_k^2 .
 \end{align*} 
 So there exists a sufficiently large constant $C_4$ such that with probability at  least $1- n^{-5}$, 
\begin{align*}
 &  \bigg|   \sum_{ i \in \I }\{ X_i^\top (\beta_\I ^* - \beta^*_i)\} ^2 -\E{\sum_{ i \in \I }\{ X_i^\top (\beta_\I ^* - \beta^*_i)\} ^2}    \bigg|
  \\ 
   \le &  C_ 4   \bigg\{\sqrt { \frac{| \J'| |\J| }{|\I | }  \log(n)  \kappa _k ^2  } + \log(n)    \frac{\max \{|\J|, |\J'|  \}  }{|\I| } \kappa _k  \bigg\}     
 \\
 \le &  C_ 4'  \bigg\{      \frac{| \J'| |\J| }{|\I | }  \kappa_k^2  + \log(n)   +  \log(n)    \frac{\max \{|\J|, |\J'|  \}  }{|\I| } \kappa _k  \bigg\}      
 \\
 \le &C_5  \bigg\{      \frac{| \J'| |\J| }{|\I | }  \kappa_k^2  + \log(n)     \bigg\}      
 \end{align*}
 where $ \kappa_k \asymp \kappa \le C_\kappa $ is used in the last inequality.
Since $ \E{\sum_{ i \in \I }\{ X_i^\top (\beta_\I ^* - \beta^*_i)\} ^2} = \sum_{i\in \I }\omega_i ^2 \le C_x \frac{| \J'| |\J| }{|\I | } \kappa_k^2$, it follows that 
$$ \p \bigg (  \bigg|   \sum_{ i \in \I }\{ X_i^\top (\beta_\I ^* - \beta^*_i)\} ^2    \bigg|  \le ( C_5 +C_x )  \frac{| \J'| |\J| }{|\I | }  \kappa_k^2  + C_5 \log(n) \bigg ) \ge 1-n^{-5}.$$
\
\\
{\bf Step 3.} For \Cref{eq:regression one change point deviation bound term 3},  it follows that with probability at least $1-2n^{-4}$
\begin{align*} 
   \frac{1}{|\I| } \sum_{i\in \I }  \epsilon_i X_i^\top ( \widehat \beta_\I   -  \beta^*_\I )
  \le  C_6 \sqrt { \frac{\log(n\vee p) }{ |\I| } } \|  \widehat \beta_\I   -  \beta^*_\I \|_1 \le C_7  \frac{\s \log(n\vee p) }{ |\I| } 
\end{align*}
where the first inequality is a consequence of \Cref{eq:independent condition 1c 1}, the second inequality follows from  \Cref{eq:lemma:interval lasso term 3} in \Cref{lemma:interval lasso}. 
\
\\
\\
{\bf Step 4.}  From {\bf Step 2}, we have  that
$ X_i^\top (\beta_\I ^* - \beta^*_i) $ is Gaussian with mean $0$ and variance 
$$ \omega_i ^2 = \begin{cases} 
\bigg(  \frac{ | \J'| (\beta^*_\J -\beta^*_{\J'}   ) }{ | \I| }  \bigg)^\top  \Sigma \bigg(  \frac{ | \J'| (\beta^*_\J -\beta^*_{\J'}   ) }{ | \I| }  \bigg)   \le   \frac{| \J'| ^2  \kappa_k^2 }{ |\I|^2 }   &\text{when }  i \in \J ,
\\
\bigg(  \frac{ | \J| (\beta^*_{\J'} -\beta^*_\J    ) }{ | \I| }  \bigg) ^\top  \Sigma \bigg(  \frac{ | \J'| (  \beta^*_{\J'}  -\beta^*_\J   ) }{ | \I| }  \bigg)   \le   \frac{| \J| ^2  \kappa_k^2 }{ |\I|^2 }   &\text{when }  i \in \J'.
\end{cases} $$
 Consequently,
$   \epsilon_i  X_i^\top (\beta_\I ^* - \beta^*_i)  $ is   centered  sub-Exponential with parameter $  \omega_i \sigma_\epsilon  $.  
By standard sub-Exponential tail bounds, it follows that 
\begin{align*}
 &\p \bigg(  \bigg|   \sum_{ i \in \I } \epsilon_i  X_i^\top (\beta_\I ^* - \beta^*_i)       \bigg| \ge C_8 \tau    \bigg)  
 \\
 \le &  \exp\bigg (-c \min\bigg \{ \frac{  \tau^2}{\sum_{i\in \I } \omega_i ^2  }  ,  \frac{ \tau }{ \max_{i\in \I }  |\omega_i| } \bigg\}  \bigg)
 \\
 \le & \exp\bigg (-c '\min\bigg \{  \tau^2  \bigg( \frac{|\I |   }{| \J'| |\J|  } \kappa_k^{-2}  \bigg)    ,    \tau   \frac{|\I| }{ \max \{|\J|, |\J'|  \}}   \kappa_k ^{-1} \bigg\}  \bigg) ,
 \end{align*}
where the last inequality follows from the observation that 
\begin{align*}
 \sum_{i\in \I} \omega_i ^2    
 \le  C_x | \J| \frac{| \J'| ^2  \kappa_k^2 }{ |\I|^2 } + C_x | \J ' | \frac{| \J| ^2  \kappa_k^2 }{ |\I|^2 }  
=   C_x \frac{| \J'| |\J| }{|\I | } \kappa_k^2 .
 \end{align*} 
 So there exists a sufficiently large constant $C_9$ such that with probability at  least $1- n^{-5}$, 
\begin{align*}
 &  \bigg|   \sum_{ i \in \I } \epsilon_i X_i^\top (\beta_\I ^* - \beta^*_i) \bigg|
  \\ 
   \le &  C_ 9   \bigg\{\sqrt { \frac{| \J'| |\J| }{|\I | }  \log(n)  \kappa _k ^2  } + \log(n)    \frac{\max \{|\J|, |\J'|  \}  }{|\I| } \kappa _k  \bigg\}     
 \\
 \le &  C_ 9'  \bigg\{      \frac{| \J'| |\J| }{|\I | }  \kappa_k^2  + \log(n)   +  \log(n)    \frac{\max \{|\J|, |\J'|  \}  }{|\I| } \kappa _k  \bigg\}      
 \\
 \le &C_9  \bigg\{      \frac{| \J'| |\J| }{|\I | }  \kappa_k^2  + \log(n)     \bigg\}      
 \end{align*}
 where $ \kappa_k \asymp \kappa \le C_\kappa $ is used in the last inequality.
\end{proof}

\bnlem \label{lemma:interval lasso} Suppose  \Cref{assp:dcdp_linear_reg} holds. Let $$ \widehat \beta _{ \mathcal I } = \arg\min_{\beta \in \mathbb R^p }  \frac{1}{ |\mathcal I |  }\sum_{i \in \mathcal I}   (y_i -X_i^\top \beta) ^2 + \lambda \|\beta\|_1$$ with $\lambda = C_\lambda (\sigma_{\epsilon} \vee 1)\sqrt { \log(n\vee p) }$ for some sufficiently large constant $C _ \lambda $. There exists a sufficiently large constant $ C_s$ such that for all   $ \mathcal I  \subset (0, n] $   such that $| \mathcal I | \ge C_s\s \log(n\vee p)$, it holds with probability at least $ 1-(n\vee p)^{-3}$ that 
\begin{align} \label{eq:lemma:interval lasso term 1}
&   \| \widehat \beta_\I -\beta^*_\I \| _2^2 \le \frac{C (\sigma_{\epsilon}^2\vee 1)\s\log(n\vee p)}{ |\I  |  } ;
 \\ \label{eq:lemma:interval lasso term 2}
 &     \| \widehat \beta_\I  -\beta^*_\I \| _1  \le     C(\sigma_{\epsilon}\vee 1) \s\sqrt {  \frac{\log(n\vee p)} {|\I  |  } }  ;
 \\ \label{eq:lemma:interval lasso term 3}
 &    \| (\widehat \beta_\I -\beta^*_\I)_{S^c} \| _1 \le  3 \| (\widehat \beta _\I  -\beta^*_\I )_{S } \| _1   .
\end{align} 
where $ \beta^*_\I  = \frac{1}{|\I|} \sum_{i\in \I } \beta^*_i $.
 \enlem 
\begin{proof}  Denote $S=\bigcup_{k=1}^K S_{\eta_k+1} $. Since $K< \infty$, 
$|S| \asymp \s.$ 
  It follows from the definition of $\widehat{\beta}_\I$ that 
	\[
	 \frac{1}{|\I | }	\sum_{ i \in I} (y_i - X_i^{\top}\widehat{\beta}_\I )^2 + \frac{ \lambda}{ \sqrt{  |\I|  }}    \|\widehat{\beta} _\I\|_1 \leq  
	 \frac{1}{|\I | } \sum_{t \in \I} (y_i - X_i^{\top}\beta^*_\I )^2 + \frac{ \lambda}{ \sqrt{  |\I|  }}  \|\beta^*_\I  \|_1.
	\]
	This gives 
	\begin{align*}
		 \frac{1}{|\I | } \sum_{i \in \I} \bigl\{X_i ^{\top}(\widehat{\beta}_\I - \beta^*_\I )\bigr\}^2 +    \frac{ 2 }{|\I | }   \sum_{ i \in \I}(y_i - X_i^{\top}\beta^*_\I)X_i^{\top}(\beta^*_\I  - \widehat{\beta}_\I )  +  \frac{ \lambda}{ \sqrt{  |\I|  }}   \bigl\|\widehat{\beta}_\I \bigr\|_1 
		\leq  \frac{ \lambda}{ \sqrt{  |\I|  }}    \bigl\|\beta^*_\I \bigr\|_1,
	\end{align*}
	and therefore
	\begin{align}
		& \frac{1}{|\I | }	  \sum_{i  \in \I} \bigl\{X_i ^{\top}(\widehat{\beta} _\I  - \beta^*_\I )\bigr\}^2 + \frac{ \lambda}{ \sqrt{  |\I|  }} \bigl\|\widehat{\beta}_\I \bigr\|_1  \nonumber \\
		 \leq  &  \frac{ 2}{|\I | }	 \sum_{i \in  \I } \epsilon_i  X_i ^{\top}(\widehat{\beta}_\I  - \beta^*_\I ) 
+ 2   (\widehat{\beta}_\I  - \beta^*_\I )^{\top}\frac{  1 }{|\I | }	 \sum_{i\in   \I} X_ i  X_i^{\top}(    \beta^*_i -\beta^*_\I   )		 
		+  \frac{ \lambda}{ \sqrt{  |\I|  }}    \bigl\|\beta^*_\I \bigr\|_1   .	 \label{eq-lem10-pf-2}
	\end{align}
To bound	
		$\left\|\sum_{ i  \in \I} X_ i  X_i ^\top (\beta^*  _\I  -\beta^*_i ) \right\|_{\infty},  
$
	note  that  for any $j \in  \{ 1, \ldots, p\}$, the $j$-th entry of 
	\\$\sum_{i \in \I} X _ i X _i^\top (\beta^*_\I   -\beta_i )$ satisfies  
	\begin{align*}
		  E \left\{\sum_{ i  \in \I}   X_i (j) X _ i ^\top (\beta^*_\I   - \beta^*_ i  )\right\} = \sum_{ i \in \I}  E \{X _ i (j  ) X_i ^\top \}\{\beta^*_\I  - \beta^*_ i \}  
		=     \mathbb{E}\{X _1( j ) X _1 ^\top \} \sum_{ i \in \I}\{\beta^*_\I    - \beta^*_ i  \} = 0.
	\end{align*}
	So   $  E\{ \sum_{ i  \in \I} X_ i  X_i ^\top (\beta^*  _\I  -\beta^*_i )\} =0 \in \mathbb R^p.$ 
By \Cref{lemma:consistency}{\bf b},
	\begin{align*}
	  \bigg|  (    \beta^*_i -\beta^*_\I   ) ^\top  \frac{  1 }{|\I | }	 \sum_{i\in   \I} X_t X_t^{\top} (\widehat{\beta}_\I  - \beta^*_\I )	 	   \bigg| \le & C_1 \big(\max_{1\le i \le n }  \|\beta^*_i -\beta^*_\I \|_2 \big) \sqrt { \frac{\log(n\vee p) }{ | \I|  }} \| \widehat{\beta}_\I  - \beta^*_\I  \|_1   
	  \\
	  \le & C_2 \sqrt { \frac{\log(n\vee p) }{ | \I| }} \| \widehat{\beta}_\I  - \beta^*_\I  \|_1  
	  \\
	  \le &  \frac{\lambda}{8\sqrt { |\I| } }   \|\widehat{\beta}_\I  - \beta^*_\I \|_1 
 	\end{align*}
 	where the second inequality follows from \Cref{lemma:beta bounded 1} and the last inequality follows from $\lambda = C_\lambda\sigma_{\epsilon} \sqrt { \log(n\vee p) }$ with sufficiently large constant $C_\lambda$.
	In addition  by \Cref{lemma:consistency}{\bf a},
	\begin{equation*}
	    \frac{ 2}{|\I | }	 \sum_{i \in  \I } \epsilon_i  X_i ^{\top}(\widehat{\beta}_\I  - \beta^*_\I )  \le C  \sigma_{\epsilon}\sqrt { \frac{ \log(n\vee p) }{ |\I| } }\|\widehat{\beta}_\I  - \beta^*_\I \|_1  \le  \frac{\lambda}{8\sqrt { |\I| } }   \|\widehat{\beta}_\I  - \beta^*_\I \|_1 .
	\end{equation*}
	So   \eqref{eq-lem10-pf-2}  gives 
	\begin{align*}
		  \frac{1}{|\I | }	  \sum_{i  \in \I} \bigl\{X_i ^{\top}(\widehat{\beta} _\I  - \beta^*_\I )\bigr\}^2 + \frac{ \lambda}{ \sqrt{  |\I|  }} \bigl\|\widehat{\beta}_\I \bigr\|_1  
		 \leq  \frac{\lambda}{4\sqrt { |\I| } }   \|\widehat{\beta}_\I  - \beta^*_\I \|_1  
		+  \frac{ \lambda}{ \sqrt{  |\I|  }}    \bigl\|\beta^*_\I \bigr\|_1   .	 
	\end{align*} 
Let $\Theta = \widehat  \beta _\I   - \beta^*_\I   $. The above inequality implies
\begin{align}
\label{eq:two sample lasso deviation 1} \frac{1}{|\I|} \sum_{i \in \I } \left( X_i^\top  \Theta \right)^2 + \frac{ \lambda}{2\sqrt{  |\I|  } }\|  (\widehat \beta _\I)   _{ S ^c}\|_1  
 \le & \frac{3\lambda}{2\sqrt{  |\I|  }  } \| ( \widehat  \beta _\I  -  \beta^*_\I  )    _{S} \| _1  ,
 \end{align}  
which also implies that 
$$ \frac{\lambda }{2}\| \Theta _{S^c}  \|_1 = \frac{ \lambda}{2 }\|   (\widehat \beta_\I)  _{ S ^c}\|_1    \le  
\frac{3\lambda}{2 } \|  ( \widehat  \beta _\I   - \beta^* _\I  ) _{S} \| _1  = \frac{3\lambda}{2 } \|  \Theta  _{S} \| _1 . $$
The above inequality and  \Cref{corollary:restricted eigenvalues 2} imply that with probability at least $1-n^{-5}$,
$$ \frac{1}{|\I| } \sum_{i \in \I }  \left( X_i^\top  \Theta \right)^2  = \Theta^\top \widehat \Sigma _\I  \Theta  
\ge 
 \Theta^\top \Sigma \Theta - C_3\sqrt{ \frac{ \s\log(n\vee p)}{ |\I| }}  \| \Theta\|_2^2   \ge  \frac{c_x}{2} \|\Theta\|_2 ^2   ,$$
 where the last inequality follows from the assumption that $| \mathcal I | \ge C_s \s \log(n\vee p)   $ for sufficiently large $C_s$.
Therefore \Cref{eq:two sample lasso deviation 1} gives
\begin{align}
\label{eq:two sample lasso deviation 2}  c'\|\Theta\|_2 ^2  + \frac{ \lambda}{2\sqrt{  |\I|  }  }\| ( \widehat \beta_\I  - \beta^*_\I ) _{ S ^c}\|_1  
 \le   \frac{3\lambda}{2\sqrt{  |\I|  }  } \| \Theta_{S} \| _1  \le \frac{3\lambda \sqrt \s }{2 \sqrt{  |\I|  } } \| \Theta  \| _2 
 \end{align}  
 and so 
 $$ \|\Theta\|_2  \le   \frac{C \lambda \sqrt  \s}{\sqrt{| \I |} } . $$ 
 The above display   gives  
 $$  \| \Theta_{S} \| _1 \le  \sqrt {\s}  \| \Theta_{S} \| _2 \le   \frac{C\lambda \s}{\sqrt{|\I| }}. $$ 
 Since 
  $ \| \Theta_{S^c } \| _1 \le 3 \| \Theta_{S} \| _1 ,$ 
it also holds that 
$$\| \Theta  \| _1 = \| \Theta_{S  } \| _1 +\| \Theta_{S^c } \| _1  \le 4 \| \Theta_{S} \| _1 \le  \frac{4C\lambda \s}{\sqrt{|\I|} } .$$
 \end{proof}

\subsection{Technical lemmas}

Throughout this section, let $\widehat { \mathcal P }  $ denote the output of  \Cref{algorithm:DCDP}.

 \bnlem[No change point]
 \label{lem:regression loss deviation no change point}
Let $\I\subset [1,T]$ be any interval that contains no change point. Then under \Cref{assp:dcdp_linear_reg}, for any interval $\J \supset \I$, it holds with probability at least $1 - (n\vee p)^{-5}$ that
\begin{equation*}
    \mclF(\beta^*_{\I},\I)\leq \mclF(\widehat{\beta}_\J,\I) + C(\sigma_{\epsilon}^2\vee 1)\s\log(n\vee p).
\end{equation*}
 \enlem
\begin{proof} 
\noindent \textbf{Case 1.} If $ |\I| < C_{\mclF} \s \log(n p)$, then by the definition of $\mclF(\beta, \mclI)$, we have $\mclF(\beta^*_{\I},\I)= \mclF(\widehat{\beta}_\J,\I) =0$ and the inequality holds automatically.
 
\noindent \textbf{Case 2.} If 
	\begin{equation}\label{eq-lem16-i-cond} 
		|\I| \geq  C_{\mclF} \s \log(n p),
	\end{equation}
	then letting $\delta_\I = \beta^*_\I  -\widehat \beta_\J$ and consider the high-probability event given in \Cref{lem:restricted eigenvalue}, we have
	\begin{align}
		& \sqrt{\sum_{t \in \I} (X_t^{\top} \delta_\I)^2} \geq {c_1'\sqrt{|\I|}} \|\delta_\I\|_2 - c_2'\sqrt{\log(p)} \|\delta_\I\|_1 \nonumber \\
		= & {c_1'\sqrt{|\I|}} \|\delta_\I\|_2 - c_2'\sqrt{\log(p)} \|(\delta_I)_{S}\|_1 - c_2'\sqrt{\log(p)} \|(\delta_\I)_{S^c}\|_1 \nonumber \\
		\geq & c_1'\sqrt{|\I|} \|\delta_\I\|_2 - c_2'\sqrt{\s\log(p)} \|\delta_\I\|_2 - c_2'\sqrt{\log(p)} \|(\delta_\I)_{S^c}\|_1 \nonumber \\
		\geq & \frac{c_1'}{2}\sqrt{|\I|} \|\delta_\I\|_2 - c_2'\sqrt{\log(p)} \|(\widehat{\beta}_\J)_{S^c}\|_1 \geq c_1\sqrt{|\I|} \|\delta_\I\|_2 - c_2\sqrt{\log(p)} \frac{\s \lambda}{\sqrt{|\I|}}, \label{eq-lem16-pf-1}
	\end{align} 
	where the last inequality follows from \Cref{lemma:interval lasso} and the assumption that $(\beta^*_t)_i = 0$, for all $t\in [T]$ and $i\in S^c$. Then by the fact that $(a - b)^2\geq \frac{1}{2}a^2 - b^2$ for all $a,b\in \mathbb{R}$, it holds that
	\begin{equation}
	    \sum_{t \in \I} (X_t^{\top} \delta_\I)^2\geq \frac{c_1^2}{2}{|\I|} \|\delta_\I\|_2^2 - \frac{c_2^2  \lambda^2\s^2\log(p)}{|\I|}.
	\end{equation}
	Notice that
	\begin{align*}
	    & \sum_{t \in \I} (y_t - X_t^{\top} \beta^*_\I)^2 - \sum_{t \in \I} (y_t - X_t^{\top} \widehat{\beta}_\J)^2  = 2 \sum_{t \in \I} \epsilon_t X_t^{\top}\delta_\I  - \sum_{t \in \I} (X_t^{\top}\delta_\I)^2 \\
	\leq & 2\|\sum_{t \in \I }  X_t \epsilon_t \|_{\infty }  \left( \sqrt {\s} \| (\delta_\I)_{S} \|_2  + \| (\widehat \beta_\J)_{S^c} \|_1  \right)- \sum_{t \in \I} (X_t^{\top}\delta_\I)^2.
	\end{align*}
	Since for each $t$, $\epsilon_t$ is subgaussian with $\|\epsilon_t\|_{\psi_2}\leq \sigma_{\epsilon}$ and for each $i\in [p]$, $(X_t)_i$ is subgaussian with $\|(X_t)_i\|_{\psi_2}\leq C_x$, we know that $(X_t)_i\epsilon_t$ is subexponential with $\|(X_t)_i\epsilon_t\|_{\psi_1}\leq C_x\sigma_{\epsilon}$. Therefore, by Bernstein's inequality (see, e.g., Theorem 2.8.1 in \cite{vershynin2018high}) and a union bound, for $\forall u\geq 0$ it holds that
	\begin{equation*}
	    \mathbb{P}(\|\sum_{t \in \I }  X_t \epsilon_t \|_{\infty }> u)\leq 2p\exp(-c\min\{\frac{u^2}{|\I|C_x^2\sigma_{\epsilon}^2}, \frac{u}{C_x\sigma_{\epsilon}}\}).
	\end{equation*}
	Take $u = cC_x\sigma_{\epsilon}\sqrt{|\I|\log(n\vee p)}$, then by the fact that $|\I|\geq C_{\mclF}\s\log(n\vee p)$, it follows that with probability at least $1 - (n\vee p)^{-7}$,
	\begin{equation*}
	    \|\sum_{t \in \I }  X_t \epsilon_t \|_{\infty } \leq CC_x\sigma_{\epsilon}\sqrt{|\I|\log(n\vee p)}\leq \lambda \sqrt{|\I|},
	\end{equation*}
	where we use the fact that $\lambda=C_{\lambda}(\sigma_{\epsilon}\vee 1)\sqrt{\log(n\vee p)}$. Therefore, we have
	\begin{align*}
	& \sum_{t \in \I} (y_t - X_t^{\top} \beta^*_\I)^2 - \sum_{t \in \I} (y_t - X_t^{\top} \widehat{\beta}_\J)^2 \\
	\leq & 2\lambda\sqrt{|\I|\s}\|\delta_\I\|_2 + 2\lambda\sqrt{|\I|}\cdot \frac{\lambda \s}{ \sqrt{|\I|}} - \frac{c_1^2 |\I|}{2}\|\delta_\I\|_2^2 + \frac{c_2^2 \lambda^2 \s^2\log(p)}{|\I|}\\  
	\leq & 2\lambda\sqrt{|\I|\s}\|\delta_\I\|_2 + {2\lambda^2 \s} - \frac{c_1^2 |\I|}{2}\|\delta_\I\|_2^2 + \frac{c_2^2 \lambda^2 \s^2\log(p)}{|\I|} \\
	\leq & \frac{4}{c_1^2}\lambda^2\s + \frac{c_1^2}{4}|\I| \|\delta_\I\|^2_2 + {2\lambda^2 \s} - \frac{c_1^2}{2}|\I|\|\delta_\I\|^2_2 + \frac{c_2^2 \lambda^2 \s^2\log(p)}{C_{\mclF}\s\log(n\vee p)} \\
	\leq & {c_3\lambda^2\s} + {2\lambda^2 \s} + \frac{c_2^2 \lambda^2 \s^2\log(p)}{C_{\mclF}\s\log(n\vee p)}\\
	\leq & c_4\lambda^2 \s.
	\end{align*}
	where the third inequality follows from $2ab \leq a^2 + b^2$. 

\end{proof}

 \bnlem[Single change point]
 \label{lemma: regression dcdp one change point}
Suppose the good events 
$\mathcal L (\tr ) $ and  $\mathcal R (\tr) $ defined  in \Cref{eq:left and right approximation of change points} hold. 
Let   $ \I=(s,e] \in \mathcal {\widehat P}  $  be such that $\I$ contains exactly  one  true  change point $ \eta_k $. 
Suppose $\gamma \geq C_\gamma K\tr \kappa^2   $.  Then  with probability  at least $1- n^{-3}$,   it holds that 
\begin{equation*}
    \min\{ \eta_k -s ,e-\eta_k \}   \lesssim \frac{\sigma_{\epsilon}^2\vee 1  }{\kappa^2 }\bigg(  \s \log(n\vee p) +\gamma \bigg)  +  \tr.
\end{equation*}
\enlem
\begin{proof}    
If either $ \eta_k -s \le  C_\mclF \s \log(n\vee p)   $ or $e-\eta_k\le C_\mclF \s \log(n\vee p)  $, then 
$$ \min\{ \eta_k -s ,e-\eta_k \}   \le C_\mclF \s \log(n\vee p)  $$
and there is nothing to show. So assume that 
$$  \eta_k -s  > C_\mclF \s \log(n\vee p)  \quad \text{and} \quad   e-\eta_k  >C_\mclF \s \log(n\vee p)  . $$
By event $\mathcal R (\tr )$, there exists  $ s_ u  \in \{ s_q\}_{q=1}^\Q$ such that 
$$0\le s_u - \eta_k \le \tr. $$

\begin{center}
 \begin{tikzpicture} 
\draw[ - ] (-10,0)--(-1,0);
 \node[color=black] at (-8,-0.3) {\small s};
 \draw[  (-, ultra thick, black] (-8,0) -- (-7.99,0);
  \draw[ { -]}, ultra thick, black] (-3.0002,0) -- (-3,0);

   \node[color=black] at (-7,-0.3) {\small $\eta_k$};
\draw(-7 ,0)circle [radius=2pt] ;

  \node[color=black] at (-6.5,-0.3) {\small $s_u$};
\draw plot[mark=x, mark options={color=black, scale=1.5}] coordinates {(-6.5,0)  }; 
  
  \node[color=black] at (-3 ,-0.3) {\small $e$};
\end{tikzpicture}
\end{center} 

{\bf Step 1.} Denote 
$$ \I_ 1 = (s,s_u] \quad \text{and} \quad \I_2 = (s_u, e] .$$
Since 
$ \eta_k-s \ge C_\mclF \s \log(n\vee p)    , $
it follows that 
   $ |\I| \ge C_\mclF \s \log(n\vee p)   $ and $ |\I_1| \ge C_\mclF \s \log(n\vee p)   $. Thus 
$$ \mathcal F  (\I)  = \sum_{ i \in \I } (y_i - X_i^\top \widehat \beta_\I )^2 \quad \text{and} \quad \mathcal F  (\I_1)  = \sum_{ i \in {\I_1}  } (y_i - X_i^\top \widehat \beta_{\I_1}  )^2  .$$
Since $\I\in \widehat { \mathcal P}$, it holds that 
\begin{align}\label{eq:regression one change basic inequality step 1} \mathcal F  (\I)  \le  \mathcal F  (\I_1)  +  \mathcal F  (\I_2)  + \gamma. 
\end{align}

{\bf Case a.} Suppose $|\I_2| <C_\mclF \s \log(n\vee p)      $.
It follows from \Cref{eq:regression one change basic inequality step 1} that 
\begin{align}\nonumber
\sum_{i \in \I }(y_i - X_i^\top \widehat \beta _\I ) ^2 \le &  \sum_{i \in \I_1 }(y_i -X_i^\top  \widehat \beta _{\I_1}  ) ^2 +  0  + \gamma 
\\ \nonumber 
\le &  \sum_{i \in \I_1 }(y_i -  X_i^\top  \beta^*  _i  ) ^2 + C_1 \big ( (s_u -\eta_k) \kappa_k^2   + \s  \log(n\vee p)  \big )   +    \gamma
\\ \nonumber 
\le &  \sum_{i \in \I _1}(y_i -  X_i^\top  \beta^*  _i ) ^2  +C_1 \big (  \tr  \kappa_k^2   +  \s \log(n\vee p )    \big ) +  \gamma
\\
\le &   \sum_{i \in \I  }(y_i -  X_i^\top  \beta^*  _i  ) ^2  +C_1 \big (  \tr  \kappa_k^2   +  \s \log(n\vee p )    \big ) +  \gamma  , \label{eq:regression one change point step 1  case a}
\end{align}
where the first inequality follows from the fact that $  \mathcal F  (\I_2)  = 0  $ when $|\I_2| < C_\mclF \s \log(n\vee p)   $,    the second inequality follows from \Cref{lem:regression one change deviation bound}   {\bf b}, the third inequality follows from  the assumption  that 
$   (s_u-\eta_k) \le  \tr   $,   and the inequality holds because $ \sum_{i \in \I _2 }(y_i -  X_i^\top  \beta^*  _i  ) ^2\ge 0 $.

{\bf Case b.} Suppose $|\I_2|  \ge C_\mclF \s \log(n\vee p)   $.
It follows from \Cref{eq:regression one change basic inequality step 1} that  
\begin{align}\nonumber
\sum_{i \in \I }(y_i - X_i^\top \widehat \beta _\I ) ^2 \le &  \sum_{i \in \I_1 }(y_i -X_i^\top  \widehat \beta _{\I_1}  ) ^2 +  \sum_{i \in \I_2  }(y_i -X_i^\top  \widehat \beta _{\I_2 }  ) ^2   + \gamma 
\\ \nonumber 
\le &  \sum_{i \in \I_1 }(y_i -  X_i^\top  \beta^*  _i  ) ^2 + C_1 \big ( (s_u -\eta_k) \kappa_k^2   + \s  \log(n\vee p)  \big )  
\\  \nonumber 
+ & \sum_{i \in \I_2 }(y_i -  X_i^\top  \beta^*  _i  ) ^2 + C_1     \s  \log(n\vee p)     + 2  \gamma
\\ 
\le &  \sum_{i \in \I }(y_i -  X_i^\top  \beta^*  _i ) ^2  +C_2 \big (  \tr  \kappa_k^2   +  \s \log(n\vee p )    \big ) +  \gamma, \label{eq:regression one change point step 1 case b}
\end{align}
where    the second inequality follows from \Cref{lem:regression one change deviation bound}  {\bf a} and {\bf b}, and the third inequality follows from  the assumption  that 
$   (s_u-\eta_k) \le  \tr   $. Combing two cases leads to 
\begin{align} 
\sum_{i \in \I }(y_i - X_i^\top \widehat \beta _\I ) ^2  
\le     \sum_{i \in \I  }(y_i -  X_i^\top  \beta^*  _i  ) ^2  +C_2 \big (  \tr  \kappa_k^2   +  \s \log(n\vee p )    \big ) +  \gamma . \label{eq:two change point step 1}
\end{align}
\\
\\
{\bf Step 2.}
Denote 
$$ \J_1  = (s,\eta_k ] \quad \text{and} \quad \J_2 = (\eta_k , e] .$$
\Cref{eq:one change point step 1}  gives 
\begin{align}  \nonumber 
& \sum_{i \in \J_1 }(y_i  -X_i^\top  \widehat \beta _\I ) ^2 +\sum_{i \in \J_2 }(y_i - X_i^\top  \widehat \beta _\I  ) ^2  
\\ 
 \le \label{eq:regression step 2 first term} & \sum_{i \in \J_1  }(y_i -  X_i^\top    \beta _{\J _1}  ^*    ) ^2   +\sum_{i \in \J_2  }(y_i - X_i^\top    \beta _{\J _1}  ^*  ) ^2  +C_2  \big (     \tr \kappa_k^2  +\s \log(n\vee p)  \big )  +   \gamma 
\end{align}
The above display  leads to 
\begin{align} \nonumber
&\sum_{i \in \J_1 } \big\{ X_i^\top  (  \widehat \beta _\I  -  \beta^* _{\J_1 }) \big\}  ^2 +\sum_{i \in \J_2  } \big\{ X_i^\top  (  \widehat \beta _\I  -  \beta^* _{\J_2 }) \big\}  ^2  
\\ \nonumber 
 \le  & 2 \sum_{i \in \J_1  } \epsilon_i  X_i^\top (  \widehat \beta _\I  -  \beta^* _{\J_1 })     +2 \sum_{i \in \J_2  } \epsilon_i  X_i^\top (  \widehat \beta _\I  -  \beta^* _{\J_2 })      +C_2  \big (     \tr   \kappa_k^2   +\s\log(n\vee p)  \big )  +  \gamma 
 \\ \label{eq:regression one change step 2 term 2}
 \le  & C_3  \sqrt { \log(n\vee p)    |\J_1| }  \|  \widehat \beta _\I  -  \beta^* _{\J_1 }\| _1  + C_3  \sqrt { \log(n\vee p)  |\J_2| }  \|  \widehat \beta _\I  -  \beta^* _{\J_2  }\| _1+ C_2  \big (     \tr \kappa_k^2  +\s \log(n\vee p)  \big ) +    \gamma 
\end{align} 
where the last inequality follows from \Cref{lemma:consistency} and  that  $|\J_1|   \ge  C_\mclF \s \log(n\vee p)   $ and $|\J_2|  \ge C_\mclF \s \log(n\vee p)   $.
Note that 
\begin{align*} 
\|  ( \widehat \beta _\I  -  \beta _{\J_1}^* )_{S^c } \|_1 = \|  ( \widehat \beta _\I )_{S^c} \|_{1} 
=   \|  ( \widehat \beta _\I -\beta _\I ^* )_{S^c} \|_{1}  \le 3 \|  ( \widehat \beta _\I -\beta _\I ^* )_{S } \|_{1}   \le C_5 \s \sqrt { \frac{ \log(n\vee p)}{| \I|} }    , 
\end{align*} 
where the last two inequalities follows from \Cref{lemma:interval lasso}.  So
\begin{align}  \label{eq:regression one change step 2 term 1}
\|   \widehat \beta _\I  -  \beta _{\J_1}^*   \|_1 = \|  ( \widehat \beta _\I  -  \beta _{\J_1}^* )_{S  } \|_1 +\|  ( \widehat \beta _\I  -  \beta _{\J_1}^* )_{S^c } \|_1  \le \sqrt {\s }  \|  \widehat \beta _\I  -  \beta _{\J_1}^*   \|_2 + C_5 \s \sqrt { \frac{ \log(n\vee p)}{| \I|} } .
\end{align} 
Therefore  \Cref{eq:regression one change step 2 term 2} gives 
\begin{align} \nonumber 
& \sum_{i \in \J_1 } \big\{ X_i^\top  (  \widehat \beta _\I  -  \beta^* _{\J_1 }) \big\}  ^2 +\sum_{i \in \J_2  } \big\{ X_i^\top  (  \widehat \beta _\I  -  \beta^* _{\J_2 }) \big\}  ^2   
\\  \nonumber 
\le & C_3  \sqrt { \log(n\vee p)    |\J_1| }   \bigg(  \sqrt {\s }  \|  \widehat \beta _\I  -  \beta _{\J_1}^*   \|_2 + C_5 \s \sqrt { \frac{ \log(n\vee p)}{| \I|} } \bigg) 
\\ \nonumber  
+ & C_3  \sqrt { \log(n\vee p)    |\J_2| }   \bigg(  \sqrt {\s }  \|  \widehat \beta _\I  -  \beta _{\J_2}^*   \|_2 + C_5 \s \sqrt { \frac{ \log(n\vee p)}{| \I|} } \bigg)  +2\gamma 
\\ \nonumber  
\le & \frac{ c_x | \J_1|}{64} \|  \widehat \beta _\I  -  \beta _{\J_1}^*   \|_2  ^2 + \frac{ c_x | \J_2|}{64} \|  \widehat \beta _\I  -  \beta _{\J_2}^*   \|_2  ^2   + C_5'   \s \log(n\vee p)
\\ +&C_2  \big (     \tr \kappa_k^2  +\s \log(n\vee p)  \big )   +2\gamma,
 \label{eq:regression one change step 2 term 3}
\end{align}
   $ |\J_1| \le |\I| , |\J_2| \le |\I|  $ are used in the last inequality. 
\\
\\
{\bf Step 3.}  Since     $|\J_1|   \ge  C_\mclF  \s \log(n\vee p)   $ and $|\J_2|  \ge C_\mclF \s \log(n\vee p)   $, for $\ell=1, 2$,
  it holds that
 \begin{align}\nonumber 
		&  \sum_{i \in \J_\ell  } \big\{ X_i^\top  (  \widehat \beta _\I  -  \beta^* _{\J_ \ell  }) \big\}  ^2 
		 \\  \nonumber  
		  \ge&  \frac{c_x|\J_\ell | }{16} \|\widehat \beta _\I  -  \beta^* _{\J_ \ell  } \|_2^2 - C_6  \log(p) \|\widehat \beta _\I  -  \beta^* _{\J_ \ell  } \|_1  ^2 
		 \\\nonumber  
		 \ge & \frac{c_x|\J_\ell | }{16} \|\widehat \beta _\I  -  \beta^* _{\J_ \ell  } \|_2^2 - C_6' \s   \log(p) \|\widehat \beta _\I  -  \beta^* _{\J_ \ell  } \|_ 2   ^2  - C_6'  \frac{  \s^2 \log(p) \log(n\vee p)  }{|\I| }
		 \\ \label{eq:regression one change step 3 term 1}
		 \ge &  \frac{c_x|\J_\ell | }{32} \|\widehat \beta _\I  -  \beta^* _{\J_ \ell  } \|_2^2 - C_7  \s \log(n\vee p),
	\end{align} 
	where the first inequality follows from   
 \Cref{lem:restricted eigenvalue},  the second inequality follows from \Cref{eq:regression one change step 2 term 1}   and the last inequality follows from the observation that 
 $$ |\I| \ge |\J_\ell|  \ge   C_\mclF \s \log(n\vee p). $$
 \Cref{eq:regression one change step 2 term 3} and \Cref{eq:regression one change step 3 term 1} together lead to 
 $$   |\J_1  |   \|\widehat \beta _\I  -  \beta^* _{\J_ 1  } \|_2^2  + |\J_2  |   \|\widehat \beta _\I  -  \beta^* _{\J_ 2  } \|_2^2  \le C_8(\s \log(n\vee p) + \tr \kappa_k^2 + \gamma) .$$
Observe  that 
$$ \inf_{  \beta  \in \mathbb R ^p   }  |\J_1 |  \| \beta    -  \beta^* _{\J_ 1  } \|_2^2  + |\J_2 | \|  \beta    -  \beta^* _{\J_ 2  } \|_2^2    = \kappa_k ^2  \frac{|\J_1| |\J_2|}{| \I| }   \ge \frac{ \kappa_k^2 }{2} \min\{ |\J_1| ,|\J_2| \}    . $$ 
This leads to 
$$  \frac{ \kappa_k^2 }{2}\min\{ |\J_1| ,|\J_2| \}    \le  C_8(\s \log(n\vee p)+\tr  + \gamma)  .$$
Since $\kappa_k \asymp \kappa $, it follows that 
$$ \min\{ |\J_1| ,|\J_2| \}   \le C_9 \bigg( \frac{\s \log(n\vee p) + \gamma  }{\kappa 	^2 } +  \tr  \bigg) .$$
 \end{proof}

 \bnlem[Two change points] 
 \label{lemma:regression two change points}
Suppose the good events 
$\mathcal L (\tr ) $ and  $\mathcal R ( \tr ) $  in  \Cref{eq:left and right approximation of change points} hold. 
Let   $ \I=(s,e] \in \mathcal {\widehat P}  $  be such that $\I$ contains  exactly two  change points $ \eta_k,\eta_{k+1}  $.  Suppose in addition that 
\begin{align} \label{eq:regression two change points snr}
\Delta_{\min} \kappa^2 \ge  C     \big( \sigma_{\epsilon}^2\s \log(n\vee p) + \gamma)    
\end{align} 
for sufficiently large constant $C   $ and that 
$\gamma \ge  C_\gamma \mclB_n^{-1}\Delta_{\min}\kappa^2   $.
 Then  with probability  at least $1- n^{-3}$,   it holds that 
\begin{align*} 
 \eta_k -s     \lesssim  \frac{\sigma_{\epsilon}^2\vee 1  }{\kappa^2 }\bigg(  \s \log(n\vee p) +\gamma \bigg) +  \tr   \quad \text{and} \quad 
 e-\eta_{k+1}   \lesssim \frac{\sigma_{\epsilon}^2\vee 1  }{\kappa^2 }\bigg(  \s \log(n\vee p) +\gamma \bigg) +  \tr  ,
 \end{align*} 
 where $C_0\ge 1 $ is some sufficiently large constant.
 \enlem 
\begin{proof} 
By symmetry, it suffices to show that $\eta_k-s\lesssim \frac{\sigma_{\epsilon}^2\vee 1  }{\kappa^2 }\bigg(  \s \log(n\vee p) +\gamma \bigg) +  \tr$.  If 
$$\eta_k-s  \le  C_\mclF \s \log(n\vee p) , $$
then the desired result follows immediately.  
So it suffices to assume that 
$$\eta_k-s > C_\mclF \s \log(n\vee p) .  $$
Since the events $\mathcal L ( \tr  ) $ and  $\mathcal R ( \tr ) $ hold, let $ s_u, s_v$ be such that 
 $\eta_k \le s_u \le s_v \le \eta_{k+1} $ and that 
 $$ 0 \le s_u-\eta_k \le   \tr ,  \quad  0\le  \eta_{k+1} - s_v \le \tr   .    $$
 \
 \\
 \begin{center}
 \begin{tikzpicture} 
\draw[ - ] (-10,0)--(1,0);
 \node[color=black] at (-8,-0.3) {\small s};
 \draw[  (-, ultra thick, black] (-8,0) -- (-7.99,0);
  \draw[ { -]}, ultra thick, black] (-1.0002,0) -- (-1,0);

   \node[color=black] at (-7,-0.3) {\small $\eta_k$};
\draw(-7 ,0)circle [radius=2pt] ;

  \node[color=black] at (-6.5,-0.3) {\small $s_u$};
\draw plot[mark=x, mark options={color=black, scale=1.5}] coordinates {(-6.5,0)  };

  \node[color=black] at (-2.3,-0.3) {\small $\eta_{k+1}$};
\draw(-2.3 ,0)circle [radius=2pt] ;
  \node[color=black] at (-3 ,-0.3) {\small $s_v$};
\draw plot[mark=x, mark options={color=black, scale=1.5}] coordinates {(-3,0)  }; 
  
   \node[color=black] at (-1,-0.3) {\small $e$};

\end{tikzpicture}
\end{center} 
\
\\
{\bf Step 1.} Denote 
 $$ \mathcal I_ 1= (  s, s _u], \quad \I_2  =(s_u, s_v] \quad \text{and} \quad \I_3  = (s_v,e]. $$
 \
 \\
Since  $ |\I| \ge \eta_{k+1} -\eta_k \ge C_\mclF \s \log(n\vee p)   $, 
$$ \mathcal F(\I ) = \sum_{i \in \I }(y_i - X_i^\top \widehat \beta _\I ) ^2  . $$
Since 
 $ |\I_1|  \ge  \eta_k-s      \ge C_\mclF \s \log(n\vee p)   , $ 
it follows that 
 $$ \mathcal F(\I_1 ) = \sum_{i \in \I_1 }(y_i - X_i^\top \widehat \beta _{\I_1}  ) ^2  . $$
In addition   since  $ |\I_1| \ge C_\mclF \s \log(n\vee p)  $, then 
 \begin{align*}\mathcal F(\I_1)  = & \sum_{i \in \I_1 }(y_i - X_i^\top \widehat \beta _{\I_1}  ) ^2 
 \\
 \le & \sum_{i \in \I_1 }(y_i - X_i^\top   \beta ^*  _i   ) ^2  + C_1  \bigg\{ \frac{( \eta_k-s) (s_u -\eta_k ) }{ ( \eta_k-s)  + (s_u -\eta_k )  } \kappa ^2 + \s\log(n\vee p)  \bigg\} 
 \\
  \le & \sum_{i \in \I_1 }(y_i - X_i^\top   \beta ^*  _i   ) ^2  +  C_1 \bigg\{ (s_u -\eta_k )   \kappa ^2 +  \s\log(n\vee p)  \bigg\} 
  \\
  \le & \sum_{i \in \I_1 }(y_i - X_i^\top   \beta ^*  _ i   ) ^2  + C_1 \bigg\{  \tr   \kappa ^2 +  \s\log(n\vee p)  \bigg\} ,
 \end{align*}
where the first  inequality follows  from \Cref{lem:regression one change deviation bound}  and that $ \kappa_{k} \asymp \kappa$.
 Similarly, since 
 $ |\I_2|  \ge \Delta_{\min} /2 \ge   C_\mclF \s \log(n\vee p)   , $ 
it follows that 
 $$ \mathcal F(\I_2 ) = \sum_{i \in \I_2 }(y_i - X_i^\top \widehat \beta _{\I_2}  ) ^2  . $$
  Since $|\I_2| \ge C_\mclF \s \log(n\vee p)   $ and  $\I_2$ contains no change points, by \Cref{lem:regression one change deviation bound},
  \begin{align*}\mathcal F(\I_2)  \le \sum_{i \in \I_2 }(y_i - X_i^\top   \beta ^*  _ i  ) ^2   + C_1\s\log(n\vee p).
  \end{align*}  
  \\
  \\
 {\bf Step 2.} If $ |\I_3 | \ge C_\mclF \s \log(n\vee p)   $, then 
  \begin{align*}\mathcal F(\I_3)  = & \sum_{i \in \I_3 }(y_i - X_i^\top \widehat \beta _{\I_3}  ) ^2 
 \\
 \le & \sum_{i \in \I_3 }(y_i - X_i^\top   \beta ^*  _i   ) ^2  + C_1  \bigg\{ \frac{( \eta_{k+1}-s_v) (e -\eta_{k+1} ) }{ ( \eta_{k+1}-s_v)+  (e -\eta_{k+1} )  } \kappa ^2 + \s\log(n\vee p)  \bigg\} 
 \\
  \le & \sum_{i \in \I_3 }(y_i - X_i^\top   \beta ^*  _i   ) ^2  +  C_1 \bigg\{ ( \eta_{k+1}-s_v)    \kappa ^2 +  \s\log(n\vee p)  \bigg\} 
  \\
  \le & \sum_{i \in \I_3 }(y_i - X_i^\top   \beta ^*  _ i   ) ^2  + C_1 \bigg\{  \tr   \kappa ^2 +  \s\log(n\vee p)  \bigg\} ,
 \end{align*}
 where the first inequality follows from \Cref{lem:regression one change deviation bound}{\bf b} and that $ \kappa_{k+1} \asymp \kappa$. 
 If $|\I_3| < C_\mclF \s \log(n\vee p)   $, then 
$\mathcal F(\I_3)  =  0 $.
 So 
both cases imply that  
 $$\mathcal F(\I_3)  \le \sum_{i \in \I_3 }(y_i - X_i^\top   \beta ^*  _ i   ) ^2  + C_1 \bigg\{  \tr   \kappa_k^2 +  \s\log(n\vee p)  \bigg\}  .$$
 \
 \\
 {\bf Step 3.} Since $\I \in \widehat{\mathcal P} $, we have
 \begin{align}\label{eq:regression two change points local min} \mathcal F(\I ) \le \mathcal F(\I_1  ) +\mathcal F(\I_2 )+\mathcal F(\I_3 ) + 2\gamma. 
 \end{align} The above display and the calculations in  {\bf Step 1} and  {\bf Step 2} implies that
\begin{align} 
 \sum_{i \in \I }(y_i -  X_i^\top \widehat \beta _\I ) ^2  
\le   \sum_{i \in \I  }(y_i -   X_i^\top \beta_i^*   ) ^2  
 +   3  C_1 \bigg\{  \tr   \kappa ^2 +  \s\log(n\vee p) \bigg\}
+2\gamma  .
  \label{eq:regression two change points step 2 term 1}
\end{align}
Denote 
$$ \J_1  = (s,\eta_k ], \quad \J_2=(\eta_k, \eta_{k} +   \eta_{k+1}    ]  \quad \text{and} \quad \J_3 = (\eta_{k+1}  , e] .$$
\Cref{eq:regression two change points step 2 term 1}  gives 
\begin{align} 
 \sum_{\ell=1}^3 \sum_{i \in \J_{\ell}  }(y_i - X_i^\top  \widehat \beta _\I ) ^2 
 \le& \sum_{\ell=1}^3 \sum_{i \in \J_ \ell  }(y_i -  X_i^\top    \beta^*  _{\J_\ell }  ) ^2    
 +  3  C_1 \bigg\{  \tr   \kappa ^2 +  \s\log(n\vee p) \bigg\} 
+2\gamma     \label{eq:regression two change points first}
\end{align} 
\
\\
{\bf Step 4.} 
Note that for $\ell\in \{1,2,3\}$, 
\begin{align*} 
\|  ( \widehat \beta _\I  -  \beta _{\J_\ell }^* )_{S^c } \|_1 = \|  ( \widehat \beta _\I )_{S^c} \|_{1} 
=   \|  ( \widehat \beta _\I -\beta _\I ^* )_{S^c} \|_{1}  \le 3 \|  ( \widehat \beta _\I -\beta _\I ^* )_{S } \|_{1}   \le C_2 \s \sqrt { \frac{ \log(n\vee p)}{| \I|} }    , 
\end{align*} 
where the last two inequalities follows from \Cref{lemma:interval lasso}.  So
\begin{align}  \label{eq:regression two change step 3 term 1}
\|   \widehat \beta _\I  -  \beta _{\J_\ell}^*   \|_1 = \|  ( \widehat \beta _\I  -  \beta _{\J_\ell }^* )_{S  } \|_1 +\|  ( \widehat \beta _\I  -  \beta _{\J_\ell}^* )_{S^c } \|_1  \le \sqrt {\s }  \|  \widehat \beta _\I  -  \beta _{\J_\ell}^*   \|_2 + C_2 \s \sqrt { \frac{ \log(n\vee p)}{| \I|} } .
\end{align} 
Note that by assumptions,
$$|\J_1|\ge C_\mclF \s \log(n\vee p)    \quad \text{and} \quad |\J_2 |\ge C_\mclF \s \log(n\vee p) .$$
 So for $\ell\in \{ 1, 2\} $,
  it holds that
 \begin{align}\nonumber 
		&  \sum_{i \in \J_\ell  } \big\{ X_i^\top  (  \widehat \beta _\I  -  \beta^* _{\J_ \ell  }) \big\}  ^2 
		 \\  \nonumber  
		  \ge&  \frac{c_x|\J_\ell | }{16} \|\widehat \beta _\I  -  \beta^* _{\J_ \ell  } \|_2^2 - C_3 \log(p) \|\widehat \beta _\I  -  \beta^* _{\J_ \ell  } \|_1  ^2 
		 \\\nonumber  
		 \ge & \frac{c_x|\J_\ell | }{16} \|\widehat \beta _\I  -  \beta^* _{\J_ \ell  } \|_2^2 - C_3' \s   \log(p) \|\widehat \beta _\I  -  \beta^* _{\J_ \ell  } \|_ 2   ^2  - C_3'  \frac{  \s^2 \log(p) \log(n\vee p)  }{|\I| }
		 \\ \label{eq:regression two change step 3 term 2}
		 \ge &  \frac{c_x|\J_\ell | }{32} \|\widehat \beta _\I  -  \beta^* _{\J_ \ell  } \|_2^2 - C_4  \s \log(n\vee p),
	\end{align} 
	where the first inequality follows from \Cref{lem:restricted eigenvalue}, the second inequality follows from \Cref{eq:regression two change step 3 term 1}   and the last inequality follows from the observation that 
 $$  |\I|\ge  |\J_\ell|   \ge C_\gamma \s \log(n\vee p). 
 $$ 
 So for $\ell\in \{ 1,2\}$, 
\begin{align*} &\sum_{i \in \J_\ell  }(y_i - X_i^\top  \widehat \beta _\I ) ^2  - \sum_{i \in \J_\ell   }(y_i - X_i^\top    \beta^*  _{\J_\ell }  ) ^2
=  \sum_{i \in \J_\ell    }\big\{ X_i^\top (\widehat \beta _\I -\beta^*  _{\J_4} )\big\}^2 -2 \sum_{i \in \J_\ell   } \epsilon_i X_i^\top (\widehat \beta _\I -\beta^*  _{\J_\ell  }  )
\\
\ge  & \sum_{i \in \J_\ell   }\big\{ X_i^\top (\widehat \beta _\I -\beta^*  _{\J_\ell  } )\big\}^2  - 2 \| \sum_{i \in \J_\ell   }  \epsilon_i X_i^\top \|_\infty \| \widehat \beta _\I -\beta^*  _{\J_\ell  }  \|_1
\\
\ge &\sum_{i \in \J_\ell   }\big\{ X_i^\top (\widehat \beta _\I -\beta^*  _{\J_\ell  } )\big\}^2  - C_5 \sqrt{ \log(n\vee p) |\J_\ell|   }\bigg(\sqrt {\s }  \|  \widehat \beta _\I  -  \beta _{\J_1}^*   \|_2 + C_2 \s \sqrt { \frac{ \log(n\vee p)}{| \I|} } \bigg) 
\\
\ge &  \frac{c_x|\J_\ell | }{32} \|\widehat \beta _\I  -  \beta^* _{\J_ \ell  } \|_2^2 - C_4  \s \log(n\vee p)  - C_5 \sqrt{ \log(n\vee p) |\J_\ell|  }\bigg(\sqrt {\s }  \|  \widehat \beta _\I  -  \beta _{\J_1}^*   \|_2 + C_2 \s \sqrt { \frac{ \log(n\vee p)}{| \I|} } \bigg)  
\\
\ge & \frac{c_x|\J_\ell | }{32} \|\widehat \beta _\I  -  \beta^* _{\J_ \ell  } \|_2^2 -\frac{c_x|\J_\ell | }{64} \|\widehat \beta _\I  -  \beta^* _{\J_ \ell  } \|_2^2 -C_6 \s \log(n\vee p)  
= \frac{c_x|\J_\ell | }{64} \|\widehat \beta _\I  -  \beta^* _{\J_ \ell  } \|_2^2     -C_6 \s \log(n\vee p)  ,
\end{align*}
where the second inequality follows from the standard sub-Exponential tail bound and \Cref{eq:regression two change step 3 term 1}, the third inequality follows from \Cref{eq:regression two change step 3 term 2}, and the fourth inequality follows from  $ \J_\ell \subset \I $ and so $ |\I | \ge |\J_\ell|$.
\\
\\
So for $\ell \in \{1,2 \}$, 
\begin{align} \label{eq:regression change point step 3 last item}\sum_{i \in \J_\ell  }(y_i - X_i^\top  \widehat \beta _\I ) ^2  - \sum_{i \in \J_\ell   }(y_i - X_i^\top    \beta^*  _{\J_\ell }  ) ^2 
\ge \frac{c_x|\J_\ell | }{64} \|\widehat \beta _\I  -  \beta^* _{\J_ \ell  } \|_2^2     -C_6 \s \log(n\vee p).
\end{align} 
\
\\
\\
{\bf Step 5.} For $\J_3$, if $|\J_3| \ge C_\mclF \s \log(n\vee p)  $,  following the same calculations as in {\bf Step 4}, 
$$\sum_{i \in \J_ 3 }(y_i - X_i^\top  \widehat \beta _\I ) ^2  - \sum_{i \in \J_ 3 }(y_i - X_i^\top    \beta^*  _{\J_3   }  ) ^2 \ge \frac{c_x|\J_3   | }{64} \|\widehat \beta _\I  -  \beta^* _{\J_ 3  } \|_2^2     -C_6 \s \log(n\vee p)  \ge -C_6 \s\log(n\vee p). 
$$
If $|\J_3 | < C_\mclF \s \log(n\vee p) $, then
\begin{align} & \sum_{i \in \J_3   }(y_i - X_i^\top  \widehat \beta _\I ) ^2  - \sum_{i \in \J_3   }(y_i - X_i^\top    \beta^*  _{\J_ 3}  ) ^2 \nonumber 
=   \sum_{i \in \J_ 3   }\big\{ X_i^\top (\widehat \beta _\I -\beta^*  _{\J_3} )\big\}^2 -2 \sum_{i \in \J_3   } \epsilon_i X_i^\top (\widehat \beta _\I -\beta^*  _{\J_3 }  )
\\  \nonumber 
\ge  & \sum_{i \in \J_3   }\big\{ X_i^\top (\widehat \beta _\I -\beta^*  _{\J_3 } )\big\}^2  -\frac{1}{2 }\sum_{i \in \J_3   }\big\{ X_i^\top (\widehat \beta _\I -\beta^*  _{\J_3 } )\big\}^2  - 4   \sum_{i \in \J_3   }\epsilon_i^2 
\\  \nonumber 
\ge & \frac{1}{2 }\sum_{i \in \J_3   }\big\{ X_i^\top (\widehat \beta _\I -\beta^*  _{\J_3 } )\big\}^2  -C_7\bigg(  \sqrt { \gamma   \log(n)} + \log(n)+ \gamma  \bigg) 
\\  \nonumber  
\ge & \frac{1}{2 }\sum_{i \in \J_3   }\big\{ X_i^\top (\widehat \beta _\I -\beta^*  _{\J_3 } )\big\}^2  -C_7' \bigg(     \log(n)+ \gamma   \bigg)  
\\ \ge 
& \frac{1}{2 }\sum_{i \in \J_3   }\big\{ X_i^\top (\widehat \beta _\I -\beta^*  _{\J_3 } )\big\}^2  -C_8    ( \s\log(n\vee p) + \gamma) 
\ge -C_8    (  \s\log(n\vee p) +\gamma)   \label{eq:regression change point step 4 last item}
\end{align}
where the second inequality follows from the standard sub-exponential deviation bound.
\
\\
\\
{\bf Step 6.} Putting \Cref{eq:regression two change points first}, \eqref{eq:regression change point step 3 last item} and \eqref{eq:regression change point step 4 last item} together, it follows that 
 $$  \sum_{\ell =1}^2 \frac{c_x|\J_\ell | }{64} \|\widehat \beta _\I  -  \beta^* _{\J_ \ell  } \|_2^2      \le C_9(\s \log(n\vee p) + \tr \kappa ^2 + \gamma) .$$
 This leads to 
  $$     |\J_1 |   \|\widehat \beta _\I  -  \beta^* _{\J_ 1  } \|_2^2     + |\J_2 |   \|\widehat \beta _\I  -  \beta^* _{\J_ 2  } \|_2^2       \le C_9(\s \log(n\vee p) + \tr \kappa ^2 + \gamma) .$$ 
Observe  that 
$$ \inf_{  \beta  \in \mathbb R ^p   }  |\J_1 |  \| \beta    -  \beta^* _{\J_ 1  } \|_2^2  + |\J_2 | \|  \beta    -  \beta^* _{\J_ 2  } \|_2^2    = \kappa_k   ^2  \frac{|\J_1| |\J_2|}{| \J_1| +|\J_2|  }   \ge \frac{ \kappa_k  ^2 }{2} \min\{ |\J_1| ,|\J_2| \}   \ge   \frac{ c \kappa   ^2 }{2} \min\{ |\J_1| ,|\J_2| \}   . $$ 
Thus
$$    \kappa ^2 \min\{ |\J_1| ,|\J_2| \}    \le  C_{10} (\s \log(n\vee p)+\tr \kappa^2  + \gamma)  ,$$
which is 
$$ \min\{ |\J_1| ,|\J_2| \}   \le C_5 \bigg( \frac{\s \log(n\vee p) + \gamma  }{\kappa ^2 } +  \tr  +\frac{ \gamma}{\kappa ^2} \bigg) .$$ 
Since  $ |\J_2  | \ge \Delta_{\min} \ge     \frac{  C     ( \s \log(n\vee p) + \gamma) }{   \kappa^{2}}   $
 for sufficiently large constant $C$, 
 it follows that
 $$ |\J_2| \ge \Delta_{\min}>  C_5 \bigg( \frac{\s \log(n\vee p) + \gamma  }{\kappa ^2 } +  \tr  +\frac{ \gamma}{\kappa ^2} \bigg) .$$  So it holds that 
 $$   |\J_1|    \le C_5 \bigg( \frac{\s \log(n\vee p) + \gamma  }{\kappa ^2 } +  \tr  \bigg) .$$ 
 \end{proof} 

 \bnlem[Three or more change points]
\label{lem:regression three or more cp}
Suppose the good events 
$\mathcal L (\tr ) $ and  $\mathcal R ( \tr ) $ defined  in  \Cref{eq:left and right approximation of change points} hold. 
 Suppose in addition that 
\begin{align} \label{eq:regression three change points snr}
\Delta_{\min} \kappa^2 \ge C      \big( \s \log(n\vee p) + \gamma)    
\end{align} 
for sufficiently large constant $C $. 
 Then  with probability  at least $1- n^{-3}$,    there is no intervals in   $  \widehat { \mathcal P}  $  containing three or more true change points.   
 \enlem

\begin{proof} For  contradiction,  suppose    $ \I=(s,e] \in \mathcal {\widehat P}  $  be such that $ \{ \eta_1, \ldots, \eta_M\} \subset \I $ with $M\ge 3$.  
\\
\\ 
Since the events $\mathcal L ( \tr  ) $ and  $\mathcal R ( \tr ) $ hold, by relabeling $\{ s_q\}_{q=1}^\Q  $ if necessary,   let $ \{ s_m\}_{m=1}^M $ be such that 
 $$ 0 \le s_m  -\eta_m  \le \tr    \quad \text{for} \quad 1 \le m \le M-1  $$ and that
 $$ 0\le \eta_M  - s_M \le \tr .$$ 
 Note that these choices ensure that  $ \{ s_m\}_{m=1}^M  \subset \I . $
 \
 \\
 \begin{center}
 \begin{tikzpicture} 
\draw[ - ] (-10,0)--(1,0);
 \node[color=black] at (-8,-0.3) {\small s};
 \draw[  (-, ultra thick, black] (-8,0) -- (-7.99,0);
  \draw[ { -]}, ultra thick, black] (-1.0002,0) -- (-1,0);

   \node[color=black] at (-7,-0.3) {\small $\eta_1$};
\draw(-7 ,0)circle [radius=2pt] ;

  \node[color=black] at (-6.5,-0.3) {\small $s_1$};
\draw plot[mark=x, mark options={color=black, scale=1.5}] coordinates {(-6.5,0)  }; 

  \node[color=black] at (-5,-0.3) {\small $\eta_2$};
\draw(-5 ,0)circle [radius=2pt] ;
  \node[color=black] at (-4.5,-0.3) {\small $s_2$};
\draw plot[mark=x, mark options={color=black, scale=1.5}] coordinates {(-4.5,0)  };

  \node[color=black] at (-2.5,-0.3) {\small $\eta_M$};
\draw(-2.5 ,0)circle [radius=2pt] ;
  \node[color=black] at (-3 ,-0.3) {\small $s_M$};
\draw plot[mark=x, mark options={color=black, scale=1.5}] coordinates {(-3,0)  }; 
  
   \node[color=black] at (-1,-0.3) {\small $e$};

\end{tikzpicture}
\end{center}
\
\\
{\bf Step 1.}
 Denote 
 $$ \mathcal I_ 1= (  s, s _1], \quad \I_m   =(s_{m-1} , s_m] \text{ for }  2 \le m \le M     \quad \text{and} \quad \I_{M+1}   = (s_M,e]. $$
 Then since $|\I| \ge \Delta_{\min} \ge  C_s \s \log(n\vee p)  $, it follows that 
 Since  $ |\I| \ge \eta_{k+1} -\eta_k \ge  C_s \s \log(n\vee p) $, 
$$ \mathcal F(\I ) = \sum_{i \in \I }(y_i - X_i^\top \widehat \beta _\I ) ^2  . $$
Since  $ | \I_m | \ge \Delta_{\min} /2 \ge   C_s \s \log(n\vee p) $ for all $ 2 \le m \le M $,  it follows  from the same argument as   {\bf Step 1} in the proof  of  \Cref{lemma:regression two change points} that 
 \begin{align*}\mathcal F(\I_m)  = & \sum_{i \in \I_m }(y_i - X_i^\top \widehat \beta _{\I_m}  ) ^2
  \le   \sum_{i \in \I_m }(y_i - X_i^\top   \beta ^*  _ i   ) ^2  + C_1 \bigg\{  \tr   \kappa ^2 +  \s\log(n\vee p)  \bigg\} \quad \text{for all } 2 
  \le m \le M.
 \end{align*}
 \
 \\
 {\bf Step 2.} It follows  from the same argument as   {\bf Step 2} in the proof  of  \Cref{lemma:regression two change points} that    
  \begin{align*}
  &\mathcal F(\I_1)  
  \le   \sum_{i \in \I_1 }(y_i - X_i^\top   \beta ^*  _ i   ) ^2  + C_1 \bigg\{  \tr   \kappa ^2 +  \s\log(n\vee p)  \bigg\} , \text{ and}
  \\
  &\mathcal F(\I_{M+1} )  
  \le   \sum_{i \in \I_{M+1} }(y_i - X_i^\top   \beta ^*  _ i   ) ^2  + C_1 \bigg\{  \tr   \kappa ^2 +  \s\log(n\vee p)  \bigg\}  
 \end{align*}
 \
 \\
 {\bf Step 3.} Since $\I \in \widehat{\mathcal P} $, we have
 \begin{align}\label{eq:regression three change points local min} \mathcal F(\I ) \le \sum_{m=1}^{M+1}\mathcal F(\I_m  )  +  M\gamma. 
 \end{align} The above display and the calculations in  {\bf Step 1} and  {\bf Step 2} implies that
\begin{align} 
 \sum_{i \in \I }(y_i -  X_i^\top \widehat \beta _\I ) ^2  
\le   \sum_{i \in \I  }(y_i -   X_i^\top \beta_i^*   ) ^2  
 +   (M +1) C_1 \bigg\{  \tr   \kappa ^2 +  \s\log(n\vee p) \bigg\}
+M\gamma  .
  \label{eq:regression three change points step 2 term 1}
\end{align}
Denote 
   $$  \J_1 =(s,  \eta_1], \ \J_m = (\eta_{m-1}, \eta_m]  \quad \text{for}\quad  2 \le m \le M , \ \J_{M+1}  =(\eta_M, e]. $$  
\Cref{eq:regression three change points step 2 term 1}  gives 
\begin{align} 
 \sum_{m=1}^{M+1}  \sum_{i \in \J_m   }(y_i - X_i^\top  \widehat \beta _\I ) ^2 
 \le& \sum_{m=1}^{M+1} \sum_{i \in \J_  m   }(y_i -  X_i^\top    \beta^*  _{\J_m  }  ) ^2    
 +  (M+1)   C_1 \bigg\{  \tr   \kappa ^2 +  \s\log(n\vee p) \bigg\} 
+M\gamma     \label{eq:regression three change points first}
\end{align}  
 \
 \\
 {\bf Step 4.} Using the same argument as in the {\bf Step 4} in the proof  of  \Cref{lemma:regression two change points},
it follows that 
\begin{align} \label{eq:regression three change point step 3 last term}\sum_{i \in \J_m   }(y_i - X_i^\top  \widehat \beta _\I ) ^2  - \sum_{i \in \J_m   }(y_i - X_i^\top    \beta^*  _{\J_m }  ) ^2 
\ge \frac{c_x|\J_m | }{64} \|\widehat \beta _\I  -  \beta^* _{\J_ m  } \|_2^2     -C_2 \s \log(n\vee p) \quad \text{for all}  \   2 \le m \le M. 
\end{align} 
 \
 \\
 {\bf Step 5.}
  Using the same argument as in the {\bf Step 4} in the proof  of  \Cref{lemma:regression two change points},  it follows that 
\begin{align} \label{eq:regression three change point step 5 first term}& \sum_{i \in \J_ 1 }(y_i - X_i^\top  \widehat \beta _\I ) ^2  - \sum_{i \in \J_ 1 }(y_i - X_i^\top    \beta^*  _{\J_1   }  ) ^2  \ge -C_3  ( \s\log(n\vee p)  +\gamma) \text{ 
 and }
\\\label{eq:regression three change point step 5 second term}
& \sum_{i \in \J_ {M+ 1} }(y_i - X_i^\top  \widehat \beta _\I ) ^2  - \sum_{i \in \J_  {M+ 1}  }(y_i - X_i^\top    \beta^*  _{\J_ {M+ 1}   }  ) ^2  \ge -C_3 ( \s\log(n\vee p)  +\gamma) 
 \end{align}
\
\\
{\bf  Step 6.} Putting \Cref{eq:regression three change points first}, \eqref{eq:regression three change point step 3 last term}, \eqref{eq:regression three change point step 5 first term} and  \eqref{eq:regression three change point step 5 second term}, 
 it follows that 
\begin{align} \label{eq:regression three change points step six} \sum_{ m  =2}^M \frac{c_x|\J_m  | }{64} \|\widehat \beta _\I  -  \beta^* _{\J_  m  } \|_2^2      \le C_4M (\s \log(n\vee p) + \tr \kappa ^2 + \gamma) .
\end{align} 
 For any $  m \in\{2, \ldots, M\}$, it holds that
\begin{align}  \label{eq:regression three change points signal lower bound} \inf_{  \beta  \in \mathbb R^p   }  |\J_{m-1} |  \|   \beta - \beta^* _{\J_ {m-1}   }  \|  ^2 + |\J_{m} | \|   \beta - \beta^* _{\J_ m  }  \|  ^2   =&     \frac{|\J_{m-1}| |\J_m|}{ |\J_{m-1}| +  |\J_m| } \kappa_m ^2  \ge \frac{1}{2} \Delta_{\min} \kappa^2,
\end{align}  
where the last inequality follows from the assumptions that $\eta_k - \eta_{k-1}\ge \Delta_{\min}  $ and $ \kappa_k \asymp \kappa$  for all $1\le k \le K$. So
\begin{align}   \nonumber &2 \sum_{ m=1}^{M }   |\J_m |\|\widehat \beta _\I  -  \beta^* _{\J_ m  } \|_2^2   
\\ 
\ge &  \nonumber    \sum_{m=2}^M  \bigg(  |\J_{m-1}  \|\widehat \beta _\I  -  \beta^* _{\J_ {m-1}   } \|_2^2      + |\J_m |\|\widehat \beta _\I  -  \beta^* _{\J_ m  } \|_2^2    \bigg) 
\\ \label{eq:regression three change points signal lower bound two} 
\ge & (M-1)  \frac{ 1}{2} \Delta_{\min} \kappa^2  \ge \frac{M}{4} \Delta_{\min} \kappa^2 ,
\end{align} 
where the second inequality follows from  \Cref{eq:regression three change points signal lower bound} and the last inequality follows from $M\ge 3$. \Cref{eq:regression three change points step six} and  \Cref{eq:regression three change points signal lower bound two} together imply that 
\begin{align}\label{eq:regression three change points signal lower bound three} 
 \frac{M}{4} \Delta_{\min} \kappa^2 \le 2  C_5  M\bigg(  \s   \log(n\vee p ) +    \tr \kappa^2
+   \gamma \bigg) .
\end{align}
Since $\Bn\to \infty $, it follows that for sufficiently large $n$,   \Cref{eq:regression three change points signal lower bound three} gives 
$$ \Delta_{\min}\kappa^2 \le C_5 \big(  \s \log(n\vee p) +\gamma),$$  
which contradicts \Cref{eq:regression three change points snr}.

 \end{proof}

 \bnlem[Two consecutive intervals]
 \label{lem:regression two intervals}
 Suppose $ \gamma \ge C_\gamma \s  \log(n\vee p) $ for sufficiently large constant $C_\gamma $. 
With probability  at least $1- n^{-3}$,    there are no  two consecutive intervals $\I_1= (s,t ] \in \widehat {\mathcal  P}  $,   $ \I_2=(t, e]  \in   \widehat  {\mathcal  P}  $    such that $\I_1 \cup \I_2$ contains no change points.  
 \enlem 
 \begin{proof} 
 For contradiction, suppose that 
 $$ \I :  =\I_1\cup  \I_2 $$
 contains no change points. 
For $\I_1$, note that if $|\I_1| \ge  C_\zeta \s \log(n\vee p) $, then 
 by \Cref{lem:regression one change deviation bound} {\bf a}, it follows that 
 \begin{align*} 
 \bigg|   \mathcal F(\I_1)   -  \sum_{i\in \I_1} (y_i - X_i^\top  \beta^*  _{i} )^2 \bigg|  = \bigg|   \sum_{i\in \I_1} (y_i - X_i^\top \widehat \beta _{\I_1} )^2- \sum_{i\in \I_1} (y_i - X_i^\top \beta_i^*   )^2    \bigg| \le  C_1 \s \log(n\vee p)  .
 \end{align*}
 If $|\I_1| <  C_\zeta \s \log(n\vee p)  $, then
\begin{align*} 
& \bigg|   \mathcal F(\I_1)   -  \sum_{i\in \I_1} (y_i - X_i^\top  \beta^*  _{i} )^2 \bigg|   =   \bigg|     \sum_{i\in \I_1} (y_i - X_i^\top  \beta^*  _{i} )^2  \bigg|  
=\sum_{i\in \I_1}   \epsilon_i^2 
 \\ 
 \le & |\I_1| E(\epsilon^2_1) + C_2 \sqrt {  |\I_1|  \log(n)  } + \log(n)  \le C_2' \s \log(n\vee p) .
 \end{align*} 
 So
 \begin{align*} 
 \bigg|   \mathcal F(\I_1)   -  \sum_{i\in \I_1} (y_i - X_i^\top  \beta^*  _{i} )^2 \bigg| \le   C_3   \s \log(n\vee p)  .
 \end{align*} 
 Similarly, 
\begin{align*} 
&\bigg|   \mathcal F(\I_2)   -  \sum_{i\in \I_2} (y_i - X_i^\top  \beta^*  _{i} )^2 \bigg| \le   C_3   \s \log(n\vee p), \quad \text{and}
\\
&\bigg|   \mathcal F(\I )   -  \sum_{i\in \I } (y_i - X_i^\top  \beta^*  _{i} )^2 \bigg| \le   C_3   \s \log(n\vee p)  .
 \end{align*} 
 So 
 $$\sum_{i\in \I_1} (y_i - X_i^\top \beta_i^*   )^2  +\sum_{i\in \I_2} (y_i - X_i^\top \beta_i^*    )^2 -2C_1  \s \log(n\vee p ) +\gamma    \le \sum_{i\in \I } (y_i - X_i^\top \beta_i^*   )^2 +C_1 \s  \log(n\vee p) .  $$
 Since $\beta^* _i$ is unchanged when $i\in \I$, it follows that 
 $$ \gamma \le 3C_1 \s \log(n\vee p).$$
 This is a contradiction when $C_\gamma> 3C_1. $
 \end{proof}

\bnlem
Let $\mathcal{S}$ be any linear subspace in $\mathbb{R}^n$ and $\mathcal{N}_{1/4}$	be a $1/4$-net of $\mathcal{S} \cap B(0, 1)$, where $B(0, 1)$ is the unit ball in $\mathbb{R}^n$.  For any $u \in \mathbb{R}^n$, it holds that
	\[
		\sup_{v \in \mathcal{S} \cap B(0, 1)} \langle v, u \rangle \leq 2 \sup_{v \in \mathcal{N}_{1/4}} \langle v, u \rangle,
	\]
	where $\langle \cdot, \cdot \rangle$ denotes the inner product in $\mathbb{R}^n$.
\enlem

\begin{proof}
Due to the definition of $\mathcal{N}_{1/4}$, it holds that for any $v \in \mathcal{S} \cap B(0, 1)$, there exists a $v_k \in \mathcal{N}_{1/4}$, such that $\|v - v_k\|_2 < 1/4$.  Therefore,
	\begin{align*}
		\langle v, u \rangle = \langle v - v_k + v_k, u \rangle = \langle x_k, u \rangle + \langle v_k, u \rangle \leq \frac{1}{4} \langle v, u \rangle + \frac{1}{4} \langle v^{\perp}, u \rangle + \langle v_k, u \rangle,
	\end{align*}
	where the inequality follows from  $x_k = v - v_k = \langle x_k, v \rangle v + \langle x_k, v^{\perp} \rangle v^{\perp}$.  Then we have
	\[
		\frac{3}{4}\langle v, u \rangle \leq \frac{1}{4} \langle v^{\perp}, u \rangle + \langle v_k, u \rangle.
	\]
	It follows from the same argument that 
	\[
		\frac{3}{4}\langle v^{\perp}, u \rangle \leq \frac{1}{4} \langle v, u \rangle + \langle v_l, u \rangle,
	\]
	where $v_l \in \mathcal{N}_{1/4}$ satisfies $\|v^{\perp} - v_l\|_2 < 1/4$.  Combining the previous two equation displays yields
	\[
		\langle v, u \rangle \leq 2 \sup_{v \in \mathcal{N}_{1/4}} \langle v, u \rangle,
	\]
	and the final claims holds.
\end{proof}

\Cref{lem:deviation piecewise constant} is an adaptation of Lemma 3 in \cite{wang2021_jmlr}.

\bnlem\label{lem:deviation piecewise constant}
	Given any interval $I = (s, e] \subset \{1, \ldots, n\}$. Let $\mclR_m := \{v \in \mathbb{R}^{(e-s)}| \|v\|_2 = 1, \sum_{t = 1}^{e-s-1} \mathbf{1}\{v_i \neq v_{i+1}\} = m\}$. Then for data generated from \Cref{assp:dcdp_linear_reg}, it holds that for any $\delta > 0$, $i \in \{1, \ldots, p\}$, 
	\[
		\mathbb{P}\left\{\sup_{v\in \mclR_m} \left|\sum_{t = s+1}^e v_t \epsilon_t (X_t)_i\right| > \Delta_{\min} \right\} \leq C(e-s-1)^m 9^{m+1} \exp \left\{-c \min\left\{\frac{\delta^2}{4C_x^2}, \, \frac{\delta}{2C_x \|v\|_{\infty}}\right\}\right\}.
	\]
\enlem

\begin{proof}
For any $v \in \mathbb{R}^{(e-s)}$ satisfying $\sum_{t = 1}^{e-s-1}\mathbbm{1}\{v_i \neq v_{i+1}\} = m$, it is determined by a vector in $\mathbb{R}^{m+1}$ and a choice of $m$ out of $(e-s-1)$ points.  Therefore we have,
	\begin{align*}
		& \mathbb{P}\left\{\sup_{\substack{v \in \mathbb{R}^{(e-s)}, \, \|v\|_2 = 1\\ \sum_{t = 1}^{e-s-1} \mathbf{1}\{v_i \neq v_{i+1}\} = m}} \left|\sum_{t = s+1}^e v_t \epsilon_t (X_t)_i\right| > \Delta_{\min} \right\} \\
		\leq & {(e-s-1) \choose m} 9^{m+1} \sup_{v \in \mathcal{N}_{1/4}}	\mathbb{P}\left\{\left|\sum_{t = s+1}^e v_t \epsilon_t (X_t)_i\right| > \delta/2 \right\} \\
		\leq & {(e-s-1) \choose m} 9^{m+1} C\exp \left\{-c \min\left\{\frac{\delta^2}{4C_x^2}, \, \frac{\delta}{2C_x \|v\|_{\infty}}\right\}\right\} \\
		\leq & C(e-s-1)^m 9^{m+1} \exp \left\{-c \min\left\{\frac{\delta^2}{4C_x^2}, \, \frac{\delta}{2C_x \|v\|_{\infty}}\right\}\right\}.
	\end{align*}

\end{proof}

\subsection{Additional Technical Results} 
\bnlem
\label{theorem:restricted eigenvalues 2}
Suppose  $\{X_{i } \}_{1\le i \le n } \overset{i.i.d.} {\sim}  N_p (0, \Sigma )  $. 
Denote $\mathcal C_S  := \{ v : \mathbb R^p : \| v_{S^c }\|_1 \le 3\| v_{S}\|_1 \}  $, where $ |S| \le \s $.
Then there exists constants $c$ and $C$ such that for all $\eta\le 1$, 
\begin{align} 
P \left( \sup_{v \in \mathcal C _S , \|v\|_2=1   } \left | v^\top ( \widehat \Sigma -\Sigma )  v \right| \ge C\eta   \Lambda_{\max} (\Sigma)   \right)
\le 2\exp( -c n\eta ^2 + 2\s \log(p) ) .
\end{align}
\enlem
\begin{proof}
This is a well known restricted eigenvalue property for Gaussian design. The proof can be found  in \cite{Basu2015} or \cite{Loh2012}.
\end{proof}  

\bnlem
\label{corollary:restricted eigenvalues 2} Suppose  $\{X_{i } \}_{1\le i \le n } \overset{i.i.d.} {\sim}  N_p (0, \Sigma )  $. 
 Denote $\mathcal C_S  := \{ v : \mathbb R^p : \| v_{S^c }\|_1 \le 3\| v_{S}\|_1 \}  $, where $ |S| \le \s $.  With probability at least $1- n ^{-5}$, it holds that
$$  \left | v^\top ( \widehat \Sigma_\I  -\Sigma )  v \right| \le C \sqrt { \frac{\s\log(n\vee p) }{|\I| } }     \|v\|_2^2   
 $$
   for all $v \in \mathcal C _S $ and all $\I \subset  (0, n]$ such that $| \I |\ge C_s \s\log(n\vee p)  $, where $ C_s$
is the constant in \Cref{lemma:consistency} which is independent of $n, p$.   
\enlem
\begin{proof} 
For any $\I \subset  (0, n]$ such that $| \I |\ge C_s\s \log(n\vee p)   $, by \Cref{theorem:restricted eigenvalues 2}, it holds that 
\begin{align*} 
P \left( \sup_{v \in \mathcal C _S , \|v\|_2=1   } \left | v^\top ( \widehat \Sigma _\I -\Sigma )  v \right| \ge C\eta   \Lambda_{\max} (\Sigma)   \right)
\le 2\exp( -c   |\I| \eta ^2 + 2\s \log(p) ) .
\end{align*}
Let  $\eta= C_1 \sqrt { \frac{\s\log(n\vee p) }{|\I | } } $ for sufficiently large constant $C_1$. Note that $\eta< 1$ if $ |\I| > C_1^2 \s\log(n\vee p) $. Then   with probability at least 
$(n\vee p)^{-7} $, 
$$\sup_{v \in \mathcal C _S , \|v\|_2=1   } \left | v^\top ( \widehat \Sigma _\I -\Sigma )  v \right| \ge C_2 \sqrt { \frac{s\log(n\vee p) }{|\I |  } } .$$
Since there are at most $n^2$ many different  choices of $ \I \subset (0,n]$, the desired result follows from a union bound argument. 
\end{proof}

\bnlem
\label{lem:restricted eigenvalue}
Under \Cref{assp:dcdp_linear_reg}, it holds that
 \begin{align*}
		 \p \bigg( \sum_{t\in \I } ( X_t^\top v )^2 \ge \frac{c_x|\I| }{4} \|v\|_2^2 - C_2 \log(n\vee p) \|v\|_1 ^2 \ \forall v \in \mathbb R^p  \text{ and } \forall |\I|\ge C_s \s \log(n\vee p)  \bigg)  \le  n^{-5} 
	\end{align*} 
	where $  C_2 > 0$ is an  absolute constant  only depending on $C_x$.
\enlem
\begin{proof}
By the well known restricted eigenvalue condition, for any $\I$, it holds that 
 \begin{align*}
		 \p \bigg( \sum_{t\in \I } ( X_t^\top v )^2 \ge \frac{c_x |\I| }{4} \|v\|_2^2 - C_2 \log(n\vee p) \|v\|_1 ^2  \ \  \forall v \in \mathbb R^p    \bigg)  \le  C_3 \exp(-c_3|\I| ).
	\end{align*} 
  Since $|\I|  \ge  C_s \s \log(n\vee p) $, 
  \begin{align*}
		 \p \bigg( \sum_{t\in \I } ( X_t^\top v )^2 \ge \frac{c_ x |\I| }{4} \|v\|_2^2 - C_2 \log(n\vee p) \|v\|_1^2 \ \ \forall v \in \mathbb R^p    \bigg)  \le  n^{-4}.
	\end{align*} 
	Since there are at most $n^2$ many subinterval $\I \subset (0,n]$, it follows from a union bound argument  that 
	 \begin{align*}
		 \p \bigg( \sum_{t\in \I } ( X_t^\top v )^2 \ge \frac{c_x |\I| }{4} \|v\|_2^2 - C_2 \log(n\vee p) \|v\|_1^2  \ \ \forall v \in \mathbb R^p  \text{ and } \forall |\I|\ge C_s \s \log(n\vee p)  \bigg)  \le  n^{-2} .
	\end{align*} 
\end{proof}

\bnlem \label{lemma:consistency}
Suppose  \Cref{assp:dcdp_linear_reg} holds. There exists  a sufficient large constant $ C_s$ such that the following conditions holds.    
\\
\\
{\bf a.} With probability at least $1-n^{-3} $, it holds that
\begin{align}\label{eq:independent condition 1c 1}
   \left |  \frac{1}{|\I | } \sum_{i \in \mathcal I }  \epsilon_i  X_i^\top \beta \right| \le C\sigma_{\epsilon}\sqrt { \frac{\log(n\vee p)}{|\I |   } } \|\beta\|_1      
 \end{align}
uniformly  for all $ \beta \in \mathbb R^p$ and all $\mathcal I \subset (0, n]$ such that $|\I|\ge C_s \s \log(n\vee p) $, 
\\
\\
 {\bf b.} Let $\{ u_i\}_{i=1}^n \subset  \mathbb R^p$  be a collection of  deterministic vectors. Then with probability at least $1-n^{-3} $, it holds that   
\begin{align}\label{eq:independent condition 1c}
       \left |  \frac{1}{|\I |   } \sum_{i \in \I }    u^\top_i   X_i   X_i^\top \beta  - \frac{1}{ |\I |  } \sum_{i \in \mathcal I }   u^\top  _i  \Sigma \beta  \right| \le C \left(  \max _{1\le i \le n }\|u_i\|_2  \right) \sqrt {  \frac{\log(n\vee p)}{|\I |   } } \|\beta\|_1    
 \end{align}
 uniformly for all $ \beta \in \mathbb R^p$ and all $\mathcal I \subset (0, n]$ such that $|\I|\ge  C_s \s \log(n\vee p) $.    
\enlem 
 \begin{proof}
   The justification   of the  \eqref{eq:independent condition 1c 1}  is   similar and simpler  than the justification  of \eqref{eq:independent condition 1c}. For conciseness, only the justification  of \eqref{eq:independent condition 1c} is presented.
   
For any $\mathcal I \subset (0, n]$  such  that $|\I|\ge  C_s \s \log(n\vee p) $, it holds  that
 \begin{align*}
 &\left|  \frac{1}{ |\I |  } \sum_{i\in \I }  u^\top  _i  X_i   X_i^\top \beta   - \frac{1}{ |\I |  }  \sum_{i\in \I }   u^\top_i  \Sigma  \beta \right| 
 \\
 =
 & \left|  \left(  \frac{1}{|\I |  } \sum_{i\in \I }   u^\top _i    X_i   X_i^\top     - \frac{1}{|\I |  }    \sum_{i\in \I }   u^\top_i  \Sigma  \right) \beta  \right| 
 \\
 \le 
 & \max_{1\le j \le p }  \left |  \frac{1}{|\I |  } \sum_{i\in \I }   u^\top  _i   X_i   X_{i,j}      - \frac{1}{|\I |  }    \sum_{i\in \I }   u^\top_i  \Sigma  (, j)  \right| \|\beta\|_1.
 \end{align*}
 Note that 
 $ E(u^\top  _i  X_i   X_{i,j} ) =u^\top _i \Sigma  (, j)    $ and in addition, 
 $$u^\top _ i   X_i   \sim N(0, u^\top _i  \Sigma u _i )  \quad \text{ and } \quad X_{i,j} \sim N(0, \Sigma({j,j}))  .$$
 So $u^\top _i   X_i X_{i,j} $  is a sub-exponential random variable such that 
 $$ u^\top _i   X_i X_{i,j} \sim SE( u^\top_i  \Sigma u _i  \Sigma ({j,j})).$$
 As a result, for $\gamma<1$ and every $j$,
$$ P \left( \left |  \frac{1}{|\I |  } \sum_{i \in \I }  u^\top _ i     X_i   X_{i,j}      - u^\top \Sigma  (, j)  \right|  \ge \gamma \sqrt {  \max_{1\le i\le n } ( u^\top_i  \Sigma u_i)  \Sigma ( {j,j} ) }\right) \le \exp(-c \gamma^2 |\I |  ). $$
Since 
$$  \sqrt { u^\top _i  \Sigma u _i \Sigma ({j,j} ) } \le C_x  \|u_i \|_2, $$
by union bound,
$$ \p \left(  \max_{1\le j \le p }\left |  \frac{1}{|\I |  } \sum_{i \in \I}  u^\top _ i     X_i   X_{i,j}      - \frac{1}{|\I |  } \sum_{i \in\I}  u^\top _i  \Sigma  (, j)  \right|  \ge \gamma 
C_x  \left(  \max _{1\le i \le n }\|u_i\|_2  \right)  \right) \le p\exp(-c \gamma^2  |\I| ). $$
Let $\gamma =  3\sqrt { \frac{\log(n\vee p) }{c  |\I| }  }$. Note that   $ \gamma<1$ if  $|\I|\ge C_s \s \log(n\vee p) $ for sufficiently large $C_s$.  Therefore 
$$ \p \left(  \max_{1\le j \le p }\left |  \frac{1}{|\I |  } \sum_{i \in \I}  u^\top _ i     X_i   X_{i,j}      - \frac{1}{|\I |  } \sum_{i \in\I}  u^\top _i  \Sigma  (, j)  \right|  \ge  
C_1\sqrt{ \frac{ \log(n\vee p)}{|\I| } } \left(  \max _{1\le i \le n }\|u_i\|_2  \right)  \right) \le p\exp(-9\log(n\vee p)  ). $$
Since there are at most $n^2$ many intervals $\I \subset(0,n]$, it follows that 
\begin{align*}
 &\p \left(  \max_{1\le j \le p }\left |  \frac{1}{|\I |  } \sum_{i \in \I}  u^\top _ i     X_i   X_{i,j}      - \frac{1}{|\I |  } \sum_{i \in\I}  u^\top _i  \Sigma  (, j)  \right|  \ge  
C_1\sqrt{ \frac{ \log(n\vee p)}{|\I| } } \left(  \max _{1\le i \le n }\|u_i\|_2   \right) \ \forall    |\I| \ge C_s \s \log(n\vee p)  \right)
\\
 \le & p n^2 \exp(-9\log(n\vee p)  ) \le n^{-3}. 
\end{align*}
This immediately  gives \eqref{eq:independent condition 1c}.
 \end{proof}

 \bnlem\label{lem-x-bound}
Uder \Cref{assp:dcdp_linear_reg}, for any interval $\I \subset (0, n]$, for any 
	\[
		\lambda \geq \lambda_1 := C_{\lambda}\sigma_{\epsilon} \sqrt{\log(n p)},
	\]
	where $C_{\lambda} > 0$ is a large enough absolute constant, it holds with probability at least $1 - n^{-5}$ that
	\[
    \|\sum_{i \in \I} \epsilon_i X_i \|_{\infty} \leq \lambda \sqrt{\max\{|\I|, \, \log(n p)\}}/8,
	\]
	where $c_3 > 0$ is an absolute constant depending only on the distributions of covariants $\{X_i\}$ and $\{\epsilon_i\}$.
\enlem

\begin{proof}
	Since $\epsilon_i$'s are sub-Gaussian random variables and $X_i$'s are sub-Gaussian random vectors, we have that $\epsilon_i X_i$'s are sub-Exponential random vectors with $\|\epsilon_iX_i\|_{\psi_1}\leq C_x \sigma_{\epsilon}$ \citep[see e.g.~Lemma~2.7.7 in][]{vershynin2018high}. It then follows from Bernstein's inequality \citep[see e.g.~Theorem~2.8.1 in][]{vershynin2018high} that for any $t > 0$,
		\[
	\mathbb{P}\left\{\|\sum_{i \in \I} \epsilon_i X_i \|_{\infty} > t\right\} \leq 2p \exp\left\{-c \min\left\{\frac{t^2}{|\I|C_x^2\sigma^2_{\epsilon}}, \, \frac{t}{C_x \sigma_{\epsilon}}\right\}\right\}.
		\]
		Taking 
		\[
			t = C_{\lambda}C_x/4 \sigma_{\epsilon} \sqrt{\log(n p)} \sqrt{\max\{|\I|, \, \log(n p)\}}
		\] 
		yields the conclusion.
\end{proof}

\bnlem \label{lemma:beta bounded 1}
Suppose   \Cref{assp:dcdp_linear_reg} holds. Let $\I \subset [1,n] $. Denote 
 $ \kappa = \min_{k\in \{1,\ldots, K\} } \kappa_k,$ 
where $ \{ \kappa_k\}_{k=1}^K$ are defined in  \Cref{assp:dcdp_linear_reg}. Then for any $i\in [T]$,
   $$\|\beta^*_\I - \beta^*_i\|_2  \le C\kappa  \le CC_\kappa,$$
  for some absolute constant $C$ independent of $n$.
\enlem 
 \begin{proof}
 It suffices to consider $\I =[1,n]$  and  $\beta_i^*= \beta_1^*$ as the general case is similar. Observe that  
 \begin{align*}
  \| \beta^*_{[1,n]}  - \beta_ 1^*  \|_2  =& 
  \|  \frac{1}{n} \sum_{i=1}^n \beta_i ^*  - \beta_1^*  
   \|_2 
 = \|  \frac{1}{n} \sum_{k=0}^{K} \Delta_k \beta_{\eta_k+ 1} ^*   -    \frac{1}{n}\sum_{k=0}^{K} \Delta_k  \beta_1 ^*
  \|_2  
  \\
 \le 
 &    \frac{1}{n} \sum_{k=0}^{K} \left\| \Delta_k  (\beta_{\eta_k+ 1} ^* -\beta_1 ^* ) \right\|_2    \le 
 \frac{1}{n}\sum_{k=0}^{K} \Delta_k  (K+1)  \kappa   \le  (K+1)\kappa . 
  \end{align*}
  By \Cref{assp:dcdp_linear_reg}, both $\kappa $ and $K $ bounded above. 
 \end{proof}
 
 \bnlem\label{eq:cumsum upper bound}
 Let $t\in \I =(s,e] \subset [1,n]$. Denote 
 $ \kappa_{\max}= \max_{k\in \{1,\ldots, K\} } \kappa_k,$ 
where $ \{ \kappa_k\}_{k=1}^K$ are defined in  \Cref{assp:dcdp_linear_reg}.  Then   
 $$ \sup_{0< s < t<e\le n  }\| \beta_{ (s, t]}  ^* - \beta_{(t,e] }  ^* \|_2  \le C\kappa  \le CC_\kappa. $$ 
for some absolute constant $C$ independent of $n$. 
 \enlem
 \begin{proof}
 It suffices to consider $(s,e]=(0,n]$, as the general case is similar.  
 Suppose that $\eta_{q} <t \le \eta_{q+1} $.
 Observe that 
 \begin{align*}
&  \| \beta^*_{(1,t]}  - \beta ^* _{(t,n]}  \|_2  
  \\
  =& 
  \left\|  \frac{1}{t } \sum_{i=1}^t \beta_i ^*  - \frac{1}{ n- t } \sum_{i=t+1}^n \beta_i ^*
  \right \|_2 
 \\
 = & \left\|  \frac{1}{t}  \left( \sum_{k=0}^{q-1} \Delta_k \beta_{\eta_k+ 1}  ^*  + (t-\eta_{q} )\beta^*_{\eta_q + 1} \right) -    \frac{1}{n-t } \left( \sum_{k=q+1}^{K}  \Delta_k \beta_{\eta_k+ 1} ^*   + ( \eta_{q+1}-t  )\beta^*_{\eta_q + 1} \right) 
  \right \|_2  
  \\
=
 & \left\|  \frac{1}{t}  \left( \sum_{k=0}^{q-1} \Delta_k (\beta ^* _{\eta_k+ 1}-\beta^*_{\eta_q+1} ) \right)  +  \beta^*_{\eta_q + 1} -    \frac{1}{n-t } \left( \sum_{k=q+1}^{K}  \Delta_k (  \beta ^* _{\eta_k+ 1}  -\beta ^* _{\eta_q+1} ) \right)   -  \beta^*_{\eta_q + 1} 
  \right \|_2 
  \\
  =
  &\left\|  \frac{1}{t}  \left( \sum_{k=0}^{q-1} \Delta_k (\beta_{\eta_k+ 1} ^* -\beta^*_{\eta_q+1} ) \right)    -    \frac{1}{n-t } \left( \sum_{k=q+1}^{K}  \Delta_k (  \beta_{\eta_k+ 1} ^*   -\beta_{\eta_q+1} ^*  ) \right)   
  \right \|_2 
  \\
  \le 
  &\frac{1}{t} \sum_{k=0}^{q-1} \Delta_k    K \kappa      +\frac{1}{n-t} \sum_{k=q+1}^{K} \Delta_k    K  \kappa 
  \le 2K\kappa  . 
  \end{align*}
  
 \end{proof}

\clearpage

\section{Gaussian graphical model}
\label{sec: main proof covariance}

In this section, we will present the proof of \Cref{thm:DCDP covariance main}. Throughout this section, we use $\Sigma$ for covariance matrices and $\Omega$ for precision matrices. For any generic interval $\I\subset [1,n]$, denote $\Omega^*_{\I} = \frac{1}{|\I|}\sum_{i\in \I}\Omega^*_i$ and
$$ \widehat{\Omega}_{\I} = \argmin_{\Omega\in \mathbb{S}^p_+} \sum_{i \in \I } {\rm Tr}[{\Omega}^\top X_iX_i^\top] - |\I|\log|\Omega|. $$
Also, unless specially mentioned, in this section, we set the goodness-of-fit function $\mathcal F  (\I)$ in \Cref{algorithm:DCDP} to be
\begin{equation}
    \mathcal F  (\widehat{\Omega}_\I,\I) 
     = \begin{cases}  0  
 & \text{ if } |\I|< C_\mclF p\log(p\vee n);\\
       \sum_{i \in \I } {\rm Tr}[\widehat{\Omega}^\top X_iX_i^\top] - |\I|\log|\widehat{\Omega}|   
       &\text{otherwise}.
    \end{cases}
\end{equation}
where $C_\mclF$ is a universal constant.

\paragraph{Additional notations.} Before presenting more details on Gaussian graphical model, we introduce some additional notations while reviewing some notations we used in the main text. We use $\mathbb{S}^p_+$ to denote the cone of positive semidefinite matrices in $\mathbb{R}^{p\times p}$. For a matrix $A\in \mathbb{R}^{m\times n}$, we use $\|A\|_F:=\sqrt{\sum_{i\in [m]}\sum_{j\in [n]}A_{ij}^2}$ to denote its Forbenius norm, $\|A\|_{op} = \sup_{v\in \mathbb{R}^p}\|Av\|_2/\|v\|_2$ as its operator norm, and ${\rm Tr}(A) = \sum_{i\in [m\wedge n]}A_{ii}$ to denote its trace. For a square matrix $A\in \mathbb{R}^{n\times n}$, denote its determinant by $|A|$. For two matrices $A, B\in\mathbb{R}^{p\times p}$, $A\preceq B$ means that $B - A\in \mathbb{S}_+^p$. For a random vector $X\in \mathbb{R}^p$, we denote $g_X$ as the subgaussian norm \citep{vershynin2018high}: $g_X:=\sup \{\|v^\top X\|_{\psi_2}: v\in \mathbb{R}^p, \|v\|_2=1\}$.

\paragraph{Assumptions.} For the ease of presentation, we combine the SNR condition we will use throughout this section and \Cref{assp:DCDP_covariance main} into a single assumption. Besides, we would like to point out that although we assume that $\{X_i\}_{i\in[n]}$ are Gaussian vectors in \Cref{assp:DCDP_covariance main}, it is actually only compulsory for the proof of the conquer step. Throughout this section for the divide step, it suffices to assume that $\{X_i\}_{i\in[n]}$ are subgaussian vectors with bounded Orlicz norm $\sup_{i\in[n]}\|X_i\|_{\psi_2}\leq g_X<\infty$ where $g_X$ is some absolute constant. Thus, we keep $g_X$ in all results in this section, although when $\{X_i\}_{i\in[n]}$ are Gaussian it holds that $g_X = C_X$.

\bnassum[Gaussian graphical model] 
\label{assp:DCDP_covariance}
Suppose that \Cref{assp:DCDP_covariance} holds. In addition, suppose that $ \Delta_{\min} \kappa^2  \geq \mclB_n   p^2\log^2(n\vee p)$ as is assumed in \Cref{thm:DCDP covariance main}.
\enassum

\bnprop\label{prop:covariance local consistency}
 Suppose  \Cref{assp:DCDP_covariance} holds. Let 
 $\widehat { \mathcal P }  $ denote the output of \Cref{algorithm:DP}. 
 Then with probability  at least $1 - C n ^{-3}$, the following conditions hold.
    \begin{itemize}
        \item [(i)] For each interval  $ \I = (s, e] \in \widehat{\mathcal{P}}$    containing one and only one true 
           change point $ \eta_k $, it must be the case that
 $$\min\{ \eta_k -s ,e-\eta_k \}  \lesssim  C_{\gamma} g_X^4\frac{C_X^2}{c_X^6}\frac{p^2\log(n\vee p)}{\kappa_k^2} +g_X^4\frac{C_X^6}{c_X^6}\tr .$$
        \item [(ii)] For each interval  $  \I = (s, e] \in \widehat{\mathcal{P}}$ containing exactly two true change points, say  $\eta_ k  < \eta_ {k+1} $, it must be the case that
            \begin{align*} 
 \eta_k -s    \lesssim \htr    \quad \text{and} \quad 
 e-\eta_{k+1}  \lesssim  \htr. 
 \end{align*} 
         
\item [(iii)] No interval $\I \in \widehat{\mathcal{P}}$ contains strictly more than two true change points; and

    \item [(iv)] For all consecutive intervals $ \I_1 $ and $ \I_2 $ in $\widehat{ \mathcal P}$, the interval 
        $ \I_1 \cup  \I_2 $ contains at least one true change point.
                
     \end{itemize}
\enprop 
\bprf
The four cases are proved in \Cref{lem:cov single cp}, \Cref{lem:cov two cp}, \Cref{lem:cov three or more cp}, and \Cref{lem:cov two consecutive interval}, respectively.
\eprf

\bnprop\label{prop:cov change points partition size consistency}
Suppose  \Cref{assp:DCDP_covariance} holds. Let 
 $\widehat { \mathcal P }  $ be the output of \Cref{algorithm:DP}. Suppose 
$ \gamma \ge C_\gamma K\tr \kappa^2 $  for sufficiently large constant $C_\gamma$. Then
 with probability  at least $1 - C n ^{-3}$,  $| \widehat { \mathcal P} | =K $.
\enprop

\begin{proof}[Proof of   \Cref{prop:cov change points partition size consistency}] 
Denote $\mathfrak{G} ^*_n = \sum_{ i =1}^n [{\rm Tr}[(\Omega_i^*)^\top X_iX_i^\top] - \log|\Omega_i^*|]$. Given any collection $\{t_1, \ldots, t_m\}$, where $t_1 < \cdots < t_m$, and $t_0 = 0$, $t_{m+1} = n$, let 
\begin{equation}\label{eq: cov sn-def}
    \G _n(t_1, \ldots, t_{m}) = \sum_{k=1}^{m} \sum_{ i = t_k +1}^{t_{k+1}} \mclF(\widehat {\Omega}_{(t_{k}, t_{k+1}]}, (t_{k}, t_{k+1}]).
\end{equation}
For any collection of time points, when defining \eqref{eq: cov sn-def}, the time points are sorted in an increasing order.

Let $\{ \widehat \eta_{k}\}_{k=1}^{\widehat K}$ denote the change points induced by $\widehat {\mathcal P}$.  Suppose we can justify that 
    \begin{align}
        \G^*_n + K\gamma  \ge  &\G _n(s_1,\ldots,s_K)   + K\gamma - C(K + 1)\frac{g_X^2}{c_{X}^4}\frac{ p^2\log(n\vee p)}{\kappa^2} - C\sum_{k\in [K]}\kappa_k^2\mclB_n^{-1}\Delta_{\min} \label{eq:cov K consistency step 1} \\ 
        \ge & \G_n (\widehat \eta_{1},\ldots, \widehat \eta_{\widehat K } ) +\widehat K \gamma - C_1'(K + 1)\frac{g_X^2}{c_{X}^4}\frac{ p^2\log(n\vee p)}{\kappa^2} - C_1'\sum_{k\in [K]}\kappa_k^2\mclB_n^{-1}\Delta_{\min}  \label{eq:cov K consistency step 2} \\ 
        \ge &  \G _n ( \widehat \eta_{1},\ldots, \widehat \eta_{\widehat K } , \eta_1,\ldots,\eta_K ) + \widehat K \gamma    - C_1(K + 1)\frac{g_X^2}{c_{X}^4}\frac{ p^2\log(n\vee p)}{\kappa^2} - C_1\sum_{k\in [K]}\kappa_k^2\mclB_n^{-1}\Delta_{\min} \label{eq:cov K consistency step 3}
    \end{align}
    and that 
    \begin{align}\label{eq:cov K consistency step 4}
        \G ^*_n   -\G _n ( \widehat \eta_{1},\ldots, \widehat \eta_{\widehat K } , \eta_1,\ldots,\eta_K ) \le C_2 (K + \widehat{K} + 2) \frac{g_X^4}{c_X^2} p^2\log(n\vee p).
    \end{align}
    Then it must hold that $| \widehat{\mclP} | = K$, as otherwise if $\widehat K \geq K+1 $, then  
    \begin{align*}
        C _2 (K + \widehat{K} + 2)  \frac{g_X^4}{c_X^2} p^2\log(n\vee p)  & \ge   \G ^*_n  -\G _n ( \widehat \eta_{1},\ldots, \widehat \eta_{\widehat K } , \eta_1,\ldots,\eta_K ) \\
        & \ge      (\widehat K - K)\gamma  -C_1 ( K +1)  \frac{g_X^4}{c_X^2} p^2\log(n\vee p).
    \end{align*} 
    Therefore due to the assumption that $| \widehat{\mclP}|=\widehat K\leq 3K $, it holds that 
    \begin{align} \label{eq:cov Khat=K} 
        [C_2 (4K + 2) + C_1(K+1)] \frac{g_X^4}{c_X^2} p^2\log(n\vee p) \geq  (\widehat K - K)\gamma \geq \gamma.
    \end{align}
    Note that \eqref{eq:cov Khat=K} contradicts the choice of $\gamma$. Therefore, it remains to show \Cref{eq:cov K consistency step 1} to \Cref{eq:cov K consistency step 4}.

{\bf Step 1.} \Cref{eq:cov K consistency step 1} holds because $\widehat{\Omega}_{\I}$ is (one of) the minimizer of $\mclF(\Omega, \I)$ for any interval $\I$.

{\bf Step 2.} \Cref{eq:cov K consistency step 2} is guaranteed by the definition of $\widehat{\mclP}$.

{\bf Step 3.} For every $ \I  =(s,e]\in \widehat{\mclP}$, by \Cref{prop:covariance local consistency}, we know that $\I$ contains at most two change points. We only show the proof for the two-change-points case as the other case is easier. Denote
    \[
         \I =  (s ,\eta_{q}]\cup (\eta_{q},\eta_{q+1}] \cup  (\eta_{q+1} ,e]  = \J_1 \cup \J_2 \cup \J_{3},
    \]
where $\{ \eta_{q},\eta_{q+1}\} =\I \, \cap \, \{\eta_k\}_{k=1}^K$. 

For each $ m=1,2,3$, by definition it holds that
\begin{equation}
    \mclF(\widehat\Omega_{\J_m},\J_m)\leq \mclF(\Omega^*_{\J_m},\J_m).
\end{equation}
On the other hand, by \Cref{lem:cov loss deviation no change point}, we have
$$
\mclF(\widehat\Omega_{\I},\J_m)   \geq \mclF(\Omega^*_{\J_m},\J_m) - C\frac{g_X^4}{c_X^2}p^2\log(n\vee p). 
$$
Therefore the last two inequalities above imply that 
\begin{align} 
\sum_{i \in \I }  \mclF(\widehat \Omega_\I,\I) \geq & \sum_{m=1}^{3} \sum_{ i \in \J_m }\mclF(\widehat \Omega_\I,\J_m) \nonumber \\
\geq & \sum_{m=1}^{3}   \sum_{ i \in \J_m}\mclF(\widehat \Omega_{\J_m},\J_m) - C\frac{g_X^4}{c_X^2}p^2\log(n\vee p). \label{eq:cov K consistency step 3 inequality 3}
\end{align}
Then \eqref{eq:cov K consistency step 3} is an immediate consequence of \eqref{eq:cov K consistency step 3 inequality 3}.

{\bf Step 4.}
Finally, to show \eqref{eq:cov K consistency step 4},   let  $\widetilde { \mathcal P}$ denote    the partition induced by $\{\widehat \eta_{1},\ldots, \widehat \eta_{\widehat K } , \eta_1,\ldots,\eta_K\}$. Then 
$| \widetilde { \mathcal P} | \le K + \widehat K+2   $ and that $\Omega^*_i$ is unchanged in every interval $\I \in \widetilde { \mathcal P}$. So \Cref{eq:cov K consistency step 4} is an immediate consequence of \Cref{lem:cov no cp}.
\end{proof}

\subsection{Fundamental lemmas}

\bnlem[Deviation, Gaussian graphical model]
\label{lem:cov one change deviation bound}
Let $\mathcal I =(s,e] $ be any generic interval, and define the loss function $\mclF(\Omega, \I)  = \sum_{i\in \I}{\rm Tr}[\Omega^\top (X_iX_i^\top)] - |\I|\log|\Omega|$.  Define $\widehat{\Omega}_{\I} = \argmin_{\Omega\in \mathbb{S}_+}L(\Omega,\I)$ and $\mclF^*(\Omega^*, \I) = \sum_{i\in \I}[{\rm Tr}((\Omega_i^*)^\top (X_iX_i^\top)) - \log|\Omega_i^*|]$.

{\bf a.} If $\I$ contains no change points. Then it holds that 
$$\mathbb{P} \bigg( |   \mclF(\widehat{\Omega}_{\I},\I)  - \mclF^*(\Omega^*, \I) |  \ge C \frac{g_X^4}{c_X^2}p^2\log(n\vee p)  \bigg)  \le  (n\vee p)^{-3}. $$

{\bf b.}  Suppose that the interval $ \I=(s,e]$ contains one and only one change point $ \eta_k $. Denote
$$  \mathcal J = (s,\eta_k] \quad \text{and} \quad  \mathcal J' =  (\eta_k, e].$$ 
Then it holds that 
$$\mathbb{P} \bigg( | \mclF(\widehat{\Omega}_{\I},\I) - \mclF^*(\I)  | \geq \frac{C_X^2p}{c_X^8}\frac{|\J||\J '| }{|\I|}\kappa_k^2 +  C\frac{g_X^4 C_X^2}{c_X^4}p^2\log(n\vee p) \bigg)   \le  (n\vee p)^{-3}. $$
\enlem
\begin{proof} 
We show {\bf b} as {\bf a} immediately follows from {\bf b} with $ |\J'| =0$. Denote 
$$  \mathcal J = (s,\eta_k] \quad \text{and} \quad  \mathcal J' =  (\eta_k, e]  .$$ 
Let $\widetilde{\Omega}_{\I} = (\frac{1}{|\I|}\sum_{i\in\I}\Sigma_i^*)^{-1}$.
Then by Taylor expansion and \Cref{lem:estimation covariance}, we have
\begin{align}
    & |\mclF(\widehat{\Omega}_{\I},\I)-\mclF(\widetilde{\Omega}_{\I},\I)| \nonumber \\
    \leq & |{\rm Tr}[(\widehat{\Omega}_{\I} - \widetilde{\Omega}_{\I})^\top (\sum_{i\in\I} X_iX_i^\top - |\I|\widetilde{\Omega}_{\I}^{-1})]| + \frac{C_X^2}{2}|\I|\|\widehat{\Omega}_{\I} - \widetilde{\Omega}_{\I}\|_F^2 \nonumber \\
    \leq & |\I|\|\widehat{\Omega}_{\I} - \widetilde{\Omega}_{\I}\|_F \|\widehat{\Sigma}_{\I} - \widetilde{\Omega}_{\I}^{-1}\|_F + \frac{C_X^2}{2}|\I|\|\widehat{\Omega}_{\I} - \widetilde{\Omega}_{\I}\|_F^2 \nonumber\\
    \leq & C\frac{g_X^4}{c_X^2}p^2\log(n\vee p) + C\frac{g_X^4 C_X^2}{c_X^4}p^2\log(n\vee p)\leq C\frac{g_X^4 C_X^2}{c_X^4}p^2\log(n\vee p). \label{eq:cov mixed deviation 1}
\end{align}
On the other hand, it holds that
\begin{align}
    & |\mclF(\widetilde{\Omega}_{\I},\I)-\mclF^*(\Omega^*, \I)| \nonumber \\
    \leq & |{\rm Tr}[(\widetilde{\Omega}_{\I} - {\Omega}_{\J})^\top (\sum_{i\in\J} X_iX_i^\top - |\J|\Omega_{\J}^{-1})]| + \frac{C_X^2}{2}|\J|\|\widetilde{\Omega}_{\I} - {\Omega}_{\J}\|_F^2 \nonumber\\
     & + |{\rm Tr}[(\widetilde{\Omega}_{\I} - {\Omega}_{\J '})^\top (\sum_{i\in\J'} X_iX_i^\top - |\J '|\Omega_{\J '}^{-1})]| + \frac{C_X^2}{2}|\J '|\|\widetilde{\Omega}_{\I} - {\Omega}_{\J '}\|_F^2.\label{eq:cov mixed deviation 2}
\end{align}
To bound $\|\widetilde{\Omega}_{\I} - {\Omega}_{\J}\|_F$ and $\|\widetilde{\Omega}_{\I} - {\Omega}_{\J '}\|_F$, notice that for two positive definite matrices $\Sigma_1,\Sigma_2\in \mathbb{S}_+$ and two positive numbers $w_1,w_2$ such that $w_1 + w_2 = 1$, we have
\begin{align*}
    & \|(w_1\Sigma_1 + w_2\Sigma_2)^{-1} - \Sigma_1^{-1}\|_F\\
    \leq &\sqrt{p}\|(w_1\Sigma_1 + w_2\Sigma_2)^{-1} - \Sigma_1^{-1}\|_{op}\\
    = &\sqrt{p}\|(w_1\Sigma_1 + w_2\Sigma_2)^{-1}[\Sigma_1 - (w_1\Sigma_1 + w_2\Sigma_2)]\Sigma_1^{-1}\|_{op}\\
    \leq & \sqrt{p}\|(w_1\Sigma_1 + w_2\Sigma_2)^{-1}\|_{op}\|\Sigma_1 - (w_1\Sigma_1 + w_2\Sigma_2)\|_{op}\|\Sigma_1^{-1}\|_{op}\\
    \leq & \|(w_1\Sigma_1 + w_2\Sigma_2)^{-1}\|_{op}\|\Sigma_1^{-1}\|_{op}\cdot\sqrt{p}w_2\|\Sigma_1 - \Sigma_2\|_{op} \\
    \leq & \|(w_1\Sigma_1 + w_2\Sigma_2)^{-1}\|_{op}\|\Sigma_1^{-1}\|_{op}\|\Sigma_1\|_{op}\|\Sigma_2\|_{op}\cdot\sqrt{p}w_2\|\Sigma_1^{-1} - \Sigma_2^{-1}\|_{op}.
\end{align*}
Therefore, under \Cref{assp:DCDP_covariance}, it holds that
\begin{equation}
    \|\widetilde{\Omega}_{\I} - {\Omega}_{\J}\|_F\leq \frac{C_X^2}{c_X^2}\frac{|\J '|}{|\I|}\sqrt{p}\kappa_k, \|\widetilde{\Omega}_{\I} - {\Omega}_{\J '}\|_F\leq \frac{C_X^2}{c_X^2}\frac{|\J|}{|\I|}\sqrt{p}\kappa_k,
\end{equation}
where in the second inequality we use the fact that $2ab\leq a^2+ b^2$. As a consequence, \Cref{eq:cov mixed deviation 2} can be bounded as
\begin{align}
     & |\mclF(\widetilde{\Omega}_{\I},\I)-\mclF^*(\Omega^*, \I)| \nonumber \\
    \leq & C\frac{g_X^2p\sqrt{\log(n\vee p)}}{c_X^2C_X^{-2}}\kappa_k(\frac{|\J '||\J|^{\frac{1}{2}}}{|\I|} + \frac{|\J '|^{\frac{1}{2}}|\J|}{|\I|}) + \frac{C_X^6 p}{2c_X^4}\kappa_k^2(\frac{|\J||\J '|^2 }{|\I|^2} + \frac{|\J '||\J|^2 }{|\I|^2}) \nonumber\\
    \leq & C\frac{g_X^4p^2{\log(n\vee p)}}{C_X^2} + \frac{C_X^6p}{c_X^4}\frac{|\J||\J '| }{|\I|}\kappa_k^2.\label{eq:cov mixed deviation 3}
\end{align}
Combine \Cref{eq:cov mixed deviation 1} and \Cref{eq:cov mixed deviation 3} and we can get
\begin{equation*}
    \begin{split}
        |\mclF(\widehat{\Omega}_{\I},\I) - \mclF^*(\Omega^*,\I)|
        \leq & |\mclF(\widehat{\Omega}_{\I},\I)-\mclF({\Omega}_{\I},\I)| +|\mclF({\Omega}_{\I},\I) - \mclF^*(\I)|\\
        \leq& C\frac{g_X^4 C_X^2}{c_X^4}p^2\log(n\vee p) + \frac{C_X^6 p}{c_X^4}\frac{|\J||\J '| }{|\I|}\kappa_k^2.
    \end{split}
\end{equation*}
\end{proof} 

\bnrmk
It can be seen later that the $p$ factor in the signal term $\frac{C_X^6 p}{c_X^4}\frac{|\J||\J '| }{|\I|}\kappa_k^2$ will require an additional $p$ factor in the number of points in the grid for DCDP, leading to an additional $p^2$ factor in the computation time.

This factor is hard to remove because it is rooted in the approximation error
$$
\|(w_1\Sigma_1 + w_2\Sigma_2)^{-1} - \Sigma_1^{-1}\|_F.
$$
We can try another slightly neater way of bounding this term. As is mentioned in \cite{kereta2021ejs}, for two matrices $\mathbf{G}, \mathbf{H}\in \mathbb{R}^{d_1\times d_2}$, it holds that
$$
\left\|\mathbf{H}^{\dagger}-\mathbf{G}^{\dagger}\right\|_F \leq \min \left\{\left\|\mathbf{H}^{\dagger}\right\|_{op}\left\|\mathbf{G}^{\dagger} (\mathbf{H}-\mathbf{G})\right\|_F,\left\|\mathbf{G}^{\dagger}\right\|_{op}\left\|\mathbf{H}^{\dagger} (\mathbf{H}-\mathbf{G})\right\|_{op}\right\},
$$
if $\operatorname{rank}(\mathbf{G})=\operatorname{rank}(\mathbf{H})=\min \left\{d_1, d_2\right\}$. Therefore, we have
\begin{align*}
    \|(w_1\Sigma_1 + w_2\Sigma_2)^{-1} - \Sigma_1^{-1}\|_F\leq & \|(w_1\Sigma_1 + w_2\Sigma_2)^{-1}\|_{op}\|\Sigma_{1}^{-1}(w_1\Sigma_1 + w_2\Sigma_2 - \Sigma_1)\|_F\\
    \leq & \|(w_1\Sigma_1 + w_2\Sigma_2)^{-1}\|_{op}\|\Sigma_{1}^{-1}\|w_2\|\Sigma_2 - \Sigma_1\|_F.
\end{align*}
However, to relate $\|\Sigma_2 - \Sigma_1\|_F$ to $\|\Sigma_2^{-1} - \Sigma_1^{-1}\|_F$, we need to proceed in the following way:
\begin{align*}
    \|\Sigma_2 - \Sigma_1\|_F\leq & \sqrt{p}\|\Sigma_2 - \Sigma_1\|_{op} \\
    \leq & \|\Sigma_1\|_{op}\|\Sigma_2\|_{op} \cdot \sqrt{p}\|\Sigma_2^{-1} - \Sigma_1^{-1}\|_{op}\\
    \leq & \|\Sigma_1\|_{op}\|\Sigma_2\|_{op} \cdot \sqrt{p}\|\Sigma_2^{-1} - \Sigma_1^{-1}\|_F,
\end{align*}
which leads to the same bound in \Cref{lem:cov one change deviation bound}.
\enrmk

\bnlem
\label{lem:estimation covariance}
Let $\{X_i\}_{i\in [n]}$ be a sequence of subgaussian vectors in $\mathbb{R}^{p}$ with Orlicz norm upper bounded by $g_X<\infty$. Suppose $\mathbb{E}[X_i] = 0$ and $\mathbb{E}[X_iX_i^\top] = \Sigma_i$ for $i\in [n]$. Consider the change point setting in \Cref{assp:DCDP_covariance} and consider a generic interval $\I\subset [1,n]$. Let $\widehat{\Sigma}_{\I} = \frac{1}{|\I|}\sum_{i \in \I} X_i X_i^\top$ and $\Sigma_{\I} = \frac{1}{|\I|}\sum_{i\in \I}\Sigma^*_i$. Then for any $u>0$, it holds with probability at least $1 - \exp(-u)$ that
\begin{equation}
    \|\widehat{\Sigma}_{\I} - \Sigma_{\I}\|_{op}\lesssim g_X^2(\sqrt{\frac{p + u}{|\mclI|}}\vee \frac{p + u}{|\mclI|}).
    \label{eq: cov concentration 1}
\end{equation}
As a result, when $n\geq C_s p\log(n\vee p)$ for some universal constant $C_s>0$, it holds with probability at least $1 - (n\vee p)^{-7}$ that
\begin{equation}
    \|\widehat{\Sigma}_{\I} - \Sigma_{\I}\|_{op}\leq Cg_X^2\sqrt{\frac{p\log(n\vee p)}{|\I|}},
    \label{eq: cov concentration 2}
\end{equation}
where $C$ is some universal constant that does not depend on $n, p, g_X$, and $C_s$. In addition let $\widehat{\Omega}_{\I} = \argmin_{\Omega\in \mathbb{S}_+}L(\Omega,\I)$ and $\widetilde{\Omega}_{\I} = (\frac{1}{|\I|}\sum_{i\in\I}\Sigma^*_i)^{-1}$. if $|\I|\geq C_s p\log(n\vee p)g_X^4 / c_X^2$ for sufficiently large constant $C_s>0$, then it holds with probability at least $1 - (n\vee p)^{-7}$ that
\begin{equation}
    \|\widehat{\Omega}_{\I} -\widetilde{\Omega}_{\I}\|_{op}\leq C \frac{g_X^2}{c_X^2}\sqrt{\frac{p\log(n\vee p)}{|\I|}}.
    \label{eq: precision concentration}
\end{equation}
\enlem
\bprf
If there is no change point in $\mclI$, then the two inequalities \eqref{eq: cov concentration 1} and \eqref{eq: cov concentration 2} are well-known results in the literature, see, e.g., \cite{kereta2021ejs}. Otherwise, suppose $\mclI$ is split by change points into $q$ subintervals $\mclI_1,\cdots, \mclI_q$. By \Cref{assp:DCDP_covariance}, we know that $q\leq C$ for some constant $C<\infty$. Thus with probability at least $1 - \exp(-u)$,
\begin{align*}
    \|\widehat{\Sigma}_{\I} - \Sigma_{\I}\|_{op} &\leq \|\frac{1}{|\mclI|}\sum_{k\in [q]}|\mclI_k|(\widehat{\Sigma}_{\I_k} - \Sigma_{\I_k})\|_{op}\\
    & \leq C_1g_X^2\frac{\sqrt{p + u}}{|\mclI|}\sum_{k\in [q]}\sqrt{|\mclI_k|\vee (p + u)} \\
    &\leq C_2g_X^2\frac{\sqrt{p + u}}{|\mclI|}\max_{k\in [q]}{|\mclI_k|\vee (p + u)}\\
    &\leq C_2g_X^2\sqrt{\frac{{p + u}}{|\mclI|}}(\sqrt{\frac{\max_{k\in [q]}{|\mclI_k|}}{|\mclI|}}\vee \sqrt{\frac{{p + u}}{|\mclI|}}) \leq C_2 g_X^2(\sqrt{\frac{{p + u}}{|\mclI|}}\vee \frac{{p + u}}{|\mclI|}).
    \end{align*}
It is then straightforward to see that \Cref{eq: cov concentration 2} holds with probability at least $1 - (n\vee p)^{-7}$ when $n\geq C_sp\log(n\vee p)$ for some sufficiently large constant $C_s>0$.

For \Cref{eq: precision concentration}, first vanish the gradient of the loss function $L(\Omega, \I)$ and we get
\begin{equation*}
    \widehat{\Omega}_{\I} = (\widehat{\Sigma}_{\I})^{\dagger}.
\end{equation*}
Then \Cref{eq: precision concentration} is implied by \Cref{eq: cov concentration 2} and the well-known property that
$$
\left\|\mathbf{H}^{\dagger}-\mathbf{G}^{\dagger}\right\|_{op} \leq C\max \left\{\|\mathbf{G}^{\dagger}\|^2_{op},\|\mathbf{H}^{\dagger}\|^2_{op}\right\}\left\|\mathbf{H}-\mathbf{G}\right\|_{op},
$$
for two matrices $G,H\in \mathbb{R}^{p\times p}$.
\eprf



\subsection{Technical lemmas}

\bnlem[No change point]
\label{lem:cov no cp}
For interval $\mclI$ containing no change point, it holds with probability at least $1 - n^{-5}$ that
\begin{equation}
    \mclF(\widehat{\Omega}_{\I}, \mclI) -\mclF({\Omega}^*, \mclI) \geq -g_X^4 p^2\log(n\vee p)\max_{k\in[K + 1]}\|\Omega^*_{\eta_k}\|_{op}^2.
\end{equation}
\enlem
\bprf
If $\I< C_s\frac{g_X^4}{c_X^2}p\log(n\vee p)$, then $\mclF (\widehat{\Omega}_{\I}, \mclI) = \mclF ({\Omega}^*, \mclI)=0$ and the conclusion holds automatically. If $\I\geq C_s\frac{g_X^4}{c_X^2}p\log(n\vee p)$, then by \Cref{lem:estimation covariance}, it holds with probability at least $1 - n^{-7}$ that
\begin{align}
\mclF(\widehat{\Omega}_{\I}, \mclI) -\mclF({\Omega}^*, \mclI) \geq & |\mclI|{\rm Tr}[(\widehat{\Omega}_{\I} - \Omega^*)^\top (\widehat{\Sigma}_{\I} - \Sigma^*)] + \frac{c|\mclI|}{2\|\Omega^*\|_{op}^2}\|\widehat{\Omega}_{\I} - \Omega^*\|_F^2.    \\
\geq & -|\mclI|\|\widehat{\Omega}_{\I} - \Omega^*\|_F \|\widehat{\Sigma}_{\I} - \Sigma^*\|_F \\
\geq & -|\mclI|p \|\widehat{\Omega}_{\I} - \Omega^*\|_{op} \|\widehat{\Sigma}_{\I} - \Sigma^*\|_{op}\\
\geq & -g_X^4 p^2\log(n\vee p)\|\Omega^*\|_{op}^2.
\end{align}
\eprf

\bnlem
\label{lem:cov loss deviation no change point}
Let $\I\subset [1,T]$ be any interval that contains no change point. Then for any interval $\J \supset \I$, it holds with probability at least $1 - (n\vee p)^{-5}$ that
\begin{equation*}
    \mclF(\Omega^*_{\I},\I)\leq \mclF(\widehat{\Omega}_\J,\I) + C\frac{g_X^4}{c_X^2}p^2\log(n\vee p).
\end{equation*}
\enlem
\begin{proof}
The conclusion is guaranteed by \Cref{lem:cov no cp}   $\widehat{\Omega}_{\I}$ is   the minimizer of $\mclF(\Omega,\I)$.
\end{proof}

\bnlem[Single change point]
\label{lem:cov single cp}
Suppose the good events 
$\mathcal L (\tr ) $ and  $\mathcal R (\tr) $ defined  in \Cref{eq:left and right approximation of change points} hold. 
Let   $ \I=(s,e] \in \mathcal {\widehat P}  $  be such that $\I$ contains exactly  one    change point $ \eta_k $. 
 Then  with probability  at least $1-(n\vee p)^{-3}$,   it holds that 
\begin{equation}
    \min\{ \eta_k -s ,e-\eta_k \} \leq CC_{\gamma} g_X^4\frac{C_X^2}{c_X^6}\frac{p^2\log(n\vee p)}{\kappa_k^2} + Cg_X^4\frac{C_X^6}{c_X^6}\tr.
\end{equation}
\enlem
\bprf
If either $ \eta_k -s \le \tr  $ or $e-\eta_k\le \tr$, then there is nothing to show. So assume that 
$$  \eta_k -s  > \tr \quad \text{and} \quad   e-\eta_k  > \tr . $$
By event $\mathcal R (p^{-1}\tr )$, there exists  $ s_ u  \in \{ s_q\}_{q=1}^\Q$ such that 
$$0\le s_u - \eta_k \le p^{-1}\tr. $$
 So 
$$ \eta_k \le s_ u \le e .$$
Denote 
$$ \I_ 1 = (s,s_u] \quad \text{and} \quad \I_2 = (s_u, e],$$
and $\mclF^*(\J) = \sum_{i\in \J}[{\rm Tr}((\Omega^*_i)^\top X_iX_i^\top) -\log|\Omega_i^*|]$. Since
$s, e,  s_u \in  \{ s_q\}_{q=1}^\Q  $, by the definition of $\widehat{P}$ and $\widehat{\Omega}$, and \Cref{lem:cov one change deviation bound}, it holds that
\begin{align}
    \mclF(\widehat{\Omega}_{\I},\I)\leq& \mclF(\widehat{\Omega}_{\I_1},\I_1) + \mclF(\widehat{\Omega}_{\I_2},\I_2) + \gamma \nonumber \\
    \leq & \mclF^*(\I_1) + \frac{C_X^6 p}{c_X^4}(s_u - \eta_k)\kappa_k^2 +  C\frac{g_X^4 C_X^2}{c_X^4}p^2\log(n\vee p) + \mclF^*(\I_2) + \gamma \nonumber \\
    \leq & \mclF^*(\I) + \frac{C_X^6}{c_X^4}\mclB_n^{-1}\Delta\kappa_k^2 +  C\frac{g_X^4 C_X^2}{c_X^4}p^2\log(n\vee p) +  \gamma,
\end{align}
where the last inequality is due to $s_u - \eta_k\leq \mclB_n^{-1}\Delta$. Let 
$$\widetilde{\gamma} = \frac{C_X^6}{c_X^4}\mclB_n^{-1}\Delta\kappa_k^2 +  C\frac{g_X^4 C_X^2}{c_X^4}p^2\log(n\vee p) +  \gamma.$$ 
Then by Taylor expansion and \Cref{lem:estimation covariance} we have
\begin{align}
    \frac{c}{\max_{k\in [K]}\|\Omega_{\I_k}^*\|_{op}^2}\sum_{t\in \mclI}\|\widehat{\Omega}_{\I} - \Omega^*_t\|_F^2 
    &\leq \widetilde{\gamma} + \sum_{i=k-1}^k|\I_i|{\rm Tr}[(\Omega_{\I_i}^* - \widehat{\Omega}_{\I})^\top (\widehat{\Sigma}_{\I_i}  - \Sigma^*_{\I_i})] \nonumber \\
    &\leq \widetilde{\gamma} + \sum_{i=k-1}^k {|\mclI_{i}|}\|\widehat{\Sigma}_{\I_i} - \Sigma^*_{\I_i}\|_F\|\widehat{\Omega}_{\I} - \Omega^*_{\I_i}\|_F \nonumber \\
    &\leq \widetilde{\gamma} + C_1g_X^2 p\log^{\frac{1}{2}}(n\vee p)\left[ \sum_{i=k-1}^k \sqrt{|\mclI_{i}|}\|\widehat{\Omega}_{\I} - \Omega^*_{\I_i}\|_F \right] \nonumber \\
    &\leq \widetilde{\gamma} + C_1g_X^2 p\log^{\frac{1}{2}}(n\vee p) \sqrt{\sum_{t\in \mclI}\|\widehat{\Omega}_{\I} - \Omega^*_t\|_F^2}.
\end{align}
The inequality above implies that
\begin{equation}
    \sum_{t\in \mclI}\|\widehat{\Omega}_{\I} - \Omega^*_t\|_F^2 \leq \frac{2}{c}\max\|\Omega_{\I_k}^*\|_{op}^2\left[\widetilde{\gamma} + \max\|\Omega_{\I_k}^*\|_{op}^2\|X\|^4_{\psi_2} p^{2}\log(n\vee p)\right].
\end{equation}
On the other hand,
\begin{equation}
    \sum_{t\in \mclI}\|\widehat{\Omega}_{\I} - \Omega^*_t\|_F^2\geq \frac{|\mclI_{k-1}||\mclI_{k}|}{|\mclI|}\|\Omega^*_{\I_{k-1}} - \Omega^*_{\I_k}\|_F^2,
\end{equation}
which implies that
\begin{equation}
    \min\{|\mclI_{k-1}|,|\mclI_{k}|\}\leq C_2C_{\gamma} g_X^4\frac{C_X^2}{c_X^6}\frac{p^2\log(n\vee p)}{\kappa_k^2} + C_2g_X^4\frac{C_X^6}{c_X^6}\tr.
\end{equation}
\eprf

Recall that we assume for $i\in [n]$, $c_XI_p \preceq \Sigma_i \preceq C_X I_p$ for some universal constants $c_X>0, C_X<\infty$.

\bnlem[Two change points]
\label{lem:cov two cp}
Suppose the good events 
$\mathcal L (\tr ) $ and  $\mathcal R ( \tr ) $ defined  in  \Cref{eq:left and right approximation of change points} hold. Let   $ \I=(s,e] \in \mathcal {\widehat P}  $  be an interval that contains exactly two change points $ \eta_k,\eta_{k+1}$. Suppose in addition that $\gamma \geq C_{\gamma}\frac{g_X^4}{c_X^2} p^2\log(n\vee p)$, and

\begin{equation}
    \Delta_{\min} \kappa^2 \geq \mathcal{B}_n \frac{g_X^4}{c_X^4}p^2\log(n\vee p),
\end{equation}
then   with probability at least $1 - n^{-5}$ it holds  that
\begin{equation}
    \max\{\eta_k - s, e - \eta_{k + 1}\}\leq \htr.
\end{equation}
\enlem
\bprf
Since the events $\mathcal L ( \tr  ) $ and  $\mathcal R ( \tr ) $ hold, let $ s_u, s_v$ be such that 
 $\eta_k \le s_u \le s_v \le \eta_{k+1} $ and that 
 $$ 0 \le s_u-\eta_k \le   \tr ,  \quad  0\le  \eta_{k+1} - s_v \le \tr   .    $$

 \begin{center}
 \begin{tikzpicture} 
\draw[ - ] (-10,0)--(1,0);
 \node[color=black] at (-8,-0.3) {\small s};
 \draw[  (-, ultra thick, black] (-8,0) -- (-7.99,0);
  \draw[ { -]}, ultra thick, black] (-1.0002,0) -- (-1,0);

   \node[color=black] at (-7,-0.3) {\small $\eta_k$};
\draw(-7 ,0)circle [radius=2pt] ;

  \node[color=black] at (-6.5,-0.3) {\small $s_u$};
\draw plot[mark=x, mark options={color=black, scale=1.5}] coordinates {(-6.5,0)  };

  \node[color=black] at (-2.3,-0.3) {\small $\eta_{k+1}$};
\draw(-2.3 ,0)circle [radius=2pt] ;
  \node[color=black] at (-3 ,-0.3) {\small $s_v$};
\draw plot[mark=x, mark options={color=black, scale=1.5}] coordinates {(-3,0)  }; 
  
   \node[color=black] at (-1,-0.3) {\small $e$};

\end{tikzpicture}
\end{center} 

Denote 
$$ \mathcal I_ 1= (  s, s _u], \quad \I_2  =(s_u, s_v] \quad \text{and} \quad \I_3  = (s_v,e]. $$
In addition, denote 
$$ \J_1  = (s,\eta_k ], \quad \J_2=(\eta_k, \eta_{k} + \frac{ \eta_{k+1}   -\eta_k  }{2}], \quad \J_3 =  ( \eta_k+ \frac{ \eta_{k+1}   -\eta_k  }{2},\eta_{k+1 } ]  \quad \text{and} \quad \J_4 = (\eta_{k+1}  , e] .$$
Since 
$s, e,  s_u ,s_v  \in  \{ s_q\}_{q=1}^\Q  $, by the event $\mclL(p^{-1}\tr)$ and $\mclR(p^{-1}\tr)$, it holds with probability at least $1 - n^{-3}$ that
$$
0\leq s_u - \eta_k\leq p^{-1}\tr,\ 0\leq \eta_{k+1} - s_v\leq p^{-1}\tr.
$$
Denote
$$\I_1 = (s ,\eta_k], \I_2 = (\eta_k, \eta_{k + 1}], \I_3 = (\eta_{k + 1}, e].$$
By the definition of DP and $\widehat{\Omega}_{\I}$, it holds that
\begin{align}
        \mclF(\widehat {\Omega}_{\I},\I)
        \leq & \sum_{i = 1}^3 \mclF(\widehat {\Omega}_{\I_i},\I_i) + 2\gamma \\
        \leq & \sum_{i = 1}^3 \mclF({\Omega}_{\I_i}^*,\I_i) +  2\gamma + \frac{C_X^6 p}{c_X^4} \frac{|\J_1|(s_u - \eta_k)}{|\J_1|+s_u - \eta_k} \kappa_k^2 + \frac{C_X^6 p}{c_X^4} \frac{|\J_4|(\eta_{k + 1} - s_v)}{|\J_4|+\eta_{k + 1} - s_v} \kappa_k^2 + C\frac{g_X^4 C_X^2}{c_X^4}p^2\log(n\vee p) \\
        \leq & \sum_{i = 1}^3 \mclF({\Omega}_{\I_i}^*,\I_i) +  2\gamma + \frac{C_X^6}{c_X^4} \tr \kappa_k^2 + C\frac{g_X^4 C_X^2}{c_X^4}p^2\log(n\vee p).
\end{align}
Let
$$
\widetilde{\gamma} = 2\frac{C_X^6}{c_X^4}\mclB_n^{-1}\Delta\kappa_k^2 +  C\frac{g_X^4 C_X^2}{c_X^4}p^2\log(n\vee p) +  2\gamma.
$$
Then by Taylor expansion and \Cref{lem:estimation covariance} we have
\begin{align}
    cc_X^2\sum_{t\in \mclI}\|\widehat{\Omega}_{\I} - \Omega^*_t\|_F^2 &\leq \widetilde{\gamma} + \sum_{i=1}^3|\I_i|{\rm Tr}[(\Omega_{\I_i}^* - \widehat{\Omega}_{\I})^\top (\widehat{\Sigma}_{\I_i}  - \Sigma^*_{\I_i})] \nonumber \\
    &\leq \widetilde{\gamma} + Cg_X^2 p\log^{\frac{1}{2}}(n\vee p)\left[ \sum_{i=1}^3 \sqrt{|\mclI_{i}|}\|\widehat{\Omega}_{\I} - \Omega^*_{i}\|_F \right]  \nonumber \\
    &\leq \widetilde{\gamma} + Cg_X^2 p\log^{\frac{1}{2}}(n\vee p) \sqrt{\sum_{t\in \mclI}\|\widehat{\Omega}_{\I} - \Omega^*_t\|_F^2}.
\end{align}
The inequality above implies that
\begin{equation}
    \sum_{t\in \mclI}\|\widehat{\Omega}_{\I} - \Omega^*_t\|_F^2 \leq \frac{C_1}{c_X^2}\left[\widetilde{\gamma} + \frac{1}{c_X^2}\|X\|^4_{\psi_2} p^{2}\log(n\vee p)\right].
\end{equation}
By the choice of $\gamma$, it holds that
\begin{equation}
    \sum_{t\in \I_1\cup\I_2}\|\widehat{\Omega}_{\I} - \Omega^*_t\|_F^2 \leq \frac{C_1C_{\gamma}}{c_X^4} \|X\|^4_{\psi_2} p^{2}\log(n\vee p).
\end{equation}
On the other hand,
\begin{equation}
    \sum_{t\in \I_1\cup\I_2}\|\widehat{\Omega}_{\I} - \Omega^*_t\|_F^2\geq \frac{|\mclI_{1}||\mclI_{2}|}{|\mclI|}\|\Omega^*_{k-1} - \Omega^*_k\|_F^2\geq \frac{1}{2}\min\{|\I_1|,|\I_2|\}\|\Omega^*_{k-1} - \Omega^*_k\|_F^2.
\end{equation}
Suppose $|\I_1|\geq |\I_2|$, then the inequality above leads to
\begin{equation*}
    \Delta_{\min} \kappa^2\leq \frac{C_1C_{\gamma}}{c_X^4} \|X\|^4_{\psi_2} p^{2}\log(n\vee p),
\end{equation*}
which is contradictory to the assumption on $\Delta$. Therefore, $|\I_1|<|\I_2|$ and we have 
\begin{equation}
    s-\eta_k = |\I_1|\leq CC_{\gamma} \frac{g_X^4}{c_X^4\|\Omega_k^* - \Omega_{k-1}^*\|_F^2}p^2\log(n\vee p).
\end{equation}
The bound for $e - \eta_{k+1}$ can be proved similarly.
\eprf

\bnlem[Three or more change points]
\label{lem:cov three or more cp}
Suppose the assumptions in \Cref{assp:DCDP_covariance} hold. Then with probability at least $1-(n\vee p)^{-3}$, there is no intervals in $\widehat { \mathcal P}$ containing three or more true change points.   
 \enlem 
\bprf
We prove by contradiction. Suppose    $ \I=(s,e] \in \mathcal {\widehat P}  $  be such that $ \{ \eta_1, \ldots, \eta_M\} \subset \I $ with $M\ge 3$. Throughout the proof, $M$ is assumed to be a parameter that can potentially change with $n$. 
Since the events $\mathcal L ( \tr  ) $ and  $\mathcal R ( \tr ) $ hold, by relabeling $\{ s_q\}_{q=1}^\Q  $ if necessary,   let $ \{ s_m\}_{m=1}^M $ be such that 
 $$ 0 \le s_m  -\eta_m  \le \tr    \quad \text{for} \quad 1 \le m \le M-1  $$ and that
 $$ 0\le \eta_M  - s_M \le \tr .$$ 
 Note that these choices ensure that  $ \{ s_m\}_{m=1}^M  \subset \I . $
 
 \begin{center}
 \begin{tikzpicture} 
\draw[ - ] (-10,0)--(1,0);
 \node[color=black] at (-8,-0.3) {\small s};
 \draw[  (-, ultra thick, black] (-8,0) -- (-7.99,0);
  \draw[ { -]}, ultra thick, black] (-1.0002,0) -- (-1,0);

   \node[color=black] at (-7,-0.3) {\small $\eta_1$};
\draw(-7 ,0)circle [radius=2pt] ;

  \node[color=black] at (-6.5,-0.3) {\small $s_1$};
\draw plot[mark=x, mark options={color=black, scale=1.5}] coordinates {(-6.5,0)  }; 

  \node[color=black] at (-5,-0.3) {\small $\eta_2$};
\draw(-5 ,0)circle [radius=2pt] ;
  \node[color=black] at (-4.5,-0.3) {\small $s_2$};
\draw plot[mark=x, mark options={color=black, scale=1.5}] coordinates {(-4.5,0)  };

  \node[color=black] at (-2.5,-0.3) {\small $\eta_3$};
\draw(-2.5 ,0)circle [radius=2pt] ;
  \node[color=black] at (-3 ,-0.3) {\small $s_3$};
\draw plot[mark=x, mark options={color=black, scale=1.5}] coordinates {(-3,0)  }; 
  
   \node[color=black] at (-1,-0.3) {\small $e$};

\end{tikzpicture}
\end{center}

{\bf Step 1.}
 Denote 
 $$ \mathcal I_ 1= (  s, s _1], \quad \I_m   =(s_{m-1} , s_m] \text{ for }  2 \le m \le M     \quad \text{and} \quad \I_{M+1}   = (s_M,e]. $$
Then since 
$ s , e,  \{ s_m \}_{m=1}^M  \subset   \{ s_q\}_{q=1}^\Q  $,   it follows that

Suppose $\I = (s,e]\in \widehat{\mclP}$ and there are $M\geq 3$ true change points $\{\eta_{q + i}\}_{i\in [M]}$ inside $\I$, and denote
\begin{equation*}
    \I_1 = (s, \eta_{q + 1}],\ \I_m = (\eta_{q + m - 1},\eta_{q + m}],\ \I_{M+1} = (\eta_{q + M},e].
\end{equation*}
Then by the definition of $\widehat{\mclP}$ and $\widehat{\Omega}_{\I_m}$, it holds that
\begin{equation*}
    \mclF(\widehat {\Omega}_{\I},\I)\leq \sum_{i = 1}^{M+1} \mclF (\widehat {\Omega}_{\I_i},\I_i) + M\gamma\leq \sum_{i = 1}^{M+1} \mclF ({\Omega}_{\I_i}^*,\I_i) + M\gamma,
\end{equation*}
which implies that
\begin{align}
    & \sum_{t\in \I}{\rm Tr}(\widehat{\Omega}_{\I}^\top (X_tX_t^\top)) - |\I|\log|\widehat{\Omega}_{\I}| \nonumber  \\
    \leq & \sum_{i=1}^{M+1} \sum_{t\in \I_i}{\rm Tr}(({\Omega}_{\I_i}^*)^\top (X_tX_t^\top)) - \sum_{i=1}^{M+1}|\I_i|\log|{\Omega}_{\I_i}^*| + M\gamma.
\end{align}
By Taylor expansion and \Cref{lem:estimation covariance} we have
\begin{align}
    cc_X^2\sum_{t\in \mclI}\|\widehat{\Omega}_{\I} - \Omega^*_t\|_F^2 &\leq M\gamma + \sum_{i=1}^{M+1}|\I_i|{\rm Tr}[(\Omega_{\I_i}^* - \widehat{\Omega}_{\I})^\top (\widehat{\Sigma}_{\I_i}  - \Sigma^*_{\I_i})] \nonumber \\
    &\leq M\gamma + Cg_X^2 p\log^{\frac{1}{2}}(n\vee p)\left[ \sum_{i=1}^{M+1} \sqrt{|\mclI_{i}|}\|\widehat{\Omega}_{\I} - \Omega^*_{i}\|_F \right]  \nonumber \\
    &\leq M\gamma + Cg_X^2 p\log^{\frac{1}{2}}(n\vee p) \sqrt{\sum_{t\in \mclI}\|\widehat{\Omega}_{\I} - \Omega^*_t\|_F^2}.
\end{align}
The inequality above implies that
\begin{equation}
    \sum_{t\in \mclI}\|\widehat{\Omega}_{\I} - \Omega^*_t\|_F^2 \leq \frac{C_1}{c_X^2}\left[M\gamma + \frac{1}{c_X^2}\|X\|^4_{\psi_2} p^{2}\log(n\vee p)\right].
\end{equation}
On the other hand, for each $i\in [M]$, we have
\begin{equation}
    \sum_{t\in \I_i\cup\I_{i +1}}\|\widehat{\Omega}_{\I} - \Omega^*_t\|_F^2\geq \frac{|\mclI_{i}||\mclI_{i + 1}|}{|\mclI|}\|\Omega^*_{\eta_{q + i + 1}} - \Omega^*_{\eta_{q + i}}\|_F^2.
\end{equation}
In addition, for each $i\in \{2,\cdots,M\}$, by definition, it holds that $\min\{|\I_i|,|\I_{i + 1}|\}\geq \Delta_{\min}$. Therefore, we have
\begin{align*}
    (M-2)\Delta_{\min} \kappa^2\leq \frac{C_1}{c_X^2}\left[M\gamma + \frac{1}{c_X^2}\|X\|^4_{\psi_2} p^{2}\log(n\vee p)\right].
\end{align*}
Since $M/(M-2)\leq 3$ for any $M\geq 3$, it holds that
\begin{equation}
    \Delta_{\min} \leq CC_{\gamma} \frac{g_X^4}{c_X^4\|\Omega_k^* - \Omega_{k-1}^*\|_F^2}p^2\log(n\vee p),
\end{equation}
which is contradictory to the assumption on $\Delta$, and the proof is complete.
\eprf

 \bnlem[Two consecutive intervals]
 \label{lem:cov two consecutive interval}
 Under \Cref{assp:DCDP_covariance} and the choice that $$\gamma\geq C_{\gamma}\frac{g_X^4}{c_X^2} p^2\log(n\vee p),$$ with probability at least $1-(n\vee p)^{-3}$, there are no  two consecutive intervals $\I_1= (s,t ] \in \widehat{\mathcal P}$, $\I_2=(t, e] \in   \widehat{\mathcal P}$ such that $\I_1 \cup \I_2$ contains no change points.  
 \enlem 
 \begin{proof} 
 We prove by contradiction. Suppose that $\I_1,\I_2\in \widehat{\mclP}$ and
 $$ \I :  =\I_1\cup  \I_2 $$
 contains no change points. 
 By the definition of $\widehat{\mclP}$ and $\widehat{\Omega}_{\I}$, it holds that
 $$ \mclF (\widehat{\Omega}_{\I_1},\I_1)  + \mclF(\widehat{\Omega}_{\I_2},\I_2) + \gamma \leq  \mclF (\widehat{\Omega}_{\I},\I)\leq \mclF ({\Omega}_{\I}^*,\I)$$
 By \Cref{lem:cov no cp}, it follows that 
\begin{align*} 
\mclF({\Omega}^*_{\I_1}, \mclI_1)\leq &  \mclF(\widehat{\Omega}_{\I_1}, \mclI_1) + C\frac{g_X^4}{c_X^2} p^2\log(n\vee p),
 \\
\mclF({\Omega}^*_{\I_2}, \mclI_2)\leq &  \mclF(\widehat{\Omega}_{\I_2}, \mclI_2) +  C\frac{g_X^4}{c_X^2} p^2\log(n\vee p) 
 \end{align*}
 So
 $$\mclF({\Omega}^*_{\I_1}, \mclI_1) +\mclF({\Omega}^*_{\I_2}, \mclI_2) -2C\frac{g_X^4}{c_X^2} p^2\log(n\vee p)  +\gamma    \leq  \mclF ({\Omega}_{\I}^*,\I).  $$
 Since $\I$ does not contain any change points, ${\Omega}^*_{\I_1} = {\Omega}^*_{\I_2} = {\Omega}^*_{\I}$, and it follows that 
 $$ \gamma \leq 2C\frac{g_X^4}{c_X^2} p^2\log(n\vee p).$$
 This is a contradiction when $C_\gamma$ is sufficiently large.
 \end{proof}

\clearpage

\section{Penalized local refinement}
\label{sec: proof local refinement}
In this section, we prove consistency results in \Cref{section:main} for penalized local refinement, or the conquer step. We also provide more details on the computational complexity of local refinement using memorization technique which is summarized in \Cref{sec:method}. In particular,
\begin{enumerate}
    \item In \Cref{sec:complexity local refine}, we analyze the complexity of the local refinement step and show that it is linear in terms of $n$, as is mentioned in \Cref{sec:method}.
    \item \Cref{sec:op1 fundations} presents some fundamental lemmas to prove other results.
    \item \Cref{sec:mean op1} prove results for the mean model, i.e., \Cref{cor:mean local refinement}.
    \item \Cref{sec:regression op1} prove results for the linear regression model, i.e., \Cref{cor:regression local refinement}.
    \item \Cref{sec:cov op1} prove results for the Gaussian graphical model, i.e., \Cref{cor:covariance local refinement main}.
\end{enumerate}

\subsection{Complexity analysis}
\label{sec:complexity local refine}
We show in \Cref{lem: complexity local refine} that the complexity of the conquer step (\Cref{algo:local_refine_general}) can be as low as $O(n\cdot \mathcal{C}_2(p))$.

\bnlem[Complexity of the conquer step]
\label{lem: complexity local refine}
For all three models we discussed in \Cref{section:main}, with a memorization technique, the complexity of \Cref{algo:local_refine_general} would be $O(n\cdot \mathcal{C}_2(p))$.
\enlem
\bprf
In \Cref{algo:local_refine_general}, for each $k\in [\widehat{K}]$, we search over the interval of length $\frac{2}{3}(\widehat{\Delta}_{k-1} + \widehat{\Delta}_k)$ where $\widehat{\Delta}_k:=\widehat{\eta}_{k+1} - \widehat{\eta}_{k}$. Without 
any algorithmic optimization, the complexity would be $O((\widehat{\Delta}_{k-1} + \widehat{\Delta}_k)\mathcal{C}_1(\widehat{\Delta}_{k-1} + \widehat{\Delta}_k, p))$ where $\mathcal{C}_1(m, p)$ is the complexity of calculating $\widehat{\theta}_{\mclI}$ and $\mclF(\widehat{\theta}_{\mclI}, \mclI)$ for an interval of length $m$,

Under the three models in \Cref{section:main}, calculating $\widehat{\theta}_{\mclI}$ involves the calculation of some sufficient statistics or gradients and a gradient descent or coordinate descent procedure which is independent of $|\mclI|$. Therefore, $\mathcal{C}_1(|\mclI|, p) = O(|\mclI|) + O(\mclC_2(p))$. For instance, solving Lasso only takes $O(p)$ time once $\sum_{i\in[n]}X_iX_i^\top$ and $\sum_{{i\in[n]}}X_iy_i$ are known. In the conquer step, each time we only update the two summations $(\sum_{i\in[n]}X_iX_i^\top,\sum_{{i\in[n]}}X_iy_i)$ by one term, so we can use memorization trick to reduce $\mathcal{C}_1(\widehat{\Delta}_{k-1} + \widehat{\Delta}_k, p)$ to $O(1) + O(\mclC_2(p))$. Consequently, the complexity at the $k-th$ step of \Cref{algo:local_refine_general} can be reduced to $O((\widehat{\Delta}_{k-1} + \widehat{\Delta}_k)\mclC_2(p))$. Taking summation over $k\in [\widehat{K}]$ and considering the fact that $\widehat{\mclP}$ is a segmentation of $[1,n]$, the total complexity of the conquer step would be
$$
\sum_{k\in [\widehat{K}]}O((\widehat{\Delta}_{k-1} + \widehat{\Delta}_k)\cdot \mclC_2(p)) = O(n\cdot \mclC_2(p)).
$$
\eprf


\subsection{Fundamental lemma}
\label{sec:op1 fundations}
 As is introduced in \Cref{sec:introduction}, the sub-gaussian norm of a random variable is defined as \citep{vershynin2018high}: $\|X\|_{\psi_2}:=\inf \{t>0: \mathbb{E} \psi_2(|X| / t) \leq 1\}$ where $\psi_2(t) = e^{t^2} - 1$.

 Similarly, for sub-exponential random variables, one can define its Orlicz norm as $\|X\|_{\psi_1}:=\inf \{t>0: \mathbb{E} \psi_1(|X| / t) \leq 2\}$ where $\psi_1(t) = e^{t}$.

\bnlem
\label{lem:subexp deviation martingale}
Suppose $\left\{z_i\right\}_{i=1}^{\infty}$ is a collection of independent centered sub-exponential random variables with $0<\sup _{1 \leq i<\infty}\left\|z_i\right\|_{\psi_1} \leq 1$. Then for any integer $d>0, \alpha>0$ and any $x>0$
$$
\mathbb{P}\left(\max _{k \in[d,(1+\alpha) d]} \frac{\sum_{i=1}^k z_i}{\sqrt{k}} \geq x\right) \leq \exp \left\{-\frac{x^2}{2(1+\alpha)}\right\}+\exp \left\{-\frac{\sqrt{d} x}{2}\right\} .
$$
\enlem
\bprf
Denote $S_n=\sum_{i=1}^n z_i$. Let $\zeta=\sup _{1 \leq i \leq \infty}\left\|z_i\right\|_{\psi_1}$. For any two integers $m<n$ and any $t \leq \frac{C_1}{\zeta}$
$$
\mathbb{E}\left(\exp \left(t\left(S_n-S_m\right)\right)\right) \leq \prod_{i=m+1}^n \mathbb{E}\left(\exp \left(t z_i\right)\right) \leq \prod_{i=m+1}^n \mathbb{E}\left(C_1^2 t^2 / 2\right)=\mathbb{E}\left[(n-m) C_1^2 t^2 / 2\right].
$$
Let $\mathcal{F}_k$ denote the sigma-algebra generated by $\left(z_1, \ldots, z_k\right)$. Without loss of generality, assume that $C_1=1$. Since $S_k$ is independent of $S_n-S_k$, this implies that when $t \geq 1 / \zeta$
$$
\mathbb{E}\left(\exp \left\{t S_n-\frac{t^2 n}{2}\right\} \mid \mathcal{F}_k\right)=\exp \left\{t S_k-\frac{t^2 k}{2}\right\} \mathbb{E}\left(\exp \left\{t\left(S_n-S_k\right)\right\}-\frac{t^2(n-k)}{2}\right) \leq \exp \left\{t S_k-\frac{t^2 k}{2}\right\}
$$
Therefore $\exp \left\{t S_n-\frac{t^2 n}{2}\right\}$ is a super-Martingale. Let $x$ be given and
$$
A=\inf \left\{n \geq d, S_n \geq \sqrt{n} x\right\} .
$$
Then $S_A \geq \sqrt{A} x \geq \sqrt{d} x$. Thus for $t \geq 1 / \zeta$,
$$
\mathbb{E}\left(\exp \left\{t \sqrt{d} x-\frac{t^2 A}{2}\right\}\right) \leq \mathbb{E}\left(\exp \left\{t S_A-\frac{t^2 A}{2}\right\}\right) \leq \mathbb{E}\left(\exp \left\{t S_1-\frac{t^2}{2}\right\}\right) \leq 1 .
$$
By definition of $A$,
$$
\mathbb{P}\left(\max _{k \in[d,(1+\alpha) d]} \frac{\sum_{i=1}^k z_i}{\sqrt{k}} \geq x\right) \leq \mathbb{P}(A \leq(1+\alpha) d) .
$$
Since $u \rightarrow \exp \left(s \sqrt{d} x-\frac{t^2 u}{2}\right)$ is decreasing, it follows that
$$
\mathbb{P}\left(\max _{k \in[d,(1+\alpha) d]} \frac{\sum_{i=1}^k z_i}{\sqrt{k}} \geq x\right) \leq \mathbb{P}\left(\exp \left\{t \sqrt{d} x-t^2 A / 2\right\} \geq \exp \left\{t \sqrt{d} x-t^2(1+\alpha) d / 2\right\}\right) .
$$
Markov's inequality implies that when $t \geq \frac{1}{\zeta}$,
$$
\mathbb{P}\left(\max _{k \in[d,(1+\alpha) d]} \frac{\sum_{i=1}^k z_i}{\sqrt{k}} \geq x\right) \leq \exp \left\{-t \sqrt{d} x+t^2(1+\alpha) d / 2\right\}
$$
Set
$$
t=\min \left\{\frac{1}{\zeta}, \frac{x}{(1+\alpha) \sqrt{d}}\right\}
$$
If $\frac{1}{\zeta} \geq \frac{x}{(1+\alpha) \sqrt{d}}$. Then $t=\frac{x}{(1+\alpha) \sqrt{d}}$ and therefore
$$
-t \sqrt{d} x+t^2(1+\alpha) d / 2=-\frac{x^2}{2(1+\alpha)}
$$
So
$$
\mathbb{P}\left(\max _{k \in[d,(1+\alpha) d]} \frac{\sum_{i=1}^k z_i}{\sqrt{k}} \geq x\right) \leq \exp \left\{-\frac{x^2}{2(1+\alpha)}\right\} .
$$
If $\frac{1}{\zeta} \leq \frac{x}{(1+\alpha) \sqrt{d}}$. Then $t=\frac{1}{\zeta} \leq \frac{x}{(1+\alpha) \sqrt{d}}$ and so
$$
-t \sqrt{d} x+t^2(1+\alpha) d / 2 \leq \frac{-\sqrt{d} x}{\zeta}+\frac{1}{\zeta} \frac{x}{(1+\alpha) \sqrt{d}} \frac{(1+\alpha) d}{2}=\frac{-\sqrt{d} x}{2 \zeta} .
$$
So
$$
\mathbb{P}\left(\max _{k \in[d,(1+\alpha) d]} \frac{\sum_{i=1}^k z_i}{\sqrt{k}} \geq x\right) \leq \exp \left\{-\frac{\sqrt{d} x}{2 \zeta}\right\} \leq \exp \left\{-\frac{\sqrt{d} x}{2}\right\},
$$
where $\zeta \leq 1$ is used in the last inequality. Putting the two cases together leads the desired result.
\eprf

\bnlem
\label{lem:subexp uniform loglog law}
Suppose $\left\{z_i\right\}_{i=1}^{\infty}$ is a collection of independent centered sub-exponential random variable with $0<\sup _{1 \leq i<\infty}\left\|z_i\right\|_{\psi_1} \leq 1$. Let $\nu>0$ be given. For any $x>0$, it holds that
$$
\mathbb{P}\left(\sum_{i=1}^r z_i \leq 4 \sqrt{r\{\log \log (4 \nu r)+x+1\}}+4 \sqrt{r \nu}\{\log \log (4 \nu r)+x+1\} \text { for all } r \geq 1 / \nu\right) \geq 1-2 \exp (-x) \text {. }
$$
\enlem
\bprf
Let $s \in \mathbb{Z}^{+}$and $\mathcal{T}_s=\left[2^s / \nu, 2^{s+1} / \nu\right]$. By \Cref{lem:subexp deviation martingale}, for all $x>0$,
$$
\mathbb{P}\left(\sup _{r \in \mathcal{T}_s} \frac{\sum_{i=1}^r z_i}{\sqrt{r}} \geq x\right) \leq \exp \left\{-\frac{x^2}{4}\right\}+\exp \left\{-\frac{\sqrt{2^s / \nu} x}{2}\right\} \leq \exp \left\{-\frac{x^2}{4}\right\}+\exp \left\{-\frac{x}{2 \sqrt{\nu}}\right\} .
$$
Therefore by a union bound,
\begin{align}
    & \mathbb{P}\left(\exists s \in \mathbb{Z}^{+}: \sup _{r \in \mathcal{T}_s} \frac{\sum_{i=1}^r z_i}{\sqrt{r}} \geq 2 \sqrt{\log \log ((s+1)(s+2))+x}+2 \sqrt{\nu}\{\log \log ((s+1)(s+2))+x\}\right) \nonumber \\
\leq & \sum_{s=0}^{\infty} 2 \frac{\exp (-x)}{(s+1)(s+2)}=2 \exp (-x) \label{tmp_eq: sub-gauss iterated log}.
\end{align}
For any $r \geq 2^s / \nu, s \leq \log (r \nu) / \log (2)$, and therefore
$$
(s+1)(s+2) \leq \frac{\log (2 r \nu) \log (4 r \nu)}{\log ^2(2)} \leq\left(\frac{\log (4 r \nu)}{\log (2)}\right)^2 .
$$
Thus
$$
\log ((s+1)(s+2)) \leq 2 \log \left(\frac{\log (4 r \nu)}{\log (2)}\right) \leq 2 \log \log (4 r \nu)+1 .
$$
The above display together with \eqref{tmp_eq: sub-gauss iterated log} gives
$$
\mathbb{P}\left(\sup _{r \geq 1 / \nu} \frac{\sum_{i=1}^r z_i}{\sqrt{r}} \geq 2 \sqrt{2 r \log \log (4 r \nu)+x+1}+2 \sqrt{r \nu}\{\log \log (4 r \nu)+x+1\}\right) \leq 2 \exp (-x) .
$$
\eprf

Next we present two analogous lemmas for sub-gaussian random variables.

\bnlem
\label{lem:subgaussian deviation martingale}
Suppose $\left\{z_i\right\}_{i=1}^{\infty}$ is a collection of independent centered sub-gaussian random variables with $0<\sup _{1 \leq i<\infty}\left\|z_i\right\|_{\psi_2} \leq \sigma$. Then for any integer $d>0, \alpha>0$ and any $x>0$
$$
\mathbb{P}\left(\max _{k \in[d,(1+\alpha) d]} \frac{\sum_{i=1}^k z_i}{\sqrt{k}} \geq x\right) \leq \exp \left\{-\frac{x^2}{2(1+\alpha)\sigma^2}\right\} .
$$
\enlem
\bprf
Denote $S_n=\sum_{i=1}^n z_i$. Let $\zeta=\sup _{1 \leq i \leq \infty}\left\|z_i\right\|_{\psi_2}$. For any two integers $m<n$,
$$
\mathbb{E}\left(\exp \left(t\left(S_n-S_m\right)\right)\right) \leq \prod_{i=m+1}^n \mathbb{E}\left(\exp \left(t z_i\right)\right) \leq \prod_{i=m+1}^n \mathbb{E}\left(\zeta^2 t^2 / 2\right)=\mathbb{E}\left[(n-m) \zeta^2 t^2 / 2\right].
$$
Let $\mathcal{F}_k$ denote the sigma-algebra generated by $\left(z_1, \ldots, z_k\right)$. Since $S_k$ is independent of $S_n-S_k$, this implies that
\begin{align*}
    \mathbb{E}\left(\exp \left\{t S_n-\frac{\zeta^2t^2 n}{2}\right\} \mid \mathcal{F}_k\right)
    &=\exp \left\{t S_k-\frac{\zeta^2t^2 k}{2}\right\} \mathbb{E}\left(\exp \left\{t\left(S_n-S_k\right)\right\}-\frac{\zeta^2 t^2(n-k)}{2}\right) \\
    &\leq \exp \left\{t S_k-\frac{\zeta^2 t^2 k}{2}\right\}
\end{align*}
Therefore $\exp \left\{t S_n-\frac{\zeta^2t^2 n}{2}\right\}$ is a super-martingale. Let $x$ be given and
$$
A=\inf \left\{n \geq d, S_n \geq \sqrt{n} x\right\} .
$$
Then $S_A \geq \sqrt{A} x \geq \sqrt{d} x$. Thus for $t >0$,
$$
\mathbb{E}\left(\exp \left\{t \sqrt{d} x-\frac{\zeta^2 t^2 A}{2}\right\}\right) \leq \mathbb{E}\left(\exp \left\{t S_A-\frac{\zeta^2 t^2 A}{2}\right\}\right) \leq \mathbb{E}\left(\exp \left\{t S_1-\frac{\zeta^2 t^2}{2}\right\}\right) \leq 1 .
$$
By definition of $A$,
$$
\mathbb{P}\left(\max _{k \in[d,(1+\alpha) d]} \frac{\sum_{i=1}^k z_i}{\sqrt{k}} \geq x\right) \leq \mathbb{P}(A \leq(1+\alpha) d) .
$$
Since $u \rightarrow \exp \left(s \sqrt{d} x-\frac{\zeta^2 t^2 u}{2}\right)$ is decreasing, it follows that
$$
\mathbb{P}\left(\max _{k \in[d,(1+\alpha) d]} \frac{\sum_{i=1}^k z_i}{\sqrt{k}} \geq x\right) \leq \mathbb{P}\left(\exp \left\{t \sqrt{d} x - \zeta^2 t^2 A / 2\right\} \geq \exp \left\{t \sqrt{d} x-\zeta^2 t^2(1+\alpha) d / 2\right\}\right) .
$$
Markov's inequality implies that,
$$
\mathbb{P}\left(\max _{k \in[d,(1+\alpha) d]} \frac{\sum_{i=1}^k z_i}{\sqrt{k}} \geq x\right) \leq \exp \left\{-t \sqrt{d} x+\zeta^2 t^2(1+\alpha) d / 2\right\}
$$
Set $t= \frac{x}{\zeta^2(1+\alpha) \sqrt{d}}$, then
$$
-t \sqrt{d} x+\zeta^2 t^2(1+\alpha) d / 2=-\frac{x^2}{2(1+\alpha)\zeta^2}
$$
So
$$
\mathbb{P}\left(\max _{k \in[d,(1+\alpha) d]} \frac{\sum_{i=1}^k z_i}{\sqrt{k}} \geq x\right) \leq \exp \left\{-\frac{x^2}{2(1+\alpha)\zeta^2}\right\} .
$$
\eprf

\bnlem
\label{lem:subgaussian uniform loglog law}
Suppose $\left\{z_i\right\}_{i=1}^{\infty}$ is a collection of independent centered sub-gaussian random variable with $0<\sup _{1 \leq i<\infty}\left\|z_i\right\|_{\psi_2} \leq \sigma$. Let $\nu>0$ be given. For any $x>0$, it holds that
$$
\mathbb{P}\left(\sum_{i=1}^r z_i \leq 4\sigma \sqrt{r\{\log \log (4 \nu r)+x+1\}} \text { for all } r \geq 1 / \nu\right) \geq 1-2 \exp (-x) \text {. }
$$
\enlem
\bprf
Let $s \in \mathbb{Z}^{+}$and $\mathcal{T}_s=\left[2^s / \nu, 2^{s+1} / \nu\right]$. By \Cref{lem:subgaussian deviation martingale}, for all $x>0$,
$$
\mathbb{P}\left(\sup _{r \in \mathcal{T}_s} \frac{\sum_{i=1}^r z_i}{\sqrt{r}} \geq x\right) \leq \exp \left\{-\frac{x^2}{4\sigma^2}\right\}.
$$
Therefore by a union bound,
\begin{align}
    & \mathbb{P}\left(\exists s \in \mathbb{Z}^{+}: \sup _{r \in \mathcal{T}_s} \frac{\sum_{i=1}^r z_i}{\sqrt{r}} \geq 2\sigma \sqrt{\log \log ((s+1)(s+2))+x}\right) \nonumber \\
\leq & \sum_{s=0}^{\infty} 2 \frac{\exp (-x)}{(s+1)(s+2)}=2 \exp (-x) \label{tmp_eq: iterated log}.
\end{align}
For any $r \geq 2^s / \nu, s \leq \log (r \nu) / \log (2)$, and therefore
$$
(s+1)(s+2) \leq \frac{\log (2 r \nu) \log (4 r \nu)}{\log ^2(2)} \leq\left(\frac{\log (4 r \nu)}{\log (2)}\right)^2 .
$$
Thus
$$
\log ((s+1)(s+2)) \leq 2 \log \left(\frac{\log (4 r \nu)}{\log (2)}\right) \leq 2 \log \log (4 r \nu)+1 .
$$
The above display together with \eqref{tmp_eq: iterated log} gives
$$
\mathbb{P}\left(\sup _{r \geq 1 / \nu} \frac{\sum_{i=1}^r z_i}{\sqrt{r}} \geq 2\sigma \sqrt{2 r \log \log (4 r \nu)+x+1}\right) \leq 2 \exp (-x) .
$$
\eprf

\clearpage
\subsection{Local Refinement in the mean model }
\label{sec:mean op1}
For the ease of notations, we re-index the observations in the $k$-th interval by $[n_0]:\{1,\cdots, n_0\}$ (though the sample size of the problem is still $n$), and denote the $k$-th jump size as $\kappa$ and the minimal spacing between consecutive change points as $\Delta$ (instead of $\Delta_{\min}$ in the main text).

By \Cref{assp: DCDP_mean} and the setting of the local refinement algorithm, we have for some $\alpha^*,\beta^*\in \mathbb{R}^p$ that
$$
y_i=\begin{cases}
\alpha^*+\epsilon_i & \text { when } i \in(0, \eta] \\
\beta^*+\epsilon_i & \text { when } i \in(\eta, n_0]
\end{cases}
$$
where $\{\epsilon_i\}$ is an i.i.d sequence of subgaussian variables such that $\|\epsilon_i\|_{\psi_2}=\sigma_\epsilon<\infty$. In addition, there exists $\theta \in(0,1)$ such that $\eta=\lfloor n \theta\rfloor$ and that $\|\alpha^*-\beta^*\|_2=\kappa<\infty$. By \Cref{assp: DCDP_mean}, it holds that $\|\alpha^*\|_0 \leq \mathfrak{s},\|\beta^*\|_0 \leq \mathfrak{s}$ and
\begin{equation}
    \frac{\mathfrak{s}^2 \log^2 (n\vee p)}{\Delta \kappa^2} \rightarrow 0.
    \label{op1_eq:mean snr}
\end{equation}
By \Cref{lem:mean refinement step 1}, with probability at least $1 - n^{-2}$, there exist $\widehat{\alpha}$ and $\widehat{\beta}$ such that
\begin{align*}
  & \|\widehat{\alpha}-\alpha^*\|_2^2\leq C\frac{\mathfrak{s} \log (n\vee p)}{\Delta} \text{ and } \|\widehat{\alpha}-\alpha^*\|_1\leq C\mathfrak{s} \sqrt{\frac{\log (n\vee p)}{\Delta}}; \\
& \|\widehat{\beta}-\beta^*\|_2^2\leq C\frac{\mathfrak{s} \log (n\vee p)}{\Delta} \text{ and } \|\widehat{\beta}-\beta^*\|_1\leq C \mathfrak{s} \sqrt{\frac{\log (n\vee p)}{\Delta}}.
\end{align*}

In fact, \Cref{lem:mean refinement step 1} shows that we are able to remove the extra $\mathcal{B}_n^{-1/2}\Delta_{\min}$ term in the localization error in \Cref{thm:DCDP mean} under the same SNR condition. In \Cref{lem:mean refinement}, we show that with slightly stronger SNR condition, the localization error can be further reduced as is concluded in \Cref{cor:mean local refinement}.

Let
$$
\widehat{\mathcal{Q}}(k)=\sum_{i=1}^k\|y_i- \widehat{\alpha}\|_2^2+\sum_{i=k+1}^{n_0}\|y_i-\widehat{\beta}\|_2^2 \quad \text { and } \quad \mathcal{Q}^*(k)=\sum_{i=1}^k\|y_i- \alpha^*\|_2^2+\sum_{i=k+1}^{n_0}\|y_i- \beta^*\|_2^2 .
$$

\bnlem[Refinement for the mean model]
\label{lem:mean refinement}
Let
$$
\eta+r=\underset{k \in(0, n_0]}{\arg \max } \widehat{\mathcal{Q}}(k) .
$$
Then under the assumptions above, for any given $\alpha\in (0,1)$, it holds with probability $1 - (\alpha\vee n^{-1})$ that
$$
\kappa^2 r \leq C \log\frac{1}{\alpha}.
$$
\enlem
\bprf
Without loss of generality, suppose $r \geq 0$. Since $\eta+r$ is the minimizer, it follows that
$$
\widehat{\mathcal{Q}}(\eta+r) \leq \widehat{\mathcal{Q}}(\eta) .
$$
If $r \leq \frac{1}{\kappa^2}$, then there is nothing to show. So for the rest of the argument, for contradiction, assume that
$$
r \geq \frac{1}{\kappa^2}
$$
Observe that
$$
\begin{aligned}
\widehat{\mathcal{Q}}(\eta + r)-\widehat{\mathcal{Q}}(\eta) &=\sum_{i=\eta+1}^{\eta+r}\|y_i- \widehat{\alpha}\|_2^2-\sum_{i=\eta+1}^{\eta+r}\|y_i-\widehat{\beta}\|_2^2 \\
\mathcal{Q}^*(\eta + r)-\mathcal{Q}^*(\eta) &=\sum_{i=\eta+1}^{\eta+r}\|y_i- \alpha^*\|_2^2-\sum_{i=\eta+1}^{\eta+r}\|y_i- \beta^*\|_2^2
\end{aligned}
$$
\textbf{Step 1}. It follows that
$$
\begin{aligned}
& \sum_{i=\eta+1}^{\eta+r}\|y_i- \widehat{\alpha}\|_2^2-\sum_{i=\eta+1}^{\eta+r}\|y_i-\alpha^*\|_2^2 \\
=& \sum_{i=\eta+1}^{\eta+r}\|\widehat{\alpha}- \alpha^*\|_2^2+2\left(\widehat{\alpha}-\alpha^*\right)^{\top} \sum_{i=\eta+1}^{\eta+r}\left(y_i-\alpha^*\right) \\
=& \sum_{i=\eta+1}^{\eta+r}\|\widehat{\alpha}- \alpha^*\|_2^2+2r\left(\widehat{\alpha}-\alpha^*\right)^{\top} \left(\beta^*-\alpha^*\right) +2\left(\widehat{\alpha}-\alpha^*\right)^{\top} \sum_{i=\eta+1}^{\eta+r}\epsilon_i
\end{aligned}
$$
By assumptions, we have
$$
\sum_{i=\eta+1}^{\eta+r}\|\widehat{\alpha}- \alpha^*\|_2^2 \leq C_1 r \frac{\mathfrak{s} \log (p)}{\Delta}.
$$
Similarly
$$
r\left(\widehat{\alpha}-\alpha^*\right)^{\top}\left(\beta^*-\alpha^*\right)\leq r\|\widehat{\alpha}-\alpha^*\|_2\|\beta^*-\alpha^*\|_2 \leq
C_1 r\kappa \sqrt{\frac{\mathfrak{s} \log (p)}{\Delta} }
$$
where the second equality follows from $\|\beta^*-\alpha^*\|_2=\kappa$, and the last equality follows from \eqref{op1_eq:mean snr}. In addition,
$$
\begin{aligned}
&\left(\widehat{\alpha}-\alpha^*\right)^{\top} \sum_{i=\eta+1}^{\eta+r} \epsilon_i \leq\|\widehat{\alpha}-\alpha^*\|_1\|\sum_{i=\eta+1}^{\eta+r}\epsilon_i\|_{\infty} \\
=& C_2\mathfrak{s} \sqrt{\frac{\log (p)}{\Delta}} \sqrt{r \log (p)}=C_2\mathfrak{s}\log(p)\sqrt{\frac{r}{\Delta}}.
\end{aligned}
$$
Therefore
\begin{align}
    \sum_{i=\eta+1}^{\eta+r}\|y_i- \widehat{\alpha}\|_2^2-\sum_{i=\eta+1}^{\eta+r}\|y_i-\alpha^*\|_2^2 &\leq C_1 r \frac{\mathfrak{s} \log (p)}{\Delta} + C_1 r\kappa \sqrt{\frac{\mathfrak{s} \log (p)}{\Delta} } + C_2\mathfrak{s}\log(p)\sqrt{\frac{r}{\Delta}}\nonumber \\
    &\leq C_1 r\kappa^2 \frac{\mathfrak{s} \log (p)}{\Delta\kappa^2} + C_1 r\kappa^2 \sqrt{\frac{\mathfrak{s} \log (p)}{\Delta\kappa^2} } + C_2\mathfrak{s}\log(p)\sqrt{\frac{r\kappa^2}{\Delta\kappa^2}}\nonumber \\
    &\leq C_3 r \kappa^2 \frac{\mathfrak{s} \log (p)}{\sqrt{\Delta\kappa^2}}.
\end{align}
\textbf{Step 2}. Using the same argument as in the previous step, it follows that
$$
\sum_{i=\eta+1}^{\eta+r}\|y_i- \widehat{\beta}\|_2^2-\sum_{i=\eta+1}^{\eta+r}\|y_i- \beta^*\|_2^2\leq C_3 r \kappa^2 \frac{\mathfrak{s} \log (p)}{\sqrt{\Delta\kappa^2}}.
$$
Therefore
\begin{equation}
    \left|\widehat{\mathcal{Q}}(\eta+r)-\widehat{\mathcal{Q}}(\eta)-\left\{\mathcal{Q}^*(\eta+r)-\mathcal{Q}^*(\eta)\right\}\right|\leq C_3 r \kappa^2 \frac{\mathfrak{s} \log (p)}{\sqrt{\Delta\kappa^2}}
    \label{tmp_eq:op1 mean main prop eq 1}
\end{equation}
Notice that $\widehat{\mathcal{Q}}(\eta+r)-\widehat{\mathcal{Q}}(\eta)\leq 0$, so our goal is to find a regime where $\mathcal{Q}^*(\eta+r)-\mathcal{Q}^*(\eta)\geq 0$, in order to get rid of the $|\cdot|$.

\textbf{Step 3}. Observe that
$$
\begin{aligned}
\mathcal{Q}^*(\eta+r)-\mathcal{Q}^*(\eta) 
=& \sum_{i=\eta+1}^{\eta+r}\|y_i- \alpha^*\|_2^2-\sum_{i=\eta+1}^{\eta+r}\|y_i- \beta^*\|_2^2 \\
=& r\| \alpha^*- \beta^*\|_2^2-2 \sum_{i=\eta+1}^{\eta+r}(y_i- \beta^*)( \alpha^*- \beta^*) \\
=& r\| \alpha^*- \beta^*\|_2^2 -2 (\alpha^*- \beta^*)^\top\sum_{i=\eta+1}^{\eta+r} \epsilon_i
\end{aligned}
$$
Let
$$
w_i=\frac{1}{\kappa} \epsilon_i^\top\left( \alpha^*-\beta^*\right)
$$
Then $\left\{w_i\right\}_{i=1}^{\infty}$ are subgaussian random variables with bounded $\psi_2$ norm. Therefore by \Cref{lem:subgaussian uniform loglog law}, uniformly for all $r \geq 1 / \kappa^2$, with probability at least $1 -\alpha/2$,
$$
\sum_{i=1}^r w_i\leq4\sqrt{r\left\{\log \log \left(\kappa^2 r\right) +\log\frac{4}{\alpha}+1\right\}}
$$
It follows that
$$
\sum_{i=\eta+1}^{\eta+r} \epsilon_i^\top \left( \alpha^*-\beta^*\right)\leq 4\sqrt{r \kappa^2\left\{\log \log \left(\kappa^2 r\right)+\log\frac{4}{\alpha}+1\right\}}.
$$
Therefore
\begin{equation}
    \begin{split}
        \mathcal{Q}^*(\eta+r)-\mathcal{Q}^*(\eta) &\geq r\kappa^2 - 4\sqrt{r \kappa^2\left\{\log \log \left(\kappa^2 r\right)+\log\frac{4}{\alpha}+1\right\}}\\
        &\geq 
        r\kappa^2 - 4\sqrt{r \kappa^2\{1\vee \log \log \left(\kappa^2 r\right)\}}-4\sqrt{r \kappa^2 \log\frac{4}{\alpha}}-4\sqrt{r \kappa^2}
    \end{split}
\label{tmp_eq:op1 mean main prop eq 2}
\end{equation}
Since $\frac{x}{144} - \log\log (x)\geq 0$ for all $x>0$, when $r\kappa^2\geq \max\{144\log\frac{4}{\alpha},144\}$, we have $\mathcal{Q}^*(\eta+r)-\mathcal{Q}^*(\eta)\geq 0$.

\textbf{Step 4}. \Cref{tmp_eq:op1 mean main prop eq 1} and \Cref{tmp_eq:op1 mean main prop eq 2} together give, uniformly for all $r$ such that $r \kappa^2 \geq 144 (1\vee \log\frac{4}{\alpha})$,
$$
 0\leq r\kappa^2 - 4\sqrt{r \kappa^2\left\{\log \log \left(\kappa^2 r\right)+\log\frac{4}{\alpha}+1\right\}} \leq C_3 r \kappa^2 \frac{\mathfrak{s} \log (p)}{\sqrt{\Delta\kappa^2}}.
$$
Since we assume that $\frac{\mathfrak{s}^2\log^2(p)}{\Delta \kappa^2}\rightarrow 0$, this either leads to a contradiction or implies that $r\kappa^2\leq C_4(1\vee \log\frac{1}{\alpha})$. 
\eprf

\bnlem[Local refinement step 1]
\label{lem:mean refinement step 1}
The output $\check{\eta}$ of step 1 of the local refinement satisfies that with probability at least $1 - n^{-3}$,
\begin{equation}
    \max_{k\in [K]}|\check {\eta}_k - \eta_k|\leq \frac{C \sigma_{\epsilon}^2\s\log(n\vee p)}{\kappa^2}.
\end{equation}
\enlem

\begin{proof}[Proof of \Cref{lem:mean refinement step 1}]
For each $k\in [K]$, let $\widehat {\mu}_t = \widehat {\mu}^{(1)}$ if $s_k<t< \check {\eta}_k$ and $\widehat {\mu}_t =\widehat {\mu}^{(2)}$ otherwise, and ${\mu}^*_t$ be the true parameter at time point $t$. First we show that under conditions $\tilde{K} = K$ and $\max_{k\in [K]}|\tilde{\eta}_k - \eta_k|\leq \Delta / 5$, there is only one true change point $\eta_k$ in $(s_k, e_k)$. It suffices to show that
\begin{equation}
    |\tilde{\eta}_k - \eta_k|\leq \frac{2}{3}(\tilde{\eta}_{k+1} - \tilde{\eta}_{k}),\ \text{and}\ |\tilde{\eta}_{k+1} - \eta_{k+1}|\leq \frac{1}{3}(\tilde{\eta}_{k+1} - \tilde{\eta}_{k}).
    \label{tmp_eq: mean goal_lr}
\end{equation}
Denote $R = \max_{k\in [K]}|\tilde{\eta}_k - \eta_k|$, then
\begin{align*}
    \tilde{\eta}_{k+1} - \tilde{\eta}_{k} &= \tilde{\eta}_{k+1} - {\eta}_{k+1} + {\eta}_{k+1} - {\eta}_{k} + {\eta}_{k} - \tilde{\eta}_{k} \\
    &= ({\eta}_{k+1} - {\eta}_{k}) + (\tilde{\eta}_{k+1} - {\eta}_{k+1}) + ({\eta}_{k} - \tilde{\eta}_{k})\in [{\eta}_{k+1} - {\eta}_{k} - 2R,{\eta}_{k+1} - {\eta}_{k} + 2R].
\end{align*}
Therefore, \Cref{tmp_eq: mean goal_lr} is guaranteed as long as
\begin{equation*}
    R\leq \frac{1}{3}(\Delta - 2R),
\end{equation*}
which is equivalent to $R\leq \Delta/5$.

Now without loss of generality, assume that $s_k<\eta_k<\check {\eta}_k<e_k$. Denote $\mclI_k = \{s_k + 1, \cdots, e_k\}$. Consider two cases:

\textbf{Case 1} If
\begin{equation*}
    \check {\eta}_k - \eta_k < \max\{C\sigma_{\epsilon}^2\s\log(n\vee p), C\sigma_{\epsilon}^2\s\log(n\vee p)/\kappa^2\},
\end{equation*}
then the proof is done.

\textbf{Case 2} If
\begin{equation*}
    \check {\eta}_k - \eta_k \geq \max\{C\sigma_{\epsilon}^2\s\log(n\vee p), C\sigma_{\epsilon}^2\s\log(n\vee p)/\kappa^2\},
\end{equation*}
then we proceed to prove that $|\check {\eta}_k - \eta_k|\leq C\sigma_{\epsilon}^2\s\log(n\vee p)/\kappa^2$ with probability at least $1 - (Tn)^{-3}$. Then we either prove the result or get an contradiction, and complete the proof in either case.

By definition, we have
\begin{equation*}
    \sum_{t\in \I}\|y_t - \widehat {\mu}_t\|_2^2 + \zeta\sum_{i = 1}^p \sqrt{\sum_{i\in \I}(\widehat {\mu}_t)_i^2}\leq \sum_{t\in \I}\|y_t - {\mu}^*_t\|_2^2 + \zeta\sum_{i = 1}^p \sqrt{\sum_{i\in \I}({\mu}^*_t)_i^2},
\end{equation*}
which implies that
\begin{equation*}
    \sum_{t\in \I}\|\mu^*_t - \widehat {\mu}_t\|_2^2 + \zeta\sum_{i = 1}^p \sqrt{\sum_{i\in \I}(\widehat {\mu}_t)_i^2}\leq 2\sum_{t\in \I}(y_t - \mu^*_t)^\top (\widehat {\mu}_t - \mu^*_t) + \zeta\sum_{i = 1}^p \sqrt{\sum_{i\in \I}({\mu}^*_t)_i^2}.
\end{equation*}
Denote $\delta_t = \widehat {\mu}_t - \mu^*_t$. Notice that
\begin{align*}
    & \sum_{i  \in [p]} \sqrt{\sum_{i\in \I}({\mu}^*_t)_i^2} - \sum_{i  \in [p]} \sqrt{\sum_{i\in \I}(\widehat {\mu}_t)_i^2} \\ 
    =& \sum_{i \in [p]} \sqrt{\sum_{i\in \I}({\mu}^*_t)_i^2} - \sum_{i \in S} \sqrt{\sum_{i\in \I}(\widehat {\mu}_t)_i^2} - \sum_{i \in S^c} \sqrt{\sum_{i\in \I}(\widehat {\mu}_t)_i^2} \\
    \leq & \sum_{i \in S} \sqrt{\sum_{i\in \I}(\delta_t)_i^2} - \sum_{i \in S^c} \sqrt{\sum_{i\in \I}(\delta_t)_i^2}.
\end{align*}
Now we check the cross term. Notice that the variance of $\sum_{t\in \I}(\epsilon_t)_i(\delta_t)_i$ is $\sum_{t\in \I}(\delta_t)_i^2$, so with probability at least $1 - (n\vee p)^{-5}$,
\begin{equation*}
    \sum_{t\in \I}(y_t - \mu^*_t)^\top (\widehat {\mu}_t - \mu^*_t)\leq C\sigma_{\epsilon}\sqrt{\log(n\vee p)}\sum_{i \in [p]}\sqrt{\sum_{t\in \I}(\delta_t)_i^2}\leq \frac{\zeta}{4}\sum_{i \in [p]}\sqrt{\sum_{t\in \I}(\delta_t)_i^2},
\end{equation*}
since $\zeta = C_{\zeta}\sigma_{\epsilon}\sqrt{\log(n\vee p)}$ with sufficiently large constant $C_{\zeta}$. Combining inequalities above, we can get
\begin{align*}
    \sum_{t\in \I}\|\delta_t\|_2^2 +\frac{\zeta}{2}\sum_{i\in S^c}\sqrt{\sum_{t\in \I}(\delta_t)_i^2}
    \leq & \frac{3\zeta}{2}\sum_{i\in S}\sqrt{\sum_{t\in \I}(\delta_t)_i^2} \\
    \leq & \frac{3\zeta}{2}\sqrt{\s}\sqrt{\sum_{t\in \I}\|(\delta_t)_S\|_2^2} \\
    \leq & \frac{3\zeta}{2}\sqrt{\s}\sqrt{\sum_{t\in \I}\|\delta_t\|_2^2},
\end{align*}
which implies that
\begin{equation}
    \sum_{t\in \I}\|\delta_t\|_2^2\leq \frac{9}{4}\s\zeta^2\leq C\s\sigma_{\epsilon}^2\log(n\vee p).
    \label{tmp_eq:upper_bound_delta_local_refine}
\end{equation}
Without loss of generality, assume that $\check {\eta}>\eta_k$ and denote
$$
\J_1 = [s_k, \eta_k), \J_2 = [\eta_k, \check {\eta}_k), \J_3 = [\check {\eta}_k, e_k),
$$
and $\mu^{(1)} = \mu^*_{\eta_k - 1}$, $\mu^{(2)} = \mu^*_{\eta_k}$. Then \Cref{tmp_eq:upper_bound_delta_local_refine} is equivalent to
\begin{equation*}
    \J_1\|\widehat {\mu}^{(1)} - \mu^{(1)}\|_2^2 + \J_2\|\widehat {\mu}^{(1)} - \mu^{(2)}\|_2^2 + \J_3\|\widehat {\mu}^{(2)} - \mu^{(2)}\|_2^2 \leq C\s\sigma_{\epsilon}^2\log(n\vee p).
\end{equation*}
Since $|\mclJ_1|=\eta_k - s_k\geq c_0\Delta$ with some constant $c_0$ under \Cref{assp: DCDP_mean}, we have
\begin{equation}
    \Delta \|\widehat {\mu}^{(1)} - \mu^{(1)}\|_2^2\leq c_0|\mclJ_1|\|\widehat {\mu}^{(1)} - \mu^{(1)}\|_2^2 \leq C \sigma_{\epsilon}^2\s\log(n\vee p)\leq c_2\Delta \kappa^2,
\end{equation}
with some constant $c_2\in (0,1/4)$, where the last inequality is due to the fact that $\Bn \rightarrow \infty$. Thus we have
\begin{equation*}
    \|\widehat {\mu}^{(1)} - \mu^{(1)}\|_2^2\leq c_2\kappa^2.
\end{equation*}
Triangle inequality gives
\begin{equation*}
    \|\widehat {\mu}^{(1)} - \mu^{(2)}\|_2\geq \|\mu^{(1)} - \mu^{(2)}\|_2 - \|\widehat {\mu}^{(1)} - \mu^{(1)}\|_2 \geq \kappa/2.
\end{equation*}
Therefore, $\kappa^2|\mclJ_2|/4\leq |\mclJ_2|\|\widehat {\mu}^{(1)} - \mu^{(2)}\|_2^2\leq C \sigma_{\epsilon}^2\s\log(n\vee p)$ and
\begin{equation*}
    |\check {\eta}_k - \eta_k| = |\mclJ_2|\leq \frac{C \sigma_{\epsilon}^2\s\log(n\vee p)}{\kappa^2}.
\end{equation*}
\end{proof}

\clearpage
\subsection{Local refinement in the regression model}
\label{sec:regression op1}
For the ease of notations, we re-index the observations in the $k$-th interval by $[n_0]:\{1,\cdots, n_0\}$ (though the sample size of the problem is still $n$), and denote the $k$-th jump size as $\kappa$ and the minimal spacing between consecutive change points as $\Delta$ (instead of $\Delta_{\min}$ in the main text).

By \Cref{assp:dcdp_linear_reg} and the setting of the local refinement algorithm, we have
$$
y_i=\left\{\begin{array}{ll}
X_i^{\top} \alpha^*+\epsilon_i & \text { when } i \in(0, \eta] \\
X_i^{\top} \beta^*+\epsilon_i & \text { when } i \in(\eta, n_0]
\end{array} .\right.
$$
In addition, there exists $\theta \in(0,1)$ such that $\eta=\lfloor n_0 \theta\rfloor$ and that $\|\alpha^*-\beta^*\|_2=\kappa<\infty$. By \Cref{assp:dcdp_linear_reg}, it holds that $\left\|\alpha^*\right\|_0 \leq \mathfrak{s},\left\|\beta^*\right\|_0 \leq \mathfrak{s}$, and
\begin{equation}
    \frac{\mathfrak{s}^2 \log ^{3}(n\vee p)}{\Delta \kappa^2} \rightarrow 0.
    \label{op1_eq:regression snr}
\end{equation}
By \Cref{lem: regression local refinement step 1}, with probability at least $1 - n^{-2}$, the output of the first step of the PLR algorithm (\Cref{algo:local_refine_general}) $\widehat{\alpha}$ and $\widehat{\beta}$ satisfies that
\begin{equation}
    \begin{split}
          & \|\widehat{\alpha}-\alpha^*\|_2^2\leq C\frac{\mathfrak{s} \log (n\vee p)}{\Delta} \text{ and } \|\widehat{\alpha}-\alpha^*\|_1\leq C\mathfrak{s} \sqrt{\frac{\log (n\vee p)}{\Delta}}; \\
& \|\widehat{\beta}-\beta^*\|_2^2\leq C\frac{\mathfrak{s} \log (n\vee p)}{\Delta} \text{ and } \|\widehat{\beta}-\beta^*\|_1\leq C \mathfrak{s} \sqrt{\frac{\log (n\vee p)}{\Delta}}.
    \end{split}
    \label{eq: regression op1 condition}
\end{equation}

In fact, \Cref{lem: regression local refinement step 1} shows that we are able to remove the extra $\mathcal{B}_n^{-1/2}\Delta_{\min}$ term in the localization error in \Cref{thm:DCDP regression} under the same SNR condition. In \Cref{lem: regression local refinement}, we show that with slightly stronger SNR condition, the localization error can be further reduced as is concluded in \Cref{cor:regression local refinement}.

Let
$$
\widehat{\mathcal{Q}}(k)=\sum_{i=1}^k\left(y_i-X_i^{\top} \widehat{\alpha}\right)^2+\sum_{i=k+1}^{n_0}\left(y_i-X_i^{\top} \widehat{\beta}\right)^2 \quad \text { and } \quad \mathcal{Q}^*(k)=\sum_{i=1}^k\left(y_i-X_i^{\top} \alpha^*\right)^2+\sum_{i=k+1}^{n_0}\left(y_i-X_i^{\top} \beta^*\right)^2 .
$$

\bnlem[Refinement for regression]
\label{lem: regression local refinement}
Let
$$
\eta+r=\underset{k \in(0, n_0]}{\arg \max } \widehat{\mathcal{Q}}(k) .
$$
Then under the assumptions above, it holds with probability at least $1 - (\alpha\vee n^{-1})$ that
$$
 r\kappa^2\leq C \log^2\frac{1}{\alpha}.
$$
where $C$ is a universal constant that only depends on $C_{\kappa}$, $\Lambda_{\min}$, $\sigma_{\epsilon}$.
\enlem
\bprf
For the brevity of notations, we denote $p_n:=n\vee p$ throughout the proof. Without loss of generality, suppose $r \geq 0$. Since $\eta+r$ is the minimizer, it follows that
$$
\widehat{\mathcal{Q}}(\eta+r) \leq \widehat{\mathcal{Q}}(\eta) .
$$
If $r \leq \frac{1}{\kappa^2}$, then there is nothing to show. So for the rest of the argument, for contradiction, assume that
$$
r \geq \frac{1}{\kappa^2}
$$
Observe that
$$
\begin{aligned}
\widehat{\mathcal{Q}}(t)-\widehat{\mathcal{Q}}(\eta) &=\sum_{i=\eta+1}^{\eta+r}\left(y_i-X_i^{\top} \widehat{\alpha}\right)^2-\sum_{i=\eta+1}^{\eta+r}\left(y_i-X_i^{\top} \widehat{\beta}\right)^2 \\
\mathcal{Q}^*(t)-\mathcal{Q}^*(\eta) &=\sum_{i=\eta+1}^{\eta+r}\left(y_i-X_i^{\top} \alpha^*\right)^2-\sum_{i=\eta+1}^{\eta+r}\left(y_i-X_i^{\top} \beta^*\right)^2
\end{aligned}
$$
\textbf{Step 1}. It follows that
$$
\begin{aligned}
& \sum_{i=\eta+1}^{\eta+r}\left(y_i-X_i^{\top} \widehat{\alpha}\right)^2-\sum_{i=\eta+1}^{\eta+r}\left(y_i-X_i^{\top} \alpha^*\right)^2 \\
=& \sum_{i=\eta+1}^{\eta+r}\left(X_i^{\top} \widehat{\alpha}-X_i^{\top} \alpha^*\right)^2+2\left(\widehat{\alpha}-\alpha^*\right)^{\top} X_i \sum_{i=\eta+1}^{\eta+r}\left(y_i-X_i^{\top} \alpha^*\right) \\
=& \sum_{i=\eta+1}^{\eta+r}\left(X_i^{\top} \widehat{\alpha}-X_i^{\top} \alpha^*\right)^2+2\left(\widehat{\alpha}-\alpha^*\right)^{\top} \sum_{i=1}^r X_i X_i^{\top}\left(\beta^*-\alpha^*\right)+2\left(\widehat{\alpha}-\alpha^*\right)^{\top} \sum_{i=\eta+1}^{\eta+r} X_i \epsilon_i
\end{aligned}
$$
By \Cref{lem:regression op1 base lemma 1}, uniformly for all $r$,
$$
\left\|\frac{1}{r} \sum_{i=1}^r X_i X_i^{\top}-\Sigma\right\|_{\infty} \leq C\left(\sqrt{\frac{\log (p_n)}{r}}+\frac{\log (p_n)}{r}\right) .
$$
Therefore
$$
\begin{aligned}
\sum_{i=\eta+1}^{\eta+r}\left(X_i^{\top} \widehat{\alpha}-X_i^{\top} \alpha^*\right)^2 &=\sum_{i=\eta+1}^{\eta+r}\left(\widehat{\alpha}-\alpha^*\right)^{\top} \sum_{i=1}^r\left\{X_i X_i^{\top}-\Sigma\right\}\left(\widehat{\alpha}-\alpha^*\right)+r\left(\widehat{\alpha}-\alpha^*\right)^{\top} \Sigma\left(\widehat{\alpha}-\alpha^*\right) \\
& \leq\|\widehat{\alpha}-\alpha^*\|_1^2\left\|\sum_{i=1}^r X_i X_i^{\top}-\Sigma\right\|_{\infty}+\Lambda_{\max } r\|\widehat{\alpha}-\alpha^*\|_2^2\\
&\leq C_1\frac{\mathfrak{s}^2 \log (p_n)}{\Delta}(\sqrt{r \log (p_n)}+\log (p_n))+C_1 r \frac{\mathfrak{s} \log (p)}{\Delta} \\
&\leq  C_1\sqrt{r}\frac{\mathfrak{s}^2 \log^{3/2} (p_n)}{\Delta}+C_1\frac{\mathfrak{s}^2 \log^2 (p_n)}{\Delta}+C_1 r \frac{\mathfrak{s} \log (p_n)}{\Delta}
\end{aligned}
$$
where the second inequality follows from \Cref{lem:regression op1 base lemma 1}. Similarly
$$
\begin{aligned}
\left(\widehat{\alpha}-\alpha^*\right)^{\top} \sum_{i=1}^r X_i X_i^{\top}\left(\beta^*-\alpha^*\right) &=\left(\widehat{\alpha}-\alpha^*\right)^{\top} \sum_{i=1}^r\left\{X_i X_i^{\top}-\Sigma\right\}\left(\beta^*-\alpha^*\right)+r\left(\widehat{\alpha}-\alpha^*\right)^{\top} \Sigma\left(\beta^*-\alpha^*\right) \\
& \leq \|\widehat{\alpha}-\alpha^*\|_1\|\left(\beta^*-\alpha^*\right)^{\top}\left\{\sum_{i=1}^r X_i X_i^{\top}-\Sigma\right\}\|_{\infty}+\Lambda_{\max } r\|\widehat{\alpha}-\alpha^*\|_2\|\beta^*-\alpha^*\|_2 \\
&\leq C_2\mathfrak{s} \sqrt{\frac{\log (p_n)}{\Delta}} (\kappa\sqrt{r \log (p_n)}+\kappa\log (p_n))+C_2 r\kappa \sqrt{\frac{\mathfrak{s} \log (p_n)}{\Delta} } .\\
&\leq C_2\mathfrak{s}\kappa\log (p_n)\sqrt{\frac{r}{\Delta}} +C_2\mathfrak{s}\kappa\sqrt{\frac{\log^3 (p_n)}{\Delta}} + C_2 r\kappa \sqrt{\frac{\mathfrak{s} \log (p_n)}{\Delta} }.
\end{aligned}
$$
where the second equality follows from $\|\beta^*-\alpha^*\|_2=\kappa$ and \Cref{lem:regression op1 base lemma 2}. In addition,
$$
\begin{aligned}
&\left(\widehat{\alpha}-\alpha^*\right)^{\top} \sum_{i=\eta+1}^{\eta+r} X_i \epsilon_i \leq\|\widehat{\alpha}-\alpha^*\|_1\|\sum_{i=\eta+1}^{\eta+r} X_i \epsilon_i\|_{\infty} \\
\leq & C_3\mathfrak{s} \sqrt{\frac{\log (p_n)}{\Delta}}(\sqrt{r \log (p_n)}+\log (p_n))\leq C_3\mathfrak{s}\log (p_n)\sqrt{\frac{r}{\Delta}} +C_3\s \sqrt{\frac{\log^3 (p_n)}{\Delta}}.
\end{aligned}
$$
where the second equality follows from \Cref{lem:regression op1 base lemma 1}. Therefore
\begin{align*}
    &\sum_{i=\eta+1}^{\eta+r}\left(y_i-X_i^{\top} \widehat{\alpha}\right)^2-\sum_{i=\eta+1}^{\eta+r}\left(y_i-X_i^{\top} \alpha^*\right)^2\\
    \leq & C_4(\kappa + 1)\mathfrak{s}\log (p_n)\sqrt{\frac{r}{\Delta}} + C_4(\kappa + 1)\s \sqrt{\frac{\log^3 (p_n)}{\Delta}} + C_4 r\kappa^2 \sqrt{\frac{\s\log (p_n)}{\Delta\kappa^2}}\\
    & \quad + C_1\frac{\s^2\log^2(p_n)}{\Delta} + C_1\sqrt{r}\frac{\s^2\log^{3/2}(p_n)}{\delta}\\
    \leq & C_4({\kappa} + 1)r\kappa^2\sqrt{\frac{\mathfrak{s}^2\log^2 (p_n)}{\Delta\kappa^2}} + C_4(\kappa^2 + \kappa) \sqrt{\frac{\s^2\log^3 (p_n)}{\Delta\kappa^2}}  + C_4 {\kappa}\sqrt{r\kappa^2} \frac{\s^2\log^{3/2}(p_n)}{\Delta \kappa^2}\\
    \leq & C_4({\kappa} + 1)r\kappa^2\sqrt{\frac{\mathfrak{s}^2\log^2 (p_n)}{\Delta\kappa^2}} + C_4\kappa(\kappa + 1 + \sqrt{r\kappa^2}) \sqrt{\frac{\s^2\log^3 (p_n)}{\Delta\kappa^2}}\leq C_5(C_\kappa^2 + 1){r\kappa^2} \sqrt{\frac{\s^2\log^3 (p_n)}{\Delta\kappa^2}}.
\end{align*}
where we use the assumption that $\Delta \kappa^2 \geq \mathcal{B}_n\s^2\log^2(p_n)$, $\kappa\leq C_{\kappa}$, and $r\kappa^2 \geq 1$.

\textbf{Step 2}. Using the same argument as in the previous step, it follows that
$$
\sum_{i=\eta+1}^{\eta+r}\left(y_i-X_i^{\top} \widehat{\beta}\right)^2-\sum_{i=\eta+1}^{\eta+r}\left(y_i-X_i^{\top} \beta^*\right)^2\leq C_5(C_\kappa^2 + 1){r\kappa^2} \sqrt{\frac{\s^2\log^3 (p_n)}{\Delta\kappa^2}}.
$$
Therefore
\begin{equation}
    \left|\widehat{\mathcal{Q}}(\eta+r)-\widehat{\mathcal{Q}}(\eta)-\left\{\mathcal{Q}^*(\eta+r)-\mathcal{Q}^*(\eta)\right\}\right|\leq C_5(C_\kappa^2 + 1){r\kappa^2} \sqrt{\frac{\s^2\log^3 (p_n)}{\Delta\kappa^2}}.
    \label{tmp_eq:op1 regression main prop eq 1}
\end{equation}
\textbf{Step 3}. Observe that
$$
\begin{aligned}
& \mathcal{Q}^*(\eta+r)-\mathcal{Q}^*(\eta) \\
=& \sum_{i=\eta+1}^{\eta+r}\left(y_i-X_i^{\top} \alpha^*\right)^2-\sum_{i=\eta+1}^{\eta+r}\left(y_i-X_i^{\top} \beta^*\right)^2 \\
=& \sum_{i=\eta+1}^{\eta+r}\left(X_i^{\top} \alpha^*-X_i^{\top} \beta^*\right)^2-2 \sum_{i=\eta+1}^{\eta+r}\left(y_i-X_i^{\top} \beta^*\right)\left(X_i^{\top} \alpha^*-X_i^{\top} \beta^*\right) \\
=& \sum_{i=\eta+1}^{\eta+r}\left(\alpha^*-\beta^*\right)^{\top}\left\{X_i^{\top} X_i-\Sigma\right\}\left(\alpha^*-\beta^*\right)+r\left(\alpha^*-\beta^*\right)^{\top} \Sigma\left(\alpha^*-\beta^*\right)-2 \sum_{i=\eta+1}^{\eta+r} \epsilon_i\left(X_i^{\top} \alpha^*-X_i^{\top} \beta^*\right)
\end{aligned}
$$
Note that
$$
z_i=\frac{1}{\kappa^2}\left(\alpha^*-\beta^*\right)^{\top}\left\{X_i^{\top} X_i-\Sigma\right\}\left(\alpha^*-\beta^*\right)
$$
is a sub-exponential random variable with bounded $\psi_1$ norm. Therefore by \Cref{lem:subexp uniform loglog law}, uniformly for all $r \geq 1 / \kappa^2$, with probability at least $1 - \alpha / 2$,
$$
\sum_{i=1}^r z_i\leq 4\left(\sqrt{r\left\{\log \log \left(\kappa^2 r\right) + \log\frac{4}{\alpha}+1\right\}}+\sqrt{r \kappa^2}\left\{\log \log \left(\kappa^2 r\right)+\log\frac{4}{\alpha}+1\right\}\right).
$$
It follows that
\begin{align*}
    \sum_{i=\eta+1}^{\eta+r}\left(\alpha^*-\beta^*\right)^{\top} &\left\{X_i^{\top} X_i-\Sigma\right\}\left(\alpha^*-\beta^*\right) \\
    &\leq 4\left(\kappa^2\sqrt{r\left\{\log \log \left(\kappa^2 r\right) + \log\frac{4}{\alpha}+1\right\}}+\kappa^3\sqrt{r}\left\{\log \log \left(\kappa^2 r\right)+\log\frac{4}{\alpha}+1\right\}\right).
\end{align*}
Similarly, let
$$
w_i=\frac{1}{\kappa} \epsilon_i\left(X_i^{\top} \alpha^*-X_i^{\top} \beta^*\right)
$$
Then $\left\{w_i\right\}_{i=1}^{\infty}$ are sub-exponential random variables with bounded $\psi_1$ norm. Therefore by \Cref{lem:subexp uniform loglog law}, uniformly for all $r \geq 1 / \kappa^2$,
$$
\sum_{i=1}^r w_i\leq 4\left(\sqrt{r\left\{\log \log \left(\kappa^2 r\right) +\log\frac{4}{\alpha}+1\right\}}+\sqrt{r \kappa^2}\left\{\log \log \left(\kappa^2 r\right)+\log\frac{4}{\alpha}+1\right\}\right)
$$
It follows that
\begin{align*}
    \sum_{i=\eta+1}^{\eta+r} \epsilon_i &\left(X_i^{\top} \alpha^*-X_i^{\top} \beta^*\right)\\
    &\leq 4\left(\sqrt{r \kappa^2\left\{\log \log \left(\kappa^2 r\right)+\log\frac{4}{\alpha}+1\right\}}+\kappa^2\sqrt{r}\left\{\log \log \left(\kappa^2 r\right)+\log\frac{4}{\alpha}+1\right\}\right).
\end{align*}
Therefore
\begin{equation}
    \begin{split}
        \mathcal{Q}^*(\eta+r)-\mathcal{Q}^*(\eta) \geq & \Lambda_{\min } r \kappa^2 - 4(\kappa + 1)\sqrt{r \kappa^2\left\{\log \log \left(\kappa^2 r\right)+\log\frac{4}{\alpha}+1\right\}}\\
        & \quad - 4(\kappa^2 + \kappa)\sqrt{r \kappa^2}\left\{\log \log \left(\kappa^2 r\right)+\log\frac{4}{\alpha}+1\right\}\\
        \geq & \Lambda_{\min } r \kappa^2 - 16\sqrt{r\kappa^2}(\kappa^2\vee 1)(1 + \log\frac{4}{\alpha} + \{1\vee \log\log (r\kappa^2)\})\\
        \geq & \Lambda_{\min } r\kappa^2  - 16\sqrt{r\kappa^2}(C_{\kappa}^2\vee 1)(1 + \log\frac{4}{\alpha} + \{1\vee \log\log (r\kappa^2)\}).
    \end{split}
\label{tmp_eq:op1 regression main prop eq 2}
\end{equation}
where $\Lambda_{\min}$ is the minimal eigenvalue of $\Sigma$. By \Cref{lem: x - c log^2log x}, for $r\kappa^2 \geq \frac{48^2(C_{\kappa}^2\vee 1)^2}{\Lambda_{\min}^2} \vee e^{2e}$, $\frac{\Lambda_{\min}}{3}r\kappa^2\geq \sqrt{r\kappa^2}\log\log (r\kappa^2)$. Thus, when $r\kappa^2\geq (\frac{48^2(C_{\kappa}^2\vee 1)^2}{\Lambda_{\min}^2}\log^2\frac{4}{\alpha}) \vee e^{2e}$, we have $\mathcal{Q}^*(\eta+r)-\mathcal{Q}^*(\eta)\geq 0$.

\textbf{Step 4}. \Cref{tmp_eq:op1 regression main prop eq 1} and \Cref{tmp_eq:op1 regression main prop eq 1} together give that, uniformly for all $r$ such that $r\kappa^2 \geq (\frac{48^2(C_{\kappa}^2\vee 1)^2}{\Lambda_{\min}^2}\log^2\frac{4}{\alpha}) \vee e^{2e}$, with probability at least $1 - (\alpha\vee n^{-1})$
$$
\Lambda_{\min } r \kappa^2 - 16\sqrt{r\kappa^2}(C_{\kappa}^2\vee 1)(1 + \log\frac{4}{\alpha} + \{1\vee \log\log (r\kappa^2)\}) \leq C_5(C_\kappa^2 + 1){r\kappa^2} \sqrt{\frac{\s^2\log^3 (p_n)}{\Delta\kappa^2}},
$$
which either leads to a contradiction or implies the conclusion.
\eprf

In what follows, we first show that the first step in the local refinement gives estimators $\hat{\alpha}, \hat{\beta}$ that satisfies \Cref{eq: regression op1 condition}, and then prove some relevant lemmas.

\bnlem[Local refinement step 1]
\label{lem: regression local refinement step 1}
For each $k\in [K]$, let $\check{\eta}_k, \widehat{\beta}^{(1)}, \widehat{\beta}^{(2)}$ be the output of step 1 of the local refinement algorithm for linear regression, with
\begin{equation*}
    R(\theta^{(1)}, \theta^{(2)},\eta; s, e) = \zeta  \sum_{i \in [p]} \sqrt{(\eta - s)(\theta^{(1)}_{i} )^2 + (e - \eta)(\theta_{i} ^{(2)}) ^2}.
\end{equation*}
and $\zeta = C_{\zeta}\sqrt{\log (n\vee p)}$. Then with probability at least $1 - n^{-3}$, it holds that
\begin{equation}
    \max_{k\in [\hat{K}]} |\check{\eta}_k - \eta_k|\leq C\frac{\s \log(n\vee p)}{\kappa^2}.
\end{equation}
\enlem

\begin{proof}[Proof of \Cref{lem: regression local refinement step 1}]
For each $k\in [K]$, let $\widehat {\beta}_t = \widehat {\beta}^{(1)}$ if $s_k<t< \widehat {\eta}_k$ and $\widehat {\beta}_t =\widehat {\beta}^{(2)}$ otherwise. Let ${\beta}^*_t$ be the true parameter at time point $t$, and $\beta^{(1)} = \beta^*_{\eta_k}$ and $\beta^{(2)} = \beta^*_{\eta_k + 1}$. First we show that under conditions $\tilde{K} = K$ and $\max_{k\in [K]}|\tilde{\eta}_k - \eta_k|\leq \Delta / 5$, there is only one true change point $\eta_k$ in $(s_k, e_k)$. It suffices to show that
\begin{equation}
    |\tilde{\eta}_k - \eta_k|\leq \frac{2}{3}(\tilde{\eta}_{k+1} - \tilde{\eta}_{k}),\ \text{and}\ |\tilde{\eta}_{k+1} - \eta_{k+1}|\leq \frac{1}{3}(\tilde{\eta}_{k+1} - \tilde{\eta}_{k}).
    \label{tmp_eq:goal_lr}
\end{equation}
Denote $R = \max_{k\in [K]}|\tilde{\eta}_k - \eta_k|$, then
\begin{align*}
    \tilde{\eta}_{k+1} - \tilde{\eta}_{k} &= \tilde{\eta}_{k+1} - {\eta}_{k+1} + {\eta}_{k+1} - {\eta}_{k} + {\eta}_{k} - \tilde{\eta}_{k} \\
    &= ({\eta}_{k+1} - {\eta}_{k}) + (\tilde{\eta}_{k+1} - {\eta}_{k+1}) + ({\eta}_{k} - \tilde{\eta}_{k})\in [{\eta}_{k+1} - {\eta}_{k} - 2R,{\eta}_{k+1} - {\eta}_{k} + 2R].
\end{align*}
Therefore, \Cref{tmp_eq:goal_lr} is guaranteed as long as
\begin{equation*}
    R\leq \frac{1}{3}(\Delta - 2R),
\end{equation*}
which is equivalent to $R\leq \Delta/5$.

Now without loss of generality, assume that $s_k<\eta_k<\widehat {\eta}_k<e_k$. Denote $\mclI_k = \{s_k + 1, \cdots, e_k\}$. Consider two cases:

\textbf{Case 1.} If
\begin{equation*}
    \widehat {\eta}_k - \eta_k < \max\{C_s(\sigma_{\epsilon}^2\vee 1)\s\log(n\vee p), C_s(\sigma_{\epsilon}^2\vee 1)\s\log(n\vee p)/\kappa^2\},
\end{equation*}
then the proof is done.

\textbf{Case 2.} If
\begin{equation*}
    \widehat {\eta}_k - \eta_k \geq \max\{C_s(\sigma_{\epsilon}^2\vee 1)\s\log(n\vee p), C_s(\sigma_{\epsilon}^2\vee 1)\s\log(n\vee p)/\kappa^2\},
\end{equation*}
then we proceed to prove that $|\widehat {\eta}_k - \eta_k|\leq C(\sigma_{\epsilon}^2\vee 1)\s\log(n\vee p)/\kappa^2$ with probability at least $1 - (n\vee p)^{-5}$. Then we either prove the result or get an contradiction, and complete the proof in either case. The first step is to prove that with probability at least $1 - (n\vee p)^{-5}$,
	\[
		\sum_{t = s_k + 1}^{e_k}\|\widehat{\beta}_t - \beta^*_t\|_2^2 \leq C_1\s \zeta^2.
	\]
By definition, it holds that
	\begin{align}\label{eq:reg local refine 1}
		& \sum_{t = s_k + 1}^{e_k} (y_t - X_t^{\top}\widehat{\beta}_t)^2 + \zeta \sum_{i = 1}^p \sqrt{\sum_{t = s_k + 1}^{e_k} \bigl(\widehat{\beta}_t\bigr)_i^2} \leq \sum_{t = s_k + 1}^{e_k} (y_{t} - X_t^{\top}\beta^*_t )^2 + \zeta \sum_{i = 1}^p \sqrt{\sum_{t = s_k + 1}^{e_k} \bigl(\beta^*_t\bigr)_{i}^2}.
	\end{align}
	Let $\delta_t = \widehat{\beta}_t - \beta^*_t$. It holds that $\sum_{t = s_k + 1}^{e_k - 1}\mathbbm{1}\left\{\delta_t \neq \delta_{t+1}\right\} = 2$.	Then \Cref{eq:reg local refine 1} implies that
	\begin{align}\label{eq:reg local refine 2}
		\sum_{t = s_k + 1}^{e_k} (\delta_t^{\top} X_t)^2 + \zeta  \sum_{i = 1}^p \sqrt{\sum_{t = s_k + 1}^{e_k} \bigl(\widehat{\beta}_t\bigr)_{i}^2}\leq 2\sum_{t = s_k + 1}^{e_k} (y_t - X_t^{\top}\beta^*_t)\delta_t^{\top} X_t + \zeta \sum_{i = 1}^p \sqrt{\sum_{t = s_k + 1}^{e_k} \bigl(\beta^*_t\bigr)_{i}^2}.
	\end{align}
Note that
	\begin{align}
		& \sum_{i = 1}^p \sqrt{\sum_{t = s_k + 1}^{e_k} \bigl(\beta^*_t\bigr)_{i}^2} - \sum_{i = 1}^p \sqrt{\sum_{t = s_k + 1}^{e_k} \bigl(\widehat{\beta}_t\bigr)_{i}^2} \nonumber \\
		=& \sum_{i \in S} \sqrt{\sum_{t = s_k + 1}^{e_k} \bigl(\beta^*_t\bigr)_{i}^2} - \sum_{i \in S} \sqrt{\sum_{t = s_k + 1}^{e_k} \bigl(\widehat{\beta}_t\bigr)_{i}^2}  - \sum_{i \in S^c} \sqrt{\sum_{t = s_k + 1}^{e_k} \bigl(\widehat{\beta}_t\bigr)_{i}^2} \nonumber \\
		\leq & \sum_{i \in S} \sqrt{\sum_{t = s_k + 1}^{e_k} \bigl(\delta_t\bigr)_{i}^2}  - \sum_{i \in S^c} \sqrt{\sum_{t = s_k + 1}^{e_k} \bigl(\delta_t\bigr)_{i}^2}.\label{eq:reg local refine 3}
	\end{align}
We then examine the cross term, with probability at least $1 - (n\vee p)^{-5}$, which satisfies the following
	\begin{align}
		& \left|\sum_{t = s_k + 1}^{e_k} (y_t - X_t^{\top}\beta^*_t)\delta_t^{\top} X_t\right| = \left|\sum_{t = s_k + 1}^{e_k} \epsilon_{t} \delta_t^{\top} X_t\right| = \sum_{i = 1}^p \left\{\left|\frac{\sum_{t = s_k + 1}^{e_k} \epsilon_{t} (\delta_t)_i (X_t)_i}{\sqrt{\sum_{t = s_k + 1}^{e_k} (\delta_t)^2_i}}\right| \sqrt{\sum_{t = s_k + 1}^{e_k} (\delta_t)^2_i}\right\} \nonumber \\
		\leq & \sup_{i = 1, \ldots, p} \left|\frac{\sum_{t = s_k + 1}^{e_k} \epsilon_{t} (\delta_t)_i (X_t)_i}{\sqrt{\sum_{t = s_k + 1}^{e_k} (\delta_t)^2_i}}\right| \sum_{i = 1}^p \sqrt{\sum_{t = s_k + 1}^{e_k} (\delta_t)^2_i} \leq (\zeta/4) \sum_{i = 1}^p \sqrt{\sum_{t = s_k + 1}^{e_k} (\delta_t)^2_i}, \label{eq:reg local refine 4}
	\end{align}
	where the second inequality follows from \Cref{lem:deviation piecewise constant}.

Combining \eqref{eq:reg local refine 1}, \eqref{eq:reg local refine 2}, \eqref{eq:reg local refine 3} and \eqref{eq:reg local refine 4} yields
	\begin{equation}\label{eq:reg local refine 5}
		\sum_{t = s_k + 1}^{e_k} (\delta_t^{\top} X_t)^2	 + \frac{\zeta}{2}\sum_{i \in S^c} \sqrt{\sum_{t = s_k + 1}^{e_k} \bigl(\delta_t\bigr)_{i}^2} \leq \frac{3\zeta}{2}\sum_{i \in S} \sqrt{\sum_{t = s_k + 1}^{e_k} \bigl(\delta_t\bigr)_{i}^2}.
	\end{equation}
Now we are to explore the restricted eigenvalue inequality. Let
	\[
		\I_1 = (s_k, \eta_k], \quad \I_2 = (\eta_k, \widehat{\eta}_k], \quad \I_3 = (\widehat{\eta}_k, e_k].
	\]
Then for $\I_1$, it holds that
	\begin{align}
		& \eta_k - s_k = \eta_k - \frac{2}{3}\widetilde{\eta}_k - \frac{1}{3}\widetilde{\eta}_k \nonumber \\
		= & \frac{2}{3}(\eta_k - \eta_{k-1}) + \frac{2}{3}(\widetilde{\eta}_k - \eta_k) -  \frac{2}{3}(\widetilde{\eta}_{k-1} - \eta_{k-1}) + (\eta_k - \widetilde{\eta}_k) \nonumber \\
		\geq & \frac{2}{3}\Delta - \frac{1}{3}\Delta = \frac{1}{3}\Delta, \label{tmp_eq:regression I_1}
	\end{align}	
where the inequality follows from \Cref{assp:dcdp_linear_reg} and \Cref{tmp_eq:goal_lr}. 

For $\I_3$, by the design of the local refinement algorithm in \Cref{algo:local_refine_general}, we have $|\I_3|\geq C_s \s\log(n\vee p)$. Since $\min\{|\I_1|,|\I_3|\}\geq C_s\s\log(n\vee p)$, by \Cref{lem:restricted eigenvalue}, it holds with probability at least $1 - (n\vee p)^{-5}$ that,
	\begin{align*}
		& \sum_{i = 1, 3}\sum_{t \in \I_i}\|\delta_{\I_i}^{\top}X_t\|_2^2 \\
		\geq & \sum_{i = 1, 3} \left(c'_1\sqrt{|\I_i|} \|\delta_{\I_i}\|_2 - c_2 \sqrt{\log(p)} \|\delta_{\I_i}\|_1\right)^2 \\
		\geq & \sum_{i = 1, 3} \left(c_1\sqrt{|\I_i|} \|\delta_{\I_i}\|_2 - c_2 \sqrt{\log(p)} \|(\delta_{\I_i})_{S^c}\|_1\right)^2,
	\end{align*}
	where the last inequality follows from $\|(\delta_{\I})_S\|_1\leq \sqrt{\s}\|\delta_{\I}\|_2$ and the fact that $\min\{|\I_1|,|\I_3|\}>C_s \s \log(n\vee p)$. Similarly, since $|\I_2| > \Delta> C_s \s \log(n\vee p)$, we have
	\[
		\sqrt{\sum_{t \in \I_2}(\delta_{\I_2}^{\top}X_t)^2} \geq {c_1\sqrt{|\I_2|}} \|\delta_{\I_2}\|_2 - c_2 \sqrt{\log(p)} \|\delta_{\I_2}(S^c)\|_1.
	\]
Denote $n_0 = C_s\s\log(n\vee p)$. We first bound the terms with $\|\cdot\|_1$. Note that
	\begin{align*}
	&\sum_{i = 1}^3\sum_{j\in S^c}|(\delta_{\I_i})_j|
		\leq \sqrt{3}\sqrt{\sum_{i = 1}^3 (\sum_{j \in S^c} |(\delta_{\I_i})_j|)^2 } \\ 
		\leq & \sqrt{3}\sqrt{\sum_{i = 1}^3{\frac{|\I_i|}{n_0}} (\sum_{j \in S^c}|(\delta_{\I_i})_j| )^2}	
		\leq  \sqrt{\frac{3}{n_0}} \sum_{i = 1}^3   \sqrt{|\I_i|}(\sum_{j \in S^c}|(\delta_{\I_i})_j| ) \\
		\leq &
		\sqrt{\frac{3}{n_0}} \sum_{j \in S^c}  \sqrt{\sum_{t = s_k + 1}^{e_k} (\delta_t)_j^2}
		\leq \frac{3\sqrt{3}}{\sqrt{n_0}}\sum_{j \in S}  \sqrt{\sum_{t = s_k + 1}^{e_k} (\delta_t)_j^2}
		\\
		\leq & \frac{3\sqrt{3}}{\sqrt{n_0}} \sqrt{\s \sum_{j \in S} \sum_{t = s_k + 1}^{e_k} (\delta_t)_j^2} \leq \frac{c}{ \sqrt{\log(n\vee p)}} \sqrt{\sum_{t = s_k + 1}^{e_k} \|\delta_t\|_2^2}.
	\end{align*}
Therefore,
	\begin{align*}
		& c_1\sqrt{\sum_{t = s_k + 1}^{e_k} \|\delta_t\|_2^2} - \frac{c_2}{ \sqrt{\log(n\vee p)}} \sqrt{\sum_{t = s_k + 1}^{e_k} \|\delta_t\|_2^2} \\
		\leq & \sum_{i = 1}^3 c_1 \|\delta_{I_i}\|_2 - \frac{c_2}{ \sqrt{\log(n\vee p)}} \sqrt{\sum_{t = s_k + 1}^{e_k} \|\delta_t\|_2^2} \leq \sqrt{3} \sqrt{ \sum_{t = s_k + 1}^{e_k} (\delta_t^{\top} X_t)^2 } \\
		\leq & \frac{3\sqrt{\zeta}}{\sqrt{2}}\s^{1/4} \left(\sum_{t = s_k + 1}^{e_k} \|\delta_t\|_2^2\right)^{1/4} \leq \frac{9\zeta \s^{1/2}}{4c_1} + \frac{c_1}{2}\sqrt{\sum_{t = s_k + 1}^{e_k} \|\delta_t\|_2^2}
	\end{align*}
	where the third inequality follows from \eqref{eq:reg local refine 5} and the fact that $\sum_{i\in S}\sqrt{\sum_{t=s_k + 1}^{e_k}(\delta_t)_i^2}\leq \sqrt{s}\sqrt{\sum_{t=s_k + 1}^{e_k}\|\delta_t\|_2^2}$. The inequality above implies that
	\[
		\frac{c_1}{4}\sqrt{\sum_{t = s_k + 1}^{e_k} \|\delta_t\|_2^2} \leq \frac{9\zeta \s^{1/2}}{4c_1}
	\]	
	Therefore,
	\[
		\sum_{t = s_k + 1}^{e_k}\|\widehat{\beta}_t - \beta^*_t\|_2^2 \leq 81\zeta^2\s/c_1^4.
	\]
  Recall that $\beta^{(1)} = \beta^*_{\eta_k}$ and $\beta^{(2)} = \beta^*_{\eta_k + 1}$. We have that
	\[
		\sum_{t = s_k + 1}^{e_k}\|\widehat{\beta}_t - \beta^*_t\|_2^2 = |I_1| \|\beta^{(1)} - \widehat{\beta}^{(1)}\|_2^2 + |I_2| \|\beta^{(2)} - \widehat{\beta}^{(1)}\|_2^2 + |I_3| \|\beta^{(2)} - \widehat{\beta}^{(2)}\|_2^2.
	\]
Since $\eta_k - s_k \geq\frac{1}{3}\Delta$ as is shown in \Cref{tmp_eq:regression I_1}. we have that
	\[
		\Delta\|\beta^{(1)} - \widehat{\beta}^{(1)}\|_2^2/3 \leq |I_1| \|\beta^{(1)} - \widehat{\beta}^{(1)}\|_2^2 \leq \frac{C_1 C_{\zeta}^2\Delta \kappa^2} {\s K \sigma^2_{\epsilon} \Bn } \leq c_3 \Delta \kappa^2,
	\]
	where $1/4 > c_3 > 0$ is an arbitrarily small positive constant.  Therefore we have
	\[
		\|\beta^{(2)} - \widehat{\beta}_1\|_2^2 \leq 3c_3 \kappa^2.
	\]
In addition we have
	\[
		\|\beta^{(2)} - \widehat{\beta}^{(1)}\|_2 \geq \|\beta^{(2)} - \beta^{(1)}\|_2 - \|\beta^{(1)} - \widehat{\beta}^{(1)}\|_2 \geq \kappa/2.
	\]	
	Therefore, it holds that 	
	\[
		\kappa^2 |I_2|/4 \leq |I_2| \|\beta^{(2)} - \widehat{\beta}^{(1)}\|_2^2 \leq C_2\s\zeta^2,
	\]
	which implies that 
	\[
		|\widehat{\eta}_k - \eta_k| \leq \frac{4C_2\s \zeta^2}{\kappa^2},
	\]
which gives the bound we want. 
\end{proof}

\bnlem
\label{lem:regression op1 base lemma 1}
Suppose $\left\{X_i\right\}_{i=1}^n \stackrel{i . i . d}{\sim} N_p(0, \Sigma)$ and $\left\{\epsilon_i\right\}_{i=1}^n \stackrel{i i . d}{\sim} N\left(0, \sigma^2\right)$. Then it holds that
$$
\begin{aligned}
&\mathbb{P}\left(\|\frac{1}{r} \sum_{i=1}^r X_i X_i^{\top}-\Sigma\|_{\infty} \geq C_1\left(\sqrt{\frac{\log (p\vee n)}{r}}+\frac{\log (p\vee n)}{r}\right) \text { for all } 1 \leq r \leq n\right) \leq (n\vee p)^{-2}, \\
&\mathbb{P}\left(\|\frac{1}{r} \sum_{i=1}^r X_i \epsilon_i\|_{\infty} \geq C_2\left(\sqrt{\frac{\log (p\vee n)}{r}}+\frac{\log (p\vee n)}{r}\right) \text { for all } 1 \leq r \leq n\right) \leq (n\vee p)^{-2}
\end{aligned}
$$
\enlem
\bprf
Proof. For the first probability bound, observe that for any $j, k \in[1, \ldots, p], X_j X_k-\Sigma_{j k}$ is subexponential random variable. Therefore for any $r>0$,
$$
\mathbb{P}\left\{\left|\frac{1}{r} \sum_{i=1}^r X_{i j} X_{i k}-\Sigma_{j k}\right| \geq x\right\} \leq \exp \left(-r c_1 x^2\right)+\exp \left(-r c_2 x\right)
$$
So
$$
\mathbb{P}\left\{\left\|\frac{1}{r} \sum_{i=1}^r X_{i j} X_{i k}-\Sigma_{j k}\right\|_{\infty} \geq x\right\} \leq p \exp \left(-r c_1 x^2\right)+p \exp \left(-r c_2 x\right) .
$$
This gives, for sufficiently large $C_1>0$,
$$
\mathbb{P}\left\{\left\|\frac{1}{r} \sum_{i=1}^r X_{i j} X_{i k}-\Sigma_{j k}\right\|_{\infty} \geq C_1\left(\sqrt{\frac{\log (p\vee n)}{r}}+\frac{\log (p\vee n)}{r}\right)\right\} \leq (n\vee p)^{-3} .
$$
By a union bound,
$$
\mathbb{P}\left\{\left\|\frac{1}{r} \sum_{i=1}^r X_{i j} X_{i k}-\Sigma_{j k}\right\|_{\infty} \geq C_1\left(\sqrt{\frac{\log (p\vee n)}{r}}+\frac{\log (p\vee n)}{r}\right) \text { for all } 1 \leq r \leq n\right\} \leq (n\vee p)^{-2}
$$
The desired result follows from the assumption that $p \geq n^\alpha$. The second probability bound follows from the same argument and therefore is omitted for brevity.
\eprf

\bnlem
\label{lem:regression op1 base lemma 2}
 Suppose $\left\{X_i\right\}_{i=1}^n \stackrel{i.i.d.}{\sim} N_p(0, \Sigma)$ and $u \in \mathbb{R}^p$ is a deterministic vector such that $|u|_2=1$. Then it holds that
$$
\mathbb{P}\left(\|u^{\top}\left\{\frac{1}{r} \sum_{i=1}^r X_i X_i^{\top}-\Sigma\right\}\|_{\infty} \geq C_1\left(\sqrt{\frac{\log (p\vee n)}{r}}+\frac{\log (p\vee n)}{r}\right) \text { for all } 1 \leq r \leq n\right) \leq (n\vee p)^{-2} .
$$
\enlem
\bprf
For fixed $j \in[1, \ldots, p]$, let
$$
z_i=u^{\top} X_i X_{i j}-u^{\top} \Sigma_{\cdot j},
$$
where $\Sigma_{\cdot j}$ denote the $j$-th column of $\Sigma$. Note that $z_i$ is a sub-exponential random variable with bounded $\psi_1$ norm. The desired result follows from the same argument as \Cref{lem:regression op1 base lemma 1}.
\eprf

\bnlem
\label{lem: x - c log^2log x}
Given a fixed constant $c>0$, for $x\geq c^2\vee e^{2e}$, it holds that
$$
x\geq c(\log\log x)^2.
$$
\enlem
\bprf
Let $f(x)  =x- c(\log\log x)^2$ for $x>1$. We have $f'(x) = 1 - \frac{2c\log\log x}{x\log x}$. Therefore, when $x\geq (2c)\vee e^e$, $f'(x)> 0$. Let $x_0= c^2 \vee e^{2e}$, and then
$$
f(x_0)\geq ce^e - c\log\log e^{2e}= c[e^e - \log 2 - 1]>0,
$$
and thus $f(x)>0$ for $x\geq x_0 = c^2 \vee e^{2e}$.
\eprf

\clearpage
\subsection{Local refinement in the Gaussian graphical model}
\label{sec:cov op1}
For the ease of notations, we re-index the observations in the $k$-th interval by $[n_0]:\{1,\cdots, n_0\}$ (though the sample size of the problem is still $n$), and denote the $k$-th jump size as $\kappa$ and the minimal spacing between consecutive change points as $\Delta$ (instead of $\Delta_{\min}$ in the main text).

By \Cref{assp:DCDP_covariance main} and the setting of the local refinement algorithm, we have for some $G^*,H^*\in \mathbb{S}_+^{p}$ that
$$
\mathbb{E}[X_iX_i^\top]=\left\{\begin{array}{ll}
G^* & \text { when } i \in(0, \eta] \\
H^* & \text { when } i \in(\eta, n_0]
\end{array} .\right.
$$
In addition, there exists $\theta \in(0,1)$ such that $\eta=\lfloor n_0 \theta\rfloor$ and that $\|G^*-H^*\|_F=\kappa_{F}<\infty$. By \Cref{assp:DCDP_covariance main}, it holds that $c_XI_d \preceq G^* \preceq C_X I_d$, $c_XI_d \preceq H^* \preceq C_X I_d$, and
\begin{equation}
    \frac{p^4\log^2(p_n)}{\Delta\kappa_F^2}\rightarrow 0, \ \frac{p^5\log^3(p_n)}{\Delta}\rightarrow 0
    \label{op1_eq:cov snr}
\end{equation}
By \Cref{lem: covariance local refinement step 1}, there exist $\widehat{G},\widehat{H}$ such that
\begin{equation}
    \begin{split}
    & \|\widehat{G}-G^*\|_{op}\leq C\sqrt{\frac{p \log (n\vee p)}{\Delta}} \text{ and } \|\widehat{G}-G^*\|_{F}\leq C p \sqrt{\frac{\log (n\vee p)}{\Delta}}; \\
& \|\widehat{H}-H^*\|_{op}\leq C\sqrt{\frac{p \log (n\vee p)}{\Delta}} \text{ and } \|\widehat{H}-H^*\|_F\leq C p \sqrt{\frac{\log (n\vee p)}{\Delta}}.
    \end{split}
\end{equation}

In fact, \Cref{lem: covariance local refinement step 1} shows that we are able to remove the extra $\mathcal{B}_n^{-1/2}\Delta_{\min}$ term in the localization error in \Cref{thm:DCDP covariance main} under the same SNR condition. In \Cref{lem: covariance local refinement}, we show that with slightly stronger SNR condition, the localization error can be further reduced as is concluded in \Cref{cor:covariance local refinement main}.

Let
$$
\widehat{\mathcal{Q}}(k)=\sum_{i=1}^k\|X_iX_i^\top- \widehat{G}\|_F^2+\sum_{i=k+1}^{n_0}\|X_iX_i^\top-\widehat{H}\|_F^2 \quad \text { and } \quad \mathcal{Q}^*(k)=\sum_{i=1}^k\|X_iX_i^\top- G^*\|_F^2+\sum_{i=k+1}^{n_0}\|X_iX_i^\top- H^*\|_F^2 .
$$
Through out this section, we use $\kappa_{F} = \|G^*-H^*\|_F$ to measure the signal.

\bnlem[Refinement for covariance model]
\label{lem: covariance local refinement}
Let
$$
\eta+r=\underset{k \in(0, n_0]}{\arg \max } \widehat{\mathcal{Q}}(k) .
$$
Then under the assumptions above, it holds that
$$
\kappa_{F}^2 r=O_P(\log(n)) .
$$
\enlem
\bprf
For the brevity of notations, we denote $n\vee p$ as $p_n$ throughout the proof. Without loss of generality, suppose $r \geq 0$. Since $\eta+r$ is the minimizer, it follows that
$$
\widehat{\mathcal{Q}}(\eta+r) \leq \widehat{\mathcal{Q}}(\eta) .
$$
If $r \leq C\frac{\log(n)}{\kappa_{F}^2}$, then there is nothing to show. So for the rest of the argument, for contradiction, assume that
$$
r \geq C\frac{\log(n)}{\kappa_{F}^2}
$$
Observe that
$$
\begin{aligned}
\widehat{\mathcal{Q}}(t)-\widehat{\mathcal{Q}}(\eta) &=\sum_{i=\eta+1}^{\eta+r}\|X_iX_i^\top- \widehat{G}\|_F^2-\sum_{i=\eta+1}^{\eta+r}\|X_iX_i^\top- \widehat{H}\|_F^2 \\
\mathcal{Q}^*(t)-\mathcal{Q}^*(\eta) &=\sum_{i=\eta+1}^{\eta+r}\|X_iX_i^\top- {G}^*\|_F^2-\sum_{i=\eta+1}^{\eta+r}\|X_iX_i^\top-{H}^*\|_F^2
\end{aligned}
$$
\textbf{Step 1}. It follows that
$$
\begin{aligned}
& \sum_{i=\eta+1}^{\eta+r}\|X_iX_i^\top- \widehat{G}\|_F^2-\sum_{i=\eta+1}^{\eta+r}\|X_iX_i^\top- {G}^*\|_F^2 \\
=& \sum_{i=\eta+1}^{\eta+r}\|\widehat{G}- G^*\|_F^2+2\left\langle G^*-\widehat{G}, \sum_{i=\eta+1}^{\eta+r}(X_iX_i^\top-G^*)\right\rangle \\
=& r\|\widehat{G}- G^*\|_F^2+2r \left\langle G^* -\widehat{G}, H^*-G^*\right\rangle +2\left\langle G^*-\widehat{G}, \sum_{i=\eta+1}^{\eta+r}(X_iX_i^\top -H^*)\right\rangle
\end{aligned}
$$
By assumptions, we have
$$
r\|\widehat{G}- G^*\|_F^2 \leq C_1 r \frac{p^2 \log (p_n)}{\Delta}.
$$
Similarly
$$
r\left\langle G^* -\widehat{G}, H^*-G^*\right\rangle\leq r\|G^* -\widehat{G}\|_F\| H^*-G^* \|_F \leq C_2 r\kappa_{F} p\sqrt{\frac{ \log (p_n)}{\Delta} },
$$
where the second equality follows from $\|G^*-H^*\|_F=\kappa_{F}$, and the last equality follows from \eqref{op1_eq:cov snr}. In addition,
$$
\begin{aligned}
&\left\langle G^*-\widehat{G}, \sum_{i=\eta+1}^{\eta+r}(X_iX_i^\top -H^*)\right\rangle \leq\| G^*-\widehat{G}\|_F\|\sum_{i=\eta+1}^{\eta+r}(X_iX_i^\top -H^*)\|_{F} \\
\leq & C_3 p \sqrt{\frac{\log (p_n)}{\Delta}} (p\sqrt{r\log (p_n)}+ p^{3/2}\log (p_n)) \leq C_3 p^2 \log (p_n) \sqrt{\frac{r}{\Delta}} + C_3 p^2  \sqrt{\frac{p \log^3 (p_n)}{\Delta}}.
\end{aligned}
$$
Therefore
\begin{align*}
    & \sum_{i=\eta+1}^{\eta+r}\|X_iX_i^\top- \widehat{G}\|_F^2-\sum_{i=\eta+1}^{\eta+r}\|X_iX_i^\top- {G}^*\|_F^2\\
    \leq & C_1 p^2\log(p_n) \frac{r}{\Delta} + C_2r\kappa_F p\sqrt{\frac{\log(p_n)}{\Delta}} + C_3 p^2 \log (p_n) \sqrt{\frac{r}{\Delta}} + C_3 p^2  \sqrt{\frac{p \log^3 (p_n)}{\Delta}}\\
    \leq & C_4 r\kappa_F^2\sqrt{\frac{p^4\log^2(p_n)}{\Delta\kappa_F^2}} + C_4 \sqrt{\frac{p^5\log^3(p_n)}{\Delta}}.
\end{align*}
\textbf{Step 2}. Using the same argument as in the previous step, it follows that
$$
\sum_{i=\eta+1}^{\eta+r}\|X_iX_i^\top- \widehat{H}\|_F^2-\sum_{i=\eta+1}^{\eta+r}\|X_iX_i^\top- {H}^*\|_F^2\leq C_4 r\kappa_F^2\sqrt{\frac{p^4\log^2(p_n)}{\Delta\kappa_F^2}} + C_4 \sqrt{\frac{p^5\log^3(p_n)}{\Delta}}.
$$
Therefore
\begin{equation}
    \left|\widehat{\mathcal{Q}}(\eta+r)-\widehat{\mathcal{Q}}(\eta)-\left\{\mathcal{Q}^*(\eta+r)-\mathcal{Q}^*(\eta)\right\}\right|\leq C_4 r\kappa_F^2\sqrt{\frac{p^4\log^2(p_n)}{\Delta\kappa_F^2}} + C_4 \sqrt{\frac{p^5\log^3(p_n)}{\Delta}}.
    \label{tmp_eq:op1 cov main prop eq 1}
\end{equation}
\textbf{Step 3}. Observe that
$$
\begin{aligned}
\mathcal{Q}^*(\eta+r)-\mathcal{Q}^*(\eta) 
=&\sum_{i=\eta+1}^{\eta+r}\|X_iX_i^\top- {G}^*\|_F^2-\sum_{i=\eta+1}^{\eta+r}\|X_iX_i^\top- {H}^*\|_F^2 \\
=& r\| G^*- H^*\|_F^2-2 \left\langle H^*- G^*, \sum_{i=\eta+1}^{\eta+r}(X_iX_i^\top- H^*) \right\rangle
\end{aligned}
$$
Denote $D^* = H^*- G^*$, then we can write the noise term as
$$
\left\langle H^*- G^*, X_iX_i^\top- H^* \right\rangle = X_i^\top D^* X_i - \mathbb{E}[X_i^\top D^* X_i].
$$
Since $X_i$'s are Gaussian, denote $\Sigma_{i} = \mathbb{E}[X_iX_i^\top] =U_i^\top \Lambda_i U_i$, then
$$
\left\langle H^*- G^*, \sum_{i=\eta+1}^{\eta+r}(X_iX_i^\top- H^*) \right\rangle = Z^\top \tilde{D} Z^\top - \mathbb{E}[Z^\top \tilde{D} Z^\top],
$$
where $Z\in \mathbb{R}^{rd}$ is a standard Gaussian vector and 
$$
\widetilde{D} = {\rm diag}\{U_1D^*U_1^\top,U_2D^*U_2^\top,\cdots, U_r D^*U_r^\top\}.
$$
Since $\|\widetilde{D}\|_F = r\kappa_F^2$, by Hanson-Wright inequality, with probability at least $1 - n^{-3}$, it holds uniformly for all $r\geq C\frac{\log(n)}{\kappa_{F}^2}$ that
$$
|\left\langle H^*- G^*, \sum_{i=\eta+1}^{\eta+r}(X_iX_i^\top- H^*) \right\rangle| \leq C_5 \|X\|_{\psi_2}^2\sqrt{r}\kappa_{F}\log(r\kappa_{F}^2).
$$
Therefore, by Hanson-Wright inequality, uniformly for all $r\geq C \frac{\log(n)}{\kappa_{F}^2}$ it holds that
\begin{equation}
    \mathcal{Q}^*(\eta+r)-\mathcal{Q}^*(\eta) \geq r \kappa_{F}^2 - C_5 \|X\|_{\psi_2}^2\sqrt{r}\kappa_{F}\log(r\kappa_{F}^2),
\label{tmp_eq:op1 cov main prop eq 2}
\end{equation}
and thus when $r\geq C (\|X\|_{\psi_2}^4\vee 1)\frac{\log(n)}{\kappa_{F}^2}$, $\mathcal{Q}^*(\eta+r)-\mathcal{Q}^*(\eta)\geq 0$.

\textbf{Step 4}. \Cref{tmp_eq:op1 cov main prop eq 1} and \Cref{tmp_eq:op1 cov main prop eq 2} together give, uniformly for all $r \geq C\log(n) / \kappa_{F}^2$,
$$
  r \kappa_{F}^2 - C_5 \|X\|_{\psi_2}^2\sqrt{r}\kappa_{F}\log(r\kappa_{F}^2) \leq C_4 r\kappa_F^2\sqrt{\frac{p^4\log^2(p_n)}{\Delta\kappa_F^2}} + C_4 \sqrt{\frac{p^5\log^3(p_n)}{\Delta}},
$$
which either leads to a contradiction or proves the conclusion since we assume that $\frac{p^4\log^2(p_n)}{\Delta\kappa_F^2}\rightarrow 0$ and $\frac{p^5\log^3(p_n)}{\Delta}\rightarrow 0$.
\eprf

\bnlem
\label{lem:estimation covariance 2}
Let $\{X_i\}_{i\in [n]}$ be a sequence of subgaussian vectors in $\mathbb{R}^{d}$ with orlitz norm upper bounded $\|X\|_{\psi_2}<\infty$. Suppose $\mathbb{E}[X_i] = 0$ and $\mathbb{E}[X_iX_i^\top] = \Sigma$ for $i\in [n]$. Let $\widehat{\Sigma}_{n} = \frac{1}{n}\sum_{i \in [n]} X_i X_i^\top$. Then for any $u>0$, it holds with probability at least $1 - \exp(-u)$ that
\begin{equation}
    \|\widehat{\Sigma}_{n} - \Sigma\|_{op}\lesssim \|X\|_{\psi_2}^2(\sqrt{\frac{d + u}{n}}\vee \frac{d + u}{n}).
\end{equation}
\enlem
\bprf
This is the same as \Cref{lem:estimation covariance}.
\eprf

\bnlem[Hanson-Wright inequality]
Let $X=\left(X_1, \ldots, X_n\right) \in \mathbb{R}^n$ be a random vector with independent, mean zero, sub-gaussian coordinates. Let $A$ be an $n \times n$ matrix. Then, for every $t \geq 0$, we have
$$
\mathbb{P}\left\{\left|X^{\top} A X-\mathbb{E} X^{\top} A X\right| \geq t\right\} \leq 2 \exp \left[-c \min \left(\frac{t^2}{K^4\|A\|_F^2}, \frac{t}{K^2\|A\|_{op}}\right)\right],
$$
where $K=\max _i\left\|X_i\right\|_{\psi_2}$
\enlem
\bprf
See \cite{vershynin2018high} for a proof and \cite{hanson-wright-general} for a generalization to random vectors with dependence.
\eprf

\bnlem[Local refinement step 1]
\label{lem: covariance local refinement step 1}
Under \Cref{assp:DCDP_covariance main}, let $\{\widetilde{\eta}_k\}_{k \in [\widetilde {K}]}$ be a set of time points satisfying
	\begin{equation}
		\max_{k \in [K]} |\widetilde{\eta}_k - \eta_k| \leq \Delta/5.
	\end{equation}
	Let $\{\check{\eta}_k\}_{k\in [\widehat{K}]}$ be the change point estimators generated from step 1 of the local refinement algorithm with $\{\widetilde{\eta}_k\}_{k\in [\widehat{K}]}$ as inputs and the penalty function $R(\cdot) = 0$. Then
 with probability at least $1 - C n ^{-3}$, $\widehat K = K$ and that
 \[ 
 \max_{ k\in [K] } |\check {\eta}_k - \eta_k|   \lesssim   \frac{\|X\|_{\psi_2}^4}{c_{X}^4}\frac{ p^2\log(n\vee p)}{\kappa^2}.
 \]
\enlem
\bprf
Denote $\I = (s_k, e_k)$ as the input interval in the local refinement algorithm. Without loss of generality, assume that
$$
\I = \I_1\cup \I_2\cup \I_3 = [s,\eta_k)\cup [\eta_k,\check {\eta}_k)\cup [\check {\eta}_k, \eta_{k + 1}).
$$
For $\I_2$, there are two cases.

\textbf{Case 1.} If
$$|\I_2| < \max\{C_s p\log(n\vee p), C_s p\log(n\vee p)/\kappa^2\},$$
then the proof is complete.

\textbf{Case 2.} If
$$|\I_2| \geq \max\{C_s p\log(n\vee p), C_s p\log(n\vee p)/\kappa^2\},$$
Then we proceed to prove that $|\check {\eta}_k - \eta_k|\leq C\frac{\|X\|_{\psi_2}^4}{c_X^4} p^2\log(n\vee p)/\kappa^2$ for some universal constant $C>0$ with probability at least $1 - (n\vee p)^{-5}$.

For $t\in \I$, let $\widehat{\Omega}_t$ be the estimator at index $t$. By definition, we have
\begin{equation}
    \sum_{t \in \I}{\rm Tr}(\widehat{\Omega}_t^\top X_tX_t^\top) - \sum_{t\in \I}\log|\widehat{\Omega}_t|\leq \sum_{t \in \I}{\rm Tr}(({\Omega}_t^*)^\top X_tX_t^\top) - \sum_{t\in \I}\log|{\Omega}_t^*|
    \label{dp_tmp_eq:cov local refine 1}
\end{equation}
Due to the property that
\begin{equation}
    \ell_t(\widehat{\Omega}) - \ell_t({\Omega}^*)\geq {\rm Tr}[(\widehat{\Omega} - \Omega^*)^\top (X_tX_t^\top - \Sigma^*)] + \frac{c}{2}\frac{1}{\|\Omega^*\|_{op}^2}\|\widehat{\Omega} - \Omega^*\|_F^2,
\end{equation}
equation \eqref{dp_tmp_eq:cov local refine 1} implies that
\begin{align}
    & \sum_{i=1}^3 \frac{|\I_i|}{\|\widehat{\Omega}_{\I_i}^*\|_{op}^2}\|\widehat{\Omega}_{\I_i} - {\Omega}_{\I_i}^*\|_F^2 \nonumber \\
    \leq & c_1\sum_{i = 1}^3 |\I_i|{\rm Tr}[({\Omega}^*_{\I_i} - \widehat{\Omega}_{\I_i})^\top (\widehat{\Sigma}_{\I_i} - {\Sigma}^*_{\I_i})] \nonumber \\
    \leq & c_1\sum_{i = 1}^3 |\I_i|\|{\Omega}^*_{\I_i} - \widehat{\Omega}_{\I_i}\|_F \|\widehat{\Sigma}_{\I_i} - {\Sigma}^*_{\I_i}\|_F \nonumber \\
    \leq & \sum_{i=1}^3 \frac{|\I_i|}{2\|\widehat{\Omega}_{\I_i}^*\|_{op}^2}\|\widehat{\Omega}_{\I_i} - {\Omega}_{\I_i}^*\|_F^2 + c_2\sum_{i = 1}^3 |\I_i|\|\widehat{\Sigma}_{\I_i} - {\Sigma}^*_{\I_i}\|_F^2\|\Omega_{\I_i}^*\|_{op}^2,\label{dp_tmp_eq:cov local refine 2}
\end{align}
where we denote $\widehat{\Omega}_{\I_1} = \widehat{\Omega}_{\I_2} = \widehat{\Omega}_{[s_k,\check {\eta}_k)}$, $\widehat{\Omega}_{\I_3} = \widehat{\Omega}_{[\check {\eta}_k,e_k)}$, ${\Omega}_{\I_1}^* = \Omega^*_{\eta_k - 1}$, and ${\Omega}_{\I_2}^* = {\Omega}_{\I_3}^* = \Omega^*_{\eta_k}$.

By the setting of local refinement, we have $\min\{|\I_1|,|\I_3|\}\geq C_s p\log(n\vee p)$. Therefore, by \Cref{lem:estimation covariance 2}, for $i=1,2,3$, it holds with probability at least $1 - (n\vee p)^{-7}$ that
$$
\|\widehat{\Sigma}_{\I_i} - {\Sigma}^*_{\I_i}\|_F^2\leq p\|\widehat{\Sigma}_{\I_i} - {\Sigma}^*_{\I_i}\|_{op}^2\leq C \|X\|_{\psi_2}^4 \frac{p^2\log(n\vee p)}{|\I_i|}.
$$
Consequently, we have
\begin{equation}
    \sum_{i=1}^3 \frac{|\I_i|}{\|\widehat{\Omega}_{\I_i}^*\|_{op}^2}\|\widehat{\Omega}_{\I_i} - {\Omega}_{\I_i}^*\|_F^2\leq c_2\sum_{i = 1}^3 \|\Omega_{\I_i}^*\|_{op}^2 \|X\|_{\psi_2}^4 {p^2\log(n\vee p)}.
    \label{dp_tmp_eq:cov local refine 3}
\end{equation}
In particular, $\Delta\kappa^2>\mclB_n \frac{\|X\|_{\psi_2}^4}{c_X^4} p^2\log(n\vee p)$, we have
\begin{align*}
    |\I_1|\|\widehat{\Omega}_{\I_1} - {\Omega}_{\I_1}^*\|_F^2\leq& c_2\|\widehat{\Omega}_{\I_1}^*\|_{op}^2\sum_{i = 1}^3 \|\Omega_{\I_i}^*\|_{op}^2 \|X\|_{\psi_2}^4 {p^2\log(n\vee p)}\\
    \leq & 3c_2\frac{\|X\|_{\psi}^4}{c_X^4} p^2\log(n\vee p)\leq \frac{1}{12}\Delta\kappa^2,
\end{align*}
for sufficiently large $n$ because $\mclB_n\rightarrow \infty$ as $n\rightarrow \infty$. Since $|\I_1|\geq \frac{1}{3}\Delta$, it follows from the inequality above that $\|\widehat{\Omega}_{\I_1} - {\Omega}_{\I_1}^*\|_F\leq \frac{\kappa}{2}$ and thus,
\begin{equation*}
    \|\widehat{\Omega}_{\I_2} - {\Omega}_{\I_2}^*\|_F\geq \|\widehat{\Omega}_{\I_2} - {\Omega}_{\I_1}^*\|_F + \|{\Omega}^*_{\I_1} - {\Omega}_{\I_2}^*\|_F\geq \frac{\kappa}{2}.
\end{equation*}
Plug this back into \Cref{dp_tmp_eq:cov local refine 3} and we can get
\begin{equation}
    \frac{\kappa^2}{4}|\I_2|\leq c_4\frac{\|X\|_{\psi_2}^4}{c_X^4}p^2\log(n\vee p),
\end{equation}
which completes the proof.
\eprf




\clearpage





\end{document}